\documentclass[bottom]{svmono-nospringer}

\usepackage{graphicx}
\usepackage{multicol}
\usepackage{footmisc}
\usepackage{hyperref}
\usepackage{pdfsync}
\usepackage{fullpage}
\usepackage{verbatim}
\usepackage{amsmath}
\usepackage{amssymb}
\usepackage{mathrsfs}
\usepackage{makeidx}
\usepackage{array}
\usepackage{bigstrut}
\usepackage{authblk}
\usepackage{mathtools}

\usepackage{wrapfig}

\usepackage[utf8]{inputenc}
\usepackage[T1]{fontenc}
\usepackage{lmodern}

\newcommand{\bea}{\begin{eqnarray}}
\newcommand{\eea}{\end{eqnarray}}
\newcommand{\be}{\begin{equation}}
\newcommand{\ee}{\end{equation}}

\def\Tr{\mbox{Tr}\,}
\def\tr{\mbox{tr}\,}
\def\sgn{\,\mbox{sgn}\,}
\newcommand{\R} {\mbox{Re}\,}

\newcommand{\eu}{{\rm e}}
\newcommand{\ii}{{\rm i}}
\newcommand{\de}{{\displaystyle\rm\mathstrut d}}
\newcommand{\ord}{
	\mathchoice
	{{\scriptstyle\mathcal{O}}}
	{{\scriptstyle\mathcal{O}}}
	{{\scriptscriptstyle\mathcal{O}}}
	{\scalebox{.7}{$\scriptscriptstyle\mathcal{O}$}}
}
\newcommand{\Ord}{{\cal O}}

\DeclareMathOperator{\arccot}{arccot}

\newcommand{\cP}{{\cal P}}
\newcommand{\bT}{{\bf {\cal T}}}

\def\XXint#1#2#3{{\setbox0=\hbox{$#1{#2#3}{\int}$}
		\vcenter{\hbox{$#2#3$}}\kern-.5\wd0}}

\makeindex

\begin{document}

\frontmatter

\author{Fabio Franchini}
{\affil{\normalsize{Ru{\dj}er Bo\v{s}kovi\'c Institute} \\
	Bijen\v{c}ka cesta 54, 10000 Zagreb, Croatia}
\affil{SISSA \\
	Via Bonomea 265, 34136, Trieste, Italy}
}

\title{An introduction to integrable techniques\\ for one-dimensional quantum systems}


\maketitle

\tableofcontents

\mainmatter

\chapter*{Preface}

These notes are the write-up and extension of the lectures I gave over a few years for a class on ``{\it Introduction to Bethe Ansatz}'' within the Ph.D. program in Statistical Physics at SISSA (Trieste). They are intended as a guidance to start the study of this extremely rich subject, by favoring a clear and physical introduction to its fundamental ideas, over many mathematical subtleties that populate its formulation. The emphasis on the physical intuition makes these notes suitable also for the scientist who mostly performs numerical simulations, but wants to compare his/her results with exact ones, and to anyone who needs to start reading the literature on Bethe Ansatz and integrable models. 

Modern physics is all about universality, but we should never forget that universal behaviors emerge from microscopic dynamics and thus solvable models have always played a pivotal role in providing concrete realizations of different phenomenologies to test hypothesis and shaping our intuition. Over the years, integrable models have helped us in constructing better numerical methods and in developing and testing general theories such as Bogoliubov theory for weakly interacting gases, Luttinger liquid, non-linear Lutting liquids and so on. 

Integrable techniques have witnessed a resurgence in popularity in recent times, mostly due to the outstanding progresses in experimental capabilities, for instance in cold atoms, which disclose the possibility of engineering virtually any desired interaction and geometry \cite{guan13}. These breakthroughs have in turn stimulated new questions, to address which exact tools are a valuable asset.

It is yet not clear which qualitative features separate integrable from non-integrable quantum systems \cite{caux2011}. While in classical physics this distinction is clear, all the proposed answers for the quantum case are somewhat unsatisfactory, although the analysis of out-of-equilibrium settings seems to provide interesting results in this respect \cite{RigolGGE,JSTATspecial}.

Regardless of these considerations, we know that a small, but interesting, subset of many-body quantum systems is amenable to exact solution. By this, we mean that each eigenstate of such systems can be uniquely characterized by a set of quantum numbers, which curiously seems to be in a one-to-one correspondence with a free fermionic system. 
This realization allows to classify the states in terms of their elementary excitations (quasi-particles). In this respect it should be stressed that the added value of these techniques does not lie on their bare efficiency (nowadays we have very powerful numerical tools at our disposal), but on the insights they provide to interpret a many-body system (and then to develop even more efficient simulations).

The fundamental ingredient responsible for such analytical solution is the fact that any scattering event can be decomposed into a sequence of two-body scatterings, and the ordering in such sequence does not alter the result. This properties means that the fundamental quasiparticle excitations cannot be created, nor destroyed in a scattering event governed by an integrable Hamiltonian. This is the realization behind Bethe's original ansatz for the solution of the Heisenberg chain and the foundation over which a beautiful mathematical physics construction has been erected.

This integrability is a peculiar property only of 1+1-dimensional quantum systems. While for many years the study of Bethe Ansatz was purely theoretically motivated, nowadays we have several systems where the degrees of freedom are effectively confined to move along a line/chain, because the transverse directions are energetically blocked. In some crystalline compounds, for instance, atoms are placed in such a way that their magnetic moments interact preeminently with neighbors in one (or two) directions. In lithography it is possible to realize conducting wires so thin that the transverse component of the electron wavefunction is frozen. Most of all, cold atoms can be manipulated with high precision with external laser beams which allow to confine them in virtually any desired geometry and probe them with remarkable accuracy.

There are several excellent sources where to learn about integrable techniques. These notes try to introduce all the basic tools and ideas, while keeping a short, but pedagogical approach.
To do so, a small number of examples were selected. We will start with the XY chain, which is essentially a free system with a non-trivial phase diagram and the prototypical model to address a variety of questions. Next we will solve the Lieb-Liniger model of one-dimensional bosons with contact interaction. This model has the merit of being very close to experimental relevance and
allows for a clear introduction of the coordinate Bethe Ansatz solution to study its zero and finite temperature thermodynamics.
We will then move to the Heisenberg spin-$1/2$ chain and the XXZ chain, which will be our reference model to introduce more advanced topics such as the issue of string solutions and the algebraic Bethe Ansatz approach. To better introduce the latter, we also briefly explain the solution of the two-dimensional classical 6-vertex model in one of the appendices, since these techniques were instrumental in the realization of the algebraic structure behind integrability. The other appendices contain a collection of results on Toeplitz determinants (which are relevant for the XY chain) and a digression on the relation between Bethe Ansatz solutions and the field theories describing the low energy properties of these models.

A noticeable absence among the topics covered is that of the nested Bethe Ansatz for systems with internal degrees of freedom. We refer the interested reader to \cite{hubbard} for an exhaustive treatment of the 1D Hubbard model, as the prototypical, and experimentally relevant, example of such systems.

I wish to thank Giuseppe Mussardo for the opportunity of teaching the class that pushed me to write these notes and deepen my understanding of the subject and to all my friends and colleagues for their advices and insights. I am also grateful to Guillaume Lang for his thorough reading of the the first version of this manuscript and for his comments and corrections.

\setlength\extrarowheight{2pt}

\chapter{The XY Chain}
\label{chap:XYModel}

\abstract{
The XY chain in a transverse magnetic field is a generalization of the 1D Ising model, with whom it shares the property of being essentially a free system. Its rich and non-trivial phase-diagram and the possibility of calculating virtually every quantity have rendered it a reference model to understand new effects or to test hypotheses.
After a brief introduction in Sec. \ref{sec:1intro}, in Sec. \ref{diagsec} we review its standard mapping to free fermions, paying particular attention on the interplay between the two parity sectors of the Hilbert space to understand the $Z_2$ symmetry breaking. In Sec. \ref{diagsec} we discuss the phase diagram and in Sec. \ref{corrsec} we show how to calculate some basic correlation functions and we discuss their behavior in the different phases. Finally, in Sec. \ref{sec:kitaev} we comment on the relation between the Ising model and the Kitaev chain, which provides a natural interpretation of the $Z_2$ symmetry in terms of Majorana boundary modes.
}

\section{Introduction and motivations}
\label{sec:1intro}

The One-Dimensional XY model in a transverse magnetic field is arguably the simplest non-trivial integrable model. 
Its simplicity derives from the fact that its excitations are non-local free fermions. This non-locality is the source of the non-trivial 2-parameters phase diagram,  characterized, at zero temperature, by two Quantum Phase Transitions (QPTs): one belonging to the universality of the anti-ferromagnetic Heisenberg chain (aka, the XX model\index{XX chain}, with conformal charge $c$ equal to $1$) and the other to the Ising model\index{Ising chain} ($c=1/2$).

The Hamiltonian of the XY model can be written as \index{XY chain}
\be
H = J  \sum_{j=1}^N \Big[
\left( {1 + \gamma } \right) S_j^x S_{j+1}^x +
\left( {1 - \gamma } \right) S_j^y S_{j+1}^y
+ h \; S_j^z \Big]
 = {J \over 2} \sum_{j=1}^N \left[
 \left( {1 + \gamma \over 2} \right) \sigma_j^x \sigma_{j+1}^x +
 \left( {1 - \gamma \over 2} \right) \sigma_j^y \sigma_{j+1}^y
 + h \; \sigma_j^z \right] \: ,
\label{spinham}
\ee 
where $\sigma_j^{\alpha}$, with $\alpha=x,y,z$, are the Pauli matrices which describe spin-$1/2$ operators on the $j$-th lattice site of a chain with $N$ sites. 
This Hamiltonian describes a one-dimensional lattice, where a 3D spin variable lives on every lattice point. The spins interact with their nearest neighbor in an anisotropic way (parametrized by $\gamma$), so that the interaction between their $z$-components (that is, the direction of an external magnetic field $h$) can be neglected. This model was first introduced and solved in the case of zero magnetic field $h$ by Lieb, Schultz and Mattis in \cite{LSM-1961} and in \cite{katsura,niemeijer} with a finite external field.

The fundamental correlation functions were calculated in \cite{mccoy}. More complicated correlators like the Emptiness Formation Probability \cite{shiroishi,abanovfran03,abanovfran05} and the Von Neumann \cite{korepinEnt,frankor07} and Renyi \cite{frankor08} entanglement entropies were calculated more recently, as well as several out-of-equilibrium properties \cite{Isingquench}. Virtually all static correlation functions of the model can be expressed as determinants of matrices with a special structure, known as Toeplitz matrices \cite{mehta2}\index{Toeplitz!matrix}. The asymptotic behavior of Toeplitz determinants\index{Toeplitz!determinant} can be studied using fairly sophisticated mathematical techniques or just by relying on known theorems, such as the Szeg\"o Theorem, the Fisher-Hartwig conjecture, Widom's theorem and so on ...\cite{Ehrhardt-2001,deift13}

The phase diagram of this model is parametrized by the {\it anisotropy
parameter} $\gamma$ capturing the relative strength of interaction in the $x$ and $y$ components and by the {\it external magnetic field} $h$, directed
along the transverse $z$-axis. We take these parameters to be dimensionless and
from now on we set the energy-scale defining parameter as $J=-1$ (that is, we will consider an easy-plane ferromagnet). The model has obvious symmetries: a rotation by $\pi/2$ along the $z$-axis interchanges the $x$ and $y$ spin interactions and is equivalent to $\gamma \to - \gamma$, while a reflection of the spin across the $x-y$ plane is compensated by $h \to - h$: thus we will concentrate only on the
first quadrant of the phase diagram ($\gamma \ge 0$, $h \ge 0$), since the rest of the phase diagram is related by the above symmetries. The two quantum phase transitions, that is, the parameters for which the spectrum becomes gapless, are located on the {\it isotropic line} $\gamma = 0$ ($|h| \le 1$), and at the {\it critical magnetic field} $|h|=1$.

We remark that for $\gamma =0$ the Hamiltonian reduces to the isotropic XX model \index{XX chain}, i.e. the $\Delta = 0$ limit of the {\bf XXZ chain} \index{XXZ chain}, which will be studied in chapter \ref{chap:XXZmodel}. For $\gamma = \pm 1$, we recover the 1D {\bf Quantum Ising model} \index{Ising chain}.

These two cases correspond to the two competing universality classes which can be realized by the XY chain. The isotropic line corresponds to free fermions hopping on a lattice and thus belong to a $c=1$ Conformal Field Theory (CFT) universality \index{CFT}. The critical magnetic field $h=\pm 1$ is an Ising transition, that is, a transition from a doubly degenerate ground state (for $|h|<1$) to a single ground state system (for $|h|>1$). It is the same as the classical phase transition occurring in the two-dimensional Ising model \cite{mussardobook}. In fact, the latter can be solved through its transfer matrix, which takes the same form as the exponential of (\ref{spinham}), with the magnetic field taking the role of the temperature \cite{LSM-1964}.
The order parameter of this transition for the classical model is the magnetization and in the quantum case it is the magnetization along the $x$-axis. Thus, consistently with the $Z_2$ symmetry of the model, $\langle \sigma^x \rangle$ goes from vanishing for $|h|>1$, to finite $\pm m_x$ for $|h|<1$.

The latter behavior is exemplified by the $(\gamma,h)=(1,0)$ point, where the two ground states are 
\be
  |GS_1\rangle =
  | \rightarrow \; \rightarrow \; \rightarrow \; \ldots \rangle
   = \prod_{j=1}^N {1 \over \sqrt{2} } \Big( | \uparrow _j \rangle + | \downarrow _j \rangle \Big) \:, 
   \quad 
  |GS_2\rangle  = 
  | \leftarrow \; \leftarrow \; \leftarrow \;  \ldots \rangle
  =  \prod_{j=1}^N {1 \over \sqrt{2} } \Big( | \uparrow _j \rangle -| \downarrow _j \rangle \Big) \: ,
\label{BellPair}
\ee
where $ | \uparrow_j \rangle$ ($ | \downarrow_j \rangle$) indicates the state with positive (negative) projection of the spin along the $z$-axis at the $j$-th lattice point. We see that these states have $\langle \sigma^x \rangle = \pm 1$.

The exact degeneracy between the two ground states is in general lifted away from the point $(\gamma,h)=(1,0)$ for finite chains, and is recovered in the whole phase only in the thermodynamic limit.
However, as noted in \cite{shrock}, the factorized structure (\ref{BellPair}) for the degenerate ground states propagates on the line $\gamma^2 + h^2 = 1$, where the two ground states can be written explicitly as 
\be
  |GS_1\rangle = \prod_{j=1}^N \Big( \cos \theta \; | \uparrow_j \rangle + \sin \theta \; | \downarrow_j \rangle \; \Big) \: ,
  \qquad
  |GS_2\rangle = \prod_{j=1}^N \Big( \cos \theta | \uparrow_j\rangle - \sin \theta \; | \downarrow_j \rangle \; \Big) \: ,
  \label{deg}
\ee
where $\cos^2 (2 \theta) =(1-\gamma)/(1+\gamma)$.
Remarkably, on this line the degeneracy is exact for any length of the chain. We see that along this line, the spins, initially aligned along the $x$-axis, progressively acquire a growing positive $z$ component and eventually merge into a perfectly polarized state at the point $(\gamma,h)=(0,1)$. This is the bicritical point of junction between the two critical lines $\gamma=0$ and $h=1$ and is quite special, in that the spectrum becomes perfectly quadratic. Thus, at this point the model is critical, but not conformal, since its dynamical critical exponent is equal to 2.


\section{Diagonalization of the Hamiltonian}
\label{diagsec}

The standard prescription to diagonalize (\ref{spinham}) assumes periodic boundary conditions $\sigma_{j+N}^\alpha = \sigma_j^\alpha$.
It is quite inconvenient to work directly with spin operators, since on each site they behave fermionically (in that they span a finite-dimensional Fock space), but between sites they obey bosonic commutation relations.
In one dimension, however, this problem can be circumvented by mapping the spins into either fermionic or bosonic operators: in the first case, one needs to introduce a strong repulsive interaction to truncate the Hilbert space, while the price for the latter choice is that the mapping is highly non-local. We pursue the latter.

Following \cite{LSM-1961}, we reformulate the Hamiltonian (\ref{spinham}) in terms of spinless fermions $\psi_j$ by means of a Jordan-Wigner transformation \index{Jordan-Wigner!transformation}:
\bea
   \sigma_j^+ & = &
   \eu^{\ii \pi \sum_{l<j} \psi_l^\dagger \psi_l} \; \psi_j =
   \prod_{l=1}^{j-1} \left( 1 - 2 \psi_l^\dagger \psi_l \right) \psi_j \: , \qquad
   \sigma_j^- =
   \psi_j^\dagger \; \eu^{-\ii \pi \sum_{l<j} \psi_l^\dagger \psi_l} =
   \prod_{l=1}^{j-1} \left( 1 - 2 \psi_l^\dagger \psi_l \right) \psi_j^\dagger \; ,
  \nonumber  \\
   \sigma_j^z & = & 1 - 2 \psi_j^\dagger \psi_j \; ,
   \label{JordanWigner} 
\eea
where, as usual, $\sigma^\pm = (\sigma^x \pm i \sigma^y)/2$. On each site, a spin up is mapped into an empty state and a spin down to an occupied one. The non-local part of this mapping is called the {\it Jordan-Wigner string}\index{Jordan-Wigner!string} and fixes the (anti)commutation relation between sites, by counting the parity of overturned spins to the left of the site on which it is applied. It should be remarked that this transformation explicitly breaks the translational invariance of the model, by singling out a particular site (site $1$) as a starting point for the string.

The Jordan-Wigner mapping transforms (\ref{spinham}) into ($J=-1$)
\bea
   H & = & - {1 \over 2} \; \sum_{j=1}^{N-1} \left( \psi_j^\dagger \psi_{j+1} +
   \psi_{j+1}^\dagger \psi_j +
   \gamma \; \psi_j^\dagger \psi_{j+1}^\dagger +
   \gamma \; \psi_{j+1} \psi_j \right)
   + h \sum_{j=1}^N \psi_j^\dagger \psi_j - {h N \over 2}
   \nonumber \\
   && \quad + {\mu_N^x \over 2} \; \left( \psi_N^\dagger \psi_{1} +
   \psi_{1}^\dagger \psi_N +
   \gamma \; \psi_N^\dagger \psi_{1}^\dagger +
   \gamma \; \psi_{1} \psi_N \right) \;  ,
   \label{realfermionH}
\eea 
where\footnote{We denote this operator as $\mu_N^x$, according to the traditional notation for the dual lattice operators of the quantum Ising Model \cite{mussardobook}.} 
\be
   \mu_N^x \equiv \prod_{j=1}^N \left(1 - 2 \psi_j^\dagger \psi_j \right) = \prod_{j=1}^N
   \sigma_j^z \; ,
   \label{muNxdef}
\ee
is the {\it parity operator}.\index{Parity operator}
This Hamiltonian describes spinless fermions hopping on a lattice, with a super\-conducting-like ``interaction'' which creates/destroys them in pairs. It is thus a simple 1D version of a BCS model\index{Superconductivity}.

The boundary terms on the second line of (\ref{realfermionH}) can often be discarded, since their effect is meant to be negligible in the thermodynamic limit. However, they are important to establish the degeneracy of the model in the ordered phase and thus we will keep them.

For non-vanishing $\gamma$, the Hamiltonian (\ref{spinham}) does not commute with $\sum_j \sigma_j^z$ and therefore (\ref{realfermionH}) does not conserve the number of fermions. Nonetheless, since fermions are created/destroyed in pairs the even/oddness of their number ({\it the parity}) is conserved, i.e. 
\be
   [\mu_N^x, H ] =0.
\ee 
This observation allows to separate the theory into two
disconnected sectors with $\mu_N^x = \pm 1$, where the plus sign
characterizes configurations with an even number of particles and
the minus the one with an odd number:
\be
   H \; = \; {1 + \mu_N^x \over 2} \; H^+ \; + \; {1 - \mu_N^x \over 2} \; H^- \; ,
   \label{hamdecomp}
\ee 
here ${1 \pm \mu_N^x \over 2}$ are the projector operators to
the states with even/odd number of particles and $H^\pm$ have the
form (\ref{realfermionH}) with $\mu_N^x = \pm 1$.

The boundary terms in (\ref{realfermionH}) can be satisfied by applying in each sector the appropriate boundary conditions to the spinless fermions: for $\mu_N^x = + 1$ (even number of particles) we
have to impose {\bf anti-periodic b.c.}
on the fermions and for $\mu_N^x = - 1$ (odd number of particles) we
require {\bf periodic b.c.}:
\bea
   \psi^{(+)}_{j+N} = - \psi^{(+)}_j & \qquad \qquad \qquad {\rm for} \qquad \qquad \qquad &
   \mu_N^x = + 1
   \nonumber \\
   \psi^{(-)}_{j+N} = \psi^{(-)}_j & \qquad \qquad \qquad {\rm for} \qquad \qquad \qquad &
   \mu_N^x = - 1.
   \label{boundarycond}
\eea 
With these definitions, we can write both Hamiltonians in
(\ref{hamdecomp}) in the compact form:
\be
   H^\pm = - {1 \over 2} \; \sum_{j=1}^N \left( \psi_j^{(\pm)\dagger} \psi^{(\pm)}_{j+1} +
   \psi_{j+1}^{(\pm)\dagger} \psi^{(\pm)}_j +
   \gamma \; \psi_j^{(\pm)\dagger} \psi_{j+1}^{(\pm)\dagger} +
   \gamma \; \psi_{j+1}^{(\pm)} \psi_j^{(\pm)} - 2 h \; \psi_j^{(\pm)\dagger} \psi_j^{(\pm)} \right) - {h N \over 2} \; .
   \label{noboundayfermionH}
\ee
Thus, each sector is governed by the same Hamiltonian, but with a different Fock space due to the different boundary conditions (\ref{boundarycond}). 

The next step towards the solution is to move into Fourier space. To account for the different boundary conditions, we sum over integer (half-integers) modes on the odd (even) particle sector:
\bea  
q \in \mathbb{N} + {1 \over 2} = {1 \over 2} \; , {3 \over 2} \ldots N - {1 \over 2} 
& \qquad \qquad & \text{for} \quad \nu_N^x = 1 \: , \quad
\text{(Even Particle Number)} \\
q \in \mathbb{N} = 0 \; , 1 \ldots N -1 
& \qquad \qquad & \text{for} \quad \nu_N^x = -1 \: , \quad
\text{(Odd Particle Number)} .
\eea
Denoting the set above to which $q$ belongs as $\Gamma_{\nu_N^x}$, we define
\footnote{We choose the asymmetric version of the Fourier transform, as it makes it easier to consider the thermodynamic limit. In fact, from $\{ \psi_j , \psi_l^\dagger \} = \delta_{j,l}$ in real space it follows that in momentum space $\{ \psi_q , \psi_k^\dagger \} = N \delta_{q,k} \stackrel{N \to \infty} \to \delta (q-k)$, where the last one is the Dirac delta function in the continuum. The additional $\pi/4$ phase is non-standard and is chosen to render explicit that the Bogoliubov transform introduced in (\ref{bogtrans}) is nothing else but a $O(2)$ rotation in Fourier space.} 
\be
  \psi^{(\pm)}_j = { \eu^{\ii \pi/4} \over N} \; \sum_{q \in \Gamma_\pm}
  \eu^{\ii {2 \pi \over N} \: q \: j} \psi_q \; ,
  \qquad \qquad
   \psi_q \equiv \eu^{-\ii \pi/4} \; \sum_{j=1}^N
  \eu^{-\ii {2 \pi \over N} \: q \: j} \psi^{(\pm)}_j \; ,
  \label{fourier}
\ee 

The Hamiltonian in Fourier space reads:
\be
   H^\pm = {1 \over N} \; \sum_{q \in \Gamma_\pm}
   \left[ h - \cos \left( \textstyle{ {2 \pi \over N} \: q } \right) \right]
   \psi_q^\dagger \psi_q
   + {\gamma \over 2 N} \; \sum_{q \in \Gamma_\pm}
   \sin \left( \textstyle{ {2 \pi \over N} \: q } \right) \left\{ \; \psi_q \psi_{-q} + \psi_{-q}^\dagger \psi_q^\dagger \; \right\}  - {h N \over 2} \; ,
   \label{fourierspinlessham}
\ee
which can also be written as a sum of $2 \times 2$ matrices
\renewcommand{\arraystretch}{1.2}
\be
   H^\pm = {1 \over 2 N} \; \sum_{q \in \Gamma_\pm}
   \left( \psi_q^\dagger ; \psi_{-q} \right)    \begin{pmatrix}
        h - \cos \left(  {2 \pi \over N} \: q \right)\: \: &
        - \gamma \; \sin \left(  {2 \pi \over N} \: q \right) \\
        - \gamma \; \sin \left(  {2 \pi \over N} \: q \right) \: \: &
        \cos \left(  {2 \pi \over N} \: q  \right) - h \cr
   \end{pmatrix} 
   \left( \begin{array}{cc} \psi_q \cr \psi^\dagger_{-q} \cr \end{array} \right) \; .
\ee
\renewcommand{\arraystretch}{1}

We can diagonalize each of these matrices (and thus the whole Hamiltonian) by means of a Bogoliubov transformation\index{Bogoliubov!transformation}, which, in our notation, is nothing but a $O(2)$ rotation in Fourier space:
\be
   \left( \begin{array}{cc} \psi_q \cr \psi^\dagger_{-q} \cr \end{array} \right) =
   \left( \begin{array}{cc}
        \cos \vartheta_q \: \: & \sin \vartheta_q \cr
        - \sin \vartheta_q \: \: & \cos \vartheta_q \cr
   \end{array} \right)
   \left( \begin{array}{cc} \chi_q \cr \chi^\dagger_{-q} \cr \end{array} \right) \; ,
   \label{bogtrans}
\ee
or   
\be
   \hskip -1cm
   \chi_{q} = \cos \! \vartheta_q \: \psi_{q}
   - \sin \! \vartheta_q \: \psi_{-q}^\dagger
   \qquad \qquad
   \chi_{-q} = \cos \! \vartheta_q \: \psi_{-q}
   + \sin \! \vartheta_q \: \psi_{q}^\dagger
\ee 
with the Bogoliubov (rotation) angle $\vartheta_q$ \index{Bogoliubov!angle}defined by
\be
   \tan \left( 2 \vartheta_q \right) = 
   { \gamma \; \sin \left( \textstyle{ {2 \pi \over N} \: q } \right) \over
   h - \cos \left( \textstyle{ {2 \pi \over N} \: q } \right) } \; ,
   \label{rotangle}
\ee
or equivalently
\be
   \eu^{\ii 2 \vartheta_q} 
   = { h - \cos \left( \textstyle{ {2 \pi \over N} \: q } \right) + 
   \ii \gamma \; \sin \left( \textstyle{ {2 \pi \over N} \: q } \right) \over 
   \sqrt{ \left( h - \cos \left( \textstyle{ {2 \pi \over N} \: q } \right) \right)^2 + 
   \gamma^2 \; \sin^2 \left( \textstyle{ {2 \pi \over N} \: q } \right) } } \; .
\ee
  
In terms of the Bogoliubov quasi-particles \index{Bogoliubov!quasi-particle} the Hamiltonian describes free fermions
\be
  H^\pm = {1 \over N} \sum_{q \in \Gamma_\pm}
  \varepsilon \left( \textstyle{ {2 \pi \over N} \: q } \right) \;
  \left\{ \chi_q^\dagger \chi_q - {N \over 2} \right\} \; ,
  \label{H+ham}
\ee with spectrum
\be
   \varepsilon (\alpha) \equiv \sqrt{ (h - \cos \alpha)^2 + \gamma^2 \; \sin^2 \alpha}.
   \label{spectrum}
\ee

Thus, we solved the model by first mapping it to spinless fermions through the Jordan-Wigner transformation (\ref{JordanWigner}), then moving into Fourier space and there performing a Bogoliubov rotation that makes the Hamiltonian diagonal. Let us now analyze the different sectors more closely.

\subsection{Even particle number}

Since the spectrum (\ref{spectrum}) is always positive, the lowest energy state $|GS\rangle_+$ in this sector is defined by
\be
   \chi_q \; | GS \rangle_+ = 0 \qquad q={1 \over 2} \ldots N - {1 \over 2}
   \label{GS+def}
\ee 
and is ``empty of quasi-particles''. To express this ground state in terms of physical fermions, we start from the vacuum state $| 0 \rangle$ defined by
\be
   \psi_q \; | 0 \rangle =0 \qquad \forall q
\ee and verify that the state
\be
   | GS \rangle_+ \equiv \prod_{q=0}^{\left[(N-1)/2\right]} \left( \cos \vartheta_{q + {1 \over 2}} + {1 \over N} \sin \vartheta_{q + {1 \over 2}} \;
   \psi_{q + {1 \over 2}}^\dagger \psi_{-q - {1 \over 2}}^\dagger \right) |0 \rangle
   \label{GS+}
\ee 
satisfies (\ref{GS+def}), where $[x]$ indicates the closest integer smaller than $x$.
The ground state energy is
\be
   E_0^+ = - {1 \over 2} \; \sum_{q=0}^{N-1} \varepsilon \left[
  \textstyle{ {2 \pi \over N} \left( q + {1 \over 2} \right) }
  \right]\stackrel{N \to \infty}{\rightarrow}
  - {N \over 2} \int_0^{2 \pi} {\de q \over 2 \pi} \;  \varepsilon (q) \; ,
  \label{E0+}
\ee where the last expression holds in the thermodynamic limit $N
\to \infty$.

The Hilbert space is generated by applying creation operators $\chi_q^\dagger$ to the ground state $| GS \rangle_+$. Each excitation adds an energy $\varepsilon \left( \textstyle{ {2 \pi \over N} \: q } \right)$. It should be remembered that physical states in this sector have an even number of excitations, and thus creation operators have to be applied in pairs.

\subsection{Odd particle number}

We define the state $|GS' \rangle$ with no quasi-particle excitations as
\be
  \tilde{\chi}_q \; | GS' \rangle_- = 0 \qquad q=0, \ldots , N-1,
\label{GS'-def}
\ee 
but this state is not allowed by the condition of odd excitations.
The lowest energy state in this sector is
\be
  |GS \rangle_- = {1 \over \sqrt{N}} \chi_0^\dagger \; | GS' \rangle_- 
  = {1 \over \sqrt{N}}\psi_0^\dagger 
  \prod_{q=1}^{[N/2]} \left( \cos \vartheta_q + {1 \over N} \sin \vartheta_q \;
  \psi_q^\dagger \psi_{-q}^\dagger \right) |0 \rangle \; .
\label{GS-def}
\ee

Note that in this case there is a zero mode\index{Zero mode} which has to be treated separately.
In fact, for $q=0$ the superconducting term in the Hamiltonian (\ref{fourierspinlessham})
vanishes\footnote{For even size lattices ($N=2M$) the same holds for the $q=M$ component
(i.e. a $\pi$-momentum particle), which is the contribution to single out for antiferromagnetic coupling $J<0$ in (\ref{spinham}). More interesting is the case of odd-size lattices with antiferromagnetic coupling: this setting introduces a frustration which renders the whole region $|h|<1$ gapless \cite{XYAFM}}. Therefore, the zero-momentum Bogoliubov particle\index{Bogoliubov!quasi-particle} coincides with the physical zero-momentum fermion (no rotation is necessary)\index{Zero mode}:
\be
   \chi_0 = \psi_0 ,
\ee
and its energy contribution is exactly $h-1$.
Thus, in this sector the diagonalized Hamiltonian is
\be
  H^- 
  = (h-1) \left\{ {1 \over N} \; \chi_0^\dagger \chi_0 - {1 \over 2} \right\} +
  \sum_{q=1}^{N-1} \varepsilon \left( \textstyle{ {2 \pi \over N} q  }  \right) \;
  \left\{ {1 \over N} \; \chi_q^\dagger \chi_q - {1 \over 2} \right\} \; .
  \label{HMinus}
\ee 

\subsubsection{Disordered Phase}
For $h>1$
\be
   h- 1 = \varepsilon (0) > 0
\ee 
and thus (\ref{HMinus}) is the same as (\ref{H+ham}).
Hence, the Hilbert space is populated like in the previous case by successive applications of pairs of operators $\chi_q^\dagger \chi_{q^\prime}^\dagger$ or $\chi_q^\dagger \chi_0$ to 
$|GS \rangle_-$.
Note that in the thermodynamic limit these states and those generated in the even excitation sector intertwine and thus one can effectively forget about the separation into the two sectors.

\subsubsection{Ordered Phase}
For $h<1$
\be
   h - 1 = - \varepsilon (0) < 0
\ee 
and thus the presence of a zero-mode\index{Zero mode} lowers the energy of the system (notice that, because of our normalization for the Fourier modes $\left( {1 \over N} \chi_0^\dagger \chi_0 - {1 \over 2} \right) \chi_0^\dagger = \chi_0^\dagger$). 
The energy of (\ref{GS-def}) for $h<1$ is
\be
  E_0^- =  \; {1 \over 2} \; (h -1) \; - {1 \over 2} \sum_{q=1}^{N-1} \varepsilon
  \left( \textstyle{ {2 \pi \over N} q } \right)
  = - {1 \over 2} \sum_{q=0}^{N-1} \varepsilon \left(
  \textstyle{ {2 \pi \over N} q } \right)
  \stackrel{N \to \infty}{\rightarrow}
  - {N \over 2} \int_0^{2 \pi} {\de q \over 2 \pi} \; \varepsilon (q) \; ,
  \label{E0-}
\ee
where the last expression holds in the thermodynamic limit $N
\to \infty$.

We see that in the thermodynamic limit the lowest energy state in this sector (with a zero mode) and the ground state with no excitation in the even particle sector become degenerate ($E_0^+ = E_0^-$).

It can be proven, see for instance \cite{damski12}, that the gap between $|GS \rangle_-$ and $|GS \rangle_+$ closes exponentially in the system size $N$ (with frequent exchanges in which out of the two $|GS\rangle_\pm$ has the lowest energy). Moreover, it is also clear that each state of the even excitation sector lies exponentially close to one state of the odd sector (for instance the states $\chi_{q+1/2}^\dagger \chi_{q^\prime +1/2}^\dagger |GS \rangle_+$ and $\chi_{q}^\dagger \chi_{q^\prime}^\dagger |GS \rangle_-$, or $\chi_{q+1/2}^\dagger \chi_{1/2}^\dagger |GS \rangle_+$ and $\chi_{q}^\dagger \chi_{0} |GS \rangle_-$).

We have thus shown that the special role of the zero mode\index{Zero mode} renders the whole spectrum of  the ordered phase doubly degenerate, while this degeneracy disappears for $h>1$.



\section{The Phase-Diagram}
\label{phasediagramsec}

As we mentioned, the zero temperature phase diagram is quite interesting, due to the presence of two different quantum phase transitions. It is easy to find them, since they are the points in the $(\gamma,h)$ plane where the minimum of the spectrum (\ref{spectrum}) is zero. Thus, at these points  the mass gap vanishes and the gapless low energy excitations determine a scale-invariant behavior.

\begin{wrapfigure}{r}{9cm}
	\vspace{-30pt}
	\begin{center}
		\includegraphics[width=8cm]{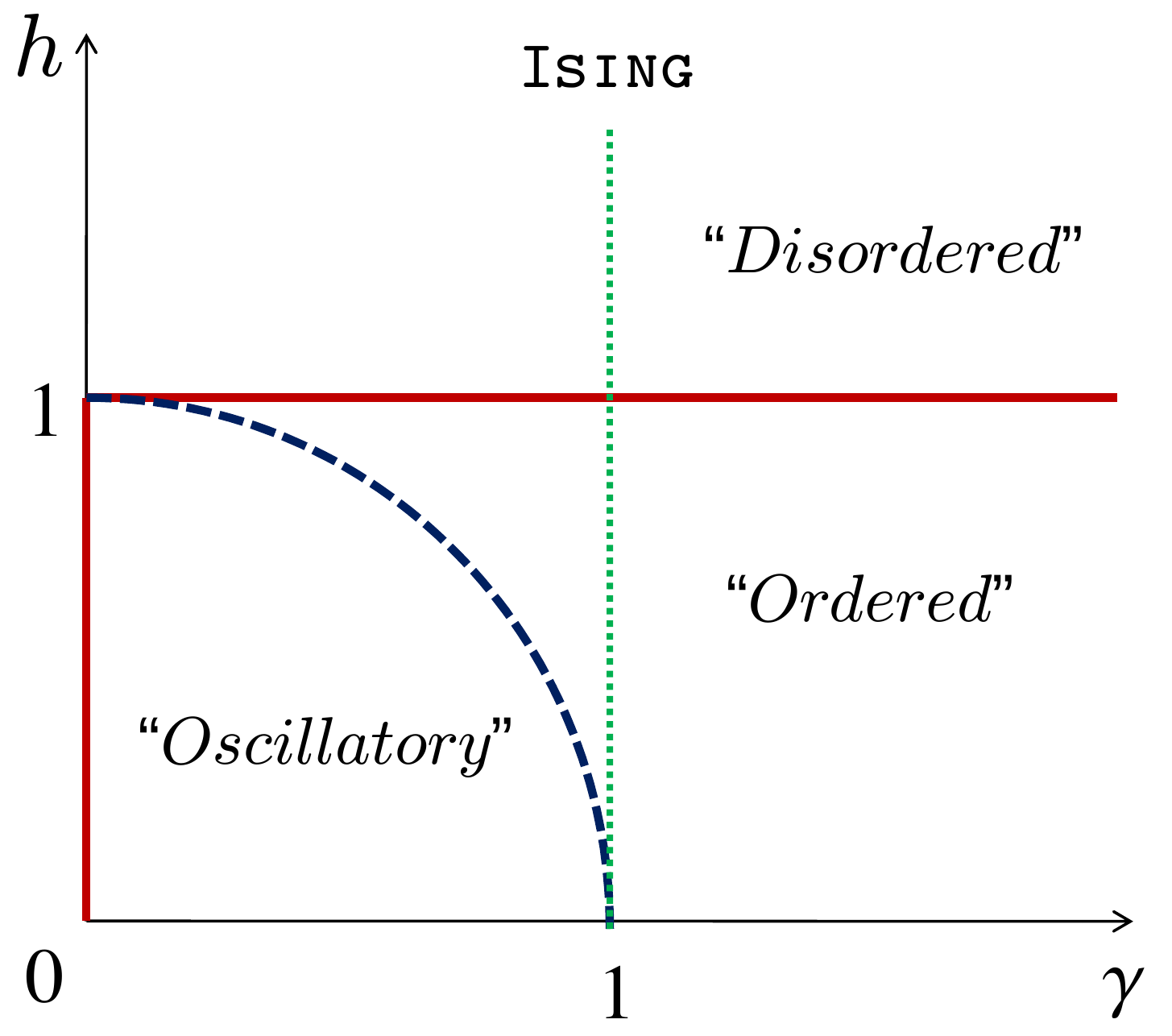}
	\end{center}
	\caption{Phase diagram of the XY Model (only $\gamma \ge
		0$ and $h \ge0$ is shown). The model is critical for $h = 1$ and
		for $\gamma = 0$ and $h < 1$ (in bold red). The line $\gamma = 1$
		is the Ising Model in transverse field (dotted line). On the line
		$\gamma^2 + h^2 = 1$ the ground states can be
		factorized as a product of single spin states (blue dashed line).}
	\label{phasediagram}
	\vspace{-20pt}	  
\end{wrapfigure}

From (\ref{spectrum}) we see that this happens for $\gamma=0$, $|h|<1$ (isotropic XX model: $c=1$ CFT)\index{XX chain} and for $|h|=1$ (critical magnetic field: $c=1/2$ CFT)\index{CFT}.
In Fig.~\ref{phasediagram} we draw the phase diagram of the XY model for
$\gamma \ge 0$ and $h \ge 0$. It shows the critical lines $\gamma = 0$ and
$h=1$ and the line $\gamma=1$ corresponding to the Ising model\index{Ising chain} in transverse
magnetic field and the line $\gamma^2 + h^2 = 1$ on which the wave function of
the ground state is factorized into a product of single spin states (\ref{deg})
\cite{shrock}.

We already determined that the $h=1$ line separates a doubly degenerate phase from a non-degenerate one and thus corresponds to the spontaneous breaking of $Z_2$. The low energy excitations close to $h=1$ have vanishing momentum. Crossing the $\gamma=0$ lines, the role of $x$ and $y$ gets inverted (the (non-)vanishing order parameters switch from $m_x$ to $m_y$). Approaching this QPT, there are two types of low energy states, with momenta approximately equal to $\pm \arccos h$. We will understand better the different phases and the nature of their low-energy excitations in the next section.

The finite temperature partition function of the XY model for $h<1$ is
\bea
    {\cal Z} = \sum \eu^{-\beta E_i}
    & = & \quad {1 \over 2} \eu^{- \beta E_0^+} \left[ \prod_{q=0}^{N-1}
    \left( 1 + \eu^{-\beta \varepsilon \left( {2 \pi \over N} \; q + {\pi \over N} \right) } \right)
    + \prod_{q=0}^{N-1}
    \left( 1 - \eu^{-\beta \varepsilon \left( {2 \pi \over N} \; q + {\pi \over N} \right) } \right)
    \right] \nonumber \\
    & & +  {1 \over 2} \eu^{- \beta E_0^-} \left[ \prod_{q=0}^{N-1}
    \left( 1 + \eu^{-\beta \varepsilon \left( {2 \pi \over N} \; q \right) } \right)
    + \prod_{q=0}^{N-1}
    \left( 1 - \eu^{-\beta \varepsilon \left( {2 \pi \over N} \; q \right) } \right)
    \right] \nonumber \\
    & = & \quad 2^{N-1} \left\{ \prod_{q=0}^{N-1} \cosh \left[
    {\beta \over 2} \varepsilon \left( \textstyle{ {2 \pi \over N} \; q + {\pi \over N} } \right) \right]
    + \prod_{q=0}^{N-1} \sinh \left[
    {\beta \over 2}\varepsilon \left( \textstyle{ {2 \pi \over N} \; q + {\pi \over N} } \right) \right]
    \right\} \nonumber \\
    & & +  2^{N-1} \left\{ \prod_{q=0}^{N-1} \cosh \left[
    {\beta \over 2} \varepsilon \left( \textstyle{ {2 \pi \over N} \; q } \right) \right]
    + \prod_{q=0}^{N-1} \sinh \left[
    {\beta \over 2} \varepsilon \left( \textstyle{ {2 \pi \over N} \; q } \right) \right]
    \right\} \; ,
    \label{degZ}
\eea
where the terms with a minus sign within each square bracket kill states with the wrong parity of excitations in each sector.
Taking the thermodynamic limit, the free energy per site is:
\be
    {\cal F} = - {1 \over \beta} \lim_{N \to \infty} {1 \over N} \ln {\cal Z}
    = - {1 \over \beta} \ln 2 - {1 \over \pi \beta} \int_0^\pi
    \ln \cosh \left[ {\beta \over 2} \varepsilon (\omega) \right] \; \de \omega
    - {1 \over \beta} \lim_{N \to \infty} {1 \over N} \ln \left[ 1 + \prod_{q=0}^{N-1}
    \tanh {\beta \over 2} \varepsilon \left( \textstyle{ {2 \pi \over N} q } \right) \right]  \; ,
\ee
where the last term, encoding the degeneracy of the model, is clearly negligible in the thermodynamic limit.

For $h>1$
\bea
    {\cal Z} & = & \quad 2^{N-1} \left\{ \prod_{q=0}^{N-1} \cosh \left[
    {\beta \over 2} \varepsilon \left( \textstyle{ {2 \pi \over N} \; q + {\pi \over N} } \right) \right]
    + \prod_{q=0}^{N-1} \sinh \left[
    {\beta \over 2}\varepsilon \left( \textstyle{ {2 \pi \over N} \; q + {\pi \over N} } \right) \right]
    \right\} \nonumber \\
    & & +  2^{N-1} \left\{ \prod_{q=0}^{N-1} \cosh \left[
    {\beta \over 2} \varepsilon \left( \textstyle{ {2 \pi \over N} \; q } \right) \right]
    - \prod_{q=0}^{N-1} \sinh \left[
    {\beta \over 2} \varepsilon \left( \textstyle{ {2 \pi \over N} \; q } \right) \right]
    \right\} \; .
    \label{nodegZ}
\eea and the free energy per site in the thermodynamic limit is
\be
    {\cal F} = - {1 \over \beta} \ln 2 - {1 \over \pi \beta} \int_0^\pi
    \ln \cosh \left[ {\beta \over 2} \varepsilon (\omega) \right] \; \de
    \omega \; .
    \label{XYFreeEn}
\ee
Clearly, from the partition function we can derive the whole thermodynamics of the model, which is essentially that of free fermions.

\section{The correlation functions}
\label{corrsec}

In this section we review the derivation of the fundamental correlators in the
ground state $|GS\rangle_+$ at zero temperature, following
McCoy and co-authors \cite{mccoy}. In the ordered phase, the ground state of the XY model breaks $Z_2$ symmetry (that is, is not an eigenstate of the parity operator (\ref{muNxdef})) and thus the true ground state of the model is $|GS\rangle = \left( |GS\rangle_+ \pm |GS \rangle_- \right)/\sqrt{2}$. However, all parity-conserving operators have the same expectation values for $|GS\rangle_+$ as with respect to $|GS\rangle$. We will thus henceforth drop the reference to the sector. 
Since the model is quadratic, all correlation functions can be expressed in terms of two-point functions using Wick's theorem. Let us thus concentrate on the latter.
Using (\ref{GS+def}) we have
\bea
   \langle GS | \chi_q \chi^\dagger_k | GS \rangle =  N \delta_{k,q} \: , 
   & \qquad \qquad &
   \langle GS | \chi^\dagger_q \chi_k | GS \rangle = 0 \: , \\
   \langle GS | \chi_q \chi_k | GS \rangle =  0 \: , 
   & \qquad \qquad &
   \langle GS | \chi^\dagger_q \chi^\dagger_k | GS \rangle = 0 \; .
\eea

As (\ref{GS+}) shows, this state is far from a vacuum, empty of fermions, as a consequence of the superconducting terms in (\ref{fourierspinlessham}). 
Using (\ref{bogtrans}), in terms of the physical fermions we have
\bea
   \langle GS | \psi^\dagger_q \psi_k | GS \rangle 
   = {1 - \cos 2 \vartheta_q \over 2} \: N \delta_{k,q} \: ,  
   & \qquad \quad &
   \langle GS | \psi_q \psi^\dagger_k | GS \rangle 
   = {1 + \cos 2 \vartheta_q \over 2} \: N \delta_{k,q} \: , \\
   \langle GS | \psi_q \psi_k | GS \rangle 
   = - {\sin 2 \vartheta_q \over 2} \: N \delta_{-k,q} \: ,
   & \qquad \quad &
   \langle GS | \psi^\dagger_q \psi^\dagger_k | GS \rangle 
   = {\sin 2 \vartheta_q \over 2} \: N \delta_{-k,q} \: .
\eea
The two-point fermionic correlators are obtained by Fourier transform. In the thermodynamic limit they read \cite{LSM-1961,mccoy}
\bea
   F_{jl} &\equiv& \ii\langle GS | \psi_j \psi_l | GS \rangle
   = - \ii\langle GS | \psi_j^\dagger \psi_l^\dagger | GS \rangle
   = \int_0^{2 \pi} {\de q \over 2\pi}\; \frac{\sin 2 \vartheta (q)}{2}
   \eu^{\ii q (j-l)} \: ,
  \label{F}
\\
   G_{jl} &\equiv& \langle GS | \psi_j \psi_l^\dagger | GS \rangle
   =   \int_0^{2 \pi} {\de q \over 2 \pi}\; \frac{1 + \cos 2 \vartheta (q)}{2}
   \eu^{\ii q (j-l)} \: ,
 \label{G}
\eea 
where the function $\vartheta(q) \equiv {1 \over 2} \arctan {\gamma \sin q \over h - \cos q}$ is the continuum limit of (\ref{rotangle}).
If our starting point was the spinless fermions Hamiltonian (\ref{noboundayfermionH}), these would be all we need. However, what we are really after are the spin-spin correlation functions, for which we have to take into account the non-local effects of the Jordan-Wigner transformation\index{Jordan-Wigner!transformation}.

Thus, we follow \cite{LSM-1961} and introduce the following expectation values
\be
   \rho^\nu_{lm} \equiv \langle GS \left| \sigma^\nu_l \: \sigma^\nu_m \right| GS \rangle
   \qquad \nu = x, y, z ,
   \label{rhocorr}
\ee 
which can be written in terms of spin lowering and raising operators as 
\bea
   \rho^x_{lm} & = & \quad \langle GS \left| \left( \sigma^+_l + \sigma^-_l \right) \:
   \left( \sigma^+_m + \sigma^-_m \right) \right| GS \rangle \; , \\
   \rho^y_{lm} & = & - \langle GS \left| \left( \sigma^+_l - \sigma^-_l \right) \:
   \left( \sigma^+_m - \sigma^-_m \right) \right| GS \rangle \; , \\
   \rho^z_{lm} & = & \quad \langle GS \left| \left( 1 - 2 \sigma^+_l \sigma^-_l \right) \:
   \left( 1 - 2 \sigma^+_m \sigma^-_m  \right) \right| GS \rangle \; .
\eea 

The key observation is that the product of two Jordan-Wigner strings\index{Jordan-Wigner!string} is the identity, since each of them measure the magnetization parity. Thus, for instance, we can use (\ref{JordanWigner}) on $\rho^x_{lm}$ to get 
\bea
   \rho^x_{lm} & = & \langle GS \left| \left( \sigma^+_l + \sigma^-_l \right) \:
   \left( \sigma^+_m + \sigma^-_m \right) \right| GS \rangle \nonumber \\
   & = &  \langle GS | \left( \psi^\dagger_l + \psi_l \right) \:
    \prod_{j=l}^{m-1} \left( 1 - 2 \psi^\dagger_j \psi_j \right) \:
   \left( \psi^\dagger_m + \psi_m \right) | GS \rangle \nonumber \\
   & = &  \langle GS |\left( \psi^\dagger_l - \psi_l \right) \:
   \prod_{j=l+1}^{m-1} \left( 1 - 2 \psi^\dagger_j \psi_j \right) \:
   \left( \psi^\dagger_m + \psi_m \right) | GS \rangle \nonumber \\
   & = &  \langle GS | \left( \psi^\dagger_l - \psi_l \right) \:
   \prod_{j=l+1}^{m-1} \left( \psi^\dagger_j + \psi_j \right)
   \left( \psi^\dagger_j - \psi_j \right) \:
   \left( \psi^\dagger_m + \psi_m \right) | GS \rangle ,
\eea 
where we have used the identities
\bea
   \sigma_j^+ 
   & = & \eu^{\ii \pi \sum_{l<j} \psi_l^\dagger \psi_l} \: \psi_j
   = \psi_j \: \eu^{-\ii \pi \sum_{l<j} \psi_l^\dagger \psi_l}  \; , \\
   \eu^{\ii \pi \psi_j^\dagger \psi_i} 
   & = & 1 - 2 \psi_j^\dagger \psi_j 
   = \left( \psi_j^\dagger + \psi_j \right) \left( \psi_j^\dagger - \psi_j \right) 
   = - \left( \psi_j^\dagger - \psi_j \right) \left( \psi_j^\dagger + \psi_j \right) \; .
\eea

Now we define the operators\footnote{Note that these operators are essentially Majorana fermions\index{Majorana fermions}, except for a missing $i$ in the definition of $B_j$ that would render it real as well.}
\be
    A_j \equiv \psi^\dagger_j + \psi_j \: , \qquad \qquad
    B_j \equiv \psi^\dagger_j - \psi_j
    \label{ABjdef}
\ee
which allow us to write the correlators (\ref{rhocorr}) as 
\bea
   \rho^x_{lm} & = & \langle GS | B_l A_{l+1} B_{l+1} \ldots
   A_{m-1} B_{m-1} A_m | GS \rangle \nonumber \\
   \rho^y_{lm} & = & (-1)^{m-1}\langle GS | A_l B_{l+1} A_{l+1} \ldots
   B_{m-1} A_{m-1} B_m | GS \rangle \nonumber \\
   \rho^z_{lm} & = & \langle GS | A_l B_l A_m B_m | GS \rangle .
   \label{rhocorr1}
\eea

We can use Wick's Theorem to expand these expectation values in terms of two-point correlation functions. By noticing that 
\be
   \langle GS | A_l A_m | GS \rangle = \langle GS | B_l B_m | GS \rangle = 0
\ee 
we write $\rho^z_{lm}$ as
\bea
    \rho^z_{lm} & = & \langle GS | A_l B_l | GS \rangle \langle GS | A_m B_m | GS \rangle
    -  \langle GS | A_l B_m | GS \rangle \langle GS | A_m B_l | GS \rangle \nonumber \\
    & = & H^2 (0) - H (m-l) H(l-m)
\eea where \be
   H (m-l) \equiv \langle GS | B_l A_m | GS \rangle = {1 \over 2}
   \int_0^{2 \pi} {\de q \over 2\pi}\; \eu^{\ii 2 \vartheta (q)} \eu^{\ii q (m-l)} .
   \label{H}
\ee

The other two correlators in (\ref{rhocorr1}) involve a string of $n=m-l$ operators from site $j$ to $m$. Their Wick's expansion can be expressed as the determinant of a $n \times n$ matrix with elements given by all non-trivial contractions \cite{mccoy,LSM-1961}:
\bea
    \rho^x_{lm} & = & \det \left| H (i-j) \right|_{i=l \ldots m-1}^{j= l+1 \ldots m} \; ,
    \label{rhoxmat} \\
    \rho^y_{lm} & = & \det \left| H (i-j) \right|_{i=l+1 \ldots m}^{j= l \ldots m-1} \: .
    \label{rhoymat}
\eea

Matrices like (\ref{rhoxmat},\ref{rhoymat}) have a special structure. Their entries
depend only on the difference between the row and column index, so that the
same elements appear on each diagonal:
\be
   \rho^x_{lm} 
   = \left| \begin{matrix}  
   		                     H(-1) & H(-2) & H(-3) & \ldots & H(-n) \\
                             H(0) & H(-1) & H(-2) & \ldots & H(1-n) \\
                             H(1) & H(0) & H(-1) & \ldots & H(2-n) \\
                             \vdots & \vdots & \vdots & \ddots & \vdots \\
                             H(n-2) & H(n-3) & H(n-4) &\ldots & H(-1)
				                     \end{matrix} \right| \: , 
   \quad
   \rho^y_{lm} 
   = \left|\begin{matrix}
   	                         H(1) & H(0) & H(-1) & \ldots & H(2-n) \\
                             H(2) & H(1) & H(0) & \ldots & H(3-n) \\
                             H(3) & H(2) & H(1) & \ldots & H(4-n) \\
                             \vdots & \vdots & \vdots & \ddots & \vdots \\
                             H(n) & H(n-1) & H(n-2) &\ldots & H(1)
		                     \end{matrix} \right| \; .
\ee

\begin{figure}[t]
	\begin{center}
		\includegraphics[width=16cm]{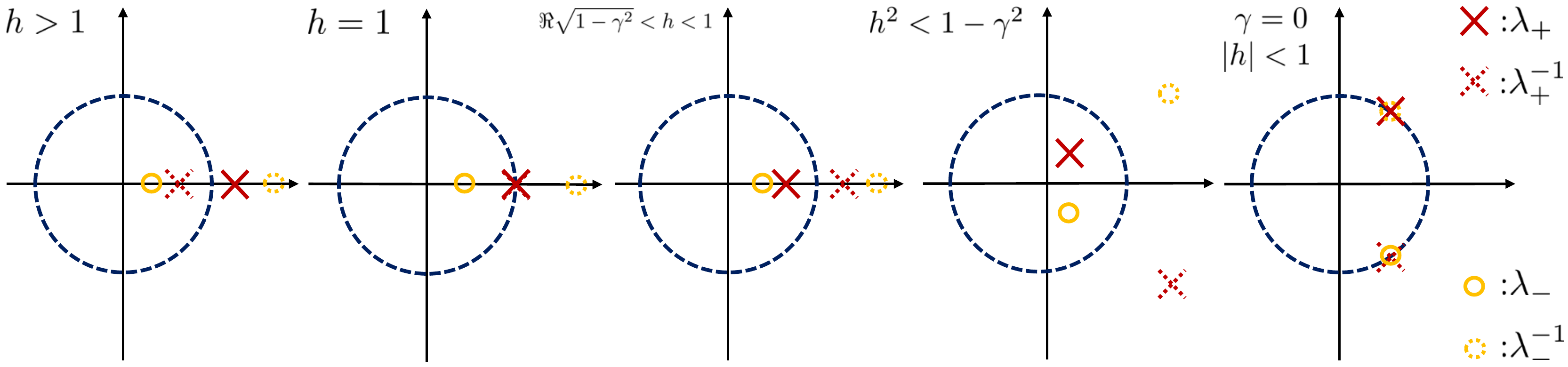}
	\end{center}
    \caption{Cartoon of the positions of the two length-scale parameters of the models $\lambda_\pm$ (\ref{lambdapmdef}) with respect to the unit circle (in dashed blue).}
    \label{fig:poles}
\end{figure}

Matrices like (\ref{rhoxmat},\ref{rhoymat}) are known as {\it ``Toeplitz
Matrices''}\index{Toeplitz!matrix} and a vast \hbox{mathematical} literature has been devoted to the study of the asymptotic behavior of their determinants ({\it ``Toeplitz Determinants''})\index{Toeplitz!determinant}. The development of the theory of Toeplitz Determinants is tightly connected with the Ising\index{Ising chain} and XY model\index{XY chain} since the seminal works by Wu, McCoy and collaborators \cite{toeplitzWu},\cite{mccoy}.
In the second paper of the series \cite{mccoy}, these techniques were applied to calculate the fundamental correlators of the XY model. It is beyond the scope of these lectures to reproduce this derivation. The main results on the asymptotic behavior of Toeplitz determinants\index{Toeplitz!determinant} are summarized in appendix \ref{ToeplitzApp} and we collected the {\bf zero-temperature} behavior of the two-point functions $\rho^\nu (n)$ in table \ref{2pointtable} and \ref{2pointpreftable} as a function of the parameters
\be
   \lambda_{\pm} \equiv {h \pm \sqrt{\gamma^2 +h^2 -1} \over 1 + \gamma } \;.
   \label{lambdapmdef}
\ee
These parameters, which are the zeros of the extension of (\ref{spectrum}) as an analytic function of the complex plane $z= \eu^{i \alpha}$, fully characterize the model: their logarithm gives the two correlation lengths of the chain (in unit of the lattice spacing $a$): $\xi_{\pm} \equiv a / |\ln \lambda_{\pm}|$. Their behavior is depicted in Fig. \ref{fig:poles}.

Looking at the asymptotic behavior of these correlation function we can derive a better interpretation of the different phases of the model.
For $h>1$ we have a ``{\it Disordered Phase}'', since there is no net magnetization along the $x$-direction. In this region the $\lambda$'s are real and $\lambda_-$ and $\lambda_+^{-1}$ are inside the unit circle, with the latter with a bigger modulus and thus providing the measurable correlation length. For $|h|<1$, the model is in an ``{\it Ordered Phase}'', with a net magnetization $m_x$. Initially, the $\lambda_\pm$ are still real and both inside the unit circle. If $h>0$ and $h^2 + \gamma^2 > 1$, $\lambda_+$ is closer to the circle and $\rho_x (n)$ approaches the saturation exponentially. For $h^2 + \gamma^2 < 1$ both $\lambda_\pm$ acquire an imaginary part and become complex conjugated. Thus, they both contribute to the asymptotic behavior and the correlation functions develop a periodic modulation (the two contributions can be traced on the existence of two distinct minima in the single particle spectrum). Hence, the name of ``{\it oscillatory phase}''. Going to negative magnetic field, the role of $\lambda_+$ and $\lambda_-$ gets inverted, while crossing the line $\gamma = 0$ one should exchange $\lambda_\pm$ with $\lambda_\pm^{-1}$.

\begin{table}
	\begin{center}
		{\small
			\centering
			\noindent\begin{tabular}{| c | l | l | l |}
				\hline
				$n \to \infty$ & $\rho^x (n) \simeq$ & $\rho^y (n) \simeq$ & $\rho^z (n) \simeq$  \\
				\hline
				\hline
				$\stackrel{\rm ``Disordered \: Phase"}{h>1}$
				& $ X_D \: {\lambda_+^{-n} \over n^{1/2} } + \ldots$
				& $ Y_D \: {\lambda_+^{-n} \over n^{3/2} } + \ldots$
				& $ {1 \over 4} - {1 \over 8 \pi} \: {\lambda_+^{-2 n} \over n^2}  + \ldots $
				\\ [2ex]
				\hline
				$\stackrel{\rm ``Ising \: Transition"}{h = 1}$
				& $ C_x \: { \gamma \over 1 + \gamma} \: {1 \over ( \gamma n)^{1/4}} + \ldots $
				& $ C_y \: { \gamma (1 + \gamma) \over ( \gamma n)^{9/4} } + \ldots$
				& $ m_z^2 - {1 \over 4 (\pi n)^2 }  + \ldots$
				\\ [2ex]
				\hline
				$\stackrel{\rm ``Ordered \: Phase"}{\Re \sqrt{1- \gamma^2} < h <1}$
				& $ m^2_x \left[ 1 + X^+_O \: {\lambda_+^{2 n} \over n^2}  + \ldots \right]$
				& $ Y_{Or} \: {\lambda_+^{2 n} \over n^3} + \ldots$
				& $ m_z^2 - {1 \over 8 \pi} \: {\lambda_+^{2n} \over n^2} + \ldots$
				\\ [2ex]
				\hline
				$\stackrel{\rm ``Factorizing \: Field"}{h^2 = 1- \gamma^2}$
				& $ {1 \over 2} \: {\gamma \over 1 + \gamma} $
				& $ 0 $
				& $ m_z^2 $
				\\ [2ex]
				\hline
				$\stackrel{\rm ``Oscillatory \: Phase"}{h^2 < 1- \gamma^2}$
				& $ m^2_x \left[ 1 + X^+_O \: {\lambda_+^{2 n} \over n^2} +
				X^-_O \: {\lambda_-^{2 n} \over n^2}  + \ldots \right]$
				& $ Y_{Os} \: {\lambda_+^n \lambda_-^n \over n}  + \ldots$
				& $ m_z^2 - {1 \over 4 \pi} \: { (\lambda_+^{n} + Z_{Os} \lambda_-^{n}) (\lambda_+^{n} + Z^{-1}_{Os} \lambda_-^{n}) \over n^2}  + \ldots$
				\\ [2ex]
				\hline
				$\stackrel{\rm ``Free \: Fermions"}{\gamma =0, \: |h| < 1}$
				& $ C \: (1 - h^2)^{1/4} \: {1 \over n^{1/2}} $
				& $ C \: (1 - h^2)^{1/4} \: {1 \over n^{1/2}} $
				& $ m_z^2 - {\sin^2 \left( n \arccos h \right) \over \pi^2 n^2} $
				\\ [2ex]
				\hline
		\end{tabular}}
		\caption{Asymptotic behavior of the fundamental two-point correlation functions (\ref{rhocorr}). $C_x = {1 \over 2 A^3} \eu^{1/4} 2^{1/12}$, $C_y = - {1 \over 32 A^3} \eu^{1/4} 2^{1/12}$, and $C = {1 \over A^6} \eu^{1/2} 2^{2/3}$,  where $A= 1.282\ldots$ is the Glaisher's constant. The other prefactors are also known from \cite{mccoy} and are listed in table \ref{2pointpreftable}. The subleading correction are all suppressed by order $1/n$.}
		\label{2pointtable}
	\end{center}
	\hskip.3cm\begin{minipage}[t]{13cm}
			\noindent\begin{tabular}{| l | l |}
				\hline
				$ m_x^2 \equiv {1 \over 4} \left[ (1 - \lambda_-^2)(1 - \lambda_+^2) (1 - \lambda_+ \lambda_-)^2 \right]^{1/4}$
				& $ m_z \equiv \int_0^\pi {p_1 \left( \eu^{\ii q} \right) + p_2 \left( \eu^{\ii q} \right) \over
					\sqrt{ p_1 \left( \eu^{\ii q} \right) p_2 \left( \eu^{\ii q} \right) }} \: {\de q \over 2 \pi} $
				\\ [2ex]
				\hline
				$ X_D \equiv {1 \over 4 \sqrt{\pi}}
				\left[ {(1- \lambda_-^2) \over (1- \lambda_+^{-2} ) } (1 - \lambda_- \lambda_+)^2 \right]^{1/4}$
				& $ Y_D \equiv - {1 \over 8 \sqrt{\pi}}
				\left[ {(1 - \lambda_+^{-2})^3 (1 - \lambda_-^2) \over (1 - \lambda_- \lambda_+)^2} \right]^{1/4}
				{1 \over 1 - \lambda_- \lambda_+^{-1}}$
				\\ [2ex]
				\hline
				$ X^+_O \equiv {1 \over 2 \pi} {\lambda_+^2 \over 1 - \lambda_+^2}$ ,
				$ X^-_O \equiv {1 \over 2 \pi} {\lambda_-^2 \over 1 - \lambda_-^2}$
				& $ Y_{Or} \equiv - {1 \over 8 \pi}
				\left[ {(1 - \lambda_-^2) \over (1 - \lambda_+^2)^3(1 - \lambda_- \lambda_+)^2} \right]^{1/4}
				{1 \over 1 - \lambda_- \lambda_+^{-1}}$
				\\ [2ex]
				\hline
				$ Z_{Os} \equiv {\lambda_+ \over \lambda_-} \sqrt{ {1 - \lambda_+^2 \over 1 - \lambda_-^2}}$
				& $ Y_{Os} \equiv { [(1 - \lambda_- \lambda_+^{-1})(1 - \lambda_-^{-1} \lambda_+)]^{1/2} \over 4 \pi
					\left[ (1 - \lambda_-^2)(1 - \lambda_+^2) (1 - \lambda_+ \lambda_-)^2 \right]^{1/4}}$
				\\ [2ex]
				\hline
		\end{tabular}
	   	\end{minipage}
   		\begin{minipage}{3cm}
    	\caption{$\lambda_\pm$-dependence of the prefactors in the asymptotic behavior of table \ref{2pointtable}.\vspace{1.5cm} $\quad$}
		\label{2pointpreftable}
	    \end{minipage}
\end{table}

Note that the spontaneous magnetization is inferred from the asymptotic behavior of the two-point function $\rho^x(\infty)$. It would be desirable to calculate directly $\langle \sigma^x \rangle$. However, a non-zero expectation value for such operator requires that the bra and the ket belong to different parity sectors and, so far, nobody has devised a method to perform such calculation directly.

\section{The Kitaev chain}
\label{sec:kitaev}

The double degeneracy in the ordered phase of the XY model is arguably its most important, defining characteristic. Nonetheless, the derivation we provided of this degeneracy is somewhat unsatisfactory, in that it relies on a subtle mathematical effect giving a ``negative mass'' to the zero mode\index{Zero mode}.

A neater derivation of the degeneracy was provided by Kitaev in \cite{kitaev2000}, focusing on the Ising line $\gamma =1$\index{Ising chain}, and highlighting the importance of the Majorana fermions representation: we introduce the operators
\be
f_{2j -1} \equiv \left[ \prod_{l<j} \sigma_l^z \right] \sigma_j^x = \psi_j^\dag + \psi_j \, , \quad 
f_{2j} \equiv \left[ \prod_{l<j} \sigma_l^z \right] \sigma_j^y =\ii \left( \psi_j^\dag - \psi_j \right) \; .
\label{majorana}
\ee
Note that $f_{2j-1} = A_j$ and $f_{2j}= \ii B_j$, where $A_j$, $B_j$ were defined in (\ref{ABjdef}). These operators are real ($f_l^* = f_l$),  satisfy anticommutation relations and square to unity ($f_l^2=1$). Thus, these fermions are their own antiparticles and are known in the literature as Majorana fermions\index{Majorana fermions}.

The Hamiltonian (\ref{spinham}) in these variables reads\index{Kitaev chain}
\be
   H = - \ii \: {J \over 2} \sum_{j=1}^N \left[
   \left( {1 + \gamma \over 2} \right) f_{2j} \: f_{2j+1} 
   - \left( {1 - \gamma \over 2} \right) f_{2j-1} \: f_{2j+2}
   + h \; f_{2j-1} \: f_{2j} \right] \; ,
   \label{majoranaham}
\ee 
where this time we take open boundary conditions.

\begin{wrapfigure}{r}{7.5cm}
	\vspace{-20pt}
	\begin{center}
		\includegraphics[width=7cm]{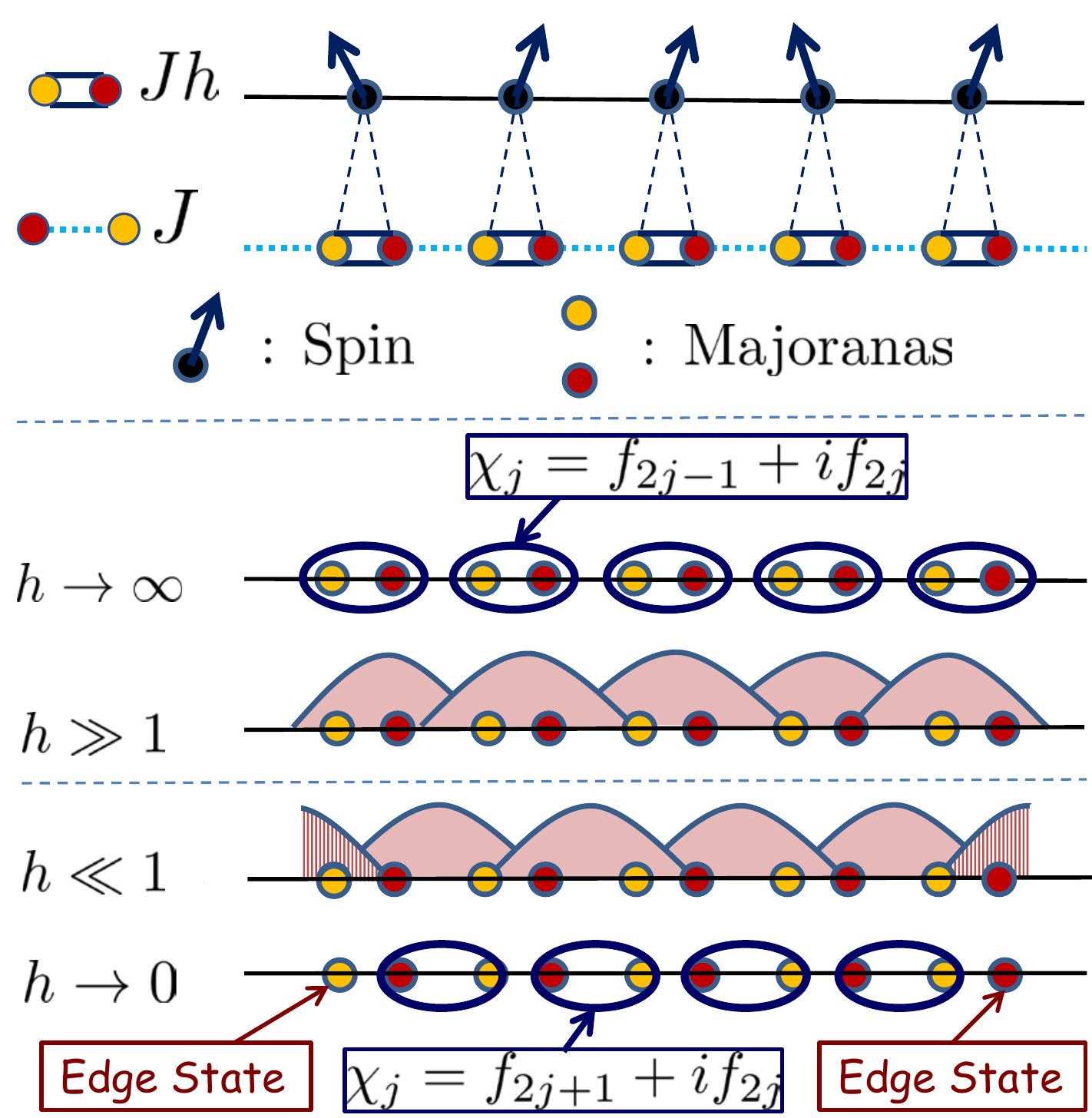}
	\end{center}
	\caption{Cartoon of the Majorana representation for the chain, with the doubling of the sites.}
	\label{fig:majorana}
	\vspace{20pt}	  
\end{wrapfigure}

The Jordan-Wigner mapping places a (complex) spinless fermion in each lattice site. Since it takes two real variables to make a complex one, the mapping (\ref{majorana}) doubles the chain, by splitting each site into two and by placing a Majorana fermion on each: in the Hamiltonian the hopping terms pair Majoranas living on formerly distinct sites, while the magnetic field couples Majoranas belonging to the same site, see Fig \ref{fig:majorana}.

For $(\gamma,h)=(1,0)$, we see that the first and last Majorana ($f_1$ and $f_{2N}$) do not appear in the Hamiltonian. It is thus possible to (mathematically) define a complex fermion out of them ($\tilde{\chi_0} \propto f_1 + \ii f_{2N}$), which can be populated without affecting the energy of the state. These Majorana are called {\it edge states}, that is, excitations localized at the beginning or at the end of the chain. They survive in the whole ordered phase, although the other terms in the Hamiltonian hybridize them so that they acquire a tail protruding inside the chain. These exponentially decaying tails are also responsible for splitting the degeneracy in finite systems, corresponding to the recombination of the two edge states at opposite boundaries \cite{kitaev2000}.

This picture provides a satisfactory explanation of the double degeneracy in terms of edge states that can be empty or filled (note that in the $h \to \infty$ limit there are no edge states and this feature is robust in the whole disordered phase). It has also sparked an interest in finding Majorana fermions\index{Majorana fermions} not as fundamental particles, but as emergent excitations in strongly-interacting systems, such as the XY model. It should be stressed, however, that the ground state of the XY model in the ordered phase is not an eigenstate of the parity operator (\ref{muNxdef}) (it has non-zero $\langle \sigma^x \rangle$) and thus does not host well-defined edge states. Thus, to observe the latter the starting point has to be the spinless fermions Hamiltonian (\ref{noboundayfermionH}), in open boundary conditions.

{\bf Ising self-duality}: One very interesting feature of the Ising model\index{Ising chain} is its self-duality. This can be seen both in the classical two-dimensional model (Kramers-Wannier duality \cite{kramerswannier}) and in the 1-D quantum. 
In the latter, the statement is that the Hamiltonian (\ref{spinham}) along the $\gamma =1$ line is invariant under the transformation between {\it order} and {\it disorder} spin operators:
\be
   \sigma_j^x = \prod_{l \le j} \mu_l^z \; , \qquad \qquad
   \sigma_j^z = \mu_j^x \mu_{j+1}^x \; ,
\ee
i.e.
\be
  \sigma_j^x \sigma_{j+1}^x - h \sigma_j^z = \mu_{j+1}^z - h \mu_j^x \mu_{j+1}^x \; .
\ee
Thus relating a system with magnetic field $h$ to one with $1/h$. Note that the duality is realized in a highly non-local way. We leave it as an exercise to the reader to determine the duality relation in terms of the Jordan-Wigner fermions:
\be
   \sigma_j^z = 1 - 2 \psi_j^\dagger \psi_j \; , \qquad
   \mu_j^z = 1 - 2 c_j^\dagger c_j \; ,
\ee
and to check that in terms of the Majorana fermion representation over the doubled chain, the duality is just a shift by one lattice site and is thus local:
\be
   f_j^\sigma = f_{j+1}^\mu \, .
\ee

\chapter{The Lieb-Liniger Model}
\label{chap:LLmodel}

\abstract{
	The first interacting problem we are going to tackle is the Lieb-Liniger model, describing bosons with contact interaction. The model is fairly realistic, although in current experimental realizations the external trapping breaks translational invariance, spoiling integrability (although in some cases the trapping can be accounted for perturbatively). The Lieb-Liniger model is also best suited to illustrate the basic ideas behind the Bethe Ansatz approach. Thus, in Sec. \ref{sec:genBA} we preview the general ideas behind the coordinate Bethe Ansatz solution. In Sec. \ref{2bodysec} we proceed in working out the two particle scattering for the LL model and in Sec. \ref{sec:BAwave} we introduce the ansatz for the eigenstates. We derive the Bethe equations and their thermodynamic limit in Sec \ref{sec:Beq} and \ref{sec:thermolimit}. After introducing some mathematical formalism on integral equations in Sec. \ref{sec:intform}, in Sec. \ref{sec:LLexcitations} we discuss the low energy excitations of the model and in Sec. \ref{sec:LLthermo} the finite temperature thermodynamics, by deriving the Yang-Yang equation. 
}

\section{Introduction}

To introduce the coordinate Bethe Ansatz approach, we will present the solution of a
model of $N$ bosons with contact interaction, which was originally solved by Lieb and Liniger in \cite{LL63}. The Hamiltonian reads\index{Lieb-Liniger model}
\be
{\cal H} = - \sum_{j=1}^N {\partial^2 \over \partial x_j^2}
+ 2 c \sum_{j<l} \delta (x_j - x_l) \; ,
\label{LLHam}
\ee
where $c$ parametrizes the interaction strength.
Physically, this is a very realistic model for 1-D particles with short range interaction \cite{kinoshita} and constitutes a powerful analytical tool to interpret experimental results \cite{jiang15}. The main limitation in the physical application of the model is that the Bethe Ansatz solution assumes translational invariance, which is spoiled by the trapping used in current cold atoms experimental implementations. In fact, due to the external potential, real systems are inhomogeneous, with a density of particles that varies in space \cite{kheruntsyan05,fabbri15,vandenberg16}.

While introducing an external potential in (\ref{LLHam}) spoils integrability, we should mention that there is a different kind of potential that admits a parabolic confinement and is also exactly solvable, namely the Calogero-Moser model \cite{CalogeroMoser}\index{Calogero-type models}:
\be
{\cal H} = - \sum_{j=1}^N {\partial^2 \over \partial x_j^2}
+ \sum_{j<l} {\lambda (\lambda -1) \over (x_j - x_l)^2} + {\omega^2 \over 2} \sum_{j=1}^N x_j^2 \; .
\label{CMModel}
\ee
The trapping prevents the repulsive potential from pushing particles toward infinity (the thermodynamic limit $N \to \infty$ is taken together with the $\omega \to 0$ limit to keep the particle density fixed). While the presence of the parabolic potential makes this model ``realistic'', its long-range potential is not (there is no known way yet to realize a inverse-square interaction in a 1-D system).
This model belongs to a family of integrable systems (for an excellent review on them \cite{sutherlandbook}), which includes a periodic version (the Calogero-Sutherland model), an elliptic interaction and even potentials with a length scale, which can be made very small to approach a short range model.
This family is solved by a different kind of ansatz, called the {\it asymptotic Bethe Ansatz}, which we will not pursue further\index{Asymptotic Bethe Ansatz}.

Let us now come back to our Lieb-Liniger model, which can be written in second quantized form
\be
   {\cal H} = 
   \int \de x \left[ \partial_x \Psi^\dagger (x) \partial_x \Psi (x)
   + c \: \Psi^\dagger (x) \Psi^\dagger (x) \Psi (x) \Psi (x) 
   - h \: \Psi^\dagger (x) \Psi (x) \right] \; ,
   \label{NLS}
\ee
where we also included a chemical potential $h$. To qualitatively understand a model, it is always convenient to extract dimensionless parameters. The strength of the interaction for this system can be captured by \cite{pethick}:
\be
   \gamma \equiv {2 m \over \hbar^2} \: {c \over n} \; , \qquad \qquad
   n \equiv {N \over L} \: ,
   \label{gammaLL}
\ee
where we provisionally restored $\hbar,m$, while, starting with (\ref{LLHam}), we set $m=\hbar^2/2$.

For $c,\gamma \to \infty$ the bosons repel so strongly that they effectively behave like (free) fermions: this is the so-called ``{\it Tonks-Girardeau regime}'' \cite{TG}.
In the regime of small interaction ($c$ much smaller than the average particle density), the bosonic field will not completely condense (it is forbidden in one-dimension, since long range order is always destroyed by fluctuations), but nonetheless, a large fraction of the particles will be in the zero momentum state and form a ``quasi-condensate'' \cite{castin2004}. Under such assumptions, we can treat (\ref{NLS}) semiclassically and take $\Psi (x)$ to be a classical, complex field. 
The Euler equation from (\ref{NLS}) is 
\be
\ii \Psi_t = - \Psi_{xx} - h \: \Psi + 2 c \: \Psi^\dagger \Psi \Psi \; ,
\label{GrossPitaevskii}
\ee
which we recognize as the 1-D Gross-Pitaevskii equation\index{Gross-Pitaevskii equation} \cite{stringari}. The Hamiltonian (\ref{NLS}) is known as the Non-Linear Schr\"odinger equation (NLS) and the Lieb-Liniger model is also called the Quantum NLS\index{Non-Linear Schr\"odinger equation}.
We will come back to the relation between the quantum and classical version of the NLS in Sec. \ref{sec:LLexcitations}.

\section{Generalities on the Bethe Ansatz approach}
\label{sec:genBA}

Let us first outline the ingredients of the coordinate BA solution, which are independent from the model:
\begin{itemize}

	\item The first step is to identify the two-particle phase-shift. In a
one-dimensional setting, conservation of energy and momentum constrains the outgoing momenta in the scattering of two identical particles to be equal to the incoming ones.
Thus, the effect of interaction is only to add a phase shift to the wavefunction.

	\item There is no consensus on a definition of what makes a quantum model integrable, but it is known that a necessary condition is that the Yang-Baxter equations\index{Yang-Baxter!equation} hold \cite{jimbo}. Although we will formally introduce it later in chapter \ref{chap:algebraic}, this condition means that a three-particle scattering can be decomposed into a sequence of two-body scatterings and that the order of this decomposition does not matter. Having determined the two-particle scattering phase\index{Scattering phase}, one checks the YBE by verifying that an ansatz wavefunction constructed as a superposition of some plane-wave modes with unknown {\it quasi-momenta}\footnote{They are called quasi-momenta because they are not observables and thus should not be confused with the physical momentum. They are just a bookkeeping way to write the eigenstates.} is an eigenstate of the Hamiltonian. This condition sets the coefficients of the superposition so that the eigenstate depends solely on the quasi-momenta.
	
	\item In order to make the wavefunction normalizable, one needs to specify the boundary conditions: for instance, we will apply periodic boundary conditions. For a system of $N$ particles, this choice generates a series of consistency relations for the quasi-momenta of the eigenstate, known as {\it Bethe Equations}. This system of $N$ algebraic equations depends on $N$ quantum numbers: they specify uniquely the quantum state of the system. For each (physical) choice of these quantum numbers one solves the set of Bethe equations to obtain the quasi-momenta (which, being algebraic, is a much lighter task than solving the original Schr\"odinger PDE) and thus the eigenstate wavefunction. These states have a fermionic nature, in that all quantum numbers have to be distinct. This is a general feature of the Bethe Ansatz solution (in a proper parametrization), valid for bosonic systems as well.

	\item A further simplification arises by considering the thermodynamic limit. Then, abandoning the precise solution for the eigenstates, one is interested in the distribution of quasi-momenta and the set of algebraic equations can be written as an integral equation for this distribution, for which both numerical and sometimes analytical solutions can be derived. The distribution function can then be used to study the thermodynamics of the model at zero and finite temperature.
	
\end{itemize}

The main limitation of the coordinate Bethe Ansatz approach is that it provides only with an implicit
knowledge of the eigenfunctions, since it is written as a superposition of an exponentially large number of terms, as a function of the particle number $N$. Therefore, the calculation of correlation functions remains a formidable task and, for many practical purposes, not attainable. In fact, there are numerical approaches valid for non-integrable systems as well that are more efficient in calculating correlation functions than those based on the coordinate BA. The Algebraic Bethe Ansatz approach we will introduce in chapter \ref{chap:algebraic} provides a more compact expression for the eigenstates. This approach is essentially a second quantized version of the Bethe Ansatz solution, in that eigenstates are generated by applying creation operators to a reference state. This formulation in turns allows to express many correlation functions in terms of Fredholm determinants \cite{ISM}. Although very elegant, these expressions require some work to provide useful results, but recent years have seen impressive progresses, both in terms of technical, analytical manipulation and in combination with numerical approaches.

\section{The two-particle problem}
\label{2bodysec}

Let us start by considering two bosons with contact interaction:
\be
  {\cal H} = - {\partial^2 \over \partial x_1^2}
  - {\partial^2 \over \partial x_2^2} + 2 \; c \; \delta (x_1 - x_2) \; ,
  \label{2bodyH}
\ee 
where the interaction can be attractive ($c<0$) or repulsive ($c>0$).

We write the generic eigenstate by dividing the $x_1 < x_2$ and $x_1 > x_2$ configurations:
\be
   \Psi (x_1, x_2) = f(x_1,x_2) \vartheta_H (x_2 - x_1) + f (x_2, x_1) \vartheta_H (x_1 - x_2) \; ,
   \label{2body1}
\ee
where 
\be
   \vartheta_H (x) = \left\{ \begin{array}{ll} 1, & x>0 \cr 0, & x<0 \cr \end{array} \right.
\ee
is the Heaviside step function.
Note that (\ref{2body1}) is completely symmetric in $x_1$ and $x_2$, as it should for bosons.

The Schr\"odinger equation connected with (\ref{2bodyH}) is a standard textbook problem, which can be solved assuming a superposition of plane-waves:
\bea
   f(x_1,x_2) & \equiv &
   A (k_1,k_2) \eu^{\ii (k_1 x_1 + k_2 x_2)} + A (k_2, k_1) \eu^{\ii (k_2 x_1 + k_1 x_2)}
   \nonumber \\
   & = & A_{12} \eu^{\ii (k_1 x_1 + k_2 x_2)} + A_{21} \eu^{\ii (k_2 x_1 + k_1 x_2)} \; .
   \label{2body2}
\eea

In solving the eigenvalue equation, one should remember that $\partial_x \vartheta_H (x) = \delta (x)$. Thus
\bea
   \partial_{x_1}^2 \Psi (x_1,x_2)
   & = & \left[ \partial_{x_1}^2 f(x_1,x_2) \right] \vartheta_H (x_2 - x_1) 
   + \left[ \partial_{x_1}^2 f(x_2,x_1) \right] \vartheta_H (x_1 - x_2)
   \nonumber \\
   && + \left[ \partial_{x_1} f(x_1,x_2) \right] \delta (x_2 - x_1) 
   - \left[ \partial_{x_1} f(x_2,x_1) \right] \delta (x_1 - x_2) \; ,
\eea
where we also used the identity (valid under integral): $f(x) \partial_x \delta (x) = - \partial_x f (x) \delta (x)$.
Therefore we have:
\be
    {\cal H} \Psi = ( k_1^2 + k_2^2 ) \Psi
    + 2 \delta (x_1 - x_2) \left[ c \left( A_{12} + A_{21} \right) +
    \ii \left( A_{12} - A_{21} \right) (k_1 - k_2) \right] \eu^{\ii (k_1 + k_2) x_1 } \; .
\ee
This eigenvalue equation is satisfied if the off-diagonal term vanishes, i.e.
\be
   {A_{12} \over A_{21} } = {\ii (k_1 - k_2) - c \over  \ii (k_1 - k_2) + c }
   \label{AmpRation}\; .
\ee
It is easy to see that this factor has unit modulus and it is therefore a
pure phase:
\be
   {A_{12} \over A_{21} } = \eu^{\ii \tilde{\theta} (k_1 - k_2)}
\ee
with\footnote{One has to pay attention to that the branch-cut of the logarithm is consistent with that of the arc-tangent.}
\be
   \tilde{\theta} (k) \equiv - 2 \arctan {k \over c} + \pi  \; .
   \label{tildephaseshift}
\ee
This is the phase shift due to the contact interaction. It is a unique signature of the potential: each integrable model is characterized by a phase shift function. We will see that this function plays a major role in the Bethe equations. Notice that in the limit $c \to \infty$ the scattering phase becomes that of free fermions.\index{Scattering phase}

Notice that, regardless of $c$, for vanishing quasi-momentum, the scattering phase also tends to $\pi$. This means that if $k_1 = k_2$ the wavefunction (\ref{2body2}) vanishes, as if it was a fermionic and not a bosonic system. It is customary to factor out this fermionic statistical phase and define the scattering phase as an odd function of its argument that vanishes for $k=0$. Thus, we make the shift $\tilde{\theta} (k) \equiv \theta (k) + \pi$, with
\be
   \theta (k) \equiv - 2 \arctan {k \over c} \; .
   \label{phaseshift}
\ee

\section{Bethe Ansatz Wavefunction}
\label{sec:BAwave}

We consider now a system with $N$ particles, with Hamiltonian (\ref{LLHam}).
We make the ansatz that the wave function can be written as a linear superposition of plane-waves as\index{Bethe!ansatz wavefunction}
\be
   \Psi (x_1, \ldots, x_N;{\cal Q}) = \sum_\cP A_\cP ({\cal Q}) 
   \eu^{\ii \sum_j k_{\cP j} x_j }
   \label{BAWF}
\ee
with $N$ quasi-momenta $k_j$ to be determined. $\cP$ is a permutation of the quasi-momenta, while ${\cal Q}$ is the permutation that specifies the particle order ({\it simplex}). Since the system is bosonic, the wavefunction must be left unchanged by any reordering of the particles. It is clear that if ${\cal Q}$ and ${\cal Q}'$ are two different permutations related by ${\cal R}$ (${\cal Q}' = {\cal R} {\cal Q}$), then consistency requires that $A_\cP ({\cal Q'}) = A_\cP ({\cal R Q}) = A_{{\cal R} \cP}({\cal Q})$. For instance, tn the two-body case considered in the previous section we saw that the connection between the two particle orderings was provided by the permutation of the quasi-momenta. Thus, knowing the wavefunction in one simplex (say the sector for which $x_1 < x_2 < \ldots < x_N$) fixes the solution completely. Thus, henceforth we will drop the explicit dependence on the coordinate permutation ${\cal Q}$, with the understanding that we are working in the given simplex and that the others can be reached by symmetry.

All permutations can be generated by exchanging two indices at a time. This is equivalent to the two-particle scattering we considered in the previous section. Therefore
\be
   {A_{\cP} \over A_{\cP'} } = -  \eu^{\ii \theta (k - k')}
\ee
where $k$, $k'$ are the momenta interchanged between permutation $\cP$ and $\cP'$ and the scattering phase\index{Scattering phase} is given by (\ref{phaseshift}). Equivalently, one could check that the wavefunction (\ref{BAWF}) is an eigenstate of the Hamiltonian (\ref{LLHam}) if:
\be
   A_\cP = \Omega_N (-1)^\cP \prod_{j<l} (k_{\cP j} - k_{\cP l} + \ii \: c )
   \label{AcPsol}
\ee
where $\Omega_N$ is a normalization constant. Notice that the wavefunction vanishes if two quasi momenta coincide.

The energy and momentum eigenvalues of (\ref{BAWF}) are
\be
   E = \sum_{j=1}^N k^2_j \: , \qquad    K = \sum_{j=1}^N k_j \; ,
   \label{PEBethe}
\ee
with the momentum operator defined as $\hat{K} \equiv - \ii \sum_{j=1}^N  {\partial \over \partial x_j}$.

\subsection{Bound States}
\label{sec:boundstates}

Before we proceed, let us analyze a limiting solution of the two-body scattering (\ref{AmpRation}), namely the case in which either $A_{12}$ or $A_{21}$ vanishes, that is
\be
k_1 - k_2 = \pm \ii c \; .
\label{imsol1}
\ee
To better understand this case, we introduce the center of mass and relative coordinates and momenta as:
\bea
X \equiv {x_1 + x_2 \over 2} \; , & \qquad \qquad &
x \equiv {x_1 - x_2 \over 2} \; , \\
K \equiv {k_1 + k_2} \; , & \qquad \qquad &
k \equiv {k_1 - k_2} \; ,
\eea
and rewrite (\ref{2body1},\ref{2body2}) as
\be
\Psi (X,x) = \eu^{\ii K X} \left\{ \begin{array}{ll}
	A_{12} \eu^{\ii k x} + A_{21} \eu^{-\ii k x} , & x>0 \cr
	A_{21} \eu^{\ii k x} + A_{12} \eu^{-\ii k x} , & x<0 \cr
\end{array} \right. \; .
\ee

For this solution to behave well at infinity we need $\Im (K) =0$. For the same reason, If $\Im (k) >0$, then $A_{21} =0$ and if $\Im (k) <0$, then $A_{12} =0$. Thus, we see that the only solutions with imaginary quasi-momenta are (\ref{imsol1}) and they are consistent with the normalizability of the wavefunction only if $c<0$. 

They represent a bound state of two particles with momenta\index{Bound state}
\be
k_{1,2} = {K \pm \ii c \over 2} \; , \qquad \qquad \Im (K) = 0 \;,
\ee
with energy $E = K^2/2 - c^2/2$ and total momentum $K$:
\be
f(x_1,x_2) = \eu^{\ii K (x_1 + x_2)/2} \eu^{c |x_1 - x_2|/2} \; ,
\label{boundstatewf}
\ee
where we recognize the boundedness in the fact that the wavefunction amplitude decays exponentially as the two particles move apart. Notice that $c$ sets the decay rate.

The same reasoning can be generalized to more particles.
With three, for $c<0$ there are two type of solutions with complex momenta:
\bea
&& k_1 = \alpha - \ii {c \over 2} \; , \qquad k_2 = \alpha + \ii {c \over 2} \; , \qquad \quad k_3 = \beta \; , \\
&& k_1 = k_3 - \ii c \; , \qquad k_2 = k_3 + \ii c \; , \qquad \Im (k_3) = 0 \; .
\eea
The former is still a two-particle bound state, scattering with a third independent particle. The second is a proper three-particle bound state.
$n$-particle bound states\index{Bound state} appear in {\it strings} of particles with the same real part of the momentum\index{String solution}. A string of length $n$ has quasi-momenta equispaced symmetrically with respect to the real axis (notice that the distance between neighboring momenta in the imaginary axis is the same for any string and is set by the interaction strength):
\be
k_j = {K \over n} - \ii \; {n + 1 - 2j \over 2} \;  c \; , \qquad j=1 \cdots n \; ,
\ee
corresponding to total momentum $K$ and energy
\be
E = {K^2 \over n} - {n(n^2-1) \over 12 } \; c^2 \; .
\ee

The Lieb-Liniger model has bound states\index{Bound state} only for attractive interactions, and we see that clearly they have lower energy than an unbound configuration.
Therefore the ground state of a system of $N$ particles is given by a zero total momentum string of length $N$: its energy diverges negatively like $N^3$. Since, for the thermodynamic limit to exist, the ground state energy should scale at most linearly in $N$, the attractive Lieb-Liniger is unstable in the thermodynamic limit. We will see in chapter \ref{chap:XXXmodel} and \ref{chap:XXZmodel} that the XXZ chain also possesses string solutions, but their energy is bounded from below and thus contribute to the thermodynamic of the model together with unbound quasi-particles.

For a finite number of particles, the attractive Lieb-Liniger is somewhat simpler than the repulsive one, because we know the ground state solution of the Bethe equations explicitly, namely the bound state\index{Bound state} (string) of all the $N$ particles with zero momentum. The first excitations are still one string state ($N_s =1$) with all $N$ particles in a bound state with finite momentum. Then one can consider two strings states ($N_s =2$), made out of two bound states with respectively $N-M$ and $M$ particles, whose momenta can be determined by a system of 2 Bethe equations (for more details, see \cite{calabresecaux} or the discussion of string solutions in chapter \ref{chap:XXXmodel}). Notice that if $M=1$ the second string collapses to a single real momentum and is not a true bound state\index{Bound state}. Multi-strings solutions can be constructed in a similar way. This ``simple'' structure for the spectrum of the attractive model allows for an explicit calculation of the basic response functions \cite{calabresecaux}. Also, the regime of infinite attraction has intriguing properties which can be experimentally studied through a sudden quench of the interaction from the Tonks-Girardeau regime $c\to \infty$ to this so called ``{\it Super-TG}'' regime $c \to -\infty$ \cite{STG}. In this way, the system is prepared in an excited state that can remain stable for suitably small density of particles. These topics are beyond the scope of these notes and we will not discuss the attractive case and its bound states\index{Bound state} any further. Henceforth, we will thus always assume $c>0$.

\section{The Bethe Equations}
\label{sec:Beq}

In order to quantize the system, we put it in a box of finite length $L$ (which we can take to infinity in the end). We impose periodic boundary conditions on the walls of this box, effectively considering a ring:
\be
   \Psi (x_1, x_2, \ldots, x_j + L, \ldots, x_N) =
   \Psi (x_1, x_2, \ldots, x_j, \ldots, x_N) \; ,
   \qquad j=1, \ldots,  N\;.
   \label{LLpbc}
\ee

Let us consider the ``first'' particle in the system (note that, on a circle, any particle can be taken as the first). The periodic boundary condition (\ref{LLpbc}) relates a given simplex with that in which the first particle has been moved to be the last one. 
From a formal point of view, this permutation can be realized as a sequence of two-body permutations, in which particle $1$ is first exchanged with particle $2$, then with particle $3$ and so on. From a physical viewpoint, taking a particle around the circle means that it has scattered across all other particles in the system. Through all these scattering events it acquires a phase equal to the sum of the scattering phases\index{Scattering phase} associated with each scattering event plus the dynamical phase accumulated through the motion (namely its quasi-momentum times $L$). To satisfy periodic boundary conditions, the sum of these phase contributions has to add up to an integer multiple of $2 \pi$.

We formalize the consistency relations imposed by the periodic boundary conditions by relating the permutation in real space with one in quasi-momenta space, as we commented above. Denoting the permutation that brings the first element to be the last as ${\cal R} \left\{ 1,2,\ldots, N \right\} = \left\{ 2,3,\dots,N,1 \right\}$, we have
\be
  A_\cP ({\cal Q}) 
  = A_\cP ( {\cal R Q}) \: \eu^{\ii k_{\cP 1} L}
  = A_{{\cal R} \cP} ( {\cal Q}) \: \eu^{\ii k_{\cP 1} L} \; .
\ee
Using (\ref{AcPsol}) this condition can be written as
\be
   \eu^{\ii k_j L} = \prod_{l \ne j} \left(
   {k_j - k_l + \ii \: c \over k_j - k_l - \ii \: c} \right)
   = (-1)^{N-1} \prod_{l=1}^N \eu^{\ii \theta (k_j-k_l)} \; ,
   \qquad j = 1, \ldots, N \; ,
   \label{LLBetheEq}
\ee
which shows how the dynamical phase on the LHS has to match the scattering phase \index{Scattering phase} on the RHS.
Taking the logarithm we get
\bea
   k_j L & = &
   2 \pi \tilde{I}_j + (N-1) \pi
   - 2 \sum_{l=1}^N \arctan \left( {k_j - k_l \over c} \right)
   \nonumber \\
   & = & 2 \pi I_j + \sum_{l=1}^N \theta ( k_j - k_l ) \; ,
   \label{betheeq}
\eea 
where the $\tilde{I}_j$ are a set of integers which define the state.
In the second line we introduced a new set $I_j$: they are integers if the number of particles $N$ is odd and half-integers if $N$ is even\footnote{This is equivalent to the different quantization of the momenta we found in the chapter \ref{chap:XYModel} for the XY model.}. The $I_j$'s are called {\it quantum} (or {\it Bethe}) {\it numbers}\index{Bethe!numbers} and they characterize uniquely the state. Notice that if two quantum numbers are equal $I_j = I_l$, their corresponding quasi-momenta also coincide $k_j = k_l$. Since, as we commented above, in such cases the Bethe wavefunction (\ref{BAWF}) vanishes, we conclude that only sets of distinct Bethe numbers correspond to physical solutions. 

The (\ref{betheeq}) are the {\it Bethe equations}\index{Bethe!equations}, a set of $N$ coupled algebraic equations in $N$ unknowns, the $k_j$. For $c \to \infty$ we have hard-core bosons: $\theta (k) \to 0$ and $k_j = 2 \pi I_j /L$. 
Since they have to be all different, the ground state which minimizes the energy and momentum (\ref{PEBethe}) is given by a symmetric distribution of quantum numbers without holes (a Fermi sea distribution):
\be
   I_j = - {N+1 \over 2} + j \; , \qquad \qquad j=1 \ldots N \; .
   \label{GSIs}
\ee

If we now decrease the coupling $c$ from infinity, we slowly turn on a scattering phase\index{Scattering phase} so that, for fixed $I_j$, the solution $k_j$ to the Bethe equations (\ref{betheeq}) will move from the original regular distribution. 
Since there cannot be a level crossing (there is no symmetry to protect a degeneracy and accidental ones cannot happen in an integrable model), the state defined by (\ref{GSIs}) will remain the lowest energy state for any interaction strength (note that, changing $c$ changes the quasi-momenta $k_j$, but cannot change the quantum number, because they are quantized).
It is possible to prove this statement, the non-degenerate condition and the uniqueness of the solution obtained through Bethe Ansatz, but we refer to \cite{ISM} for such rigorous proofs.  

Each choice for the quantum numbers yields an eigenstate, provided that all Bethe numbers are different. This rule confers a fermionic nature to the Bethe Ansatz solution (in quasi-momentum space!), even if the underlying system is composed by bosons (in the real space) as in this case.

Finally, please note that since $\theta (-k) = - \theta(k)$, the momentum of the system is $K = {2 \pi \over L} \sum_j I_j$. So the momentum is quantized and does not vary as the coupling constant is varied.

\section{The thermodynamic limit}
\label{sec:thermolimit}

If we order the Bethe numbers $I_j$'s (and therefore the momenta $k_j$'s) in increasing order, we can write the Bethe Equations ({\ref{betheeq}) as
\be
   k_j - {1 \over L} \sum_{l=1}^N \theta (k_j-k_l) = y (k_j)
   \label{countingfdef}
\ee where we defined the {\it ``counting function''} $y(k)$, as an arbitrary function constrained by two properties: 1) to be monotonically increasing; 2) to take the value of a quantum number at the corresponding quasi-momentum $y (k_j) \equiv {2 \pi I_j \over L }$. By definition,
\be
   y (k_j) - y (k_l) = {2 \pi \over L} ( I_j - I_l) \; .
\ee

We now take the limit $N, L \to \infty$, keeping the density $N/L$ fixed and
finite. We introduce a density of quasi momenta as
\be
   \rho (k_j) = \lim_{N,L \to \infty} {1 \over L (k_{j+1} - k_j) } >0 \; ,
   \label{rhokdef}
\ee 
and we replace sums with integrals over $k$ as
\be
   \sum_j \to L \int \rho(k) \de k \; .
\ee

It is easy to prove that
\be
   y' (k_j) = \lim_{N,L \to \infty} {y(k_j) - y(k_{j-1}) \over k_j - k_{j-1} }
   = \lim_{N,L \to \infty} {2 \pi \over L ( k_j - k_{j-1}) }
   = 2 \pi \; \rho (k_j)
   \label{yprimekj}
\ee
and therefore
\be
   {1 \over 2 \pi} y (k) = \int^k \rho (k') \de k' \: ,
\ee establishing a direct connection between the distribution of the integers
and of the quasi-momenta.

With these definitions, in the thermodynamic limit, the system of algebraic
equations (\ref{betheeq}) can be written as an integral equation for the
counting function and the quasi-momentum distribution:
\be
   y (k) = k - \int_{k_{min}}^{k_{max}} \theta (k - k') \rho (k') \de k'
   \label{yIntEq}
\ee 
and, by taking the derivative with respect to $k$:
\bea
   \rho (k) & = &
   {1 \over 2 \pi} - {1 \over 2 \pi} \int_{k_{min}}^{k_{max}} \theta' (k - k') \rho (k') \de k'
   \nonumber \\
   & = & {1 \over 2 \pi} - {1 \over 2 \pi} \int_{k_{min}}^{k_{max}}
   {\cal K} (k - k') \rho (k') \de k'
   \label{intbetheeq}
\eea 
where we introduced the kernel\index{Kernel} of the integral equation as the derivative
of the scattering phase\index{Scattering phase}:
\be
   {\cal K} (k) \equiv {\de \over \de k} \; \theta (k)
   = - {2c \over c^2 + k^2} \; .
   \label{Kdef}
\ee
The integral equation (\ref{intbetheeq}) with this kernel is known as the
Lieb-Liniger equation \cite{LL63} and it is a Fredholm type linear integral
equation\index{Integral equation!linear}.

Equation (\ref{intbetheeq}) determines the distribution of the quasi-momenta, which depends on the support of the kernel, that is, by the limits of integration $k_{min}$ and $k_{max}$.
In turn, this support is a reflection of the choice in the Bethe numbers in (\ref{betheeq}). For the ground state, the limits of integration are symmetric to minimize the momentum ($k_{max} = - k_{min} = q$).  We have
\be
p = {K \over L} = \int_{-q}^{q} k \; \rho (k) \de k = 0 \; , \qquad
e \equiv {E \over L} = \int_{-q}^{q} k^2 \; \rho(k) \de k \; ,
\label{BethePE}
\ee 
where we used the fact that, because (\ref{intbetheeq}) is even, $\rho(-k) = \rho(k)$. Notice that in taking the thermodynamic limit we lost the exact knowledge of the Bethe equation solution and thus of the eigenstates, but we are advancing toward the understanding of the macroscopic properties of the system, such as through (\ref{BethePE}).

The limits of integration in (\ref{intbetheeq}) is determined through the number of particles: $ N = L \int_{-q}^{q} \rho (k) \de k$. Inverting this relation, one has $q$ as a function of $N$. However, this equation depends on the density of quasi-momenta as the solution of (\ref{intbetheeq}), which, in turn, depends on $q$. It is thus convenient to perform the following rescaling \cite{LL63}
\be
   k \equiv q \: x \; , \qquad \qquad c \equiv q \: g \; ,
   \label{LLunivrescaling}
\ee
so that ($\tilde{\rho}(x) \equiv \rho (q x) $)
\bea
   \tilde{\rho} (x) & = & {1 \over 2 \pi} 
   + {1 \over \pi} \int_{-1}^{1} {g \over g^2 + (x-y)^2} \tilde{\rho} (y) \de y \; ,
   \label{reducedinteq} \\
   n & \equiv & {N \over L} = q \: \: G(g) \; , \qquad 
   \: G (g) \equiv \int_{-1}^1 \tilde{\rho} (x) \de x \; ,
   \label{BetheN} \\
   e & \equiv & {E \over L} =  q^3 \: F(g) \; , \qquad 
   F (g) \equiv \int_{-1}^1 x^2 \tilde{\rho} (x) \de x \; .
   \label{BetheE}
\eea
Notice that we can eliminate the $q$-dependence by considering the ratios ${e \over n^3} = {F(g) \over G^3(g)}$ and ${c \over n} = {g \over G (g)} \equiv \tilde{G} (g)$. The latter is also dimensionless, when the appropriate factors are reintroduced, see (\ref{gammaLL}), which means that systems with the same value of $\gamma = c/n$ have the same physics and, in particular, the same ground state energy:
\be
   e =  n^3 u (\gamma) \; , \qquad \qquad
   u(\gamma) \equiv {F\left(\tilde{G}^{-1} (\gamma)\right) \over G^3\left(\tilde{G}^{-1} (\gamma)\right)}  \; .
\ee 
Thus, with every solutions of (\ref{reducedinteq}) for a given $g$, one can calculate $G(g)$ and $F(g)$, and thus $e$ and $\gamma$. 

\begin{itemize}
	
	\item {\bf Strong repulsion:} We already discussed how in the Tonks–Girardeau regime $g \to \infty$ ($c \to \infty$) the system behaves essentially like free fermions. In fact, the kernel\index{Kernel} vanishes  ${\cal K} \to 0$ and the quasi-momenta distribution is given by
   \be
      \tilde{\rho} (x) = \left\{ \begin{array}{ll}
      	{1 \over 2 \pi} \; , & |x| \le 1 \;, \\
      	0 \; , & |x| > 1 \; .
      \end{array} \right. \; , \qquad \quad \Rightarrow \qquad \quad
      \rho (k) = \left\{ \begin{array}{ll}
      {1 \over 2 \pi} \; , & |k| \le q \;, \\
      0 \; , & |k| > q \; .
      \end{array} \right. \; .
      \label{FFDistsol}
    \ee
        Corrections to this result for large, but finite, $g$ can be calculated by expanding the kernel\index{Kernel} and the density in powers of $1/g$: at each order, the kernel is convoluted with the solution obtained at the previous order and thus the problem is reduced to integrating a rational function. In this way, one eventually reaches:
    \be
       u(\gamma) = {\pi^2 \over 3} \left[ 1 - {4 \over \gamma} + {12 \over \gamma^2} + \Ord \left( 1 \over \gamma^3 \right)\right] \; ,
    \ee
    which is convergent if $\gamma >2$ \cite{samaj}. One can also approach this asymptotic expansion following \cite{ristivojevic04}.

   \item {\bf Weak interaction:} The $g \to 0$ limit is tricky because, as ${\cal K} (x) \to - 2 \pi \delta (x)$, (\ref{reducedinteq}) gives $\tilde{\rho} (x) = {1 \over 2 \pi} + \tilde{\rho} (x)$, which indicates that $\tilde{\rho} (x)$ is, in fact, diverging. This problem was solved in \cite{hutson62} in a different context. In fact, eq. (\ref{reducedinteq}) is known in electrostatic theory as the Love equation for disk condensers \cite{sneddon,samaj} and the asymptotic analysis performed for this case translates for the Lieb-Liniger as
   \bea
      \tilde{\rho} (x) & = & {1 \over 2 \pi g} \sqrt{1 - x^2} +
      {1 \over 4 \pi^2} {1 \over \sqrt{1 - x^2}} 
      \left[ x \ln \left( {1 - x \over 1 +x}\right) + \ln {16 \pi \eu \over g} \right]
      + \ord (1) \; , 
      \label{weakLLrho} \\
      u(\gamma) & = & \gamma - {4 \over 3 \pi} \gamma^{3/2} +
      \left( {1 \over 6} - {1 \over \pi^2} \right) \gamma^2 + \Ord \left( \gamma^{5/2}\right) \; .
      \label{weakLLu}
   \eea
   Notice that the first two terms of (\ref{weakLLu}) are common in Bogolioubov theory \cite{stringari}, while the leading order in (\ref{weakLLrho}) is a {\it semi-circle law}, typical of Gaussian ensembles in Random Matrix Theory \cite{mehta2}. This can be understood through a ``duality'' transformation based on the identity
   \be
     \arctan x + \arctan {1 \over x} = \sgn(x) \; {\pi \over 2}\;,
   \ee
   which allows to rewrite (\ref{betheeq}) as
   \be
      k_j L =  2 \pi \left[ I_j - j + {1 \over 2} \left( N+1 \right) \right]
       + 2 \sum_{l \ne j} \arctan \left( {c \over k_j - k_l} \right) \; ,
       \qquad j=1,2,\ldots, N \; .
      \label{weakLLkj}
   \ee
   For the ground state, using (\ref{GSIs}) we see that the term in square parenthesis is identically zero. For $c \simeq 0$ we rescale the quasi-momenta as $k_j = \sqrt{2c \over L} \chi_j$ and expand (\ref{weakLLkj}) to leading order to get
   \be
      \chi_j = \sum_{l \ne j} {1 \over \chi_j - \chi_l} \; , \qquad \qquad 
      \qquad j=1,2,\ldots, N \; .
   \ee
   This equation gives the equilibrium positions of particles with Calogero interaction (\ref{CMModel}) \cite{sutherlandbook}, which are also the zeros of Hermite polynomials and their density is known to follow the Wigner semi-circle law.

   While (\ref{weakLLrho}) is sufficient to determine the ground state energy (\ref{weakLLu}), certain macroscopic properties, such as those we will study in Appendix \ref{app:CFT}, depend strongly on the behavior at the boundaries $x \to \pm 1$, where (\ref{weakLLrho}) turns out to be quite inaccurate.

\end{itemize}

\section{Some formalities on Integral Equations}
\label{sec:intform}

Linear integral equations\index{Integral equation!linear} like (\ref{intbetheeq}) are called inhomogeneous Fredholm equation of the second kind and are a subject of a vast mathematical literature that has developed advanced ways to deal with them \cite{musk,inthandbook}. Integral equations are in a sense the inverse of differential equations. In general, exact analytical solutions are not available for these equations, but there are often efficient approximation or perturbative schemes and one can resort quite effectively to numerical approaches. Moreover, often formal manipulations can shed light on the physical properties of the solution.

The linear integral operator $\hat{{\cal K}}_q$ is associated with a positive kernel\index{Kernel} ${\cal K} (k,k')$\index{Kernel} and the support $(-q,q)$ as:
\be
   \Big( \hat{{\cal K}}_q \rho \Big) (k) \equiv
   \int_{-q}^q {\cal K} (k,k') \rho(k') \de k'
\ee
and equation (\ref{intbetheeq}) can be written compactly as
\be
   \rho + {1 \over 2 \pi} \hat{{\cal K}}_q \: \rho =
   \left( \hat{\cal I} + {1 \over 2 \pi} \hat{{\cal K}}_q \right) \rho
   = {1 \over 2 \pi} \; ,
\ee
where $\hat{\cal I}$ has a Dirac delta as kernel\index{Kernel}.

One can then define the {\it resolvent} $\hat{{\cal L}}_q$\index{Resolvent} of $\hat{{\cal K}}_q$ as the operator that satisfies
\bea
   \left( \hat{\cal I} - \hat{{\cal L}}_q \right) \;
   \left( \hat{\cal I} + {1 \over 2 \pi} \hat{{\cal K}}_q \right) = \hat{\cal I} \; ,
   \label{resolvent} \\
   \left( \hat{\cal I} - \hat{{\cal L}}_q \right) \;
   \hat{{\cal K}}_q = 2 \pi \; \hat{{\cal L}}_q \; .
   \label{resolvent1}
\eea

One can also introduce the Green's function\index{Green's function} associated to a linear operator as the symmetric function (${\cal U}_q (k,k') = {\cal U}_q (k',k)$) satisfying:
\be
   {\cal U}_q (k,k')
   + {1 \over 2 \pi} \int_{-q}^q {\cal K} (k,k'') \: {\cal U}_q (k',k'') \de k'' =
   \delta (k-k') \quad \iff  \quad
   \left( \hat{\cal I} + {1 \over 2 \pi} \hat{{\cal K}}_q \right) {\cal U}_q = 
   \hat{\cal I} \; .
   \label{Greenf}
\ee

Knowing either the Green's function or the resolvent\index{Resolvent}, we can write the density of quasi-momenta as 
\be
    \rho (k) = {1 \over 2 \pi} \int_{-q}^q {\cal U}_q (k,k') \de k' 
    = {1 \over 2 \pi} - {1 \over 2 \pi} \int_{-q}^q {\cal L}_q (k,k') \; \de k' \; .
   \label{rhoU}
\ee
Even when the Green's function/resolvent cannot be calculated analytically, often formal manipulations in terms of these operators provide useful physical results, as we will see the next sections and chapters.

\section{Elementary excitations}
\label{sec:LLexcitations}

One of the fundamental advantages of the Bethe Ansatz solution is that it provides an interpretation of a many-body wavefunction in terms of individual excitations. For an integrable system, these excitations can be regarded as stable {\it quasi-particles}\index{Quasi-particle}, in terms of which we can interpret the interacting many-body state. In one dimension, the Fermi liquid description breaks, which means that even the low energy excitations are emergent degrees of freedom, possibly very different from the bare constituents of the system. Over the years we have understood how to describe them in terms of Conformal Field Theories, but historically this was accomplished also with the insights provided by integrable models.

We consider two kinds of elementary excitations over the ground state: we can add a new particle with momentum $|k_p| > q$ ({\bf Type I} excitation)\index{Excitation!Type I}; or we can remove a particle and create a hole with $|k_h| \le q$ ({\bf Type II} excitation)\index{Excitation!Type II}. A third option is to excite one of the momenta inside the Fermi sea $|k| \le q$ and move it above the $q$-threshold, but we can realize this as a combination of the first two. In general, low energy excitations can be all be constructed using type I \& II excitations, on which we now concentrate.

Let us start with the ground state, given by (\ref{GSIs}):
\be
   \{ I_j \} = \left\{ -{N-1 \over 2}, -{N -3 \over 2} , \ldots, {N -1 \over 2}\right\} \; ,
   \label{Ic}
\ee
and consider the excitation that adds one particle (say with positive momentum), taking the number of particles from $N$ to $N+1$. The new state is realized starting from the ground state of a system with $N+1$ particles\footnote{Notice that adding a particle turns integer Bethe numbers into half-integers and vice-versa.} by boosting the quantum number at the Fermi edge to a higher value:
\be
   \{ I'_j \} = \left\{ -{N \over 2}, -{N\over 2} + 1, \ldots, {N \over 2} -1, {N \over 2} +  m \right\} \; ,
   \label{I'c}
\ee
with $m > 0$. Such excitation has total momentum
\be
   K = {2 \pi \over L} \; m \; .
   \label{totmom}
\ee 
This momentum is realized through a complete rearrangement of the particle quasi-momenta in the system. If the original ground state configuration has quasi-momenta $\{k_1, k_2, \ldots, k_N \}$, this excited state is characterized by $\{k'_1, k'_2, \ldots, k'_N, k_p \}$, solution of a system like (\ref{betheeq}) but with a set of integers given by (\ref{I'c}), see Fig. \ref{fig:Type12}.

\begin{figure}[t]
	\noindent\begin{minipage}[t]{8.5cm}
		\includegraphics[scale=.34]{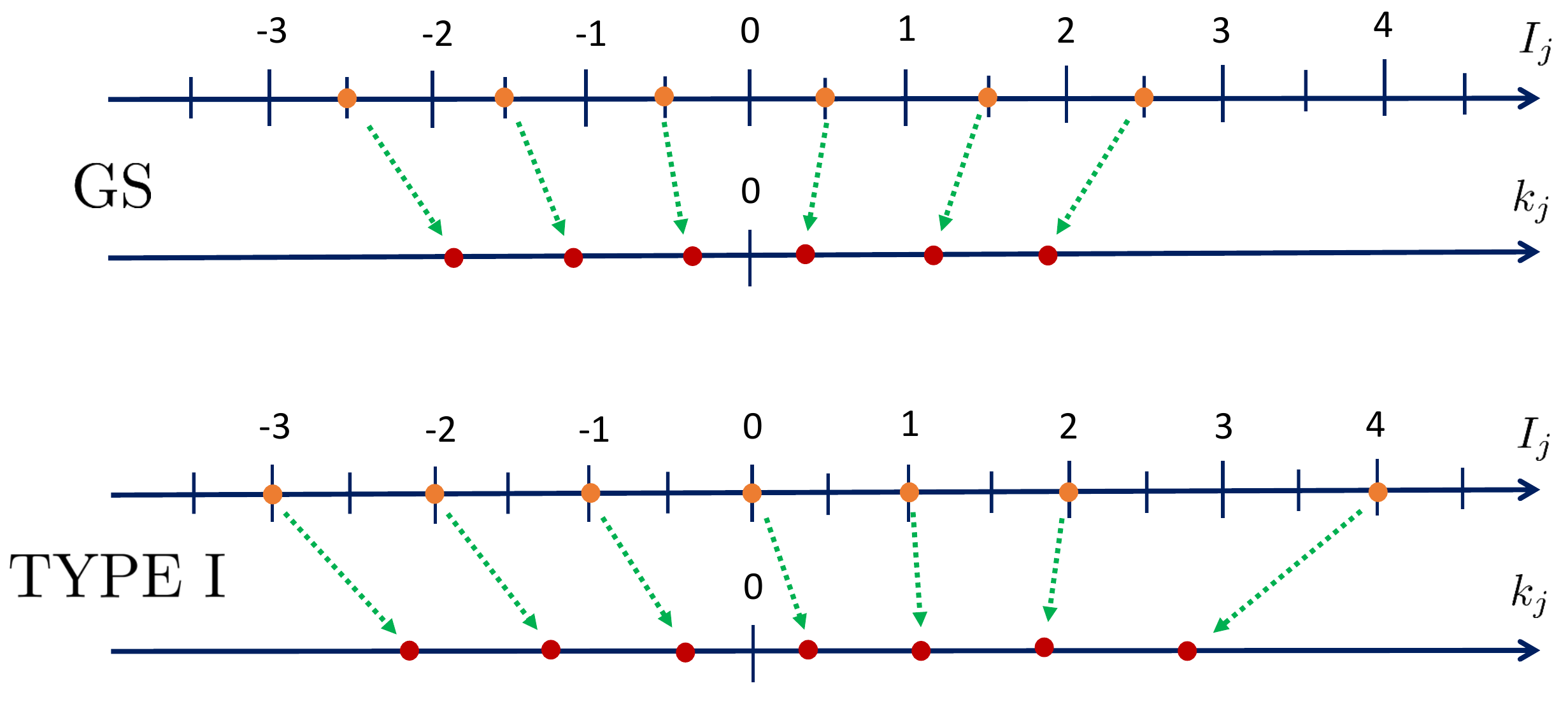}
	\end{minipage}
	\hfill
	\begin{minipage}[t]{8.5cm}
		\includegraphics[scale=.34]{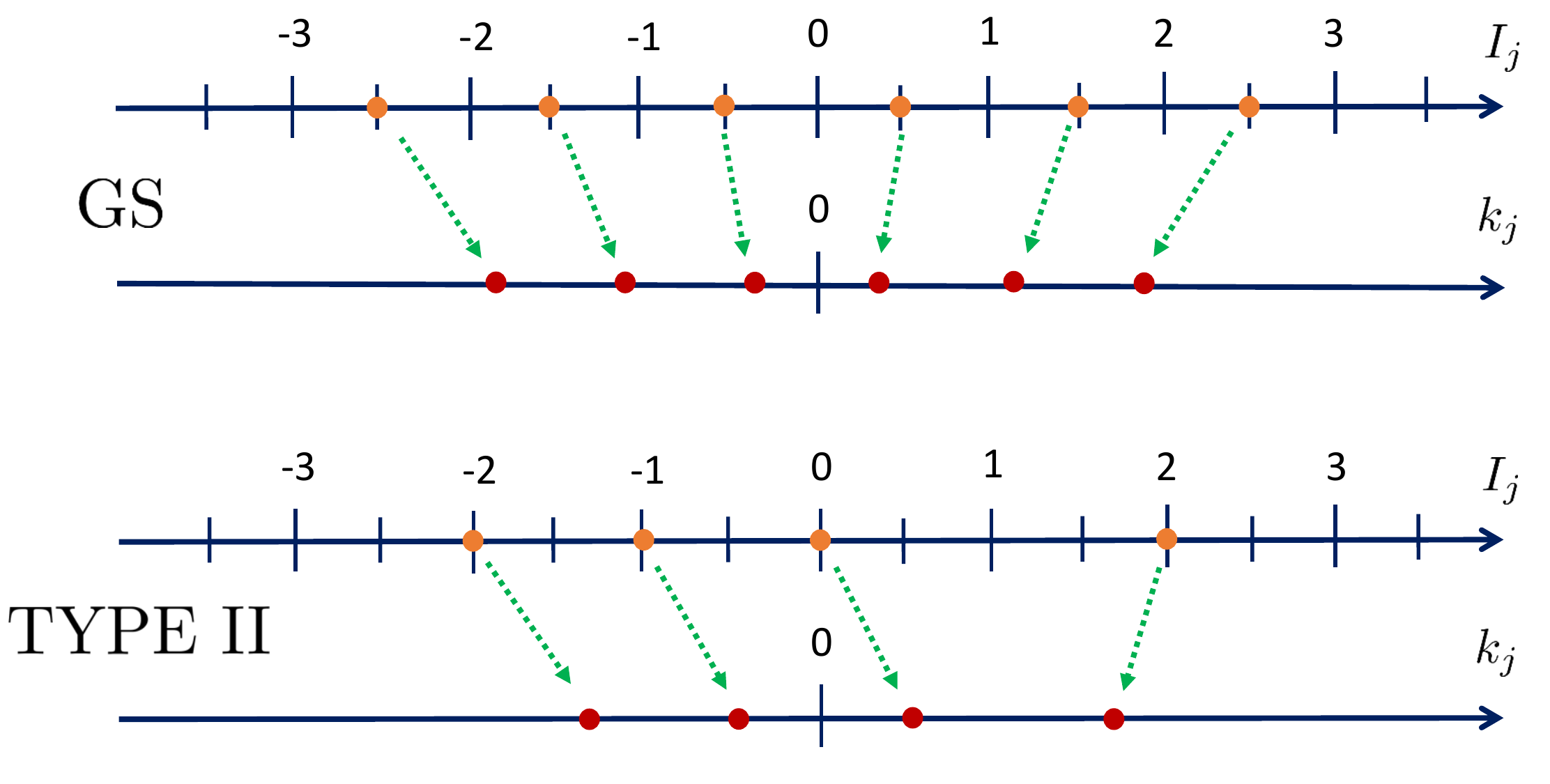}
	\end{minipage}
	\caption{Cartoon of Type I \& II excitations. On top, the ground state configuration in terms of the Bethe numbers and of the corresponding quasi-momenta. On the bottom, the quantum number and quasi-momenta configurations for a Type I (left) and a Type II (right) excitations.}
	\label{fig:Type12}
\end{figure}

We see that, while the momentum of the new particle is just $k_p$, the momentum gained by the whole system is different and given by (\ref{totmom}). The former is referred to as the {\it bare momentum} of the particle, in contrast with the latter, the observed or {\it dressed momentum}\index{Dressed!momentum}, due to the rearrangement of the whole system in reaction to the insertion of a new particle. This is a sign of the intrinsic non-local nature of a one-dimensional system excitation and provides a concrete example of the concept that one-dimensional systems are intrinsically strongly interacting, regardless of the actual strength of the coupling constant.

To calculate the reaction of the system to the addition of this extra particle, let us calculate the quantity $\Delta k_j = k'_j - k_j$ by subtracting the
Bethe equations for the two configurations:
\be
   \Delta k_j \; L = \pi +
   \sum_{l=1}^N \left[ \theta (k'_j - k'_l) - \theta (k_j - k_l) \right]
   + \theta (k'_j - k_p) \; .
\ee
The $\pi$ contribution in the right-hand-side appears as a consequence of the quantum numbers shifting from integers to half-integers (or viceversa.) Since $\Delta k_j$ if of order $\Ord(L^{-1})$ (or equivalently $\Ord(N^{-1})$), we expand the right-hand side to the same order and, remembering the definition of the kernel\index{Kernel} as the derivative of the scattering phase (\ref{Kdef})\index{Scattering phase}, we obtain
\be
   \Delta k_j \; L = \pi + \sum_{l=1}^N K(k_j - k_l)
   \left( \Delta k_j - \Delta k_l \right)
   + \theta (k_j - k_p) \; .
\ee 
Collecting the terms in the following way:
\be
   \Delta k_j \left[ 1 - {1 \over L} \sum_{l=1}^N {\cal K} (k_j - k_l) \right] = {1 \over L} \left[ \pi + \theta (k_j - k_p) \right]
   - {1 \over L} \sum_{l=1}^N {\cal K} (k_j - k_l) \Delta k_l \; ,
\ee 
we can go to the thermodynamic limit and with the help of (\ref{intbetheeq}) write
\be
   2 \pi \; \Delta k \; \rho (k) =
   {1 \over L} \left[ \pi + \theta (k -k_p) \right]
   - \int_{-q}^{q} {\cal K} (k-k') \; \Delta k' \; \rho (k') \; \de k' \; .
\ee

We introduce the {\it back-flow} or {\it shift function}\index{Back-flow}
\be
   J(k|k_p) \equiv L \; \Delta k \; \rho (k) =
   \lim_{k \to k_j} \lim_{N,L \to \infty} {k_j - k'_j \over k_{j+1} - k_j } \; ,
   \label{Jdef}
\ee 
which satisfies the integral equation\index{Integral equation!linear} above
\be
   J (k|k_p)
   + {1 \over 2 \pi} \int_{-q}^{q} {\cal K} (k-k') \; J (k'|k_p) \; \de k'
   = {1 \over 2 \pi} \tilde{\theta} (k -k_p) \; ,
   \label{jkp}
\ee 
where we remember the definition of the scattering phase $\tilde{\theta} (k)$ in (\ref{tildephaseshift})\index{Scattering phase}. Using the Green's function introduced in (\ref{Greenf}), this equation can be solved as
\be
   J(k|k_p) = {1 \over 2 \pi} \int_{-q}^q {\cal U}_q (k,k')
   \tilde{\theta}(k'-k_p) \; \de k' \; .
   \label{jkkp1}
\ee

The back-flow helps in calculating the changes in the macroscopic quantities under the addition of an excitation with momentum $|k_p| \ge q$, namely
\bea
  \Delta K (k_p) & = & {2 \pi \over L} \; m 
  = k_p + \sum_{j=1}^N \Delta k_j
  = k_p + \int_{-q}^q J(k|k_p) \de k
  \label{dressedpp} \\
  & = & k_p + {1 \over 2\pi} \int_{-q}^q \de k \int_{-q}^q \de k' {\cal U}_q (k,k')
  \tilde{\theta} (k' - k_p)
  = k_p + \int_{-q}^q \rho(k) \tilde{\theta} (k - k_p) \; \de k \; ,
  \label{deltapkp}
\eea
where we used (\ref{Jdef}) in the first line and (\ref{jkkp1},\ref{rhoU}) in the second. Similarly, for the energy
\be
  \Delta e (k_p) = k_p^2 + \sum_{j=1}^N \left[ k^{\prime 2}_j - k_j^2 \right]
  = k_p^2 + \sum_{j=1}^N \left[ 2 k_j \Delta k_j + (\Delta k_j)^2 \right]
  \simeq k^2_p +  \int_{-q}^q 2 k \; J(k|k_p) \de k \; ,
  \label{dressedep}
\ee
remembering that the term $(\Delta k_j)^2$ is suppressed like $1 /N$ in comparison with the leading one.

These equations really show that this excitation has a collective nature and cannot be assigned simply at the single bosons we added. This is the difference between the bare and dressed quantities. We have added a particle with bare momentum $k_p$ and bare energy $k^2_p$, but the whole system rearranges itself and acquires the {\it dressed momentum}\index{Dressed!momentum} (\ref{dressedpp}) and the {\it dressed energy}\index{Dressed!energy} (\ref{dressedep}).


As we mentioned, the creation of a particle excitation is referred to as a {\bf Type I} excitation\index{Excitation!Type I}. Comparing its dispersion relation to that of the classical Non-Linear Schr\"odinger (Gross-Pitaevskii)\index{Gross-Pitaevskii equation}\index{Non-Linear Schr\"odinger equation} equation in the $\gamma \to 0$ limit, Type I excitations are identified as Bogoliubov quasi-particles, i.e. purely quadratic excitations \cite{solitons1980}.

Let us now consider the other kind of excitation, a hole, obtained by removing a particle from the Fermi sea. The quantum number configuration is obtained starting from the ground state of a system with $N-1$ particles and by displacing one of the quantum numbers within the Fermi sea to the nearest empty space, just over the Fermi point:
\be
   \{ I''_j \} = \left\{ -{N \over 2} + 1 , -{N\over 2} + 2, \ldots,
   {N \over 2} - m - 1, {N \over 2} -m +1 , \ldots, {N \over 2} \right\} \; ,
   \label{I''c}
\ee
giving this state a (positive) momentum $K = {2 \pi \over L} \; m$.
As before we can consider the reaction of the system as the quasi-momenta change to accommodate for the absence of a particle with  momentum $|k_h|<q$, corresponding to the missing quantum number, see figure \ref{fig:Type12}. Proceeding in the same way, we can introduce the back-flow\index{Back-flow} for this hole excitation, which in this case satisfies the following integral equation:\index{Integral equation!linear}
\be
   J (k|k_h) + {1 \over 2 \pi} \int_{-q}^{q} {\cal K} (k-k') \; J (k'|k_h) \; \de k'
   = - {1 \over 2 \pi} \tilde{\theta} (k -k_h) \; .
   \label{jkh}
\ee
The change in energy and momentum for the whole system are
\bea
  \Delta K (k_h) & = & -k_h - \int_{-q}^q J(k|k_h) \de k 
  = - k_k - \int_{-q}^q \rho(k) \tilde{\theta} (k - k_h) \; \de k \; \; , \\
  \Delta e (k_h) & = & -k_h^2 + \int_{-q}^q 2 k \; J(k|k_h) \de k \; .
\eea

One can show \cite{solitons1980} that in the $\gamma \to 0$ limit these {\bf Type II} excitations\index{Excitation!Type II} are not simple sound waves, but have non-linear corrections to a simple relativistic dispersion relation. In fact, in the weakly interacting limit ($\gamma \ll 1$) it has been argued that they correspond to the dark solitons of the Gross-Pitaevskii equation\index{Gross-Pitaevskii equation}, as the dispersion relations of the latter matches that of Type II excitations \cite{solitons1980}. 

Due to the linear nature of the integral equations defining the back-flows\index{Back-flow} for Type I and II excitations\index{Excitation!Type I}\index{Excitation!Type II} (\ref{jkp}, \ref{jkh}), all low energy states can be constructed from these two fundamental ones. In particular, taking a particle from the Fermi sea to an excited level can be seen as the combination of a Type II (hole) and Type I (particle). After each operation of this type the whole system goes through a rearrangement, that dresses the particles and the final configuration is given by the back-flow defined by an integral equation like (\ref{jkp}, \ref{jkh}), but with the source term given the sums of each contribution:\index{Integral equation!linear}
\bea
   \left[ \left( \hat{\cal I} + {1 \over 2 \pi} \hat{{\cal K}}_q \right) 
   J \right] \left( k|k_{p1} \ldots k_{pM^+}; k_{h1} \ldots k_{hM^-} \right) & = &
   {1 \over 2 \pi} \sum_{j=1}^{M^+} \tilde{\theta} (k - k_{pj})
   - {1 \over 2 \pi} \sum_{j=1}^{M^-} \tilde{\theta} (k - k_{hj})
   \nonumber \\
   \Rightarrow \qquad J \left( k|k_{p1} \ldots k_{pM^+}; k_{h1} \ldots k_{hM^-} \right)
   & = & \sum_{j=1}^{M^+} J(k|k_{pj}) + \sum_{j=1}^{M^-} J(k|k_{hj}) \; .
\eea

So far, we described the excitations in terms of density of quasi-momenta. Let us now introduce a function $\varepsilon (k)$ as the solution of the linear integral equation\index{Integral equation!linear}
\be
   \varepsilon (k)
   + {1 \over 2 \pi} \int_{-q}^q {\cal K} (k,k') \; \varepsilon (k') \; \de k'
   = k^2 - h \equiv \epsilon_0 (k) \; ,
   \label{epsilonen}
\ee
with the boundary condition
\be
   \varepsilon (q) = \varepsilon (-q) = 0 \; .
   \label{epsilonBC}
\ee
This is the same integral equation satisfied by the momentum density $\rho (k)$ and back-flow\index{Back-flow}, but with the bare energy as source. We added a chemical potential $h$, for later convenience. So far, we implicitly worked with a micro-canonical ensemble. If we relax the fixed number of particle condition and allow a grand-canonical approach, we need $h$ in (\ref{epsilonen}), which is the Lagrange multiplier appearing for the particle density. The relation between the chemical potential and the number of particles is given by the boundary condition (\ref{epsilonBC}), that implicitly relates $h$ to the support of the integral equation $q$.

Eq. (\ref{epsilonen}) will be derived as the zero-temperature limit of the Yang-Yang equation (\ref{YangYangEq})\index{Yang-Yang equation}.
Physically, the function $\varepsilon (k)$ defined by (\ref{epsilonen}) is the dressed energy\index{Dressed!energy} of a particle with quasi-momentum $k$ . Condition (\ref{epsilonBC}) means that the theory is gapless. The solution of (\ref{epsilonen}) also satisfies the following properties:
\bea
    && \varepsilon' (k) >0 \qquad {\rm for} \qquad k>0 \label{varepsilon1} \\
    && \varepsilon (k) = \varepsilon (-k) \; , \label{varepsilon2} \\
    && \varepsilon (k) <0 \qquad {\rm for} \qquad |k|<q \; , \label{varepsilon3} \\
    && \varepsilon (k) >0 \qquad {\rm for} \qquad |k|>q \; , \label{varepsilon4}
\eea
which reflect the fact that all excitations must bring a positive energy contribution over the ground state.

To support our interpretation of the function $\varepsilon (k)$, let us
calculate the change in energy due, for instance, to the insertion of a new
particle and the removal of another (hole). From (\ref{BethePE}):
\be
   \Delta e (k_p, k_h) 
   = \epsilon_0 (k_p) - \epsilon_0 (k_h)
   + \int_{-q}^q \epsilon'_0 (k') \; J(k'|k_p, k_h) \; \de k' \; .
   \label{deltae1}
\ee
We wish to prove that
\be
   \Delta e (k_p, k_h) = \varepsilon (k_p) - \varepsilon (k_h) \; .
\ee

We note that (\ref{jkp},\ref{jkh}) can be written as
\be
  \left[ \left( \hat{\cal I} + {1 \over 2 \pi} \hat{{\cal K}}_q \right) J \right] 
  \left( k|k_p,k_h \right) =
  {1 \over 2 \pi} \int_{k_h}^{k_p} {\de \over \de k'} \tilde{\theta} (k - k') \; \de k' =
  - {1 \over 2 \pi} \int_{k_h}^{k_p} {\cal K} (k-k') \; \de k' \; .
\ee
By acting on this with the operator $\hat{\cal I} - \hat{{\cal L}}_q$ and using the defining property of the resolvent\index{Resolvent} in (\ref{resolvent}):
\be
 J \left( k|k_p,k_h \right) = - \int_{k_h}^{k_p} {\cal L}_q (k,k') \; \de k' \; .
\label{JLrel}
\ee

We also note that the derivative by $k$ of (\ref{epsilonen}) satisfies the integral equation\index{Integral equation!linear}
\be
   \left[ \left( \hat{\cal I} + {1 \over 2 \pi} \hat{{\cal K}}_q \right) \varepsilon' \right] (k) =   \epsilon'_0 (k) \; ,
\ee
obtained by integrating by parts. Acting on this with $\hat{\cal I} - \hat{{\cal L}}_q$ we get
\be
   \varepsilon' (k) - \epsilon'_0 (k) = -
   \int_{-q}^q {\cal {\cal L}}_q (k,k') \; \epsilon'_0 (k') \; \de k' \; .
   \label{varepsilon'}
\ee

By combining (\ref{JLrel}) and (\ref{varepsilon'}) with (\ref{deltae1}) we have
\bea
   \Delta e (k_p, k_h) & = &
   \epsilon_0 (k_p) - \epsilon_0 (k_h)
   - \int_{k_h}^{k_p} \de k \int_{-q}^q {\cal L}_q(k,k') \; \epsilon'_0(k') \; \de k'
   \nonumber \\
   & = & \epsilon_0 (k_p) - \epsilon_0 (k_h)
   +\int_{k_h}^{k_p} \left[ \varepsilon' (k) - \epsilon'_0(k) \right] \de k
   = \varepsilon (k_p) - \varepsilon (k_h) \; ,
\eea as we set out to prove.

In conclusion, (\ref{epsilonen}) defines the one-particle dressed energy, while its momentum is (\ref{deltapkp}).
As we add particles with $|k_p| \ge q$ and holes with $|k_h|<q$, the total
change in energy and momentum of the system is
\bea
   \Delta e & = & \sum_{\rm particles} \varepsilon (k_p) - \sum_{\rm holes} \varepsilon
   (k_h) \; , 
   \label{DeltaELL} \\
   \Delta K & = & \sum_{\rm particles} \Delta K (k_p) - \sum_{\rm holes} \Delta K (k_h) \; .
   \label{DeltaKLL}
\eea 
Out of these equations one can determine the dispersion relation of the various excitations, whose leading part is always relativistic, that is, $\varepsilon \simeq v_S \Delta K $ (solitons have higher momentum corrections).  In appendix \ref{sec:LLbosonization} we derive the sound velocity $v_S$ and describe these excitations as a Luttinger liquid.

In this section we have seen that the same kind of linear integral equations\index{Integral equation!linear}, with the same kernel\index{Kernel} but with different source terms (and appropriate boundary conditions) generates the various physical quantities that characterize each state, from their bare to the dressed form. In doing so, we lost touch with the original Bethe Ansatz construction and with the form of the eigenstates, but we learned that macroscopic observables can be calculated by dressing their bare (free) expressions with the same integral equation, which encodes the effects of the interaction.

\section{Thermodynamics of the model: the Yang-Yang equation}
\label{sec:LLthermo}

We now want to describe the system at finite temperatures, which requires considering general excited states. As each eigenstate of the system is characterized by a set of Bethe numbers $\{ I_j \}$, we can write the finite-temperature partition function as
\be
   {\cal Z} = {1 \over N!} \sum_{\{I_j\}} \exp \left[ - {E_N \over T} \right] 
   = \sum_{I_1 < I_2 < \ldots < I_N} \exp \left[ - {E_N \over T} \right] 
   = \sum_{n_1 = 1}^\infty \sum_{n_2 =1}^\infty \cdots \sum_{n_{N-1}=1}^\infty \eu^{-E_N /T}  \; ,
\label{Zsum}
\ee
where $E_N = \sum_{j=1}^N k_j^2$ and the quasi-momenta $k_j$ are the solutions of the Bethe equations with the given set of quantum numbers $\{ I_j \}$. In the last passage we introduced $n_j = I_{j+1} - I_j$ for later convenience.
We know that in general it is not easy to calculate the energy of the state directly from its quantum numbers. It is thus desirable to convert the sums into a functional integration over rapidity densities. Of course, in doing so we will lose some information of the microscopics of the state, as it is customary in any thermodynamic approach.

A central role in this ``change of variables'' is played by the counting function $y(k)$ that we introduced in (\ref{countingfdef}), since it connects the density of quasi-momenta with the corresponding quantum numbers. Let us take the point of view that we know the quasi-momenta $\{ k_j \}$ that are solution of the Bethe equations (\ref{betheeq}) for a given set of Bethe numbers $\{ I_j \}$.
We then define the counting function as
\be
   y(k) \equiv k - {1 \over L} \sum_{j=1}^N \theta \left( k - k_j \right)
   \label{yeq}
\ee
for generic $k$.  By construction, $y(k_j)= {2 \pi \over L} \: I_j$. We also look for the other values of $k$ for which the counting function takes a ``quantized'' value ${2 \pi \over L}n$ for some $n$, (half-)integer as for the $I_j$. We call these $k_n^v$ {\it vacancies}:\index{Vacancy}
\be
   y \left( k_n^v \right) = {2 \pi \over L} \: n \; ,
\ee
The vacancies are sort of ``placeholders'' for the quantum numbers: each quantum number $n$ is mapped by (\ref{yeq}) into a $k_n^v$. The subset $\{ k_j \}$ of vacancies that correspond to the Bethe numbers of the the state are callled {\it particles}. The remaining solutions $\{ k_j^h \} = \{ k_n^v \} \setminus \{ k_j\}$ are the {\it holes} \index{Hole} and are the images of the missing quantum numbers. If we consider a state generated from the ground states by removing some quantum numbers from inside the Fermi sphere, those $k$ are the holes.

We define the densities of quasi-momenta for the particles, holes and vacancies as we did in (\ref{rhokdef}):
\be
   \rho (k_j) = \lim_{N,L \to \infty} {1 \over L \left( k_{j+1} - k_j \right) }
   \; , \qquad
   \rho_v (k^v_j) = \lim_{N,L \to \infty} {1 \over L \left( k^v_{j+1} - k^v_j
   \right) } \; , \qquad
   \rho_h (k^h_j) = \lim_{N,L \to \infty} {1 \over L \left( k^h_{j+1} - k^h_j
   \right) } \; .
   \nonumber
\ee
Since $L \rho (k) \de k$ and $L \rho_h (k) \de k$ are the number of particles and holes in an interval $\de k$, we have $L \rho_v (k) \de k = L \left[ \rho (k) + \rho_h (k) \right] \de k$. As for the zero-temperature case (\ref{yprimekj}) we have
\be
   y' \left( k^v_j \right) = \lim_{N,L \to \infty} 
   {y \left(k^v_j \right) - y \left( k^v_{j-1} \right) \over
   k^v_j - k^v_{j-1} }
   = \lim_{N,L \to \infty} {2 \pi \over L \left( k^v_j - k^v_{j-1} \right)}
   = 2 \pi \: \rho_v \left( k^v_j \right) \; ,
\ee
and thus
\be
   y(k) = 2 \pi \int^k \left[ \rho(k') + \rho_h(k') \right] \de k' \; .
   \label{ykrhov}
\ee

This equation can be used in conjunction with the thermodynamic limit of (\ref{yeq})
\be
y(k) = k - \int_{- \infty}^\infty \theta \left(k - k' \right) \rho(k') \de k' \; ,
\ee
to equate both their RHS and to take their derivative with respect to $k$ like we did in (\ref{yIntEq}). One gets
\be
\rho(k) + \rho_h(k) = {1 \over 2 \pi}
- {1 \over 2 \pi} \int_{- \infty}^\infty {\cal K} (k, k') \; \rho (k') \; \de k' \; .
\label{phrel}
\ee
Note that, compared to the integral equations we dealt with so far, the introduction of the density of holes pushes the support of the integral to extend over the whole real axis. Most of all, compared to the zero-temperature case, this integral equation in not closed, as $\rho(k)$ depends on the, yet undetermined, density of holes $\rho_h (k)$. We will use it soon as a relation between $\rho_h(k)$ and $\rho (k)$.

Now, we use (\ref{ykrhov}) to relate the variables $n_j$ in (\ref{Zsum}) to the densities and thus estimate the integration measure to be used in(\ref{Zsum}) when converting sums into integrals:
\be
   n_j = I_{j+1} - I_{j} 
	= {L \over 2 \pi} \Big[ y \left( k_{j+1} \right) - y \left( k_j \right) \Big]
	= {L \over 2 \pi} \int_{k_j}^{k_{j+1}} \rho_v (k') \de k'
	= {L \over 2 \pi} \int_{k_j}^{k_j + {1 \over L \rho (k_j)}} \rho_v (k') \de k' 
	\simeq {1 \over 2 \pi} \: {\rho_v (k_j) \over \rho (k_j)} \; .
	\nonumber
\ee
Before we completely switch to densities in (\ref{Zsum}), we need to estimate the number of microstates which are not distinguishable in our macroscopic description. This entropy can be calculated as one does for free fermions, by counting in how many ways one can distribute a set of consecutive quantum numbers in an interval between particles and holes. We use the fact that the counting function maps the quantum numbers into the quasi-momenta to write the differential entropy as the ways to distribute $L \rho (k) \de k$ particles and $L \rho_h (k) \de k$ holes in an interval $\de k$:
\bea
   \de {\cal S} & = &
   \ln { \big[ L \big( \rho(k) + \rho_h (k) \big) \de k \big] ! \over
	\big[ L \rho(k) \de k \big]! \: \big[ L \rho_h (k) \de k \big]! }
   \label{LLentropy} \\
  & \approx &  L \: \Big[
  \big( \rho(k) + \rho_h (k) \big) \ln \big( \rho(k) + \rho_h (k) \big)
  - \rho(k) \ln \rho (k) - \rho_h (k) \ln \rho_h (k) \Big] \: \de k \; ,
  \nonumber
\eea
where in the last line we used Stirling's approximation formula ($\ln n! \approx n \ln n - n$).

Finally, we can write (\ref{Zsum}) in terms of the macroscopical variables $\rho(k)$ and $\rho_h (k)$ as
\be
   {\cal Z} = {\rm const} \int {\cal D}
   \left( {\rho_v (k) \over \rho (k)} \right) \:
   \delta \left( \int \rho(k) \de k - n \right) \:
   \eu^{{\cal S} - L e/T} \; ,
   \label{part}
\ee 
where ${\cal S}$ is the entropy from (\ref{LLentropy}) and $e = E_N/L = \int k^2 \rho (k) \de k$ is the energy of the state. We also introduced a delta-function to enforce the particle number conservation in a macro-canonical ensemble. By using the representation
\be
   \delta (x) = {1 \over 2 \pi \ii} \int_{-\ii \infty}^{\ii \infty} \eu^{h x}
   \de h \; ,
\ee 
we can write (\ref{part}) as
\be
   {\cal Z} = {\rm const} \int \de h \int {\cal D}
   \left( {\rho(k) + \rho_h (k) \over \rho (k)} \right) \:
   \eu^{W[\rho,\rho_h;h]} \; ,
\ee 
where
\bea
    {\cal W} [\rho,\rho_h;h] & \equiv & -{L \over T} \int \de k \Big\{
   k^2 \rho (k) + h \big[ \rho (k) - n \big]
   \\
   && \qquad \qquad - T \Big[
   \left( \rho(k) + \rho_h (k) \right) \ln \left( \rho(k) + \rho_h (k) \right)
   - \rho(k) \ln \rho (k) - \rho_h (k) \ln \rho_h (k) \Big] \Big\} \; .
   \nonumber
\eea
The Lagrange multiplier $h$ has the physical interpretation of a chemical
potential.

As $L \to \infty$, we can employ a saddle-point approximation to find the configuration that extremizes the action and gives the most relevant contribution to the partition function:
\be
   \delta {\cal W} [\rho,\rho_h;h] = - {L \over T} \int \de k \left\{
   \left[ k^2 - h
   - T \ln \left( {\rho(k) + \rho_h (k) \over \rho (k)}
   \right) \right] \delta \rho (k) 
   - T \ln \left( {\rho(k) + \rho_h (k) \over \rho_h (k)}
   \right) \delta \rho_h (k) \right\} = 0\; .
   \label{part1}
\ee

Using (\ref{phrel}) as
\be
   \delta \rho_h(k) =
   - \delta \rho (k) + {1 \over 2 \pi}
   - \int_{- \infty}^\infty {\cal K}(k, k') \; \delta \rho (k') \; \de k' \; ,
\ee
we can eliminate $\rho_h$ from (\ref{part1}) to get
\be
   \int \de k \left\{ k^2 - h
   - T \ln \left( {\rho_h (k) \over \rho (k)} \right)
   + {T \over 2 \pi} \int {\cal K} (k,k') \ln \left( 1 + {\rho (k') \over \rho_h (k')}
   \right) \de k' \right\} \delta \rho (k) = 0 \; .
\ee
For this condition to hold for any $\delta \rho$, we demand
\be
   \varepsilon (k) = k^2 - h
   + {T \over 2 \pi} \int_{-\infty}^\infty {\cal K}(k,k') \ln \left( 1 + \eu^{-\varepsilon (k')/T} \right) \de k'\; ,
   \label{YangYangEq}
\ee
where we defined
\be
   \varepsilon (k) \equiv T \ln \left( {\rho_h (k) \over \rho (k)} \right)
   \qquad \to \qquad
   {\rho_h (k) \over \rho (k)} = \eu^{\varepsilon(k)/T} \; .
\ee
Equation (\ref{YangYangEq}) is a non-linear integral equation\index{Integral equation!non-linear} whose solution gives the dressed energy\index{Dressed!energy} per particle excitation, using which the thermodynamic quantities are at hand. Equation (\ref{YangYangEq}) is known as the {\it Yang-Yang equation}\index{Yang-Yang equation}. The interpretation of the function $\varepsilon (k)$ is supported by noting that the number of excitations over the number of available states is
\be
   {\rho (k) \over \rho (k) + \rho_h (k) } =
   {1 \over 1 + \eu^{\varepsilon(k) / T} } \; ,
\ee 
where we recognize the RHS as the usual Fermi weight distribution.

The entropy (\ref{LLentropy}) evaluated at this saddle point gives
\bea
   {\cal S} & = & L \int  \Big[
   \big( \rho(k) + \rho_h (k) \big) \ln \big( 1 + \eu^{-\varepsilon (k) /T} \big)
   + {1 \over T} \: \rho(k) \varepsilon (k) \Big] \: \de k 
   \nonumber \\
   & = & L \int \left[ {1 \over 2 \pi} \ln \left( 1 + \eu^{-\varepsilon (k)/T} \right)
   + {1 \over T} \left( k^2 -h \right) \rho(k) \right] \: \de k \; ,
\eea
where we eliminated $\rho(k) + \rho_h (k)$ using (\ref{phrel}) and simplified the resulting expression using the Yang-Yang equation\index{Yang-Yang equation} (\ref{YangYangEq}). Most of all, we get the leading contribution to the Helmholtz free energy by evaluating the partition function (\ref{part}) at the saddle:
\be
   {\cal F} = - T \ln {\cal Z} = L \int k^2 \rho (k) \de k - T \, {\cal S} =
   N h - {T L \over 2 \pi} \int \de k \ln \left( 1 + \eu^{-\varepsilon (k)/T}
   \right) \; .
   \label{partitionfunction}
\ee

We see now why the Bethe Ansatz construction is so powerful in addressing the thermodynamics of an integrable model: equation (\ref{partitionfunction}) looks like the partition function of a system of non-interacting particles with single-particle spectrum $\varepsilon (k)$, similar to (\ref{XYFreeEn}). That is, once the Yang-Yang equation\index{Yang-Yang equation} (\ref{YangYangEq}) has been solved (maybe numerically, or by a series expansion...) and the dressed energies\index{Dressed!energy} have been calculated, the strongly interacting problem of the integrable theory is reduced to the partition function of a free theory with a non-trivial spectrum.

It is worth stressing that, while in free systems the decomposition of the many-body wavefunction into excitations is real and, in principle, measurable, in an interacting system such as those solved by Bethe Ansatz this is not true. Bethe Ansatz allows us to characterize each state in terms of its quasi-particle content\index{Quasi-particle}, but this should be intended only as a bookkeeping trick, since removing or changing one of the quasi-particles modifies all the state constituents, through the Yang-Yang equation\index{Yang-Yang equation}. In the same way, the quasi-momenta are a good way to characterize the excitations, but are not observable: only the dressed quantities are.

To conclude, from the knowledge of the partition function (\ref{partitionfunction}), the whole thermodynamics of the model can be calculated. The pressure is
\be
   P  = - \left( {\partial {\cal F} \over \partial L } \right)_T =
   {T \over 2 \pi} \int \de k \ln \left( 1 + \eu^{-\varepsilon (k)/T} \right) \; .
   \label{LLPressure}
\ee
(it satisfies $\de P = {\cal S}/L \de T + n \de h$) and
\be
   n = - { \partial \over \partial h} \left( {\cal F} - Nh \right) \; , \qquad
   {\cal S} = - {\partial {\cal F} \over \partial T} \; , \qquad
   e = {\cal F} + T {\cal S} \; , \ldots
\ee

The density of quasi momenta can be determined from the energy per particle
using (\ref{phrel}):
\be
   2 \pi \rho(k) \left[ 1 + \eu^{\varepsilon(k)/T} \right] =
   1 - \int {\cal K} (k,k') \rho(k') \de k' \; ,
\ee 
while particle density is always given by $n = \int \rho(k) \de k$.

The Yang-Yang equation\index{Yang-Yang equation} has been the first example of what has become known as the {\it Thermodynamic Bethe Ansatz} (TBA).\index{Thermodynamic Bethe Ansatz} Although the basic idea is the same when applied to other integrable models, often complications arise because the system develops different type of excitations. For instance, in chapters \ref{chap:XXXmodel} and \ref{chap:XXZmodel} we will see how complex (string) solutions enrich the description of the Hilbert space of the XXZ chain. Due to lack of space, we will not pursue TBA further: the interested reader can find the finite temperature physics of the XXZ chain in \cite{takahashi} and learn about other models in \cite{samaj}. We notice once more that the finite temperature description is quite independent from the original ansatz on the eigenstates. In fact, the only memory of the original model is encoded in the kernel, which means that this formalism is quite general and abstract. We should thus mention that TBA has developed into a fascinating mathematical subject with several applications \cite{tongeren16}, from the study of finite-size effects in 1+1-dimensional field theories \cite{klassen90,zamolodchikov90,klassen91}, to the calculation of the excitation spectra of string theories in the AdS/CFT correspondence \cite{AdSCFTreview}.

Before leaving this chapter, let us consider some limiting cases of the Lieb-Liniger model:

\subsection{$T \to 0^+$}

For $h < 0$, one can show that $n=0$.

For $h > 0$, one can show that the function $\varepsilon (k)$ has two zeros on
the real axis for
\be
   \varepsilon (\pm q) = 0 \; , \qquad \qquad h>0 \; ,
\ee 
and it satisfies (\ref{varepsilon1}--\ref{varepsilon4}).

This means that in the zero-temperature limit, $\ln (1+\eu^{-\varepsilon(k)/T})$ tends to zero for $|k|>q$ and to $-\varepsilon(k)/T$ inside the Fermi sea (where $\varepsilon(k)<0$. Thus, the Yang-Yang equation\index{Yang-Yang equation} (\ref{YangYangEq}) becomes linear and reduces to (\ref{epsilonen}).
Consequently, $\rho (k) =0$ for $|k|>q$, $\rho_h (k) = 0$ for $|k|<q$ and (\ref{phrel}) reduces to (\ref{intbetheeq}).
The zero-temperature limit of the pressure (\ref{LLPressure}) gives
\be
   {\cal P} = - {E \over L} = - e = - {1 \over 2 \pi} \int_{-q}^q \varepsilon (k) \de k \; .
   \label{T0P}
\ee

\subsection{$c \to \infty$}

In the Tonks–Girardeau limit the kernel\index{Kernel} vanishes, therefore
\be
   \varepsilon (k) = k^2 - h
\ee and
\bea
   \rho (k) & = & {1 \over 2 \pi} {1 \over 1 + \eu^{(k^2-h)/T} } \; , \\
   {\cal F} & = & N h - {T \over 2 \pi} \int \de k \ln
   \left( 1 + \eu^{-(k^2-h)/T} \right) \; .
\eea This is equivalent to free fermions.

Corrections can be accounted perturbatively for large, but finite $c$, by expanding the kernel and the solution in powers of $1/c$ and equating the different powers. For instance, at first order we have
\bea
 \varepsilon (k) & = & k^2 - h - {2 \over c} P + \Ord \left( {1 \over c^3} \right) \; , \\
 \rho (k) & = & {1 \over 2 \pi} \left( {1 + {2 \over c} \; n \over 1 + \eu^{\varepsilon(k)/T}} \right) \; , \\
 \rho_h (k) & = & {1 \over 2 \pi} \left( {1 + {2 \over c} \; n \over 1 + \eu^{-\varepsilon(k)/T}} \right) \; , \\
 \rho_v (k) & = & {1 \over 2 \pi} \left( 1 + {2 \over c} \; n \right) \; ,
\eea
where $P$ is given by (\ref{LLPressure}).

\subsection {$c \to 0^+$}

In this limit
\be
   {\cal K} (k,k') \to - 2 \pi \delta(k-k') \; .
\ee Therefore
\be
   \varepsilon (k) = T \ln \left[ \eu^{(k^2-h)/T} -1 \right]
\ee and \bea
   \rho (k) & = & {1 \over 2 \pi} {1 \over \eu^{(k^2-h)/T} -1} \; , \\
   \rho_h (k) & = & {1 \over 2 \pi} \; , \\
   {\cal F} & = & N h + {T \over 2 \pi} \int \de k \ln
   \left( 1 - \eu^{-(k^2-h)/T} \right) \; .
\eea This is coherent with what we know as free bosons.

\chapter{The Heisenberg chain}
\label{chap:XXXmodel}

\abstract{
Historically, the Heisenberg chain has been the first exactly solved (interacting) model. 
It is a fairly realistic model for a one-dimensional quantum magnet and it has been instrumental in moving beyond the classical Ising-type models. 
We will solve this chain following the same steps we introduced in the previous chapter for the Lieb-Liniger model: in Sec. \ref{sec:XXXmagnons} we introduce the chain's fundamental excitation, the {\it magnon}, in Sec. \ref{sec:XXX2body} we study the two-body problem and in Sec. \ref{sec:XXXBetheSol} we write the coordinate Bethe Ansatz solution and the Bethe equations.
We will see that the latter admit a mixture of complex and real solutions and discuss the challenges this implies. For infinitely long chains, it is believe that the string hypothesis allows to organize the different bound solutions: this is the topic of Sec. \ref{sec:XXXstrings}. Having organized the Hilbert space of the model, we discuss the ground state and the low energy excitations for the ferromagnetic and antiferromagnetic order in Sec. \ref{sec:XXXFM} and \ref{sec:XXXAFM}, respectively. For the latter, we introduce and discuss the role of spinons as emergent quasi-particles. Finally, in Sec. \ref{sec:XXXh} we study the effect of switching on an external magnetic field.
}

\section{Definition of the model}

The Hamiltonian of the Heisenberg spin-$1/2$ chain with $N$ sites and periodic boundary conditions ${\bf S}_{j+N}={\bf S}_j$ is\index{XXX chain}\index{XXX chain}\index{Heisenberg chain} \cite{Heisenberg28}
\be
  \label{H}
  {\cal H} = -J\sum_{n=1}^N {\bf S}_n \cdot {\bf S}_{n+1} 
    = -J\sum_{n=1}^N \biggl [
   \frac{1}{2}\bigl( S_n^+S_{n+1}^- + S_n^-S_{n+1}^+ \bigr )
     +S_n^zS_{n+1}^z \biggr ] \; ,
\ee
where $S_n^\pm\equiv S_n^x\pm i S_n^y$ are spin flip operators. ${\cal H}$ acts on a Hilbert space of dimension $2^N$ spanned by the orthogonal basis vectors $|\sigma_1 \ldots \sigma_N\rangle$, where $\sigma_n=\uparrow$ represents an up spin and $\sigma_n=\downarrow$ a down spin at site $n$.  The $SU(2)$ spin commutation relations (with $\hbar=1$) are\index{$SU(2)$ algebra}
\begin{equation}
   \label{SSS}
   [S_n^z,S_{n'}^\pm]=\pm S_n^\pm\delta_{n n'}, \quad
   [S_n^+,S_{n'}^-] = 2 S_n^z \delta_{nn'}.
\end{equation}

The coupling $J$ sets the energy scale, thus the Hamiltonian (\ref{H}) has the same eigenstates, independently of $J$. However, the order of the states is reversed by changing the sign of the coupling: $J >0$ favors ferromagnetic alignment, while $J<0$ gives an antiferromagnet. The Bethe Ansatz diagonalization gives the same result for any $J$, but the ground state nature (and hence the low-energy excitations) will differ greatly in the FM and AFM case.

The Heisenberg (or XXX)\index{XXX chain}\index{Heisenberg chain} chain is the original model solved by Hans Bethe in 1931 \cite{bethe31} using the intuition that will become the Bethe Ansatz. At the time, Bethe was very intrigued by the success of his approach, namely that simple superpositions of plane-waves would be exact eigenstates of the system, and intended to investigate it further. But he never did. In his career, Bethe contributed to virtually all fields of physics and in many of them he brought innovative ideas \cite{bethestory}. His creativity was such that he never had time to get involved in the development of the Bethe Ansatz techniques and eventually lost track of the most advanced progresses in them.

Looking for the solution of the model, we will take advantage of its symmetries. The (lattice) translational invariance will be used in constructing the eigenstates as superpositions of plane waves (same as for the Lieb-Liniger model\index{Lieb-Liniger model}). The Heisenberg chain\index{XXX chain}\index{Heisenberg chain} also possesses full $SU(2)$ rotational invariance. However, since the model remains integrable after the application of a magnetic field (say, in $z$ direction) we will use only the $U(1)$ rotational symmetry about the $z$-axis, which implies the conservation of the $z$-component of the total spin $S^z\equiv\sum_{n=1}^N S_n^z$: $[{\cal H},S^z]=0$. Since the magnetization is conserved, we can consider separately sectors defined by the quantum number $S^z=N/2-R$, where $R$ is the number of down spins. The full $SU(2)$ invariance renders the spectrum degenerate in states belonging to the same multiplets. These degeneracies are lifted in the XXZ chain\index{XXZ chain} we will consider in chapter \ref{chap:XXZmodel}.

\section{The vacuum state and the magnon basis}
\label{sec:XXXmagnons}

The $R=0$ sector consists of a single vector $|0\rangle$, which is an eigenstate, ${\cal H}|0\rangle=E_0|0\rangle$, with energy 
\be
  |0 \rangle \equiv |\uparrow \ldots \uparrow\rangle 
  \qquad \Rightarrow \qquad 
  E_0 \equiv - {J \over 4} \: N \; .
 \label{E0def}
\ee

The $N$ natural basis vectors in the $R=1$ invariant subspace (one down spin) are labeled by the position of the flipped spin:
\begin{equation}
   \label{vecn}
    |n\rangle= S_n^-|0\rangle \qquad n=1,\ldots,N.
\end{equation}
These states are clearly not eigenstates of ${\cal H}$, but out of them we can construct $N$ linear combinations that respect translational symmetry, i.e., the invariance of ${\cal H}$ with respect to discrete translations:
\begin{equation}
  \label{sws}
  |\psi\rangle = \frac{1}{\sqrt{N}}\sum_{n=1}^N \eu^{\ii k n} |n\rangle \; ,
\end{equation}
for wave numbers $k=2\pi m/N,\; m=0,\ldots,N-1$. (The lattice spacing has been set equal to unity.)  The vectors $|\psi\rangle$ are eigenstates of the translation operator with eigenvalues $\eu^{\ii k}$ and also of ${\cal H}$
with eigenvalues
\begin{equation}
  \label{e1k}
   E = E_0 + J(1-\cos k) \; ,
\end{equation}
as can be verified by inspection. The vectors (\ref{sws}) represent {\it magnon} excitations ($\Delta S =1$ excitations)\index{Magnon}, in which the complete spin alignment of the polarized vacuum state $|0\rangle$ is periodically disturbed by a spin wave with wavelength $\lambda=2\pi/k$.
Note that the $k=0$ state is degenerate with $|0\rangle$. It is easy to see that this state is the $S^z = {N \over 2} -1$ component of the $S = {N \over 2}$ multiplet. Thus, its degeneracy with the fully ferromagnetic state is a consequence of the $SU(2)$ invariance of the Heisenberg chain\index{XXX chain}\index{Heisenberg chain}.

\section{The two-body problem}
\label{sec:XXX2body}

The invariant subspace with $R>1$ is not a simple superposition of magnons, as can be immediately inferred from comparing the number of basis states. For $R=2$, for instance, we write a generic eigenstate as
\begin{equation}
   \label{psi2}
   |\psi\rangle = \! \sum_{1\leq n_1<n_2\leq N} f(n_1,n_2)  |n_1,n_2\rangle,
\end{equation}
where $|n_1,n_2\rangle\equiv S_{n_1}^-S_{n_2}^-|F\rangle$ are the basis vectors in this subspace of dimension $N(N-1)/2$. 
The eigenvalue equation translates into:
\begin{eqnarray}
   2[E-E_0] f(n_1,n_2) &=& J \: \big[ 
   4  f(n_1,n_2)\!-\!f(n_1\!-\!1,n_2)
   -f(n_1\!+\!1,n_2)  -f(n_1,n_2\!-\!1)-f(n_1,n_2\!+\!1) \big] \, ,
   \nonumber \\   
   && \qquad \qquad \qquad \qquad \qquad \qquad \qquad \qquad \qquad \qquad \qquad 
   \, \text{for} \: \:  n_2>n_1\!+\!1
    \label{b1}, \\
   2[E-E_0] f(n_1,n_2) &=& J \: 
   \big[2f(n_1,n_2)-f(n_1\!-\!1,n_2) -f(n_1,n_2\!+\!1) \big]
   \; , \qquad \: \text{for} \: \: n_2=n_1\!+\!1. \label{b2}
\end{eqnarray}

Bethe's preliminary ansatz to determine the coefficients $f(n_1,n_2)$ has been\index{Bethe!ansatz wavefunction}
\begin{equation}
   \label{a1a2}
   f(n_1,n_2) = A \eu^{\ii(k_1 n_1 + k_2 n_2)} + A' \eu^{\ii(k_1 n_2 + k_2 n_1)} \; ,
\end{equation}
which automatically satisfies (\ref{b1}) with energy
\begin{equation}
  \label{Ek}
   E = E_0 + J\! \sum_{j=1,2}(1-\cos k_j) \: .
\end{equation}
Condition (\ref{b1}) translates into the scattering phase\index{Scattering phase} relation for the coefficients of the ansatz (\ref{a1a2}) 
\begin{equation}
  \label{AA}
  \frac{A}{A'}\equiv \eu^{\ii\theta}=
  -\frac{\eu^{\ii(k_1+k_2)} + 1 - 2 \eu^{\ii k_1}}{\eu^{\ii(k_1+k_2)} + 1 - 2 \eu^{\ii k_2}} \; .
\end{equation}
Note that this equation can be rewritten as
\begin{equation}
  \label{ba2}
  2\cot\frac{\theta}{2} = \cot\frac{k_1}{2} - \cot\frac{k_2}{2}.
\end{equation}
The quasi-momenta $k_1,k_2$ of the Bethe Ansatz wave function can be determined by  requiring that the wave function (\ref{psi2}) satisfies the periodic boundary conditions: $f(n_1,n_2)=f(n_2,n_1+N)$:
\be
  \eu^{\ii k_1 N} = \eu^{\ii \theta} \; , \qquad
  \eu^{\ii k_2 N} = \eu^{-\ii \theta} \; .
\ee
Equivalently, we can write (after taking their logarithm)\index{Bethe!equations}
\begin{equation}
   \label{bapbc2}
   Nk_1 = 2\pi I_1 + \theta, \qquad
   Nk_2 = 2\pi I_2 - \theta,
\end{equation}
where the $I_j\in\{0,1,\ldots,N-1\}$ are integer quantum numbers.\index{Bethe!numbers} Note that, due to the lattice (and hence to the existence of the Brillouin zone), the range of inequivalent quantum numbers is restricted. This was not the case for the Lieb-Liniger model\index{Lieb-Liniger model}.

The total momentum of this state is
\begin{equation}\label{k12}
 K = k_1+k_2= \frac{2\pi}{N}(I_1 + I_2) \; .
\end{equation}
The magnons interaction is reflected by the phase shift $\theta$ and in the deviation of the quasi-momenta $k_1,k_2$ from the single (free) magnon wave numbers. This is because the magnons either scatter off each other or form bound states\index{Bound state}. Note that the momenta $k_1,k_2$ specify the Bethe Ansatz wave function (\ref{psi2}) but are not observable, while the wave number $K$ is the quantum number associated with the translational symmetry of ${\cal H}$ and exists independently of the Bethe Ansatz.

The allowed $(I_1, I_2)$ pairs are restricted to $0 \leq I_1 \le I_2 \leq N-1$. Switching $I_1$ with $I_2$ simply interchanges $k_1$ and $k_2$ and produces the same solution. There are $N(N+1)/2$ pairs that meet the ordering restriction, but only $N(N-1)/2$ of them yield a solution of Eqs.~(\ref{ba2}) and (\ref{bapbc2}). 
Note that, compared to the Bethe equations we found for the Lieb-Liniger, the scattering phase (\ref{ba2}) does not depend on the difference between the momenta of the scattering particles. This means that equal Bethe numbers\index{Bethe!numbers} $I_1=I_2$ do not imply $k_1=k_2$ (which would make the wavefunction (\ref{a1a2},\ref{AA}) vanish): thus, we cannot exclude solutions with equal quantum numbers and we lack of a good criterion to exclude the $N$ spurious choices of Bethe numbers. The solutions can be determined analytically or computationally. Some of them have real $k_1,k_2$, and others yield complex conjugate momenta, $k_2=k_1^*$.

If $I_1 = 0$ all solutions are real and $k_1 = 0$, $k_2 = 2 \pi I_2 /N$, $\theta =0$ and $I_2 = 0, 1, \ldots, N-1$.
These states are degenerate with the single magnon states and they belong to the same multiplet.

The majority of solutions are real and different from zero. It turns out \cite{mullernotes} that they correspond to $I_2 - I_1 \ge 2$.
They can be determined by combining (\ref{ba2}), (\ref{bapbc2}), and (\ref{k12}) into a single equation for $k_1$:
\begin{equation}
  \label{r2ss}
   2\cot \frac{N k_1}{2} = \cot \frac{k_1}{2} - \cot \frac{K-k_1}{2} .
\end{equation}
Considering that the total momentum of the state is quantized ($K= 2\pi n/N$), we can substitute for different $n$ in (\ref{r2ss}) to determine $k_1$ and $k_2 = K - k_1$.

The remaining choice of quantum numbers (differing by no more than $1$) either yield complex solutions or no solution. It is actually hard to numerically find the complex solutions of a system of equations. It is thus better to turn everything into a real equation by parameterizing
\begin{equation}
  \label{k1k2c}
   k_1\equiv \frac{K}{2} + \ii k \; , \qquad
   k_2\equiv \frac{K}{2} - \ii k \; , 
\end{equation}
which, substituted into (\ref{bapbc2}) yields
\be  
   \theta  = \pi (I_2 - I_1) + \ii N k \; .
\ee
Eq. (\ref{ba2}) becomes
\begin{equation}
   \label{r2bs}
   \cos\frac{K}{2} \sinh (N k) = \sinh [(N-1)k] + \cos \left[ \pi(I_1-I_2)\right] \sinh k,
\end{equation}
which gives $k$ as a function of the total momentum $K = 2 \pi/N (I_1+I_2)$. 
This solution represents a bound state\index{Bound state} in which the two flipped spins cannot be more than order of $k$ sites apart.\index{Bound state} 
Substituting (\ref{k1k2c}) into (\ref{Ek}) we find the energy of this complex solution to be 
\begin{equation}
  \label{Ekc}
   E = E_0 + 2J \biggl(1-\cos \frac{K}{2} \cosh k \biggr).
\end{equation}
For $N \to \infty$, (\ref{r2bs}) gives
\be
k_{1,2} = {K \over 2} \pm \ii \ln \cos {K \over 2} \; ,
\ee
which means that for large systems the energy of the bound state\index{Bound state} is
\be
   E \stackrel{N \to \infty}{\rightarrow}= E_0 + {J \over 2} \left( 1 - \cos K \right) \; .
   \label{EknNlarge}
\ee

This behavior should be contrasted with that of real solutions (\ref{e1k}). In the large $N$ limit, the real solutions (\ref{bapbc2}) are not too different from a simple superpositions of two magnons, as the quasi-momentum of each excitation differs from the ``free'' quantization as
\be
  k_{1,2} = {2 \pi \over N} I_{1,2} + \Ord \left( {1 \over N^2} \right) \; .
\ee
Thus, the dispersion relations of these states form a continuum with boundaries
\be
  E = E_0 + 2 J \left( 1 \pm \cos {K \over 2} \right) \; .
  \label{2partcont}
\ee

We stress once more that the quasi-momenta assigned to the different magnons for real solution are just bookkeeping artifacts of the Bethe Ansatz technique. However, the difference between real and complex solutions is physical and important. In fact, while the dispersion relation of real solutions (\ref{2partcont}) form a continuum, indicating the existence of internal degrees of freedom (the relative quasi-momentum of the two magnons), the complex solution's dispersion relation is just a line (\ref{Ekc},\ref{EknNlarge}), showing that this is a bound-state of two magnons, which behave like a single entity, with no additional internal dynamics. Note that the dispersion relation of such bound state\index{Bound state} is also different from that of a simple magnon (\ref{e1k}). Substituting the complex solution into the Bethe Ansatz wavefunction, one also sees that its amplitude vanishes exponentially as the distance between the flipped spins grows.

\section{The Bethe Solution}
\label{sec:XXXBetheSol}

Having discussed the basic features of the two-body problem, we proceed with the construction of the eigenstates with generic $R$ overturned spins. As in (\ref{psi2}), we expand the states into the natural (computational) basis
\begin{equation}
  \label{psir}
  |\psi\rangle = \sum_{1\leq n_1<\ldots<n_R\leq N} f(n_1,\ldots,n_R)
  |n_1,\ldots,n_R\rangle.
\end{equation}
In analogy with the ansatz we employed for the Lieb-Liniger model in the previous chapter and knowing the two-body solution (\ref{a1a2}, \ref{AA}), we write the (non-normalized) wavefunction as
\be
  \label{bar}
   f(n_1,\ldots,n_R) =  \sum_{\cal P} \exp \left[ \ii \sum_{j=1}^R k_{{\cal P} j}n_j
   + \frac{\ii}{2}\sum_{l<j} \theta \left( k_{{\cal P}l}, k_{{\cal P}j} \right) \right] \; ,
\ee
where the sum extends over all $R!$ permutations ${\cal P}$ of the assignments of the quasi-momenta to each overturned spin and where we introduced the antisymmetric phase shift $\theta \left( k_l, k_j \right) = - \theta \left( k_j, k_l \right)$.
The consistency equations for the coefficients $f(n_1,\ldots,n_R)$ are extracted from the eigenvalue equation ${\cal H}|\psi\rangle=E|\psi\rangle$. They are a straightforward generalization of the two-particle case (\ref{b1}, \ref{b2}). The energy eigenvalue equation becomes
\begin{equation}
  \label{Ekr}
   E = E_0 + J\sum_{j=1}^R (1-\cos k_j) \; ,
\end{equation}
and the eigenstate condition can be written as
\be
  \label{annr}
   2f(n_1, \ldots , n_j , n_j + 1, \ldots ,n_R) =
   f(n_1, \ldots , n_j , n_j , \ldots ,n_R)
   + f(n_1, \ldots , n_j + 1 , n_j + 1, \ldots ,n_R) \; ,
\ee
for $j=1,\ldots,R$. These conditions fix the phase shift $\theta \left( k_j, k_l \right)$ to be:\index{Scattering phase}
\begin{equation}
  \label{ethetaij}
  \eu^{\ii \theta \left( k_j, k_l \right)} = -\frac{ \eu^{\ii(k_j+k_l)} + 1 - 2 \eu^{\ii k_j}} {\eu^{\ii(k_j+k_l)} + 1 - 2 \eu^{\ii k_l}} \; ,
\end{equation}
which can be cast in real form as
\begin{equation}
  \label{ba2r}
   2\cot\frac{\theta \left( k_j, k_l \right)}{2} = \cot\frac{k_j}{2} - \cot\frac{k_l}{2} \; ,  \qquad \qquad j,l=1,\ldots,R \; .
\end{equation}

The periodicity of the chain for translations by $N$ sites $f(n_1,\ldots,n_R)=f(n_2,\ldots,n_R,n_1+N)$ gives
\be
  \sum_{j=1}^R k_{{\cal P}j}n_j +
  \frac{1}{2} \sum_{l<j}\theta \left( k_{{\cal P}l}, k_{{\cal P}j} \right)
  = \frac{1}{2} \sum_{l<j}\theta \left( k_{{\cal P}'l}, k_{{\cal P}'j} \right) - 2\pi \tilde{I}_{{\cal P}'R}
  +\sum_{j=2}^R k_{{\cal P}'(j-1)}n_j + k_{{\cal P}'R}(n_1+N) \; ,
\ee
where the permutations on the LHS are defined as ${\cal P}'(j-1)={\cal P}j, \; j=2,\ldots,R;\;{\cal P}'R={\cal P}1$ and the $\tilde{I}_j$ are integers. All terms not involving the index ${\cal P}'R={\cal P}1$ are identically equal, thus we are left with $R$ relations between the phase shifts and the quasi-momenta:\index{Bethe!equations}
\be
   \label{ba1r}
   N k_j = 2\pi \tilde{I}_j + \sum_{l\neq j} \theta \left(k_j,k_l\right) \; , 
   \qquad \qquad \qquad j=1,\ldots,R,
\ee
where $\tilde{I}_j\in\{0,1,\ldots,N-1\}$ as in (\ref{bapbc2}). 

As we saw in the previous section, this is not the end of the story, since not all choices of quantum numbers\index{Bethe!numbers} $I_j$ produce solutions and among the others some do yield complex quasi-momenta, which increase the computational complexity of solving these equations. For an excellent account on these issues we refer the interested reader to \cite{mullernotes}, which also provided the material for preparing these sections. Indeed, the identification and classification of complex solutions is still an open problem. The situation simplifies somewhat if we take the thermodynamic limit $N \to \infty$. In this case, it is possible to assume that all complex solutions organize themselves into strings, similar to those we discussed for the attractive case of the Lieb-Liniger model. In the next section we will show how these structures help in constructing the Hilbert space of the Heisenberg\index{XXX chain}\index{Heisenberg chain} chain and thus in studying its thermodynamic properties and we comment on the validity of this {\it string hypothesis}.

Before we proceed further, one unpleasant feature of (\ref{ba1r}) can be readily fixed. Namely, the scattering phase (\ref{ethetaij}, \ref{ba2r}) does not depend on the difference of the particle quasi-momenta and thus we have more possible choices of quantum numbers than states in the Hilbert space. To fix this problem and restore ``translational invariance'' to the Bethe equations, we introduce the {\it rapidities} $\lambda_j$ to parametrize the quasi-momenta:
\be
  \cot {k_j \over 2} = \lambda_j \; , \quad {\rm or} \qquad
  k_j = {1 \over \ii} \ln {\lambda_j + \ii \over \lambda_j - \ii} = \pi - \theta_1 (\lambda_j) \; ,
  \label{rapdef}
\ee
where
\be
   \theta_n (\lambda) \equiv 2 \arctan {\lambda \over n} \; .
\ee

The (bare) energy and momentum of an individual magnon, characterized by a quasi-momentum $k$, is
\bea
  p_0 (\lambda) & = & {1 \over \ii} \ln {\lambda + \ii \over \lambda - \ii} = k \; , \\
  \epsilon_0 (\lambda) & = & - J {\de k \over \de \lambda} = {2 J \over \lambda^2 + 1} = J (1 - \cos k) \; .
  \label{XXXepsilon0}
\eea
We introduced the subscript $0$ to indicate that it corresponds to a single (real) particle, that is, what later we will call a {\it $0$-type string}.\index{String solution}

In terms of these rapidities, the scattering phase is\index{Scattering phase}
\be
   \theta \left( k_j, k_l \right) = - \theta_2 (\lambda_j - \lambda_l) + \pi \sgn \left[ \Re (\lambda_j - \lambda_l) \right] \; ,
   \label{thetarap}
\ee
where $\Re (x)$ is the real part of $x$ and $\sgn (y) = \pm 1$ denotes the sign of $y$.
The Bethe equations (\ref{ba1r}) in terms of the rapidities become\index{Bethe!equations}
\be
  N \theta_1 (\lambda_j) = 2 \pi I_j + \sum_{l=1}^R \theta_2 (\lambda_j - \lambda_l) \; , \qquad \qquad \qquad
  j=1, \ldots, R \; .
  \label{XXXBE1}
\ee
The state is now defined by these ``new'' Bethe numbers\index{Bethe!numbers}  $\{ I_j \}, j=1, \ldots, R$. It is not easy to relate them to the $\tilde{I}_j$ in (\ref{ba1r}), because of the second term in (\ref{thetarap}), but we do not need to. Because (\ref{XXXBE1}) is translational invariant, two equal $I_j$ produce the same rapidities and thus a non valid solution\footnote{We shall see however that states with differing quantum numbers can have equal quasi-momenta, if these lie at the edge of the Brillouin zone.}. Therefore, it is more convenient to work with the $I_j$, which have the same ``fermionic'' properties we found for the Lieb-Liniger model\index{Lieb-Liniger model} and thus gives the proper counting of the states (each choice of increasing and non-repeating quantum numbers produces a physical state). Using the rapidities to parametrize the eigenstates, their energies and momenta are
\bea
   E & = & E_0 + J \sum_{j=1}^R \epsilon_0 (\lambda_j) \; , \\
   K & = & \left[ \sum_{j=1}^R p_0 (\lambda_j) \right] \: {\rm mod} \: 2 \pi
   = \left[ \pi \: R - {2 \pi \over N} \sum_{j=1}^R I_j \right] \: {\rm mod} \: 2 \pi \; .
\eea

Originally we expanded the wavefunction coefficients in (\ref{psir}) as a sum of plane waves parametrized by the quasi-momenta (\ref{bar}), but found that in this way the scattering phase\index{Scattering phase} was not explicitly translational invariant. The change of variable (\ref{rapdef}) shows that the basis $\left( {\lambda + \ii \over \lambda - \ii} \right)^n$ is a more appropriate choice for the wavefunction ansatz. It also shows that any complete single-particle basis can be used for the ansatz.

We still have the problem of identifying and separating real and complex solutions, since turning to rapidities does not improve the computational complexity of finding the latter. However, by embracing the string hypothesis for the complex solutions, we can account for them in a remarkably elegant way.

\subsection{String solutions}
\label{sec:XXXstrings}

We have seen that the Bethe equations admit complex solutions, a fact already noticed by Bethe in \cite{bethe31}. In general, these have to be found numerically, which can be computationally hard. However, a simple structure emerges if we take the thermodynamic limit $N \to \infty$. This structure is known as {\it string hypothesis} \cite{takahashithemo}, as it is not yet clear whether the string solutions we are about to describe exhaust the whole Hilbert space (especially for the XXZ chain we will study in chapter \ref{chap:XXZmodel}). However, there is consensus about the fact that the string hypothesis provides an accurate description of the thermodynamics of the chain, indicating that solutions\index{String solution} that do not conform to the string structure are relevant only for certain response functions or out-of-equilibrium  \cite{hagemans07}.

Following \cite{faddeev96}, let us look again at the $R=2$ case (i.e. two overturned spins).
The Bethe Equations written in terms of the rapidities are:
\bea
   \left( {\lambda_1 + \ii \over \lambda_1 - \ii} \right)^N & = &
   {\lambda_1 - \lambda_2 + 2 \ii \over \lambda_1 - \lambda_2 - 2 \ii} \; ,
   \label{l1BE} \\
   \left( {\lambda_2 + \ii \over \lambda_2 - \ii} \right)^N & = &
   {\lambda_2 - \lambda_1 + 2 \ii \over \lambda_2 - \lambda_1 - 2 \ii} \; .
\eea
If $\Im (\lambda_1) \ne 0$, the LHS in (\ref{l1BE}) will grow (or decrease) exponentially in $N$. Therefore, in the thermodynamic limit the LHS is strictly zero or infinity and the RHS will have to do the same. Thus, we must have
\be
   \lambda_1 - \lambda_2 = \pm 2 \ii \; , \qquad {\rm i.e.} \qquad
   \lambda_{1,2} = \lambda \pm \ii\footnote{Note that, as proven in \cite{vladimirov86}, complex solutions to the Bethe equations always appear in conjugated pairs: $\{ \lambda_j \} = \{ \lambda_j^*\}$} \; .
\ee
The energy and momentum of this state are real:
\bea
   p_{1/2} (\lambda) & = & p_0(\lambda + \ii) + p_0(\lambda - \ii) =
   {1 \over \ii} \ln {\lambda + 2 \ii \over \lambda - 2 \ii} \; , \\
   \epsilon_{1/2} (\lambda) & = & \epsilon_0 (\lambda + \ii) + \epsilon_0 (\lambda -\ii) = {4 J \over \lambda^2 + 4} \; ,
\eea
which gives the dispersion relation
\be
   \epsilon_{1/2} (p) = {J \over 2} \left( 1 - \cos p_{1/2} \right) \; .
\ee
We see that for $J>0$ $\epsilon_{1/2} (p) < \epsilon_0 (p-p') + \epsilon_0 (p')$ for every $0 \le p, p' < 2 \pi$, and thus in the ferromagnetic regime these bound states\index{Bound state} are energetically favored compared to real solutions.

\begin{figure}[t]
	\noindent\begin{minipage}[t]{12cm}
		\includegraphics[width=12cm]{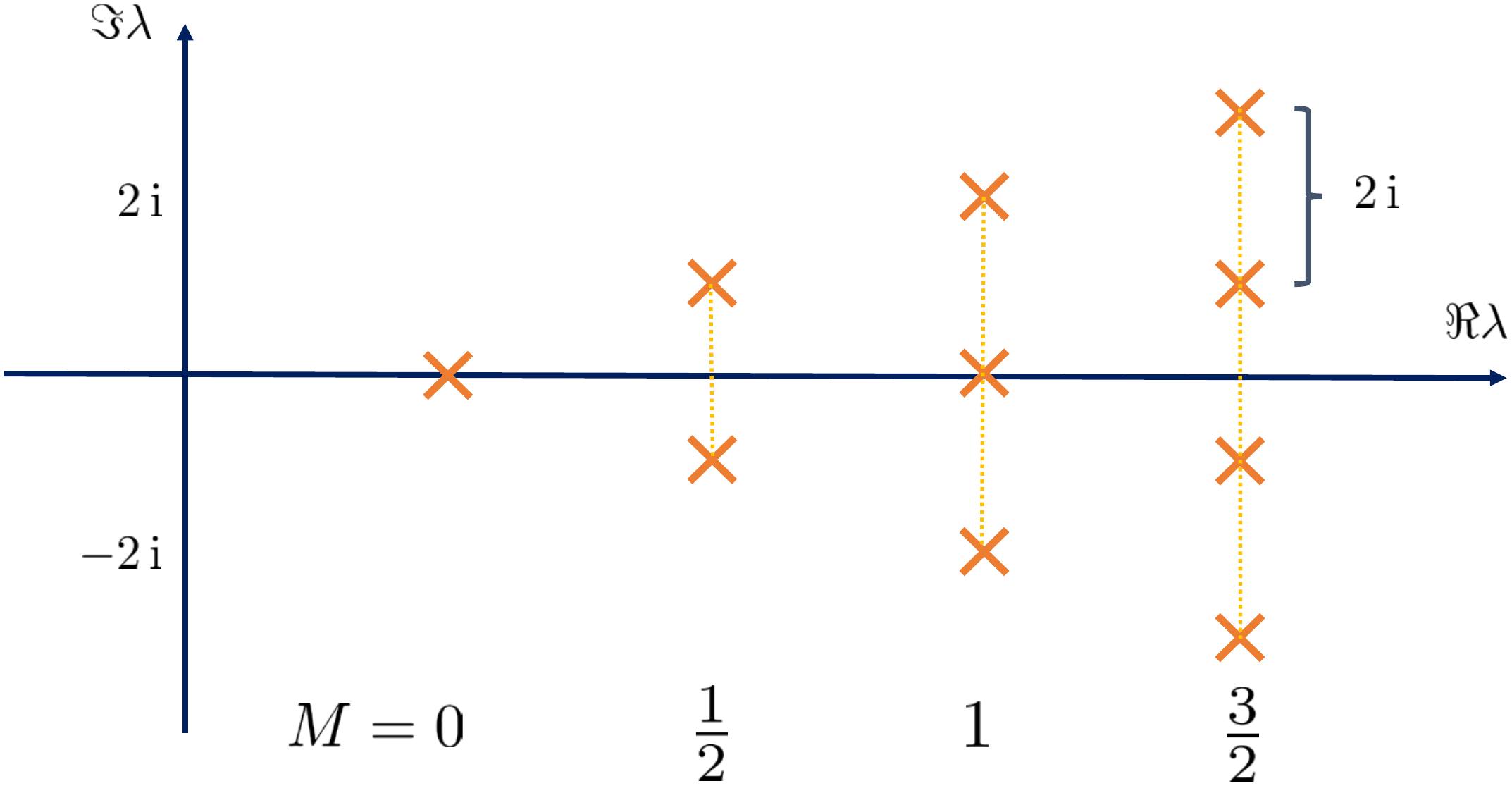}
	\end{minipage}
	\hskip.5cm
	\begin{minipage}[b]{3.5cm}
		\caption{Cartoon of the string hypothesis. We assume that complex solutions to the Bethe Equations (\ref{XXXBE1}) can be grouped in strings with the same real part and equidistant imaginary components, which we call ``complexes''. Single, real, solutions make $0$-complexes, a pair of complex rapidities a $1/2$-complex and so on. \vspace{.2cm} $\quad$}
		\label{fig:strings}
	\end{minipage}
\end{figure}

For $R>2$, we assume that complex solutions can be organized into {\it complexes} (or {\it strings})\index{M-complex} of $2M +1$ rapidities characterized by the same real value $\lambda_M$ and different, equidistant, imaginary parts, see figure \ref{fig:strings}.\index{String solution} Here $M=0, 1/2, 1, \ldots $ and the rapidities have the structure\footnote{The reader must have noticed that we use notations reminiscent of the representations of $SU(2)$.}
\be
   \lambda^{(M)}_m = \lambda_M + 2 \ii m \; , \qquad \qquad \qquad
   m = -M, -M +1, \ldots, M-1, M \; .
\ee
Denoting by $\nu_M$ the number of complexes of length $M$, a state with a given magnetization satisfies
\be
   R = \sum_M \big( 2M+1 \big) \: \nu_M \; .
\ee
We expect the results obtained through the string hypothesis to be a good approximation of reality as long as $\nu_0$, i.e. the number of single-particle solutions, is dominating over all other complexes in this sum. 

Since these rapidities represent a group of $2M+1$ spins that move together with the same real rapidity and that want to stay close to one-another (otherwise penalizing the wavefunction with an exponential decay as they are taken apart), we will treat them as a single entity. In fact, all interactions between the individual rapidities of a complex can be factorized and summed over separately in the interactions between the complexes.
The energy and momentum of a $M$-complex are obtained by summing over all the rapidities within one string
\bea
   p_M (\lambda_M) & = & 
   \sum_{m=-M}^M p_0 (\lambda_M + 2 \ii m) =
   {1 \over \ii} \ln
   {\lambda_M + \ii (2 M + 1) \over \lambda_M - \ii (2 M +1) }= \pi - \theta_{2M+1} (\lambda_M) \; ,
   \label{pMdef}\\
   \epsilon_M (\lambda_M) & = & 
   \sum_{m=-M}^M \epsilon_0 (\lambda_M + 2 \ii m) =
   {2 J (2 M + 1) \over \lambda_M^2 + (2M+1)^2}
   = {J \over 2M +1} (1 - \cos p_M ) \; .
   \label{epsilonMdef}
\eea
Due to their regular structure we have a lot of cancellations: taking them into account we see that we can consider the scattering phase\index{Scattering phase} of a $M$-complex with a simple magnon\index{Magnon} ($0$-complex) again by taking the product with respect to all the particles in a given complex, obtaining
\be
   S_{0,M} (\lambda_0 - \lambda_M) = S_{0,M} (\lambda) =
   {\lambda + \ii 2 M \over \lambda - \ii 2 M} \;
   {\lambda + \ii 2 (M+1) \over \lambda - \ii 2 (M+1)} \; ,
\ee
and the scattering of two complexes of length $M$ and $M'$ is
\be
   S_{M,M'} (\lambda) = \prod_{L=|M-M'|}^{M+M'} S_{0,L} (\lambda) \; ,
\ee
which is reminiscent of the Clebsh-Gordan coefficients. This is not surprising, since the whole structure we develop is reminding of an $SU(2)$ algebra.

With these notations, we want to describe an eigenstate of the Heisenberg chain\index{XXX chain}\index{Heisenberg chain} in terms of the number of complexes $\nu_M$ for each type $M$ and by the rapidities of their center of mass $\lambda_{M,j}$, where $j=1, \ldots, \nu_M$\footnote{To be clear in the notation used, $\lambda_{M,j}$ is the real part of the $j$-th complex of length $M$.}. 
The Bethe equations\index{Bethe!equations} for the complexes is obtained by grouping all the rapidities $\lambda^{(M)}_j$ belonging to the same complex and performing fist the products within each complex so to be left only with consistency conditions on their real centers $\lambda_{M,j}$:
\be
   \eu^{\ii p_M (\lambda_{M,j}) N} = 
   \prod_{M'} \prod_{\stackrel{j'}{(M',j') \ne (M,j)}}^{\nu_{M'}}
   S_{M,M'} \left( \lambda_{M,j} - \lambda_{M',j'} \right) \; , \qquad \qquad
   \forall M; \quad j=1, \ldots, \nu_M \; .
   \label{stringBE1}
\ee
As usual, we take the logarithm of (\ref{stringBE1}), introduce (half-)integer quantum numbers $I_{M,j}$ to take into account the branches of the logarithms for each complex type and, using the familiar identity
\be
   {1 \over \ii} \ln {\lambda + \ii n \over \lambda - \ii n}
   = \pi - 2 \arctan {\lambda \over n}
   = \pi - \theta_n (\lambda) \; ,
\ee
we get
\be
  N \theta_{2M+1} (\lambda_{M,j}) = 2 \pi I_{M,j}
  + \sum_{(M',j') \ne (M,j)} \theta_{M,M'}
  \left( \lambda_{M,j} - \lambda_{M',j'} \right) \; ,
  \label{BEMj}
\ee
where
\be
   \theta_{M,M'} (\lambda) \equiv \sum_{L=|M-M'|}^{M+M'}
   \left[ \theta_{2L} (\lambda) + \theta_{2L+2} (\lambda) \right] \; ,
   \label{thetaMM}
\ee
and the $L=0$ is intended to be omitted. Eq. (\ref{BEMj}) are called the Bethe-Gaudin-Takahashi equation.\index{Bethe!equations}

In the string hypothesis, each state is thus characterized by the number of complexes $\nu_M$\index{M-complex} and by the Bethe numbers\index{Bethe!numbers} $I_{M,j}$ of each complex type. Since a spin chain's Hilbert space is limited, not all quantum numbers are allowed. First, within each complex, $I_{M,j} \ne I_{M,j'}$ in order to have a non-vanishing solution. Moreover, since momenta are constrained within a Brillouin zone (due to the existence of a lattice in real space), the Bethe numbers are bounded.\index{Bethe!numbers}
We notice that a diverging rapidity $\lambda_M^{(\infty)} = \infty$ (corresponding to a quasi-momentum at the edge of the Brillouin zone) has a fixed scattering phase\index{Scattering phase} with all other particles, since $\arctan \pm \infty = \pm \pi / 2$.
Thus, inverting (\ref{BEMj}) we have that $\lambda_M^{(\infty)}$ is given by the Bethe number
\be
   I_M^{(\infty)} = -\sum_{M'\ne M} \left[ 2 \min(M,M') +1 \right] \nu_{M'}
   - \left( 2M + {1 \over 2} \right) \left(\nu_M -1 \right) + {N \over 2} \; .
   \label{IMinfty}
\ee
Since adding a $M$-complex shifts this boundary by ${1 \over 2 \pi} \theta_{M,M} (\infty) = 2M + {1 \over 2}$, the maximum quantum number that characterizes a finite rapidity (before it joins the rapidities at the edges) is
\be
   I_M^{\rm max} = I_M^{(\infty)} - \left( 2M + {1 \over 2} \right) - {1 \over 2}
   = {N - 1 \over 2}  - \sum_{M'} J(M,M') \nu_{M'} \; ,
   \label{IMMax}
\ee
where
\be
   J (M,M') \equiv \left\{ \begin{array}{ll}
                    2 \min(M,M') +1 & M \ne M' \cr
                    2 M + {1 \over 2} & M = M' \cr
                    \end{array} \right. \; ,
\ee
and where the additional shift of $1/2$ in (\ref{IMMax}) takes into account that with each rapidity the Bethe numbers\index{Bethe!numbers} shift from integers to half-integers and vice-versa.
Since all the scattering phases\index{Scattering phase} are odd functions of their argument, we have that
\be
   I_{M,{\rm min}} = - I_{M,{\rm max}} \; ,
\ee
which means that there are
\be
   P_M = 2 I_M^{\rm max} +1
   =N - 2 \sum_{M'} J(M,M') \nu_{M'}
   \label{numvacancies}
\ee
vacancies for a $M$-complex.\index{M-complex}
We notice that the range of allowed values becomes narrower for complexes of any size if any string is added to the system. Using these results, one can estimate the number of states accessible within the string hypothesis and it can be shown that it scales like $2^N$ as one would desire \cite{takahashi,faddeev84}, meaning that only few states are possibly neglected in this framework.
Such states typically involve a large number of complex rapidities (a finite fraction of the number of sites $N$) which are not organized in strings, but are still somewhat able to satisfy the Bethe equations (\ref{XXXBE1}) because the exponential growth/decay on the LHS is (``accidentally'') properly compensated on the RHS \cite{nostrings,hagemans07}.
As we wrote, these spurious states do not contribute significantly to the thermodynamics of the model, but are important to determine the completeness of the Bethe solution and for other investigations, such as for correlation functions, dynamical responses, or in working with finite systems.

We have introduced the elementary excitations\index{Quasi-particle} of the Heisenberg chain\index{XXX chain}\index{Heisenberg chain}, providing a classification of its states. While both the ferromagnetic and antiferromagnetic regimes share the same eigenstates, their order in energy space is reversed in the two cases: in the next two sections we will discuss the properties of the low energy states for each regime and we will see that the AFM case is best understood in terms of an additional emergent quasi-particle: the {\it spinon}.\index{Spinon}

\section{The Ferromagnetic case: $J=1$}
\label{sec:XXXFM}

For a ferromagnetic coupling, the completely polarized state $|0\rangle$ can be taken as the ground state. In fact, it is degenerate with all the other members of the $S = N/2$ multiplet, which can be generated from $|0\rangle$ by adding zero-momentum magnons. The lowest energy states are individual long wavelength magnons as well as bound states\index{Bound state} complexes, that have lower energy compared to multiple magnon excitations (clearly, a ferromagnetic coupling favors the clustering of flipped spins).
The ground state can thus be characterized as a magnon-vacuum with quadratic dispersion relation for the excitations. Thus, in the scaling limit the ferromagnetic Heisenberg chain is not described by a conformal field theory and, due to its high ground state degeneracy, it is a somewhat singular point in the phase diagram of the XXZ/XYZ chain \cite{FMXXXentanglement}.

\section{The Anti-Ferromagnetic case: $J=-1$}
\label{sec:XXXAFM}

The anti-ferromagnetic regime is the most relevant one for physical application, since ferromagnetic couplings are more rare in nature. Moreover, while ferromagnetism admits a semi-classical description, the AFM Heisenberg chain shows a truly quantum nature.
It can be proven, and it makes intuitive sense, that the ground state of the AFM regime has to be found in the $S^z = 0$ ($R=N/2$) sector\footnote{We assume that $N$ is even. For the odd case there are two degenerate ground states in the $S^z = \pm 1/2$ sectors, but we will not discuss this case further.} \cite{yangyangXXZ}.
Since bound states\index{Bound state} in this regime have higher energy compared to unbound magnons, the ground state configuration must be composed by $0$-type complexes, i.e. single quasi-particle excitations
\be
   \nu_0 = {N \over 2} \; : \qquad \qquad
   \nu_M = 0, \quad M \ge {1 \over 2} \; , \qquad \qquad \rightarrow \qquad
   R= {N \over 2} \; .
\ee
Using (\ref{numvacancies}) we find that the number of vacancies for this configuration is
\be
   P_0 = N - 2 J(0,0) \nu_0 = N - {N \over 2} = {N \over 2} \; ,
\ee
which equals the number of particle states. Hence, the quantum numbers occupy all the allowed vacancies:
\be
   -{N \over 4} + {1 \over 2} \le I_{0,k} \le {N \over 4} - {1 \over 2} \; ,
\ee
and are integer (half-integer) for N/2 odd (even).
Thus there is only one state with $N/2$ real magnons\index{Magnon} and is the antiferromagnetic ground state $|{\rm AFM}\rangle$.\footnote{Note that negative magnetization states cannot be reached in this formalism: one starts from the completely negatively polarized states and excites magnons out of it. On general ground, the model is invariant under the reversal of every spin across the $x-y$ plane (a particle/hole duality) and thus any positive magnetization state is related by this symmetry to one with a negative one.}

Excited states over this ground state are constructed by progressively taking away quasi-particles from the single state and moving them into complexes, i.e. we will characterize the excited states by $\kappa$, with
\be
   \nu_0 = {N \over 2} - \kappa \; .
\ee
For $\kappa =1$, we cannot excite any complexes, thus we have $R = N/2 -1$, which corresponds to total spin $S^z = 1$.
The number of vacancies in this case is
\be
   P_0 = N - 2 \cdot {1 \over 2} \left( {N \over 2} -1 \right)
   = {N \over 2} +1
\ee
which exceeds the number of particles by two. This means that the Bethe numbers of a state in this sector are all the quantum numbers in the allowed range but two: the choice of these two holes\index{Hole} characterizes the state.

For $\kappa=2$ we have two possibilities: we can keep $\nu_M = 0$ for $M \ge 1/2$ like before and have a state with magnetization $S^z = 2$. The physics is similar to that of the $\kappa=1$ sector, except that the state is described by four missing quantum numbers (4 holes).
The second possibility is to have $\nu_{1/2} = 1$ (and $\nu_M = 0$ for $M \ge 1$), which keeps $R=N/2$ and $S^z = 0$.
The vacancies are
\bea
   P_0 & = & N - 2 \left( {N \over 2} -2 \right) {1 \over 2}
   - 2 J \left( 0,{1 \over2} \right) = {N \over 2} \; , \\
   P_{1/2} & = & N - 2 \left( {N \over 2} -2 \right)
   J \left( {1 \over2}, 0 \right)
   - 2 J \left( {1 \over 2} ,{1 \over2} \right) = 4 -3 = 1 \; .
\eea
Once more, the number of vacancies for real quantum numbers allows for two holes\index{Hole}, while there is no freedom for the $1/2$-complex, whose state is therefore fixed.

For generic $\kappa$, we can have configurations with
\be
   \nu_0 = {N \over 2} - \kappa \; : \qquad \qquad
   \nu_M = 0, \quad M \ge {1 \over 2} \; , \qquad \qquad \rightarrow \qquad
   R= {N \over 2} - \kappa \, ,
\ee
with $P_0 = {N \over 2} + \kappa$ vacancies, which give rise to	$2\kappa$ holes (characterizing the state) and a total spin $S^z = \kappa$\footnote{These excitations are holes\index{Hole} with respect to the description we have been using, but they should be considered as particle excitations on top of the vacuum state.}.
In addition to these solutions, we can have states with smaller magnetization (all the way to $0$) and a proliferation of complexes.

Before we proceed further, let us analyze better the states we introduced in the previous examples. As we did for the Lieb-Liniger model\index{Lieb-Liniger model} in Sec. \ref{sec:thermolimit}, in the thermodynamic limit $N \to \infty$ we can approximate the solutions of the Bethe equation\index{Bethe!equations} for $0$-complexes by the (continuous) density distribution of the state's rapidities.
We start with the ground state, for which the Bethe numbers\index{Bethe!numbers} fill the allowed interval of vacancies without holes\index{Hole}. Let us assume that $N/2$ is odd (the even case requires just minor modifications) so that
\be
   I_{0,j} = j \; , \qquad \qquad
   j = -{N \over 4} + {1 \over 2} , -{N \over 4} + {3 \over 2},
   \ldots , {N \over 4} - {1 \over 2} \; .
\ee
The Bethe Equations can be written as
\be
   \arctan \lambda_j = \pi \; {j \over N}
   + {1 \over N} \sum_k \arctan
   \left( {\lambda_j - \lambda_k \over 2} \right) \; .
   \label{BAEarc}
\ee
In the $N \to \infty$ limit, the variable $x={j \over N}$ becomes continuous and limited in the range $-1/4 \le x \le 1/4$. The set of roots $\lambda_j$ turns into a function $\lambda (x)$ and (\ref{BAEarc}) becomes
\be
   \arctan \lambda (x) = \pi x +
   \int^{1/4}_{-1/4} \arctan \left( {\lambda(x) - \lambda (y) \over 2} \right) \de y \; .
   \label{BAEint}
\ee
As observables depend on (are best expressed in terms of) the rapidities $\lambda_j$ and not on the integers $I_{0,j}$, we like to perform a change of variables and integrate over $\lambda$ rather than $x$:
\be
   {1 \over N} \sum_j f(\lambda_j) 
   = \int^{1/4}_{-1/4} f \big( \lambda(x) \big) \de x
   = \int^\infty_{-\infty} f(\lambda) \rho_0(\lambda) \de \lambda \; ,
\ee
where the change of variables $x \to \lambda(x)$ maps the interval $-1/4 \le x \le 1/4$ into the whole real line $-\infty < \lambda < \infty$. More explicitly, the density $\rho_0 (\lambda)$ of real rapidities can be written as the Jacobian of the change of variable, that is
\be
   \rho_0(\lambda) = {\de x \over \de \lambda} =
   \left. {1 \over \lambda'(x)}	\right|_{x=\lambda^{-1}(\lambda)} \; .
\ee
Finally, differentiating (\ref{BAEint}) with respect to $\lambda$ we obtain a linear integral equation\index{Integral equation!linear} for the density $\rho_0 (\lambda)$:
\be
   \rho_0 (\lambda) = {1 \over \pi} {1 \over 1 + \lambda^2}
   - {1 \over \pi} \int^\infty_{-\infty}
   {2 \over (\lambda-\mu)^2 + 4} \; \rho_0 (\mu) \de \mu \; .
   \label{rho0inteq}
\ee
Notice that this integral equation is of the same type as the one we found for the Lieb-Liniger model\index{Lieb-Liniger model} (\ref{intbetheeq}) and can be cast in the same form by remembering the definition of the scattering phase\index{Scattering phase} (\ref{thetandef})
\be
   \rho_0 (\lambda) + {1 \over 2 \pi}
   \int_{-\infty}^\infty {\cal K} (\lambda - \mu) \; \rho_0 (\mu) \de \mu
   = {1 \over 2 \pi} \theta'_1 (\lambda) \; ,
   \label{XXXinteeq}
\ee
where we defined the kernel\index{Kernel}
\be
   {\cal K} (\lambda) \equiv {\de \over \de \lambda} \; \theta_2 (\lambda) =
   {4 \over \lambda^2 + 4} \; .
\ee

Since the support of this integral equation is over the whole real axis, it can be solved by Fourier transform:
\be
   \tilde{\rho}_0 (\omega) = \int_{-\infty}^\infty \eu^{-\ii \omega \lambda}
   \rho_0 (\lambda) \de \lambda \; .
\ee
Using
\be
   {1 \over \pi} \int {n \over \lambda^2 + n^2} \eu^{-\ii \lambda \omega} \de \lambda = \eu^{-n|\omega|} \; ,
\ee
we can turn the integral equation (\ref{rho0inteq}) into
\be
  \tilde{\rho}_0 (\omega) \; \left( 1 + \eu^{-2|\omega|} \right) = \eu^{-|\omega|} \; ,
\ee
which yields
\be
   \rho_0 (\lambda) = {1 \over 2 \pi} \int_{-\infty}^\infty
   \eu^{\ii \omega \lambda} \tilde{\rho}_0 (\omega) \de \omega
   = {1 \over 4 \cosh \left( {\pi \lambda \over 2} \right)} \; .
\ee
The momentum and energy of the ground state are then given by
\bea
   K & = & N \int p_0 (\lambda) \rho_0(\lambda) \de \lambda = {\pi \over 2} \; N \: {\rm mod} \: 2 \pi \equiv K_{\rm AFM}  \; , \\
   E & = & E_0 +  N \int \epsilon_0 (\lambda) \rho_0(\lambda) \de \lambda
   = N \left( {1 \over 4} - \ln 2 \right) \equiv E_{\rm AFM} \; ,
\eea
where $\rho_0 (\lambda)$ and $\epsilon_0 (\lambda)$ where defined in (\ref{pMdef}, \ref{epsilonMdef}). This result was originally derived by Hulthen \cite{hulthen38}.

We now look at states with $\nu_0=N/2-1$ and $\nu_M=0$ for $M \ge 1/2$.
They are characterized by two holes\index{Hole}: let us say that the empty quantum numbers are $j_1$ and $j_2$:
\be
   I_{0,j} = j + \vartheta_H (j-j_1) + \vartheta_H (j-j_2) \; ,
\ee
where $\vartheta_H (x)$ is the Heaviside step-function.
The integral equation \index{Integral equation!linear} for the real roots rapidity density $\rho_{\rm t}(\lambda)$ (where t stands for triplet) is
\be
   \rho_{\rm t}(\lambda) + {1 \over 2 \pi} \int^\infty_{-\infty}
   {\cal K} (\lambda - \mu) \; \rho_{\rm t} (\mu) \de \mu
   = {1 \over \pi} {1 \over 1 + \lambda^2}
   - {1 \over N} \left[ \delta (\lambda - \lambda_1)
   + \delta (\lambda - \lambda_2) \right] \; ,
   \label{linint}
\ee
where $\lambda_{1,2}$ are the images of	$x_1=j_1/N$ and $x_2=j_2/N$
under the map $x \to \lambda(x)$.
Since we are dealing with linear equations, we can write the solution of (\ref{linint}) as
\be
   \rho_{\rm t} (\lambda) = \rho_0 (\lambda)
   + {1 \over N} \left[ \tau ( \lambda - \lambda_1 )
   + \tau (\lambda - \lambda_2) \right] \; ,
   \label{rhotsol}
\ee
where $\tau (\lambda)$ solves the equation\index{Integral equation!linear}
\be
   \tau (\lambda) + {1 \over 2 \pi} \int^\infty_{-\infty}
   {\cal K} (\lambda-\mu) \; \tau (\mu) \de \mu
   = - \delta (\lambda) \; ,
   \label{taueq}
\ee
whose solution, in Fourier space, reads:
\be
  \tilde{\tau} (\omega) = {1 \over 1 + \eu^{-2 |\omega|}} \; .
\ee
Its real space form is a bit convoluted, but we can evaluate its contribution to the momentum and energy of the states by working in the $\omega$ space ($\tilde{p}_0 (\omega) = 2 \pi {\eu^{-|\omega|} \over \ii \omega}$):
\bea
   \int p_0 (\lambda) \tau (\lambda - \lambda') \de \lambda  & = &
   {1 \over 2 \pi} \int \tilde{p}_0 (\omega) \tilde{\tau} (-\omega)
   \eu^{\ii \omega \lambda'}
   =  \int {\eu^{-|\omega|} \over 1 + \eu^{-2|\omega|}} \;
   {\eu^{\ii \omega \lambda'} \over \ii \omega} \; \de \omega
   = \int \tilde{\rho}_0 (\omega)
   \left( \int^{\lambda'} \eu^{\ii \omega \lambda'} \de \lambda \right)
   \de \omega
   \nonumber \\
   & = & {\pi \over 2} \int^{\lambda'} \rho_0 (\lambda)
   = \arctan \left[ \sinh {\pi \lambda' \over 2} \right] \; ,
   \\
   \int \epsilon_0 (\lambda) \tau (\lambda - \lambda') \de \lambda & = &
   - \int p'_0 (\lambda) \tau (\lambda - \lambda' ) \de \lambda
   = - {\ii \over 2 \pi} \int \omega \; \tilde{p}_0 (\omega)
   \tilde{\tau} (-\omega) \eu^{\ii \omega \lambda'}
   =  \int \tilde{\rho}_0 (\omega) \; \eu^{\ii \omega \lambda'} \de \omega
   \nonumber \\
   & = & - {\pi \over 2} \; {1 \over \cosh {\pi \lambda \over 2}} \; .
\eea
Hence the total momentum and energy of the state given by (\ref{rhotsol}) are
\bea
   K & = & N \int p_0 (\lambda) \; \rho_{\rm t} (\lambda) \de \lambda
   = K_{\rm AFM} + k (\lambda_1) + k (\lambda_2) \; ,
   \label{Pklam} \\
   E & = & N \int \epsilon_0 (\lambda) \; \rho_{\rm t} (\lambda) \de \lambda
   = E_{\rm AFM} + \varepsilon (\lambda_1) + \varepsilon (\lambda_2) \; ,
   \label{Ehlam}
\eea
where
\be
   k (\lambda) \equiv {\pi \over 2} -  \arctan \sinh {\pi \lambda \over 2} \; , \qquad \qquad
   \varepsilon (\lambda) \equiv {\pi \over 2 \cosh {\pi \lambda \over 2} } \; .
   \label{klam}
\ee
The state we constructed was first considered in \cite{pearson62} and has two excitations over the ground state (spinons\index{Spinon}). The spinon's dressed energy\index{Dressed!energy} and momentum\index{Dressed!momentum} are (\ref{klam}). Combining the two, we find that each of these excitations are characterized by the dispersion relation
\be
   \varepsilon (k) = {\pi \over 2} \sin k \; , \qquad \qquad \qquad
   -{\pi \over 2} \le k \le {\pi \over 2} \; .
  \label{dispe}
\ee

\begin{wrapfigure}{r}{9cm}
	\vspace{-10pt}
	\begin{center}
		\includegraphics[width=8.5cm]{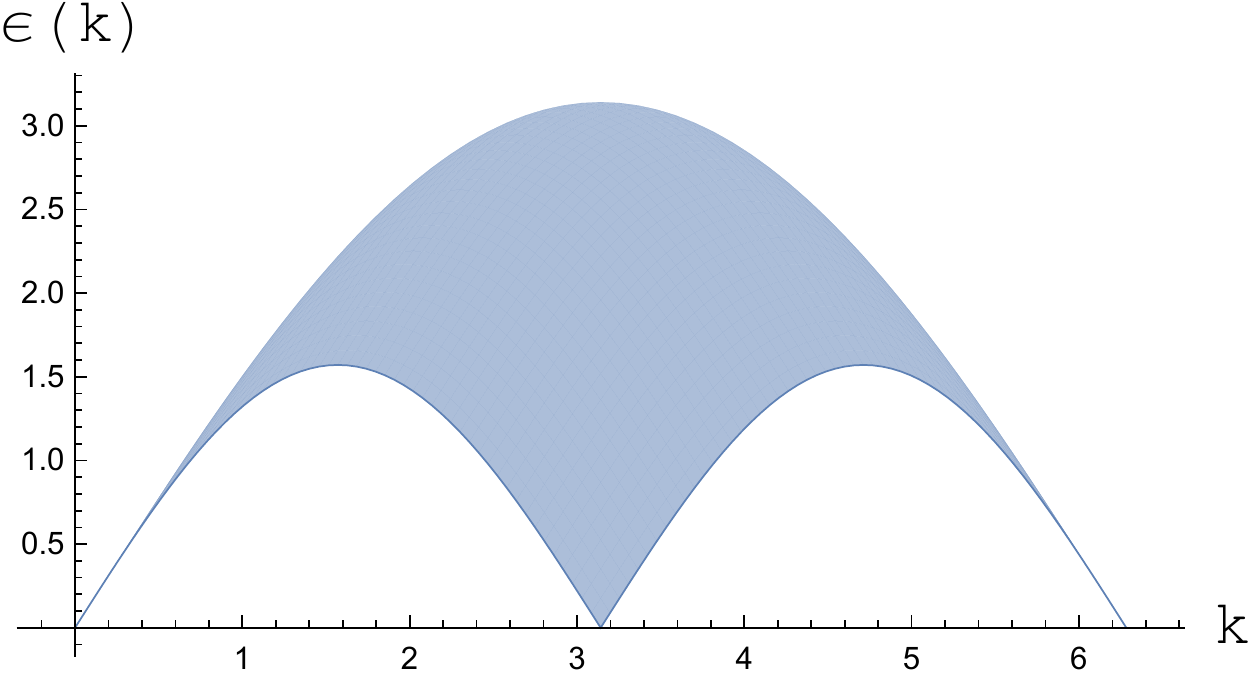}
	\end{center}
	\caption{Low energy dispersion relation emerging from (\ref{Pklam},\ref{Ehlam}), reflecting the two-spinon continuum (\ref{dispe}).}
	\label{fig:2spinons}
	\vspace{-10pt}
\end{wrapfigure}
Each hole\index{Hole} in the quantum numbers generates a quasi-particle\index{Quasi-particle} excitation, which is called a {\it spinon}\index{Spinon}, i.e. an spin-$1/2$ excitation. Spinons only exist as collective excitations (since flipping a spin-$1/2$ creates a spin-$1$ excitation) and they are an example of {\it fractionalization} that commonly happens in one dimension. Spinons are thus emergent excitations that exist over a ground state which is far from a vacuum state: they carry a signature on top of a structured ground state and thus would be hard to understand and visualize without the help of explicit examples like the Heisenberg chain\index{XXX chain}\index{Heisenberg chain} (the simplest example of spinon is in fact the {\it domain wall} excitation of the N\'eel state in an AFM Ising chain, created by flipping every spin after a given reference point\footnote{Such configuration is also a beautiful physical proof of the mathematical identity $\sum_{n=1}^\infty (-1)^n = - {1 \over 2}$, which is otherwise obtained through analytical continuation.}).

Individual spinons\index{Spinon} cannot be excited in a chain with an even number of sites (while they can be present with an odd number of sites, due to the degeneracy between states with $R ={N-1 \over 2}$ and $R = {N+1 \over 2}$). However, the  dispersion relations of a pair of spinons is very different from the one of a pure spin-$1$ excitation made by one magnon, since the latter is a simple line, while the former shows its composite nature in that it makes a band, as shown in Fig. \ref{fig:2spinons} (in the same way, we distinguished two-magnon excitations from their bound state\index{Bound state} in section \ref{sec:XXX2body}).
Also, notes that while the dispersion relation of long-wavelength magnons is quadratic, that of spinons is linear, see (\ref{dispe}) as well as Fig. \ref{fig:2spinons}.

Let us now look into the last state we considered before, i.e. the one with $\nu_0=N/2-2$, $\nu_{1/2}=1$, and $\nu_M=0$ for $M\geqslant1$.
For the density of real roots $\rho_{\rm s}(\lambda)$ (s is for singlet) we get the integral equation\index{Integral equation!linear}
\be
   \rho_{\rm s} (\lambda) + {1 \over 2\pi} \int^\infty_{-\infty}
   {\cal K} (\lambda-\mu) \; \rho_{\rm s} (\mu) \de \mu
   = {1 \over \pi} {1 \over 1 + \lambda^2}
   - {1 \over N} \left[ \delta (\lambda - \lambda_1)
   + \delta (\lambda - \lambda_2)
   + {1 \over \pi} \; \theta'_{0,1/2} (\lambda - \lambda_{1/2} ) \right] \; ,
   \label{denseq}
\ee
where $\lambda_{1,2}$ stand for the holes and the last term	in the RHS is the contribution from the scattering off the type $1/2$-complex with rapidity $\lambda_{1/2}$, which is the solution of the Bethe equations (see \ref{BEMj}):\index{Bethe!equations}
\be
   2 \arctan {\lambda_{1/2} \over 2} = {1 \over N} \sum_j
   \theta_{1/2,0} \left( \lambda_{1/2} - \lambda_{0,j} \right)
   = \int_{-\infty}^\infty
   \theta_{1/2,0} \left( \lambda_{1/2} - \lambda \right)
   \rho_{\rm s} (\lambda) \de \lambda \; ,
   \label{arctgF}
\ee
where, using (\ref{thetaMM}), we have
\be
   \theta_{1/2,0} \left( \lambda \right) =
   2 \arctan \lambda + 2 \arctan {\lambda \over 3}
\ee
and in the second line we took the continuous limit for $N \to \infty$.
As explained before, in (\ref{arctgF}) the type $1/2$-complex's Bethe number is $I_{1/2,1}=0$, since its allowed range is limited to just one point.
The solution of (\ref{denseq}) is
\be
   \rho_{\rm s} (\lambda) = \rho_0 (\lambda)
   + {1 \over N} \left[ \tau (\lambda - \lambda_1)
   + \tau (\lambda - \lambda_2)
   + \sigma (\lambda - \lambda_{1/2}) \right] \;\; ,
\ee
where $\tau(\lambda)$ is given by (\ref{taueq}) and $\sigma (\lambda)$ is the solution of
\be
   \sigma (\lambda) + {1 \over 2\pi} \int^\infty_{-\infty}
   {\cal K} (\lambda-\mu) \; \sigma (\mu) \de \mu
   = - {1 \over \pi} \; \theta'_{0,1/2} (\lambda) \; ,
   \label{sigmaeq}
\ee
which, in Fourier space, reads
\be
   \tilde{\sigma} (\omega) \left( 1 + \eu^{-2 |\omega|} \right) =
   - \left( \eu^{-|\omega|} + \eu^{-3 |\omega|} \right) 
   \qquad \Rightarrow \qquad 
   \tilde{\sigma} (\omega) = - \eu^{-|\omega|} \; .
   \label{sigmasol}
\ee

To evaluate $\lambda_{1/2}$ we can rewrite (\ref{arctgF}) as
\bea
   2 \arctan {\lambda_{1/2} \over 2} & = &
   \int^\infty_{-\infty} \theta_{1/2,0} \left( \lambda_{1/2} - \lambda \right) \rho_0(\lambda) \de \lambda
   \label{lambda12int}
   \\
   && + {1 \over N} \int^\infty_{-\infty}
   \theta_{1/2,0} \left( \lambda_{1/2} - \lambda \right)
   \left[ \tau (\lambda - \lambda_1) + \tau (\lambda - \lambda_2)
   + \sigma ( \lambda - \lambda_{1/2}) \right] \de \lambda \; .
   \nonumber
\eea
The last term in the RHS vanishes due to the oddness of the integrand $\theta_{1/2,0} (\lambda) \sigma (\lambda)$. Moreover, we have
\bea
    \int \theta_{1/2,0} \left( \lambda_{1/2} - \lambda \right) \rho_0(\lambda) \de \lambda & = &
    {1 \over 2 \pi} \int \tilde{\theta}_{1/2,0} (\omega) \;
    \tilde{\rho}_0 (\omega) \; \eu^{\ii \omega \lambda_{1/2}} \de \omega
    =  \int \left[ {\eu^{-|\omega|} \over \ii \omega} +
    {\eu^{-3|\omega|} \over \ii \omega} \right]
    {\eu^{-|\omega|} \over 1 + \eu^{-2|\omega|}} \;
    \eu^{\ii \omega \lambda_{1/2}} \de \omega
    \nonumber \\
    & = & \int {\eu^{-2|\omega|} \over \ii \omega} \;
    \eu^{\ii \omega \lambda_{1/2}} \de \omega
    = \int^{\lambda_{1/2}} \de \lambda' \int
    \eu^{-2|\omega|} \; \eu^{\ii \omega \lambda'} \de \omega
    = \int^{\lambda_{1/2}} {4 \over \lambda'^2 +4 } \; \de \lambda
    \nonumber \\
    & = & 2 \arctan {\lambda_{1/2} \over 2} \; .
\eea
This means that (\ref{lambda12int}) reduces to
\bea
   \int^\infty_{-\infty}
   \theta_{1/2,0} \left( \lambda_{1/2} - \lambda \right)
   \left[ \tau (\lambda - \lambda_1) + \tau (\lambda - \lambda_2)
   \right] \de \lambda
   & = &  {1 \over 2 \pi} \int \tilde{\theta}_{1/2,0} (\omega) \;
   \tilde{\tau} (\omega) \; \eu^{\ii \omega \lambda_{1/2}}
   \left( \eu^{-\ii \omega \lambda_1} + \eu^{-\ii \omega \lambda_2} \right)
   \de \omega
   \nonumber \\
    =  \int {\eu^{- |\omega|} \over \ii \omega}
   \left[ \eu^{\ii \omega (\lambda_{1/2} - \lambda_1)} +
   \eu^{\ii \omega (\lambda_{1/2} - \lambda_2)} \right]
   \de \omega
   & = & \arctan (\lambda_{1/2} - \lambda_1)
   + \arctan (\lambda_{1/2} - \lambda_2) =0  \; ,
\eea
i.e.
\be
	\lambda_{1/2} ={\lambda_1 + \lambda_2 \over 2} \; .
\ee
Thus, the rapidity of the type $1/2$-complex is fixed and determined by the rapidities of the holes\index{Hole}. This is a consequence of the lack of freedom in choosing a quantum number for this excitation.

Moreover, if we evaluate the momentum for this state we find:
\bea
   K & = & N \int p_0 (\lambda) \; \rho_{\rm s} (\lambda) \de \lambda
   + p_{1/2} \left( \lambda_{1/2} \right)
   = K_{\rm AFM} + k (\lambda_1) + k (\lambda_2)
   + \int p_0 (\lambda) \; \sigma \left( \lambda - \lambda_{1/2} \right)
   + 2 \arctan {\lambda_{1/2} \over 2}
   \nonumber \\
   & = & K_{\rm AFM} + k (\lambda_1) + k (\lambda_2) \; ,
\eea
since, using (\ref{sigmasol}),
\be
   \int p_0 (\lambda) \; \sigma \left( \lambda - \lambda_{1/2} \right)
   =  {1 \over 2 \pi} \int \tilde{p}_0 (\omega) \; \tilde{\sigma} (-\omega) \; \eu^{\ii \omega \lambda_{1/2}} \de \omega
   = - \int {\eu^{-2|\omega|} \over \ii \omega} \;
   \eu^{\ii \omega \lambda_{1/2}} \de \omega
   = - 2 \arctan {\lambda_{1/2} \over 2} \; .
\ee
Similarly, for the energy
\be
   E = E_{\rm AFM} + N \int \epsilon_0 (\lambda) \; \rho_{\rm s} \de \lambda
   + \epsilon_{1/2} \left( \lambda_{1/2} \right)
   = E_{\rm AFM} + \varepsilon (\lambda_1) + \varepsilon (\lambda_2) \; ,
\ee
which can be easily derived from the previous result remembering that $\epsilon_M (\lambda) = - {\de \over \de \lambda} k (\lambda)$.
Hence, we see that the contributions from the string cancel out and this state has exactly the same momentum, energy (and dispersion relation) as the one without complexes that we calculated before (\ref{Pklam}, \ref{Ehlam}, \ref{dispe}). In particular, the two excitations in both cases obey (\ref{klam}).

Thus, we saw that these two families of states with two holes\index{Hole} in the distribution of purely real roots have the same energy and momentum (when the same holes are taken in the two cases) and they only differ in their total spin, being $S^z=1$ in the first case and $S^z = 0$ in the latter.
One notices that, since applying the operator $S^+ \equiv \sum_{n=1}^N S_n^+$ to any of these states kills it, these are highest-weight states. This supports the interpretation of each hole\index{Hole} excitation as a spin-$1/2$ excitation (spinon)\index{Spinon}. In the first case we described the combination of two excitations into a triplet (in its highest-weight state $S^z =1$), while in the second we got a singlet ($S^z = 0$).

For general $\kappa$ the same picture holds: the states with $\nu_0 = N/2 - \kappa$ and $\nu_M = 0$ for $M \ge 1/2$ are $2 \kappa$-spinons\index{Spinon} states in the highest-weight state of spin $S^z = \kappa$. All other states with the same $\kappa$ have lower magnetization, entering into multiplets with a number of particles non exceeding $2 \kappa$. In all these cases, the contribution of $M$-complexes\index{M-complex} to the energy and momentum identically vanishes and so the energy/momentum depend only on the number of particles, i.e. on the holes\index{Hole} in the purely real solutions. Note that these multiplets are exactly degenerate only at the Heisenberg point (in zero external magnetic field) and will get split in the general XXZ model.

Even if only spin-$1$ excitations are observed in the Heisenberg chain\index{XXX chain}\index{Heisenberg chain}, we see that these excitations are not pure {\it magnons}\index{Magnon}, but a combination of an even number (since the number of particles is $2 \kappa$) of spin-$1/2$ excitations ({\it spinons})\index{Spinon} with dispersion relation (\ref{dispe}). Note that the dispersion relation for each spinon is defined only on half of the Brillouin zone, while the dispersion for the integer spin collective excitation is defined for $-\pi \le k \le \pi$.

All the states we described so far are highest-weight states. To lower the magnetization in a multiplet by one we place an extra rapidity at infinity, which corresponds to an excitation with zero-momentum. This amounts to adding a quantum number at $I_M^{(\infty)}$ (remember that by adding one particle, one has to shift all the quantum numbers by $1/2$), which leaves the existing rapidities unaffected. Additional complexes at infinity generate all the members in a given multiplet.

We have sketched how all excitations can be constructed in terms of spinons \index{Spinon} over the anti-ferromagnetic ground state $|{\rm AFM}\rangle$, which can thus be deemed the spinon-vacuum. This point of view is particularly suitable to describe the anti-ferromagnetic phase. We saw that in the ferromagnetic one, instead, the ground state is a magnon-vacuum and excitations are  magnons and their bound states\index{Bound state}. As the whole Hilbert space can be described in either pictures, this is a reminder of how powerful the Bethe Ansatz construction is, but also that the classification of a many-body state in terms of elementary excitations should be used as a tool to understand its properties and not as an actual ``factorization'' into quasi-particles (as it was the case for free fermionic systems such as the XY chain studied in chapter \ref{chap:XYModel}). Moreover, while certain states are easily decomposed in terms of spinons and others in terms of magnons, the reverse is often complicated and some states will lie in between.

\section{Interaction with a magnetic field}
\label{sec:XXXh}

In the presence of a magnetic field $h$, the Hamiltonian (\ref{H}) is supplemented by a Zeeman energy:
\begin{equation}
   \label{eq:Hh}
   {\cal H} = -J\sum_{n=1}^N {\bf S}_n \cdot {\bf S}_{n+1} -  2 h \sum_{n=1}^N S_n^z \; .
\end{equation}
As the Hamiltonian commutes with the total magnetization, the magnetic field does not affect the eigenstates and only alters their eigenenergies. For the ferromagnetic case $J>0$, the ground state remains the fully polarized one and the magnetic field only splits the energy of the elements in each multiplets.

In the anti-ferromagnetic case $J<0$, the two parts of ${\cal H}$ are in competition. Spin alignment in the positive $z$-direction is energetically favored by the magnetic field, while the interaction penalizes any aligned nearest-neighbor pair of spins.
Clearly, for $h \to \infty$ the magnon vacuum $|0 \rangle$ with maximal magnetization $S_z = N/2$ is the lowest energy state, while at $h =0$ we constructed the ground state as the spinon vacuum $|{\rm AFM} \rangle$ with $S_z = 0$, out of the $R=N/2$ real rapidities. Switching on a positive magnetic field does not affect the energy of $|{\rm AFM} \rangle$, but progressively lowers the energies of higher magnetization states. In particular, for a given magnetization $S_z$, the highest weight state of $S = S_z$ minimizes the interaction energy and is thus the energetically favored one. Thus, for any value of $h$, the ground state is constructed out of real magnon rapidities. 

For a given magnetization $S^z=N/2-R$, the lowest energy state is given by $\nu_0 = R$ and $\nu_M = 0$ for $M > 0$ and the Bethe quantum numbers for the $0$-complexes of this state are
\be
  I_{0,j} ={1 \over 2} \left( 2 j - R- 1 \right) \: ,  
  \qquad j=1,\ldots,R \; .
  \label{finitehIs}
\ee
Starting with $|{\rm AFM}\rangle$ ($R=N/2$) at $h=0$ and increasing $h$ we will have a first level crossing with the $R=N/2-1=\nu_0$ state, which will become the ground states until at a higher $h$ it will be taken over by the $R=N/2-2=\nu_0$ state and so on until at $h_{\rm s}$ the $R=0=\nu_0$ state $|0\rangle$ becomes the lowest energy state and remains so for ever larger $h$, having saturated the possible magnetization. As the Zeeman term contributes linearly in the $S^z$ to the energy of each state, the level crossings between the lowest energy states in sectors of progressive magnetization is equal to (half) the energy gap between these states in the absence of a magnetic field $\Delta E_R = E_{GS}^{(R)} - E_{GS}^{(R+1)}$, which increases with $S^z$. Thus, the last level crossing, which is between the $R=1$ state with one $\pi$-momentum magnon and $|0\rangle$, happens for $h_{\rm s} = \Delta E_0/2$. From (\ref{e1k}) we have $\Delta E_0 = 2J$ (which is the biggest gap in this staircase) and thus the saturation field at which the polarized state becomes the absolute ground state is $h_{\rm s} = J$

Although the magnetization changes at finite intervals at each level crossing, in the thermodynamic limit this staircase structure can be approximated by a smooth line, whose derivative gives the magnetic susceptibility.
At finite $h$, in the $N \to \infty$ limit the integral equation (\ref{XXXinteeq})\index{Integral equation!linear} becomes
\be
  \rho_0 (\lambda) 
  + {1 \over 2 \pi}
  \int_{-\Lambda}^\Lambda {\cal K} (\lambda - \nu) \; \rho_0 (\mu) \de \mu 
  = {1 \over 2 \pi} \theta'_1 (\lambda)\; ,
\label{XXXinteeqh}
\ee
where the finite support of the rapidity density reflects (\ref{finitehIs}).
A change in $h \to h + \delta h$ changes the support $\Lambda \to \Lambda + \delta \Lambda$ and thus the density $\rho_0 (\lambda)  \to \rho_0 (\lambda) + \delta \rho_0 (\lambda)$.
The energy and magnetization of the state defined by (\ref{XXXinteeqh}) is
\bea
  e & = & {E \over N} = {|J| - 4 h \over 4} +
  2 \int_{-\Lambda}^{\Lambda} \left[ h - {|J| \over \lambda^2 +1}\right] 
  \rho_0 (\lambda) \de \lambda \; , 
  \label{XXXeh} \\
  S_z & = & {N \over 2} - N \int_{-\Lambda}^{\Lambda} \rho_0 (\lambda) \de \lambda \; .
  \label{XXXSz}
\eea

The ground state condition is ${\partial e \over \partial \Lambda}=0$.
In section \ref{sec:XXZparah} we show that the ground state condition is equivalent to the following relation between $h$ and $\Lambda$:
\be
   h = \pi |J| {\rho_0 (\Lambda) \over Z(\Lambda)} \; ,
   \label{hlambda}
\ee
where $Z(\lambda)$ is the {\it dressed charge}\index{Dressed!charge} defined by (\ref{Zdef1}) and where the prefactor difference compared to (\ref{XXZhlambda}) is due to the difference in the dispersion relation between (\ref{XXXepsilon0},\ref{XXXeh}) and (\ref{XXZepsilon0}). The ground state magnetization for a given $h$ is obtained inverting (\ref{hlambda}) and inserting it into (\ref{XXXSz}) with the solution of (\ref{XXXinteeqh}).

For $h << 1$, $\Lambda >>1$, one can take advantage of the Fourier transform solution available for $\Lambda \to \infty$ and extract the small magnetic field result through the Wiener-Hopf factorization \cite{yangyangXXZ,takahashi}
\be
   h = h_0 \: \eu^{-{\pi \over2} \Lambda } 
   \left[ 1 - {1 \over 2 \pi \Lambda} + \Ord \left( {1 \over \Lambda^2} \right) \right] \; , \qquad \qquad h_0 \equiv |J| \: \sqrt{\pi^3 \over 2 \eu}  \; ,
   \label{XXXhLambda}
\ee
and the susceptibility (i.e. the proportionality coefficient between the applied magnetic field and the system's magnetization) is
\be
   \chi (h) = {S_z \over h} = {2 \over \pi^2 |J|} \left[ 1 + 
   {1 \over 2 \ln {h_0 / h} } 
   + \Ord \left( {\ln |\ln h| \over (\ln h)^2 } \right) \right] \; ,
   \label{XXXchi}
\ee
which shows that the susceptibility starts at a finite value with a diverging (vertical) slope, due to the logarithmic singularity in (\ref{XXXchi}). This finite initial value is quite elusive to numerical simulation because of the singularity in the approach. Comparison with (\ref{dispe}) shows that it obeys the general formula
\be
   \chi (0) = {1 \over \pi \:  v_F} \; ,
\ee
where $v_F = {\pi |J| \over 2}$ is the {\it Fermi velocity}, that is, the velocity of a low energy spinon.

Starting from this initial value, the susceptibility grows monotonically with $h$, and
finally diverges at the saturation field $h_{\rm s}$ as \cite{mullernotes}
\begin{equation}
  \label{eq:chi-sat}
  \chi (h) \stackrel{h\to h_{\rm s}}{\longrightarrow}
  \frac{1}{\pi}\frac{1}{\sqrt{J(h_{\rm s}-h)}} \; .
\end{equation}

\chapter{The XXZ Chain}
\label{chap:XXZmodel}

\abstract{
The XXZ spin chain is an integrable generalization of the Heisenberg chain that accounts for a uni-axial anisotropy in the spin interaction. Its Bethe Ansatz solution is a ``straightforward'' generalization of the one employed in the previous chapter, but the classification of complex roots is more involved and the nature of the low energy excitations changes with the anisotropy.
After previewing the phase diagram of the chain in Sec. \ref{sec:XXZGeneralities}, we recap the coordinate Bethe Ansatz solution in Sec. \ref{sec:XXZBetheSol}. We then analyze the different phases in Sec. \ref{sec:XXZFerro}, \ref{sec:XXZplanar}, and \ref{sec:XXZAFM}, by focusing on the physical properties and skipping some technical derivations.
}

\section{Generalities}
\label{sec:XXZGeneralities}

We consider the spin chain defined by the Hamiltonian:\index{XXZ chain}
\be
   {\cal H} = - J \sum_{n=1}^N \left[
   S_n^x S_{n+1}^x + S_n^y S_{n+1}^y +
   \Delta \; S_n^z S_{n+1}^z \right]
   - 2h \sum_{n=1}^N S_n^z ,
   \label{XXZham}
\ee
with the usual periodic boundary conditions.
Here, $S_n^\alpha = {1\over2} \sigma_n^\alpha$, where $\sigma_n^\alpha$ are the Pauli matrices.
For $\Delta =1$ we recover the Heisenberg chain we discussed in the previous chapter.

For $J>0$ the {\it ferromagnetic} order is preferred along the $x-y$ plane, while when $J<0$ we have an {\it antiferromagnet} in the plane. The parameter $\Delta$ sets the strength of the uniaxial anisotropy along the $z$ direction competing with the planar $x-y$ term: it distinguishes a planar regime ($|\Delta|<1$) from the axial ones ($|\Delta|>1$). For $|\Delta|>1$ we have a {\it ferromagnet} along the $z$ direction for $J \Delta > 0$ and an {\it antiferromagnet} when $J \Delta < 0$. In the literature, sometimes phases are named after the planar ordering and other times after the axial one and one should pay attention in order to avoid ambiguities.
We will henceforth assume $J>0$, remembering that the anti-ferromagnetic case ($J<0$) can be obtained through a $\pi$-rotation around the $z$ axis of every-other spin, followed by the transformation $\Delta \to - \Delta$.

The model has three phases, depicted in Fig. \ref{fig:XXZPhaseDiag}:
\begin{itemize}

	\item For $\Delta>1$ we have a gapped ferromagnet. This phase is best understood starting from the $\Delta \to \infty$ limit, where the ground state is $ |0 \rangle \equiv \prod_{j=1}^N | \uparrow_j \rangle$ (for $h=0$ degenerate with the one obtained by flipping every spin) and low energy excitations are individual magnons\index{Magnon} or their bound states\index{Bound state}.
	
	\item For $\Delta =1$ we have an isotropic ferromagnet (described in chapter \ref{chap:XXXmodel}), with a gapless spectrum. 
	
	\item For $h=0$, the chain remains critical in the interval $|\Delta|\le 1$: the ground state has zero magnetization and it is a paramagnet characterized as a spinon vacuum. The low energy states are thus spinons\index{Spinon}, but for $\Delta >0$ magnons\index{Magnon} also become stable and for $\cos\big( \pi/(m+1)\big) < \Delta <1$ bound states\index{Bound state} of $m$ magnons appear in the low energy sector. At $\Delta =0$ the model becomes the XX chain\index{XX chain}, the isotropic limit of the XY chain described in chapter \ref{chap:XYModel}. It is therefore no longer interacting and can be described, through the Jordan-Wigner transformation\index{Jordan-Wigner!transformation}, as a system of free spinless fermions hopping on a lattice. For non-zero magnetic fields, the ground state acquires a finite magnetization, but the systems remains in a gapless phase up to the critical value $h_{\rm s} = {J \over 2} \left( 1 - \Delta \right)$, past which the ground state has saturated to $|0 \rangle$. For higher $h$ we reenter the gapped ferromagnetic phase in continuity with $\Delta >1$.
	
	\item At $\Delta=-1$ the model is equivalent (up to the rotation of every other spin around the $z$ axis) to the AFM Heisenberg chain analyzed in the previous chapter.
	
	\item For $\Delta <-1$, at zero magnetic field, a gap opens again and this phase is dominated by the Ising AFM of the $\Delta \to -\infty$ limit, with two degenerate N\'eel ground states $|N_1 \rangle \equiv |\uparrow \downarrow \uparrow \downarrow \uparrow \ldots \rangle$ and $|N_2 \rangle \equiv |\downarrow \uparrow \downarrow \uparrow \downarrow \ldots \rangle$. In this limit, the low-energy excitations are constructed in terms of {\it domain walls}, i.e. regions where one type of N\'eel order changes into the other, thus creating two consecutive ferromagnetically aligned spins. These states are created by flipping a potentially macroscopical number of spins (that is, the number of sites between two consecutive domain walls), but their energy cost only lies at the boundaries and does not depend on the number of flipped spins. These are collective (fractionalized) excitations: each domain wall carries spin $S=1/2$ and is the simplest example of a spinon\index{Spinon}. This picture is qualitatively valid for finite $\Delta <-1$, although the structure of the ground state and of the gapped spinon is more complicated. The ground state has zero magnetization and remains gapped up to a critical value of the magnetic field $h_{\rm c}$ (\ref{hcdef}), past which the model becomes gapless and the ground state magnetization starts increasing as a function of $h$. This region is in continuity with the paramagnetic one that starts for $h=0$ at $|\Delta|<1$. At $h_{\rm s}$ the ground state becomes fully polarized and for larger magnetic fields the phase is again the gapped ferromagnet.

\end{itemize}

\begin{figure}[t]
	\noindent\begin{minipage}[t]{8.5cm}
		\includegraphics[width=8.5cm]{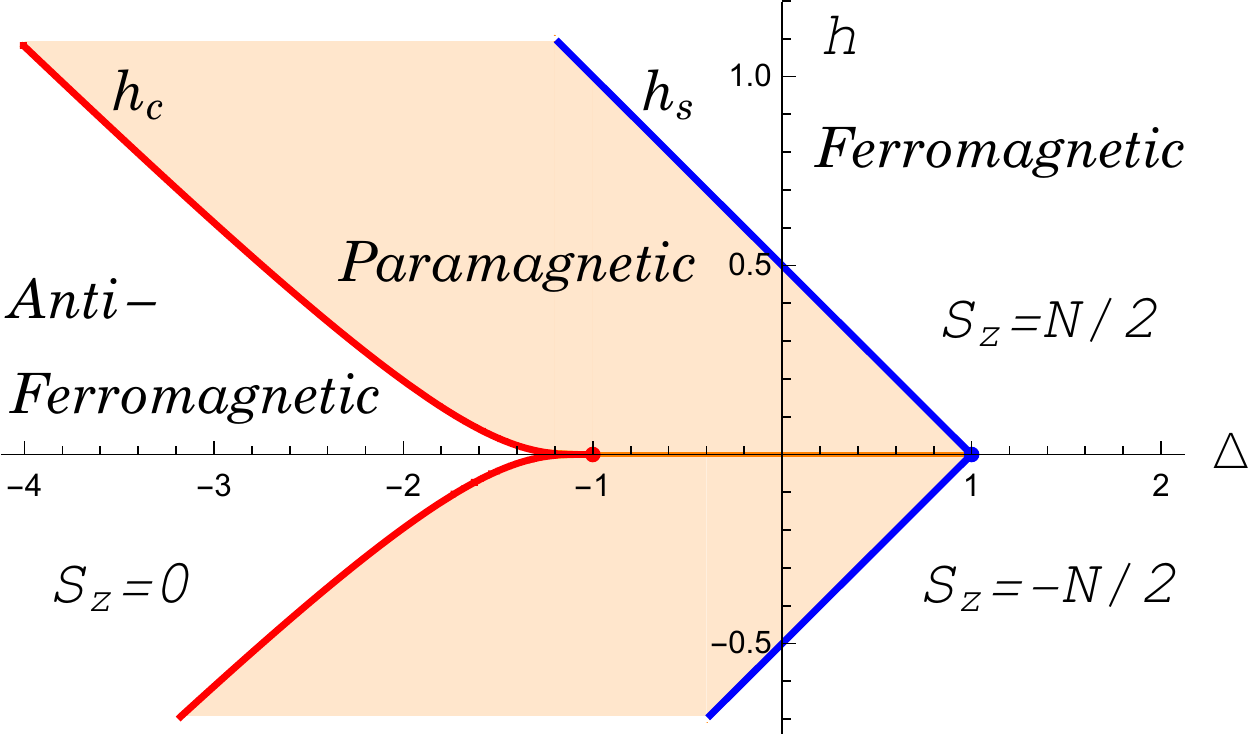}
	\end{minipage}
	\hfill
	\begin{minipage}[b]{7.5cm}
		\caption{Phase diagram of the XXZ chain. The shaded region is gapless and described by a $c=1$ conformal field theory (with varying compactification radius) in the scaling limit. To its left there is an anti-ferromagnetic phase: the ground state has zero magnetization along the $z$-axis of the external magnetic field $h$ and excitations are gapped spinons. To the right, the phase is ferromagnetic: the ground state has maximal magnetization and the low energy excitations are gapped magnons. In the paramagnetic phase, the ground state magnetization varies from zero at vanishing magnetic field and on the critical $h_{\rm c}$ line (\ref{hcdef}) separating this phase from the AFM one, to maximal at the saturation field $h_{\rm s}$ (\ref{hFM}), above which the system becomes ferromagnetic.}
		\label{fig:XXZPhaseDiag}
	\end{minipage}
\end{figure}

The phase transition at $\Delta = -1$ is of infinite order, of the Berezinsky–Kosterlitz–Thouless (BKT) type: although the excitation spectrum goes from massive to gapless, the ground state energy and all its derivatives can be shown to be continuous across the point. At $\Delta =1$ the ground state energy is continuous, but 
this transition has elements similar to a first order one, since the isotropic ferromagnet has a massively degenerate ground state (corresponding to the multiplets of the $S=N/2$ states), which is immediately removed away from this point.

\section{Bethe Ansatz solution}
\label{XXZBetheSec}
\label{sec:XXZBetheSol}

The solution of the XXZ Hamiltonian (\ref{XXZham}) proceeds exactly as for the isotropic case. Thus, we consider states with a given magnetization along the $z$-axis 
\be
   S^z \equiv \sum_{n=1}^N S_n^z\; ,
   \label{mag}
\ee
which is conserved by (\ref{XXZham}): $\left[ {\cal H}, S^z \right] = 0$. Since a rotation along the $x$-axis followed by the transformation $h \to -h$ leaves the Hamiltonian unchanged, we can consider only the case $0 \le S^z < N/2$.

We can start with the reference state $|0\rangle$ with all spins up ($S^z = N/2$). As the only state in this magnetization sector, is also an eigenstate:
\be
   {\cal H} |0\rangle = E_0 \; |0\rangle \; , \qquad
   E_0 = - \left( {J \Delta \over 4} + h \right) N \; .
\ee

If we flip one spin, we have $N$ possible states in this sector with magnetization $S^z= {N \over 2} -1$, corresponding to all the sites where the spin can be flipped. As we flip more spins, the dimension of the Fock space of states with a given magnetization increases very quickly (exponentially). We can write the generic state with $R$ spin-flips as
\be
  | \Psi \rangle = \sum_{\{n_l\}} f \left( n_1, n_2, \ldots , n_R \right)
  |n_1, n_2, \ldots , n_R \rangle \; ,
  \label{betheans}
\ee
where the sum is over all the choices of $R$ lattice sites out of $N$ and
\be
  |n_1, n_2, \ldots , n_R \rangle \equiv S_{n_1}^- S_{n_2}^- \ldots
  S_{n_R}^- | 0 \rangle
\ee
is the state with $R$ spins flipped at the lattice sites $\{n_l\}$. We order the coordinates such that $1 \le n_1 < n_2 < \ldots < n_R \le N$. This state has magnetization $S^z = {N \over 2} - R$. For the above-mentioned symmetry, we can take $R \le N/2$ and we notice that only for {\it even} $N$ we can have a $SU(2)$ invariant state, i.e. $S^z= 0$, while for odd $N$ the magnetization is a half integer.

Instead of determining the two-body scattering phase\index{Scattering phase} by considering a system of just two particles interacting like we did in the previous chapters, let us apply the Hamiltonian (\ref{XXZham}) directly to the state (\ref{betheans}) to determine the eigenstate consistency conditions, which, as usual, can be factorized as a sequence of pairs of permutation events.

The eigenvalue equation for the Hamiltonian (\ref{XXZham}) using the wave function (\ref{betheans}) is
\bea
    \left( {\cal H} - E  \right) \Psi & = &
    - {J \over 2} \sum_{j=1}^R \left(1 - \delta_{n_j+1,n_{j+1}} \right) \Big[
    f \left( n_1, \ldots , n_j +1, n_{j+1}, \ldots , n_R \right) 
    + f \left( n_1, \ldots , n_j, n_{j+1} -1, \ldots , n_R \right) \Big]
    \nonumber  \\
    && + \left[ E_0 - E + (J \Delta + 2 h) R - J \Delta \sum_{j=1}^R \delta_{n_j +1,n_{j+1}}
    \right]  f \left( n_1, n_2, \ldots , n_R \right) = 0 \; ,
    \label{eigeneq}
\eea
where we dropped writing the spin part of the wave-function, as it is assumed to be paired in an obvious way to the coordinate part.
Notice that acting with the Hamiltonian on the state, leaves a diagonal part and a series of terms involving only two-particle (nearest neighbor) interaction.

As usual, we write the ansatz for the coordinate wave function as a superposition of plane-waves\footnote{As we saw in the previous chapter, the assumptions of using plane-waves as a basis is not restrictive, as later we will find a change of variable that gives a more appropriate basis for the expansion.}:\index{Bethe!ansatz wavefunction}
\be
   f \left( n_1, n_2, \ldots , n_R \right) \equiv
   \sum_\cP^{R!} A [\cP] \eu^{\ii \sum_{j=1}^R k_{\cP j} n_j}
   = \Omega_R  \sum_\cP^{R!} \exp \left[ \ii \sum_{j=1}^R k_{\cP j} n_j +
   {\ii \over 2} \sum_{j<l}^R \tilde{\Theta} (k_{\cP j}, k_{\cP l}) \right] \; ,
   \label{bethewf}
\ee
where the sum is over the permutation $\cP$ of the quasi-momenta $k_j$ and where in the second equality we factorized a (yet undetermined) normalization constant $\Omega_R$ and wrote the expansion as to emphasize the scattering phase\index{Scattering phase}.

The Bethe wave function (\ref{bethewf}) has total lattice momentum $K = \left( \sum_{j=1}^R k_j \right) \: {\rm mod} \: (2 \pi)$,
and is an eigenfunction of (\ref{eigeneq}) with eigenvalue
\be
   E = E_0 + \sum_{l=1}^R \left[ J ( \Delta - \cos k_l ) + 2 h \right]
   = E_0 + (J \Delta + 2 h) R - \sum_{l=1}^R \cos k_l
\ee
if
\be
   A[\cP] \left( \eu^{\ii k_{\cP j}} + \eu^{-\ii k_{\cP (j+1)}} - 2 \Delta \right)
   \eu^{\ii k_{\cP (j+1)} } +
   + A[\cP (j,j+1)] \left( \eu^{\ii k_{\cP (j+1)}} + \eu^{-\ii k_{\cP j}}
   - 2 \Delta \right) \eu^{\ii k_{\cP j} }=0 \; ,
\ee
i.e. for
\be
   A[\cP] \propto (-1)^\cP \prod_{j<l} \left( \eu^{\ii (k_{\cP j}+k_{\cP l})} + 1 - 2 \Delta \eu^{\ii k_{\cP j}}
   \right)
\ee
or, equivalently, by fixing the scattering phases as\index{Scattering phase}
\bea
  \eu^{\ii \tilde{\Theta} (k,k')} = -
  { \eu^{\ii (k+k')} + 1 - 2 \Delta \eu^{\ii k} \over
  \eu^{\ii (k+k')} + 1 - 2 \Delta \eu^{\ii k'} }
  \label{bigthetatilde}
\eea
which can also be written as 
\be
   \Theta (k,k') \equiv
   \tilde{\Theta} (k,k') - \pi
   = 2 \arctan { \Delta \sin {1 \over 2} (k-k') \over
   \cos {1 \over 2} (k+k') - \Delta \cos {1 \over 2} (k-k') } \; .
   \label{bigtheta}
\ee

By imposing periodic boundary conditions, we get the following quantization
relations:
\be
  \eu^{\ii k_j N} =
  \prod_{j \ne l} \eu^{\ii \tilde{\Theta} (k_j,k_l)}
  = (-1)^{R-1} \prod_{j \ne l}
  { \eu^{\ii (k_{j}+k_{l})} + 1 - 2 \Delta \eu^{\ii k_{j}} \over
  \eu^{\ii (k_{j}+k_{l})} + 1 - 2 \Delta \eu^{\ii k_{l} } } \; , \qquad
  j= 1, \ldots, R \; ,
  \label{XXZBE}
\ee
and by taking their logarithm we get the Bethe equations\index{Bethe!equations}
\be
   k_j N = 2 \pi \tilde{I}_j - \sum_{l=1}^R \Theta(k_j,k_l) \; ,
   \qquad j=1,\ldots, R \; ,
   \label{bigTBE}
\ee
where the $\{ \tilde{I}_j \}$ are the integer/half-integer quantum numbers defining the state.

The two-body scattering phase\index{Scattering phase} (\ref{bigtheta}) has the unpleasant property of not being translational invariant for shifts of the momenta and this makes the counting of the states harder, as well as to show the factorizations of the scattering matrix.
It is then convenient to introduce the {\it rapidities} $\tilde{\lambda}_j$ to parametrize the quasi-momenta $k_j$:
\be
   \eu^{\ii k_j} = { \sin {\phi \over 2} \left( \tilde{\lambda}_j + \ii \right) \over
   \sin {\phi \over 2} \left(\tilde{\lambda}_j - \ii \right) } \; ,
   \qquad \qquad \text{or} \qquad \qquad
   \cot {k_j \over 2} = \coth {\phi \over 2} \;
   \tan \left( {\phi \tilde{\lambda}_j \over 2} \right) \; .
   \label{paramet}
\ee
The parameter $\phi$ is determined by requiring the scattering phase\index{Scattering phase} to be a function of rapidity difference only:
\be
    { \eu^{\ii (k_{j}+k_{l})} + 1 - 2 \Delta \eu^{\ii k_{j}} \over
	\eu^{\ii (k_{j}+k_{l})} + 1 - 2 \Delta \eu^{\ii k_{l} } }
    = {\cos \left[ {\phi \over 2} \left( \tilde{\lambda}_j + \tilde{\lambda}_l \right) \right]
	\left( \cosh \phi - \Delta \right) +
	\Delta \cos \left[ {\phi \over 2} \left( \tilde{\lambda}_j - \tilde{\lambda}_l + 2 \ii \right) \right]
	- \cos \left[ {\phi \over 2}  \left( \tilde{\lambda}_j - \tilde{\lambda}_l \right) \right] \over
	\cos \left[ {\phi \over 2} \left( \tilde{\lambda}_j + \tilde{\lambda}_l \right) \right]
	\left( \cosh \phi - \Delta \right) +
	\Delta \cos \left[ {\phi \over 2} \left( \tilde{\lambda}_l - \tilde{\lambda}_j + 2 \ii \right) \right]
	- \cos \left[ {\phi \over 2} \left( \tilde{\lambda}_l - \tilde{\lambda}_j \right) \right] } \; ,
    \label{scat1}
\ee
which sets
\be
\cosh \phi \equiv \Delta \; .
\ee
Note that the change of variable (\ref{paramet}) also defines a different (more convenient) basis for the Bethe Ansatz expansion (\ref{bethewf}).
Simplifying (\ref{scat1}) we get
\be
   { \eu^{\ii (k_{j}+k_{l})} + 1 - 2 \Delta \eu^{\ii k_{j}} \over
   \eu^{\ii (k_{j}+k_{l})} + 1 - 2 \Delta \eu^{\ii k_{l} } }
   = {\sin \left[ {\phi \over 2} \left( \tilde{\lambda}_j - \tilde{\lambda}_l + 2 \ii \right) \right] \over
   \sin \left[ {\phi \over 2} \left( \tilde{\lambda}_j - \tilde{\lambda}_l - 2 \ii \right) \right] } \; .
\ee

Using the second expression in (\ref{paramet}) we write the Bethe equations (\ref{bigTBE}) as:
\be
   N \tilde{\theta}_1 \left( \tilde{\lambda}_j \right) =
   2 \pi I_j 
   + \sum_{l \ne j}^R \tilde{\theta}_2 \left( \tilde{\lambda}_j - \tilde{\lambda}_l \right) \, ,
   \qquad \qquad j=1 \ldots R \; ,
   \label{theta12BE}
\ee
with
\be
   \tilde{\theta}_n \left( \tilde{\lambda} \right) \equiv 
   2 \arctan \left[ \coth \left( {n \phi \over 2} \right) \; \tan \left( {\phi \tilde{\lambda} \over 2} \right) \right] \; .
   \label{thetandef}
\ee
In terms of the rapidities, the energy and momentum are given by:
\be
  E = E_0 + 2 h R + \sum_{j=1}^R \tilde{\epsilon} \left( \tilde{\lambda}_j \right) \; , \qquad \qquad \qquad
  K = 2 \sum_{j=1}^R \cot^{-1} {\tan \left( \phi \tilde{\lambda}_j /2 \right) \over \tanh \left( \phi /2 \right) } \; ,
  \label{EKrap}
\ee
with
\be
  \tilde{\epsilon} \left( \tilde{\lambda} \right) \equiv
  - {J \sinh^2 \phi \over \cosh \phi - \cos \left( \phi \tilde{\lambda} \right) }
  \label{epsilonlam}
\ee
the quasi-particle energy.
The phase $\tilde{\theta}_1 \left( \tilde{\lambda}_j \right)$ is actually the original quasi-momentum $k_j$. This means that the energy momentum of the quasi-particle can be written in terms of the rapidities as\footnote{The minus sign arises as a consequence of the different branch cut between the cotangent defining $k_j$ in (\ref{paramet}) and the tangent used for the phase in (\ref{thetandef}).}
\be
   p \left( \tilde{\lambda}_j \right) \equiv \tilde{\theta}_1 \left( \tilde{\lambda}_j \right) = k_j \; ,
   \qquad \qquad \qquad
   {1\over J} \: \tilde{\epsilon} \left( \tilde{\lambda} \right)  = -
   {\sinh \phi \over \phi} \; {\de \over \de \tilde{\lambda}} p \left( \tilde{\lambda} \right) \; \;
   \Big( = \Delta - \cos k \Big) \; .
\ee

The parametrization for the rapidities $\tilde{\lambda}$ (\ref{paramet}) has the merit of covering with evident continuity the whole phase diagram of the model. In fact, while $0\le \phi < \infty$ directly gives $1 \le \Delta < \infty$, at the isotropic point $\Delta =1$ ($\phi = 0$) the limit reached by continuity yields the parametrization we used in the previous chapter:
\be
\cot {k_j \over 2} = \tilde{\lambda}_j \; , \quad {\rm or} \qquad
\eu^{\ii k_j} = {\tilde{\lambda}_j + \ii \over \tilde{\lambda}_j - \ii} \, \qquad {\rm and} \qquad
\tilde{\theta}_n \left( \tilde{\lambda} \right) = 2 \arctan {\tilde{\lambda} \over n} \; .
\ee
After that, in the planar regime $|\Delta| \le 1$, $\phi = \ii \gamma$ moves in the complex plane along the imaginary axis, with $0 \le \gamma \le \pi$. The limit $\phi \to \ii \pi$ gives again an isotropic parametrization ($\Delta = -1$) and is the point where $\phi = \ii \pi + \phi'$ makes a turn to run over the line parallel to the real axis ($0 \le \phi' < \infty$) to cover the remaining axial phase $-\infty < \Delta \le 1$.

To study the distinct phases better, however, we will use a different definition of rapidity, due originally to Orbach \cite{orbach58}, which essentially amounts to $\lambda = \phi \tilde{\lambda}$. With this parametrization, the continuity of the solution across the quantum phase transition is hidden, but this choice has some merit within the study of complex (string) solutions. The relevant definitions for this parametrization in the different phases of the model are summarized in Table \ref{table:XXZparam} and allow to write the logarithmic form of the Bethe equations as\index{Bethe!equations}
\be
   N \theta_1 (\lambda_j) = 
   2 \pi I_j + \sum_{l} \theta_2 \left(\lambda_j - \lambda_l \right) \; , 
   \qquad \qquad j=1, \ldots, R \; .
   \label{XXZBE1}
\ee
Let us now look at the different phases separately.

\begin{table}[t]
\begin{center}
\renewcommand{\arraystretch}{1.5}
\setlength\extrarowheight{2pt}
\begin{tabular}{|l|c|c|c|}
\hline
&{\bf Uni-axial Ferromagnet}& {\bf Planar Paramagnet} & {\bf Uni-axial Anti-Ferromagnet} \\[2pt]
\hline
\hline
$\Delta=$ & $ \cosh \phi$ ($0<\phi<\infty$) & $- \cos \gamma$  ($0<\gamma<\pi$) &
$-\cosh \phi$ ($0<\phi<\infty$) \bigstrut \\[.2cm]
\hline
$ \eu^{\ii k}=$ & 
  $ \displaystyle{ \sin {1 \over 2} \left( \lambda + \ii \phi \right) \over 
	\sin {1 \over 2} \left( \lambda - \ii \phi \right)}$ & 
   $ - \displaystyle{ \sinh {1 \over 2} \left( \lambda - \ii \gamma \right) \over 
   	\sinh {1 \over 2} \left( \lambda + \ii \gamma \right) }$ & 
   $ - \displaystyle{ \sin {1 \over 2} \left( \lambda - \ii \phi \right) \over 
   	\sin {1 \over 2} \left(\lambda + \ii \phi \right) }$  \bigstrut \\[.2cm]
\hline
\renewcommand{\arraystretch}{2}
$ \eu^{\ii \tilde{\Theta} (k,k')} =$ & 
   $ - \displaystyle{\sin {1 \over 2} 
   	\displaystyle{\left( \lambda - \lambda' + 2 \ii \phi \right)} 
   	\over \sin {1 \over 2} 
   	\displaystyle{\left( \lambda - \lambda' - 2 \ii \phi \right) }}$ & 
   $ - \displaystyle{\sinh {1 \over 2} 
   	\displaystyle{\left( \lambda - \lambda' - 2 \ii \gamma \right)} 
   	\over \sinh {1 \over 2} 
   	\displaystyle{\left( \lambda - \lambda' + 2 \ii \gamma \right) }}$ & 
   $- \displaystyle{\sin {1 \over 2} 
   	\displaystyle{\left( \lambda - \lambda' - 2 \ii \phi \right)}
   	\over \sin {1 \over 2} 
   	\displaystyle{\left( \lambda - \lambda' + 2 \ii \phi \right) }}$ \bigstrut \\[.2cm]
\hline
\renewcommand{\arraystretch}{1.5}
$ \theta_n (\lambda) \equiv $ &
   $ { 2 \arccot \left[ \coth \left( {n \phi \over 2} \right) \; \tan \left( { \lambda \over 2} \right) \right] }$ &
   $ { 2 \arctan \left[ \cot \left( {n \gamma \over 2} \right) \; \tanh \left( { \lambda \over 2} \right) \right] }$ &
   $ { 2  \arctan \left[ \coth \left( {n \phi \over 2} \right) \; \tan \left( { \lambda \over 2} \right) \right] }$ \bigstrut \\[.2cm]
\hline
\hline
\end{tabular}
\caption{Orbach rapidity parametrization for the three phases of the XXZ chain}
\vspace{-.2cm}
\label{table:XXZparam}
\end{center}
\renewcommand{\arraystretch}{1}
\end{table}
\setlength\extrarowheight{1pt}

\section{Uni-axial Ferromagnet: $\Delta >1$}
\label{sec:XXZFerro}

With the definitions in Table \ref{table:XXZparam}, a real rapidity $\lambda \in \left(-\pi,\pi\right)$ gives a real quasi-momentum $k \in \left(-\pi,\pi\right)$.

The ground state (for $h \ge 0$) is the fully polarized one $|0\rangle$. The low energy states are thus magnons\index{Magnon} (and their bound states\index{Bound state}). Note that spinons\index{Spinon} are not a natural way to interpret the many-body states in this phase.

In Sec. \ref{sec:XXXstrings} we introduced the {\it string hypothesis} to organize the complex rapidities. The structure of the Bethe solutions in this phase is similar to that of the isotropic point and thus we will employ the same assumption (although, as we discussed, deviations exist both because of finite-size effects and possibly in the thermodynamic limit). Thus, we assume that each solution of (\ref{XXZBE1}) belongs to a $M$-type complex\index{M-complex} of rapidities
\be
\lambda_{M,j} = \lambda_M + \ii 2 (M - j) \phi \; , \qquad
j= 0, \ldots, 2M \; ,
\label{lambdacomplexes}
\ee
where in this phase the center of mass rapidity varies in the range $\lambda_M \in \left[ -\pi, \pi \right]$ and strings of arbitrary lengths are allowed. Each complex can be regarded as an elementary excitation (that captures the behavior of a bound state\index{Bound state} of $2M+1$ magnons\index{M-complex}) with momentum and energy:
\bea
   p_M \left( \lambda_M \right) & = & 
   {1 \over \ii} \sum_{j=0}^{2M} \ln 
   \left[ { \sin {1 \over 2} \left( \lambda_{M,j} + \ii \phi \right) \over
   	\sin {1 \over 2} \left( \lambda_{M,j} - \ii \phi \right) } \right]
   = {1 \over \ii} \ln 
   { \sin {1 \over 2} \left[ \lambda_M + \ii (2M+1) \phi \right] \over
   \sin {1 \over 2} \left[ \lambda_M - \ii (2M+1 ) \phi  \right] }
   \nonumber \\
   & = & \arccos {1 - \cos \lambda_M \cosh \left[ (2M+1) \phi  \right] \over
   \cosh \left[ (2M+1) \phi \right] - \cos \lambda_M } 
   \cdot \sgn\big( \Re (\lambda_M) \big) \; , \\
   \epsilon_M \left( \lambda_M \right) & = &
   J \sum_{j=0}^{2M} {\sinh^2 \phi \over \cosh \phi - \cos \lambda_{M,j} }
   = J \: { \sinh \phi \: \sinh \left[ (2 M +1) \phi \right] \over
	\cosh \left[ (2 M +1) \phi \right] - \cos \lambda_M } \; ,
\eea
which can be combined to give the dispersion relation
\be
   \epsilon_M (p_M)  = J { \sinh \phi \over \sinh \left[ (2 M +1) \phi \right] } 
   \Big[ \cosh \left[ (2 M + 1) \phi \right] - \cos p_M \Big] \; .
\label{Mdisprel}
\ee

We see that these excitations are gapped: their lowest energy is $\epsilon_M (0) = J \sinh \phi \tanh \left[ {(2M+1) \phi / 2}\right]$ and the single (free) magnon\index{Magnon} ($M=0$) is the lightest one, yielding $2 J \sinh^2 \left( \phi/2 \right)$, which is thus the energy gap in this phase. 
Note that as $\Delta$ increases, the spacings in the imaginary parts in (\ref{lambdacomplexes}) increase. In the $\Delta \to \infty$ ($\phi \to \infty$) limit, the rapidities within a complex get stretched along the imaginary axis, indicating that these states have large complex components and are thus ever tighter bound states\index{Bound state}, approaching a domain wall of a series of $2M+1$ overturned spins ({\it stretched string}). In this Ising limit, the energy cost of a domain wall is practically independent from the number of spins involved, since it lies only at the boundaries: (\ref{Mdisprel}) shows that the energy gaps of complex of arbitrary length converge to the same value as $\phi \to \infty$.

\subsection{Effect of the magnetic field on the ferromagnet}

In this phase, a non-zero magnetic field does not alter the physics significantly. Beside splitting the degeneracy between the two fully polarized states (up and down), yielding a unique ground state, its main effect is just to increase the energy cost of configurations with a higher number of magnons\index{Magnon} or of higher complexes\index{M-complex}.

\section{Paramagnetic/Planar Regime: $|\Delta| < 1$}
\label{sec:XXZplanar}

\begin{wrapfigure}{r}{7cm}
	\vspace{-20pt}
	\begin{center}
		\includegraphics[width=7cm]{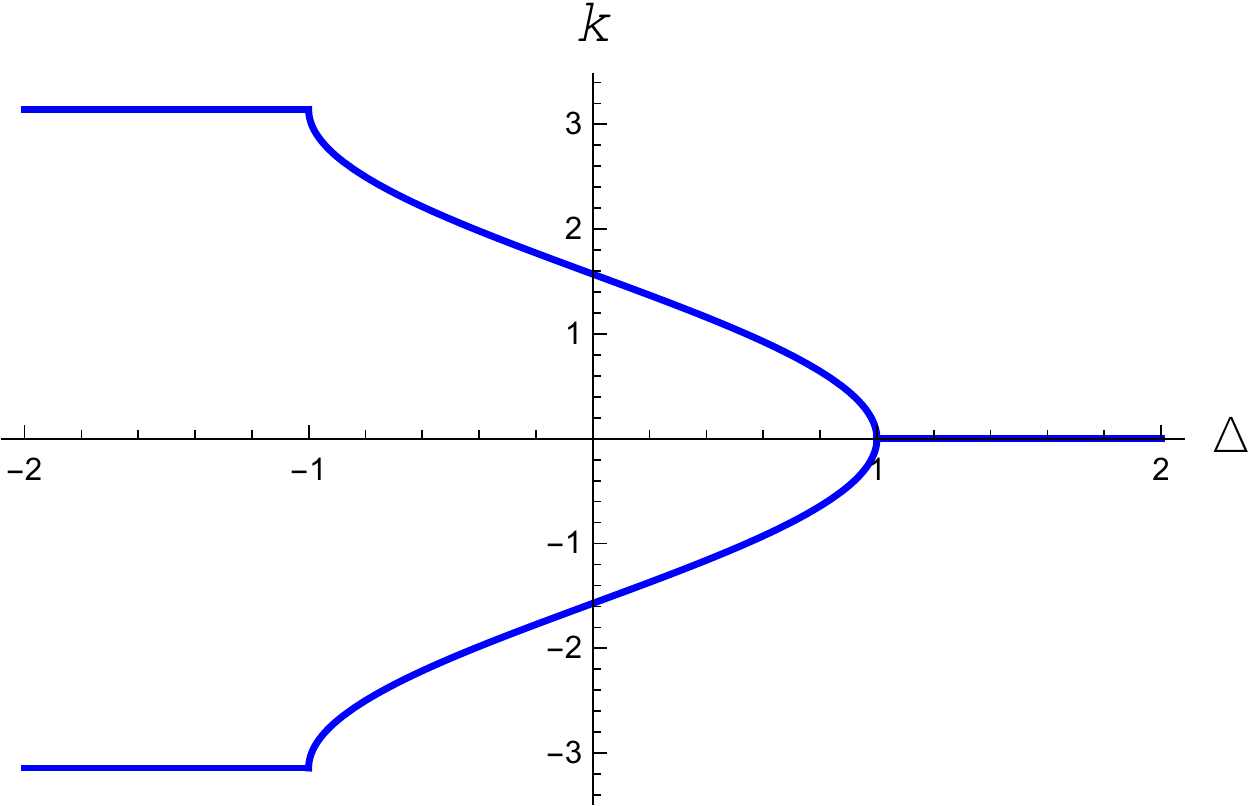}
	\end{center}
	\caption{Quasi-momentum range in the ground state as a function of $\Delta$.}
	\label{fig:XXZkrange}
	\vspace{-20pt}
\end{wrapfigure}
This is arguably the most interesting phase, as it hosts gapless excitations throughout.
Using $k(\lambda) = \theta_1 (\lambda)$, we see that real rapidities ($-\infty < \lambda < \infty$) generate real quasi-momenta constrained on a $\gamma$-dependent interval: $k \in \left[ -(\pi-\gamma), \pi - \gamma \right]$, which shrinks to zero at $\Delta=1$, see Fig. \ref{fig:XXZkrange}. Quasi-momenta outside of this interval correspond to rapidities lying on the $\ii \pi$ horizontal axis, i.e. $\lambda + \ii \pi$, with $\Im(\lambda) =0$.
Mathematically, the main difference between this phase and the others is that the function $k(\lambda)$ is periodic in the imaginary axis and this renders the classification of complex solutions more challenging. Moreover, as we saw, the center of mass of complex solution can lie on the real axis or on $\ii \pi$.

The ground state is given by pure, real rapidities, filling the Fermi sea, as the one we discussed in Sec. \ref{sec:XXXAFM}. In the thermodynamic limit, the rapidity density is the solution of the familiar integral equation (\ref{XXXinteeq})\index{Integral equation!linear}:
\be
  \rho_0 (\lambda) + {1 \over 2 \pi}
  \int_{-\Lambda}^\Lambda {\cal K} (\lambda - \mu) \; \rho_0 (\mu) \de \mu
  = {1 \over 2 \pi} \theta'_1 (\lambda) \; ,
\label{XXZinteeq}
\ee
which descends from (\ref{XXZBE1}) and where 
\be
   {\cal K} (\lambda) \equiv {\de \over \de \lambda} \; \theta_2 (\lambda) =
   {\sin \left( 2 \gamma \right) \over \cosh \lambda - \cos \left( 2 \gamma \right)} \; .
   \label{XXZinteq3}
\ee
The integral support is consistently determined by the magnetization:
\be 
    S_z = {N \over 2} - N \int_{-\Lambda}^{\Lambda} \rho_0 (\lambda) \de \lambda \; .
    \label{XXZSZ}
\ee

The energy of this state is
\be
  E = E_0 + N \int_{-\Lambda}^\Lambda \epsilon_0 (\lambda) \rho_0(\lambda) \de \lambda \: ,
  \label{paramE}
\ee
where the (bare) single magnon\index{Magnon} energy is
\be
   \epsilon_0 (\lambda) \equiv 2h - J \sqrt{1-\Delta^2} \theta_1^\prime (\lambda)
   = 2h - J {\sin^2 \gamma \over \cosh \lambda - \cos \gamma} \; .
   \label{XXZepsilon0}
\ee

At $h=0$, the lowest energy state has zero magnetization (we assume $N$ is even), and is thus given by $N/2$ real rapidities (this sector is called {\it half filled}), which exhausts the allowed vacancies. Hence, $\Lambda = \infty$ and the integral equation\index{Integral equation!linear} can be solved by Fourier transform, similarly to what we did in section \ref{sec:XXXAFM}, yielding
\bea
   \tilde{{\cal K}} (\omega|\gamma) & = & 
   \int_{-\infty}^{\infty} \de \lambda \: \eu^{- \ii \omega \lambda} \:
   {\sin \left( 2 \gamma \right) \over \cosh \lambda - \cos \left( 2 \gamma \right)}
   = 2\pi \: {\sinh \left[ \left( \pi - 2 \gamma \right) \omega \right] 
   \over \sinh \left( \pi \omega \right) } \; ,
   \label{KtildeXXZ} \\
   \rho_0 \left( \lambda \right) & = & 
   \int_{-\infty}^{\infty} {\de \omega \over 2 \pi} \: \eu^{\ii \omega \lambda} \:
   {\tilde{{\cal K}} (\omega|\gamma/2) \over 2\pi + \tilde{{\cal K}} (\omega|\gamma) }
   = \int_{-\infty}^{\infty} {\de \omega \over 2 \pi} \: \eu^{\ii \omega \lambda} \:
   {1 \over 2 \cosh \left( \gamma \omega \right)} \: 
   = {1 \over 4 \gamma} \left[ \cosh \left( {\pi \over 2} \: {\lambda \over \gamma}  \right) \right]^{-1}\: , 
   \label{rho0XXZ} \\
   E & = &  E_0
   - J N \sin \gamma \int_0^\infty {\sinh \left[(\pi - \gamma) \omega \right]
   \over \sinh \left( \pi \omega \right) \: \cosh \left(\gamma \omega \right)} \: \de \omega \: .
\eea

\subsection{Spinon and magnon excitations in the paramagnetic phase}

Similarly to what we discussed in section \ref{sec:XXXAFM}, low energy excitations are created by removing $\kappa$ real rapidities from the ground state contribution. Each removal leaves two holes\index{Hole} among the allowed vacancies: these holes are {\it spinons}. At $h=0$, the energy and momentum of the excitations can be determined exactly as in (\ref{klam}), because the integral equations\index{Integral equation!linear} extend over the whole real axis and can be solved through Fourier transform. Using the kernel\index{Kernel} (\ref{XXZinteq3}), each spinon contribution is found to be\index{Spinon}
\be
k (\lambda) \equiv {\pi \over 2} -  \arctan \sinh {\pi \lambda \over 2 \gamma} \; , \qquad \qquad
\varepsilon (\lambda) \equiv J \: {\pi \over 2} \:  {\sin \gamma \over \gamma} \:
{1 \over \cosh {\pi \lambda \over 2 \gamma} } \; ,
\label{klamXXZ}
\ee
yielding the dispersion relation
\be
\varepsilon (k) = J \: {\pi \over 2} \: { \sin \gamma \over \gamma} \: \sin k \;  .
\label{dispeXXZ}
\ee
For the isotropic case, we found that pairs of spinon\index{Spinon} excitations are degenerate with solutions with other $M$-complexes\index{M-complex} which realize the additional representations of the spin algebra. This degeneracy is due to the full $SU(2)$ symmetry enjoyed by the Heisenberg chain, which is broken in the XXZ chain. Thus, in the latter for each state one has to proceed as we did in Sec. \ref{sec:XXXAFM} to explicitly determine the contributions from each excitation separately.

Additional real excitations can be generated by placing some of the $\kappa$ rapidities removed from the real axis on the $\ii \pi$ axis. The spectrum of these excitations can be calculated through the usual techniques \cite{takahashi} and is
\be
   \varepsilon (k) = J \: \pi \: {\sin \gamma \over \gamma} 
   \left| \sin {k \over 2} \right| 
   \sqrt{1 + \cot^2\left[ \left( {\pi \over \gamma } -1 \right) {\pi \over 2} \right]  \sin^2 {k \over 2}} \;.
   \label{XXZmagnons}
\ee
Each of these excitations carries spin $S_z=1$ and in the $\Delta \to 1$ ($\gamma \to \pi$) limit they approach the magnonic dispersion\index{Magnon} (\ref{e1k}). These excitations are physical only for $0 < \Delta <1$, while for $\Delta<0$ their contribution vanishes in the linear approximation (meaning that the back-flow\index{Back-flow} on the other particles cancels it) and they only appear in high energy states \cite{takahashi}.
Note that at low momentum (\ref{XXZmagnons}) they have the same dispersion relation as individual spinons (\ref{dispeXXZ})\index{Spinon}. However, the two types of excitations are clearly distinguishable, because spinons appear in pairs (and thus each spinon has a halved momentum range), while these magnonic states are not composite excitations. Moreover, their momentum range is complementary to that of the spinon excitations and thus low momentum magnons are possible only close to $\Delta = 1$.

\subsection{String solutions in the paramagnetic phase}

String solutions in the paramagnetic phase are higher energy states compared to unbound states, but due to their number and structure they contribute significantly to the dynamics and thermodynamics of the model.
Compared to the Heisenberg case, not all complexes are allowed in the paramagnetic phase of the XXZ chain, and the issue of properly accounting for all states remains controversial to date.
The fundamental constraint comes from the normalizability of the wave-function: let us consider a $M$-complex\index{M-complex}, with the overturned spins situated at the sites $n_1 < n_2 < \ldots < n_{2M+1}$. The relevant parts of the wave-function can be written as 
\be
   f(n_1, \ldots, n_{2M+1}) = (z_1 z_2 \ldots z_{2M+1})^{n_1}
    \sum_\cP (-1)^\cP 
    \left[ \prod_{j<l} \sinh {1 \over 2}
    \left( \lambda_{\cP j} - \lambda_{\cP l} + 2 \ii \gamma \right) \right]
    \prod_{n=1}^{2M} \left( \prod_{l=n+1}^{2M+1} z_{\cP l} \right)^{n_{j+1} - n_j} \; ,
\label{BAWFXXZ}
\ee
where $z_j \equiv \eu^{\ii k_j} = {\sinh {1 \over2} \left(\ii \gamma - \lambda_j \right)  \over \sinh {1 \over2} \left(\ii \gamma + \lambda_j \right) }$.
When $\Im (\lambda_j) \ne 0$, $|z_j| \ne 1$: assuming $\Im (k_j) \ge \Im (k_{j+1})$, for the wavefunction (\ref{BAWFXXZ}) not to explode we require
\be
   \left| z_1 z_2 \ldots z_{2M+1}\right| =1 \, 
   \qquad \qquad \text{and} \qquad \qquad
   \left| \prod_{l=n+1}^{2M+1} z_l \right| < 1 \; , \quad n=1,\ldots, 2M \; .
   \label{normcond1}
\ee
These conditions guarantee that one of the permutations (the identity $\cP = {\bf 1}$ in (\ref{BAWFXXZ})), yields a normalizable wave-function. All other permutations necessarily have components that explode as the distance between flipped sites increases, and thus have to vanish (an equivalent way to say this is to require that the scattering phase\index{Scattering phase} of all other permutations vanishes). These considerations yield the string structure for bound states\index{Bound state}
\be
   \lambda_{M,j} = \lambda_M + {1 - \eta \over 2} \pi + \ii 2 (M-j) \gamma \, \qquad 
   j=0,\ldots,2M \: ,
\ee
where $\eta = \pm 1$ is called the ``{\it parity}'' of the strings, and classifies the two types of complexes, with the center of mass lying either on the real axis or on the $\ii \pi$ axis.
The energy and momentum of these $M$-complexes\index{M-complex} are
\bea
   p_M \left( \lambda_M \right) & = & 
   {1 \over \ii} \ln 
   { \sinh {1 \over 2} \left[ \ii (2M+1) \gamma - \lambda_M - \ii {1 -\eta \over 2} \pi \right] \over
	\sinh {1 \over 2} \left[ \ii (2M+1) \gamma + \lambda_M + \ii {1 -\eta \over 2} \pi \right] } \\
   \epsilon_M \left( \lambda_M \right) & = &
   - J {\sin \gamma \: \sin \left[ (2 M +1) \gamma \right] \over
	\eta \cosh \lambda_M - \cos \left[ (2 M +1) \gamma \right] } \; .
\eea
Using the identity
\be
  \cos p_M (\lambda_M) = 
  { 1- \eta \cosh \lambda_M \: \cos \left[ (2M+1) \gamma \right] \over
  \eta \cosh \lambda_M - \cos \left[ (2M+1) \gamma \right] } \: ,
  \label{pMparaXXZ} 
\ee
we can write their dispersion relation as
\be
   \epsilon_M (p_M)  = - J { \sin \gamma \over \sin \left[ (2 M +1) \gamma \right] } 
   \Big[ \cos \left[ (2 M + 1) \gamma \right] + \cos p_M \Big] \; .
\ee
Eq. (\ref{pMparaXXZ}) indicates that parity $\eta=1$ strings are low momentum states satisfying $\epsilon_M (p_M)<0$:
\bea
   \cos p_M > - \cos \left[ (2M+1) \gamma \right]
   & \qquad \qquad & \text{for} \: \: \eta=1 \: , 
   \label{pconstraint1}  \\
   \cos p_M < - \cos \left[ (2M+1) \gamma \right]
   & \qquad \qquad & \text{for} \: \: \eta=-1 \: ,
   \label{pconstraint2} 
\eea
which shows that the allowed momentum range depends on the length and parity of the complex and on $\Delta$.

The second condition in (\ref{normcond1}) can be proven \cite{samaj,takahashi,takahashithemo} to be equivalent to
\be 
   \eta \: \sin \left[ (2M+1 - n) \gamma \right] \: \sin \left[ n \: \gamma  \right] > 0 
   \qquad \qquad n=1, \ldots, 2M \: .
   \label{stringconstraint}
\ee
These last constraints are quite tricky. At {\it roots of unity}, that is points at which $\gamma = {p \over q} \pi$ ($p$ and $q$ co-primes), they roughly mean that only strings shorter than $q$ are allowed. At rational $\gamma / \pi$, only a finite set of strings can satisfy (\ref{stringconstraint}). Irrational points can be approximated by continued fractions to work out the selection rules: this was done in \cite{takahashi,takahashithemo} to construct the Hilbert space of the XXZ chain. With that they proceeded in working out the finite temperature thermodynamics of the model, similarly to the Yang-Yang construction\index{Yang-Yang equation} for the Lieb-Liniger model we presented in section \ref{sec:LLthermo}. The existence of the different types of excitations, their structure and condition of stability complicate the derivation of the thermodynamics Bethe Ansatz \cite{takahashi}, but the results are in good agreement with the actual behavior of the model. This success corroborates the validity of the string hypothesis. 

However, a direct check of this construction is still lacking, because, from one side, these are many-body states and thus individual excitations are hardly accessible, and from another point of view, direct numerical solutions of the the Bethe equations (in finite systems) always show significant deviations from the regular structures described here: these discrepancies are hard to study \cite{hagemans07,nostrings}, but they are indeed important when considering the (out of equilibrium) dynamics of the system.
Moreover, it is rather unpleasant that the existence of certain string solutions seems to depend on the precise value of $\gamma$ and on its rational or irrational character.

There is also a somewhat more ``physical'' point of view to approach these complex solutions \cite{sutherlandbook}.
We start remarking that the scattering phase\index{Scattering phase} (\ref{bigthetatilde}) has a branch point at $k_0 \equiv \arccos \Delta = \pi - \gamma$: 
\be
  \eu^{\ii \tilde{\Theta} (k, \pm k_0)}
   = \eu^{\mp 2 \ii k_0} \equiv - \eu^{\mp \ii \theta_0}  \; ,
\ee
meaning that the scattering phase of a $k_0$ magnon\index{Magnon} is constant $\theta_0 \equiv \pi - 2 k_0$ and independent from the rapidity of the other particle.
Thus a magnon with quasi-momentum $k_0$ factorizes out in the Bethe equations and its only effect is to introduce an overall phaseshift for the whole system (as if a flux was threading the system imposing an Aharonov-Bohm phase). Note that $k \to \pm k_0$ for $\lambda \to \mp \infty$.

Moreover, this critical $k_0$ also corresponds to a {\it threshold} state, i.e. a complex\index{M-complex} that has just coalesced into real momenta. To see this, let us take once more the two-body scattering phase\index{Scattering phase} in its original form (\ref{bigthetatilde}) and consider a $1/2$-string made out of two complex rapidities $k_{1,2} = k \pm \ii \kappa$ ($\kappa>0$):
\be
   \eu^{\ii \tilde{\Theta} (k + \ii \kappa, k - \ii \kappa)} = -
   {\cos k - \Delta \: \eu^{- \kappa}
   \over \cos k - \Delta \: \eu^{\kappa} } \; .
   \label{boundtheta}
\ee
For the bound state\index{Bound state} to be normalizable, the coefficient of the ``exploding'' part of the wavefunction has to vanish (see sections \ref{sec:boundstates} and \ref{sec:XXXstrings}), which means that (\ref{boundtheta}) should either vanish or diverge: $\eu^{\pm \kappa} \: \cos k = \Delta$.
This condition implies that for $|\Delta| \le 1$, the total momentum of a parity $\eta=1$, $1/2$-complex\index{M-complex} has to satisfy $\cos k \le \Delta$ (with a decay factor $\eu^{\kappa} = {\Delta \over \cos k}$). On threshold($\kappa =0$) we have $k = \pm k_0$.
For a $1/2$-complex, this threshold is $\Delta =0$ ($\gamma = k_0 = \pi/2$): approaching this point $\kappa \to 0$, the quasi-momenta of the complex get progressively closer to the real axis and eventually merge on it.

This exercise provides the fundamental ingredients to interpret the evolution of the string solutions in the paramagnetic phase. Starting from the isotropic point $\Delta=-1$ where strings of arbitrary length and center of mass momentum are allowed, moving toward $\Delta =0$ we progressively lose the longest complexes. Qualitatively, we can say that approaching $\Delta_M \equiv - \cos {\pi \over 2M +1}$, the momentum of the parity $\eta=1$, $M$-complex\index{M-complex} becomes confined towards zero $p_M \simeq 0$ (\ref{pconstraint1}). All quasi-momenta $k_{M,j} = \theta_1 \left( \lambda_{M,j} \right)$ of the complex converge towards the real axis at $k_0$ ($p_M \simeq (2M+1) k_0 \: {\rm mod} \: 2 \pi$). Past this point, for irrational values of ${\gamma \over \pi} > {1 \over (2M+1)}$, $\eta=1$ $M$-complexes\index{M-complex} are not allowed and are dissolved into real solutions \cite{sutherlandbook}. Exceptions exist at rational values, but they can be regarded more as mathematical features than as physically relevant states.

Note that at the point $\Delta = 0$ we recover the isotropic XY (or XX) model\index{XX chain}, i.e. one of the critical lines of the XY model, which is non-interacting. The Jordan-Wigner transformed\index{Jordan-Wigner!transformation} Hamiltonian becomes just that of free fermions on a lattice\footnote{As we saw in chapter \ref{chap:XYModel}, on this line the Bogoliubov angle vanishes and the Bogoliubov quasi-particles coincide with the physical fermions.}. Consistently with our picture, at this point all complexes have progressively disappeared. However,  approaching this point in continuity still yields pairs of imaginary solutions for the rapidities, but these are in fact not bound states\index{Bound state} and yield real quasi-momenta \cite{mullerXXZ}.

We can determine what happens for $\Delta >0$ by rotating every other spin by $\pi$ about the $z$ axis and effectively performing the transformation $\Delta \to - \Delta$ (and $J \to - J$). In this way, $1/2$-complexes\index{M-complex} reemerge and thus exist in the whole paramagnetic phase (except at $\Delta=0$). Then, for $\Delta \ge \cos {\pi \over 2M +1}$ we have the progressive reappearance of (parity $\eta =-1$) $M$-complexes. This time, however, they are not high energy states (because $J \to - J$). In fact, the $M$-complexes\index{M-complex} become stable as they cost less energy than a state made of $2M+1$ real magnons\index{Magnon} \cite{sutherlandbook}. Note that this picture is consistent with the results of \cite{johnson73} on the XYZ chain. Introducing a small anisotropy in the coupling along the $y$ direction opens a gap. In the scaling limit, the physics is well captured by a sine-Gordon model \cite{johnson73,luther76} and the thresholds $-\Delta_M$ correspond to those at which bound states\index{Bound state} of $2M+1$ solitons (breathers) become stable. Finally, approaching $\Delta \to 1$ we have recovered strings of arbitrary length.

In closing, we stress once more that, exceptionally, it is possible to have states with multiple instances of the same quasi-momentum, if this is $\pm k_0$. We can interpret these particles as remnants of a string solution: the rapidities of such solutions are at infinity ($k (\lambda \to \pm \infty) \to \pm k_0$) and they correspond to the maximum allowed integer $I_M^{(\infty)}$ (\ref{IMinfty}). It should be noticed that these excitations do not contribute to the energy, since $\epsilon (\pm k_0) = \Delta - \cos (\pm k_0) = 0$ and they only contribute to the total momentum (i.e. the flux) and magnetization, since they have a constant scattering phase\index{Scattering phase}. These states act as reservoir to change the magnetization of a state without changing its energy, thus generating the non-highest weight state in each representation.

\subsection{Effect of the magnetic field on the paramagnet}
\label{sec:XXZparah}

A finite magnetic field lowers the energy of sectors with finite magnetization, 
causing progressive level crossings.
The ground state for a given $h$ is the state which minimizes the energy.
Similarly to what we did in section \ref{sec:LLexcitations}, it is possible to introduce the dressed single magnon\index{Magnon} energy\index{Dressed!energy} $\varepsilon_0 (\lambda)$ through the integral equation\index{Integral equation!linear}
\be
\varepsilon_0 (\lambda) + {1 \over 2 \pi} \int_{-\Lambda}^\Lambda {\cal K}
(\lambda,\mu) \varepsilon_0 (\mu) \de \mu = \epsilon_0 (\lambda) \; .
\label{varepsiloneq}
\ee
The support of the density is then determined by the condition that at the Fermi points the dressed energy vanishes: $\varepsilon_0 (\pm \Lambda)=0$. Solving this condition yields the desired relation between $h$ and $\Lambda$ \cite{ISM}.

We present a different route, which we already introduced in section \ref{sec:XXXh}, by minimizing the energy (\ref{paramE}) as  ${\partial E \over \partial h} \propto {\partial E \over \partial \Lambda}=0$.
We consider the change $\Lambda \to \Lambda + \delta \Lambda$, which also triggers $\rho_0 (\lambda) \to \rho_0 (\lambda) + \delta \rho_0 (\lambda)$: expanding (\ref{XXZinteeq}) we obtain the integral equation for $\delta \rho_0 (\lambda)$\index{Integral equation!linear}
\be
\delta \rho_0 (\lambda) + {1 \over 2 \pi}
\int_{-\Lambda}^\Lambda {\cal K} (\lambda - \mu) \; \delta \rho_0 (\mu) \de \mu =  - {1 \over 2 \pi} \Big[ {\cal K} (\lambda - \Lambda) + {\cal K} (\lambda + \Lambda) \Big] \rho_0(\Lambda) \: \delta \Lambda  \; .
\label{XXXdeltarho0}
\ee
We (re-)introduce the back-flow\index{Back-flow} (shift) function $J(\lambda|\nu)$ through the same integral equation as in (\ref{jkp}):\index{Integral equation!linear}
\be  
  J (\lambda|\nu) + {1 \over 2 \pi}
  \int_{-\Lambda}^\Lambda {\cal K} (\lambda - \mu) \; J (\mu|\nu) \de \mu 
  = {1 \over 2 \pi} \theta_2 (\lambda - \nu)  \; ,
\label{JXXXdef}
\ee
and define
\be  
   D (\lambda) \equiv \left. {\partial \over \partial \nu} J (\lambda|\nu) \right|_{\nu=\Lambda} \qquad \Rightarrow \qquad \quad
   D (\lambda) + {1 \over 2 \pi}
   \int_{-\Lambda}^\Lambda {\cal K} (\lambda - \mu) \; D (\mu) \de \mu 
   = - {1 \over 2 \pi} {\cal K} (\lambda - \Lambda)  \; .
   \label{Ddef}
\ee
We have 
\be
   \delta \rho_0 (\lambda) = \rho_0 (\Lambda) \: \delta \Lambda \big[ D(\lambda) + D(-\lambda) \big] \: , 
   \label{deltarho0D}
\ee
as can be checked substituting (\ref{deltarho0D}) into (\ref{XXXdeltarho0}), and using (\ref{Ddef}) and the fact that ${\cal K} (-\lambda) = {\cal K} (\lambda)$.

The change in energy (\ref{paramE}) is
\be
  \delta E = 2 \: \epsilon_0 (\Lambda) \: \rho_0 (\Lambda) \: \de \Lambda
  + \int_{-\Lambda}^\Lambda \epsilon_0 (\lambda) \: \delta \rho_0 (\lambda) \: \de \lambda
  = 2 \rho_0 (\Lambda) \: \de \Lambda \left[ \epsilon_0 (\Lambda)
  + \int_{-\Lambda}^\Lambda \epsilon_0 (\lambda) \: D (\lambda) \: \de \lambda 
   \right] \; ,
  \label{deltaELambda}
\ee
where we used that $\int_{-\Lambda}^\Lambda f(\lambda) \delta \rho_0 (\lambda) \de \lambda = 2 \rho_0 (\Lambda) \de \Lambda \int_{-\Lambda}^\Lambda f(\lambda) D (\lambda) \de \lambda$ if $f(x)$ is an even function, which follows from (\ref{deltarho0D}).
Remembering the definition of $\epsilon_0 (\lambda)$ (\ref{XXZepsilon0}), we use (\ref{XXZinteeq}) to get
\bea
   \int_{-\Lambda}^\Lambda \epsilon_0 (\lambda) D(\lambda) \de \lambda  & = &  
   \int_{-\Lambda}^\Lambda \left[ 2 h - J \sqrt{1-\Delta^2} \theta_1^\prime (\lambda) \right] D(\lambda) \: \de \lambda
   \nonumber \\
   & = & 2 h \int_{-\Lambda}^\Lambda  D(\lambda) \de \lambda
   - J \sqrt{1-\Delta^2} \int_{-\Lambda}^\Lambda \de \lambda \left[
   2 \pi \rho_0 (\lambda) + \int_{-\Lambda}^\Lambda {\cal K} (\lambda - \mu) \rho_0(\mu) \de \mu \right] D(\lambda)
   \nonumber \\
   & = & 2 h \int_{-\Lambda}^\Lambda  D(\lambda) \de \lambda
   - J \sqrt{1-\Delta^2} \int_{-\Lambda}^\Lambda \de \lambda \left[
   2 \pi D (\lambda) + \int_{-\Lambda}^\Lambda {\cal K} (\lambda - \mu) D (\mu) \de \mu \right] 
   \rho_0 (\lambda)
   \nonumber \\
   & = & 2 h \int_{-\Lambda}^\Lambda  D(\lambda) \de \lambda
   + J \sqrt{1-\Delta^2} \int_{-\Lambda}^\Lambda \de \lambda \:  \rho_0 (\lambda) \: {\cal K} (\Lambda - \lambda) \; .
\eea
Inserting this into (\ref{deltaELambda}) we get
\bea
   \delta E & = & 2 \: \rho_0 (\Lambda) \: \de \Lambda
   \left[2h - J \sqrt{1-\Delta^2} \theta_1^\prime (\Lambda)
   + 2 h \int_{-\Lambda}^\Lambda  D(\lambda) \de \lambda
   + J \sqrt{1-\Delta^2} \int_{-\Lambda}^\Lambda \de \lambda \:  \rho_0 (\lambda) \: {\cal K} (\Lambda - \lambda) \right]
   \nonumber \\
   & = & 2 \: \rho_0 (\Lambda) \: \de \Lambda \left[ 
   2 h \left( 1 + \int_{-\Lambda}^\Lambda D(\lambda) \de \lambda \right) 
   - 2 \pi J \sqrt{1-\Delta^2} \rho_0 (\Lambda) \right] \; .
\eea
Thus, the ground state condition is
\be
   h (\Lambda) \left( 1 + \int_{-\Lambda}^\Lambda D(\lambda) \de \lambda \right) 
   = \pi J \sqrt{1-\Delta^2} \rho_0 (\Lambda) \; ,
   \label{GScond1}
\ee
which can be recast in terms of the more important ``{\it dressed charge}'' function $Z(\lambda)$ defined by\index{Dressed!charge}
\be
Z (\lambda) + {1 \over 2 \pi} 
\int_{-\Lambda}^\Lambda {\cal K} (\lambda - \mu) \; Z (\mu) \de \mu = 1 \; .
\label{Zdef1}
\ee
The dressed charge plays an important role in characterizing the thermodynamics and the low energy excitations of the model \cite{ISM}: in appendix \ref{app:CFT} we will use it do calculate the Luttinger parameter and the conformal dimensions of the operators describing the low energy particles. We have
\be
   Z (\Lambda) = 1 + \int_{-\Lambda}^\Lambda D(\lambda) \de \lambda  \;, 
\label{ZJrel}
\ee
as can be verified by plugging (\ref{ZJrel}) into (\ref{Zdef1}) and using (\ref{Ddef}),
Inserting (\ref{ZJrel}) into (\ref{GScond1}), the relation between the support of the ground state integral equation (given by $\Lambda$) and the magnetic field is
\be 
h (\Lambda) = \pi J \sqrt{1-\Delta^2} {\rho_0 (\Lambda) \over Z (\Lambda)} \: .
\label{XXZhlambda}
\ee

\subsubsection{Behavior close to saturation}

In the $\Lambda \to 0$ limit, both integral equations\index{Integral equation!linear} for $\rho_0 (\lambda)$ and $Z(\lambda)$ become trivial, yielding $\rho_0 (\lambda) = {1 \over 2 \pi} \: \theta_1^\prime (\lambda) $ and $Z(\lambda) = 1$. Thus, the critical magnetic field at which the lowest energy state is the fully polarized one is 
\be
h (\Lambda=0) = h_{\rm s} \equiv {J \over 2} \sqrt{1-\Delta^2} \theta_1^\prime (0) = 
{J \over 2} \left( 1 - \Delta \right) \: .
\label{hFM}
\ee
For $h > h_{\rm s}$ we re-enter the ferromagnetic phase described in the previous section.

Close but below saturation field, we can solve the integral equations\index{Integral equation!linear} (\ref{XXZinteeq},\ref{Zdef1}) perturbatively as
\be
\rho_0 (\Lambda) \left[ 1 + {1 \over 2 \pi} \; {\cal K} (0) \; 2 \Lambda \right] \simeq {1 \over 2 \pi} \theta'_1 (\Lambda) \; ,
\qquad \qquad \qquad
Z (\Lambda) \left[ 1 + {1 \over 2 \pi} \; {\cal K} (0) \; 2 \Lambda \right] 
\simeq 1 \; ,
\label{rhoZsat}
\ee
which can be inserted into (\ref{XXZhlambda}) to find $h(\Lambda)$. This is a transcendental equation: to invert it we can expand $\theta_1^\prime(\Lambda)$ for small $\Lambda$, turning $h(\Lambda)$ into a quadratic equation, whose solution is
\be
\Lambda \stackrel{h \to h_{\rm s}}{\simeq} 
2 \tan {\gamma \over 2} \: \sqrt{h_{\rm s} - h \over J} \: , 
\qquad \qquad \qquad \qquad \qquad \qquad h \le h_{\rm s} \; .
\label{Lambdasat}
\ee
Notice that this condition can be extracted also from (\ref{varepsiloneq}): requiring $\varepsilon_0 (\Lambda)=0$ for a vanishing support integral equation amounts to $\epsilon_0 (\Lambda)=0$ which gives (\ref{Lambdasat}). Inserting this in (\ref{XXZSZ}) the magnetization behaves as
\be
{S_z \over N} \stackrel{h \to h_{\rm s}}{\simeq} 
{1 \over 2} - 2 \Lambda \rho(0) 
={1 \over 2} - {2 \over \pi} 
\sqrt{ h_{\rm s} - h \over J} \: , 
\qquad \qquad \qquad h \le h_{\rm s} \: .
\label{saturationmag}
\ee 
which indicates that the susceptibility diverges approaching the saturation point.

\subsubsection{Behavior close to zero field}

For small magnetic fields (large $\Lambda$), the integral equations\index{Integral equation!linear} can be solved through the Wiener-Hopf method\index{Wiener-Hopf method} \cite{ISM, takahashi}: one takes advantage of the fact that the solution for $\Lambda = \infty$ is known through Fourier transform to build a perturbative solution in $1/\Lambda$. Looking for the solution of $ \left( \hat{\cal I} + {1 \over 2 \pi} \hat{{\cal K}}_\Lambda \right) f (\lambda) = f^{(0)} (\lambda)$ we write
\be
    \left( \hat{\cal I} + {1 \over 2 \pi} \hat{{\cal K}}_\infty \right) f(\lambda)
    = f^{(0)} (\lambda) + {1 \over 2 \pi} \left\{ \int_{-\infty}^{-\Lambda} + \int_\Lambda^\infty
    {\cal K} (\lambda - \mu) \: f(\mu) \; \de \mu \right\} \; ,
    \label{WH0}
\ee
and apply the resolvent\index{Resolvent} (\ref{resolvent1}) $\hat{\cal I} - \hat{{\cal L}}_\infty$ on both side and use the symmetry $f(\lambda) = f(-\lambda)$ to shift the variable by introducing $g(x) \equiv f(\Lambda + x)$ to get 
\be
  g(x) = f_\infty (\Lambda + x) 
  + \int_0^\infty \Big[ {\cal L}_\infty (x-y) + {\cal L}_\infty (2\Lambda + x + y) \Big] g(y) \: \de y \; ,
  \label{WH1}
\ee
where $f_\infty (\lambda)$ is the solution of the integral equation\index{Integral equation!linear} for $\Lambda=\infty$, found by Fourier transform. At this point, one notices that the second term in the integral on the RHS is small compared to the rest of the equation if the interaction is local and the resolvent decays toward zero for large $\Lambda$ (in this, ${\cal L}_\infty (2 \Lambda)$ provides a natural expansion parameter). In this way, the solution of (\ref{WH1}) is constructed as a series $g(x) = \sum_n g_n (x)$ where
\bea
   g_n (x) & = & 
   g_n^{(0)} (x) + \int_0^\infty {\cal L}_\infty (x-y) \: g_n (y) \: \de y \; , 
   \label{WH2} \\
   g_n^{(0)} (x) & \equiv & 
   \int_0^\infty {\cal L}_\infty (2 \Lambda + x + y) \: g_{n-1} (y) \: \de y \; , 
   \qquad \qquad g_0^{(0)} (x) \equiv f_\infty (\Lambda + x) \; .
\eea
Eq. (\ref{WH2}) is of the Wiener-Hopf type and can be solved (order by order) by Fourier transform and by decomposing its elements in components that are analytic in the upper and lower Fourier plane \cite{WienerHopfMethod}. 

Let us find the first order corrections for large but finite $\Lambda$ to (\ref{XXZinteeq}) and (\ref{Zdef1}) as the solution of
\be
   g(x) = g_0 (x) 
   + \int_0^\infty {\cal L}_\infty (x-y) \: g(y) \: \de y
  \qquad \quad \rightarrow \qquad \quad
  \big[ 1 - \tilde{\cal L}_\infty (\omega) \big] \tilde{g}_- (\omega) 
  + \tilde{g}_+ (\omega) = \tilde{g}_0 (\omega) \; ,
  \label{WHF1}
\ee 
where in the last expression we moved into Fourier space and defined
\be
   \tilde{f}_\pm (\omega) \equiv 
   \int_{-\infty}^\infty \eu^{-\ii \omega x} \; f(x) \;
   \vartheta_H (\mp x) \; \de x \; .
   \label{WHdec}
\ee
We also decompose the integral operator as the product:
\be
   1 + {1 \over 2 \pi }\tilde{\cal K} (\omega)
   = {1 \over 1 - \tilde{\cal L}_\infty (\omega)}
   = \tilde{\cal G}_+ (\omega) \tilde{\cal G}_- (\omega) \; ,
\ee
where function $\tilde{\cal G}_+ (\omega)$ ($\tilde{\cal G}_- (\omega)$) is analytic and non-vanishing for $\Im (\omega) >0$ ($\Im (\omega) <0$) and normalized as $\lim_{|\omega| \to \infty }\tilde{\cal G}_\pm (\omega) = 1$. Notice that, since ${\cal K} (x) = {\cal K} (-x)$, $\tilde{\cal G}_+ (\omega) = \tilde{\cal G}_- (-\omega)$ and thus $1 + \tilde{\cal K} (0)/2 \pi= \tilde{\cal G}_+^2 (0)$. Note that the expressions for $\tilde{\cal G}_\pm (\omega)$ will not be needed (they can  be found in \cite{takahashi}) and only the latter property will be used.
With these definitions, we can write (\ref{WHF1}) as
\be
   {\tilde{g}_- (\omega) \over \tilde{\cal G}_- (\omega)} +
   \tilde{\cal G}_+ (\omega) \: \tilde{g}_+ (\omega) = 
   \tilde{\cal G}_+ (\omega) \: \tilde{g}_0 (\omega) \; ,
\ee
which can be solved by equating the components in the different half planes:
\be
   \tilde{g}_- (\omega) = \tilde{\cal G}_- (\omega) 
   \left[ \tilde{\cal G}_+ (\omega) \: \tilde{g}_0 (\omega) \right]_- \; , 
   \qquad \qquad \qquad
   \tilde{g}_+ (\omega) = {1 \over \tilde{\cal G}_+ (\omega)}
   \left[ \tilde{\cal G}_+ (\omega) \: \tilde{g}_0 (\omega) \right]_+ \; .
   \label{WHsol1}
\ee
We are after the value of the solution of (\ref{WH0}) at the support boundary, that is $g(0) = f(\Lambda)$. This can be computed by noticing that since $\tilde{g}_- (\omega)$ is analytic over the negative half plane, the closed integral $\left\{ \int_{\underset{-r<x<r}{\omega= x}} + \int_{\underset{\pi<\phi<2\pi}{\omega= r \eu^{\ii \phi}}} \right\} \tilde{g}_- \left(\omega \right) \; \de \omega = 0$ vanishes for any value of $r$. Thus, taking $r=\infty$ we have
\be
   f(\Lambda) = \lim_{x \to 0^+ }g(x) 
   = 2 \int_{-\infty}^\infty \tilde{g}_- (\omega) {\de \omega \over 2 \pi} 
   = \ii \lim_{|\omega| \to \infty} \omega \; \tilde{g}_- (\omega) 
   = \ii \lim_{|\omega| \to \infty} \omega \; 
   \left[ \tilde{\cal G}_+ (\omega) \: \tilde{g}_0 (\omega) \right]_-\; ,
   \label{WHsol2}
\ee
where the factor of $2$ is due to the discontinuity of $g(x)$ at $x=0$ and we used $\tilde{\cal G}_- (\infty)=1$. 

Applying this construction to (\ref{XXZinteeq}), in (\ref{WHF1}) we can use the asymptotic behavior of the solution (\ref{rho0XXZ})
\be
   g_0 (x) \coloneqq \rho_0^{(\infty)} (\Lambda + x) \vartheta_H (x) 
   \stackrel{\Lambda \gg 1}{\simeq}
   { \eu^{- {\pi \over 2} {\Lambda + x \over \gamma}} \over 2 \gamma} \vartheta_H (x)
   \quad \rightarrow \quad
   \tilde{g}_0 (\omega) \coloneqq {\eu^{-{\pi \over 2} {\Lambda \over \gamma}} \over 2 \gamma} \int_0^\infty \eu^{-\ii \omega x - {\pi \over 2} {x \over \gamma}} \de x =
  {\eu^{-{\pi \over 2} {\Lambda \over \gamma}} \over \pi + 2 \ii \gamma \omega} \; .
  \label{rho0as}
\ee
Since $\tilde{g}_0 (\omega)$ has a simple pole at $\omega = {\ii \pi \over 2 \gamma}$, the separation in components is achieved as
\be
   \left[ \tilde{\cal G}_+ (\omega) \: \tilde{g}_0 (\omega) \right]_+ \coloneqq
   \left[ \tilde{\cal G}_+ (\omega) - \tilde{\cal G}_+ \left( {\ii \pi \over 2 \gamma} \right) \right] \tilde{g}_0 (\omega) \; , \qquad \qquad
   \left[ \tilde{\cal G}_+ (\omega) \: \tilde{g}_0 (\omega) \right]_- \coloneqq
   \tilde{\cal G}_+ \left( {\ii \pi \over 2 \gamma} \right) \tilde{g}_0 (\omega) \; .
   \label{rhoWHdec}
\ee
Inserting this and (\ref{rho0as}) in (\ref{WHsol2}) we obtain
\be
   \rho_0 (\Lambda) \stackrel{\Lambda \gg 1}{\simeq} 
   \tilde{\cal G}_+ \left( {\ii \pi \over 2 \gamma} \right) 
   {1 \over 2 \gamma} \; \eu^{-{\pi \over 2} {\Lambda \over \gamma}} \; .
   \label{rhoLambdaas}
\ee

For the dressed charge (\ref{Zdef1}) the source is a constant which means that the $\Lambda = \infty$ solution is
\be
   \tilde{g}_0 (\omega) \coloneqq \tilde{Z}^{(\infty)} (\omega) 
   = {2 \pi \delta(\omega) \over 1 + {1 \over 2\pi}\tilde{\cal K} (\omega) }
   \quad \rightarrow \quad
   g_0 (x) \coloneqq Z^{(\infty)} (\Lambda + x) 
   = \lim_{\omega \to 0} {2\pi \over 2 \pi + \tilde{\cal K} (\omega)} 
   = {\pi \over 2 (\pi - \gamma)} \; .
\ee
Using Plemelj representation $ \delta(\omega) = {\ii \over 2 \pi} \lim_{\epsilon \to 0^+} \left[ {1 \over \omega + \ii \epsilon} - {1 \over \omega - \ii \epsilon} \right]$, we decompose as (limit over $\epsilon \to 0$ intended)
\be
  \left[ \tilde{\cal G}_+ (\omega) \: \tilde{g}_0 (\omega) \right]_+ \coloneqq
  {\ii \over \omega + \ii \epsilon} \; {1 \over \tilde{\cal G}_- (-\ii \epsilon) } \; , 
  \qquad
  \left[ \tilde{\cal G}_+ (\omega) \: \tilde{g}_0 (\omega) \right]_- \coloneqq
  {\ii \over \omega + \ii \epsilon} \left[ {1 \over \tilde{\cal G}_- (\omega) } - 
  {1 \over \tilde{\cal G}_- (-\ii \epsilon) } \right]
  - {\ii \over \omega - \ii \epsilon} \; {1 \over \tilde{\cal G}_- (\omega) } \; ,
\ee
which inserted in (\ref{WHsol2}) yields
\be
  Z(\Lambda) \stackrel{\Lambda \gg 1}{\simeq} 
  \lim_{\epsilon \to 0^+} {1 \over \tilde{\cal G}_- (-\ii \epsilon) }
  = \lim_{\epsilon \to 0^+} \sqrt{2 \pi \over 2\pi +  \tilde{\cal K} (-\ii \epsilon)}
  = \sqrt{\pi \over 2 (\pi - \gamma)} \; .
  \label{ZLambdaas}
\ee

We can now evaluate (\ref{XXZhlambda}) for $h \ll 1$ as
\be
   h = J {\pi \over 2} \; {\sin \gamma \over \gamma} \: 
   \tilde{\cal G}_- \left( 0 \right) \; 
   \tilde{\cal G}_+ \left( {\ii \pi \over 2 \gamma} \right) 
   \eu^{-{\pi \over 2} {\Lambda \over \gamma}} \; ,
\ee
and use it to calculate the magnetization.
We start with the integral equation for the density of rapidity written as (\ref{WH0}) with $f(\lambda) = \rho_0 (\lambda)$ and apply to it the resolvent\index{Resolvent} $\hat{\cal I} - \hat{\cal L}_\infty$ to get
\be
   \rho_0 (\lambda) = \rho^{(\infty)}_0 (\lambda) + \left\{ \int_{-\infty}^{-\Lambda} + \int_\Lambda^\infty
   {\cal L}_\infty (\lambda - \mu) \: \rho_0(\mu) \; \de \mu \right\} \; .
   \label{WHrhoint}
\ee
We compute $\int_{-\infty}^{\infty} \rho_0 (\lambda) \: \de \lambda$ in two ways:
\bea
   \int_{-\infty}^{\infty} \rho_0 (\lambda) \: \de \lambda
   & = &  \int_{-\Lambda}^{\Lambda} \rho_0 (\lambda) \: \de \lambda +
   \left\{ \int_{-\infty}^{-\Lambda} + \int_\Lambda^\infty
   \rho_0(\lambda) \; \de \lambda \right\} \; ,
   \nonumber \\
   \int_{-\infty}^{\infty} \rho_0 (\lambda) \: \de \lambda
   & = &   \int_{-\infty}^{\infty} \rho_0^{(\infty)} (\lambda) \: \de \lambda +
   \int_{-\infty}^{\infty} {\cal L}_\infty (\lambda) \de \lambda 
   \left\{ \int_{-\infty}^{-\Lambda} + \int_\Lambda^\infty
   \rho_0(\mu) \; \de \mu \right\} \; ,
\eea
where in the last term we used (\ref{WHrhoint}) and shifted the argument of the resolvent by taking advantage of the unbound limits of integration in $\lambda$. Now, noticing that $\int_{-\infty}^{\infty} \rho_0^{(\infty)} (\lambda) \: \de \lambda =1/2$,
we use the identities above to calculate the magnetization as 
\bea
  {S_z \over N} = {1 \over 2} - \int_{-\Lambda}^\Lambda \rho_0 (\lambda) \de \lambda 
  & = &
  \left[ 1 - \int_{-\infty}^\infty {\cal L}_\infty (\lambda) \de \lambda \right] \left\{ \int_{-\infty}^{-\Lambda} + \int_\Lambda^\infty
  \rho_0(\mu) \; \de \mu \right\} 
  \nonumber \\
  & = & {2 \pi \over 2\pi + \tilde{\cal K} (0)} \: 
  2 \int_0^\infty \rho_0 (\Lambda +x) \de x 
  = {\pi \over \pi - \gamma} \; \tilde{\cal G}_- (0) \;
  \tilde{\cal G}_+ \left( {\ii \pi \over 2 \gamma}\right)
  { \eu^{-{\pi \over 2} {\Lambda \over \gamma}} \over \pi} \; ,
\eea
where in the last passage we use the result for the Fourier Transform (\ref{WHdec}) of $\rho_0 (\Lambda +x)$, eq. (\ref{rhoWHdec}) and (\ref{rho0as}). Comparing this expression with (\ref{ZLambdaas}) we conclude that the (ground state) magnetization grows linearly following the general law
\be 
{S_z \over N} \stackrel{h \to 0}{\simeq} \chi (0) \: h \: , \qquad \qquad
\chi (0) = {1 \over \left( \pi - \gamma \right) v_F} \: , 
\ee 
where $v_F = J {\pi \over 2} {\sin \gamma \over \gamma}$ is the (Fermi) velocity of the low energy (spinon)\index{Spinon} excitations (\ref{dispeXXZ}). Higher order corrections in the Wiener-Hopf expansion show that for $-1 < \Delta < -0.8$ the zero field finite value of the susceptibility is approached with an infinite slope, due to an algebraic singularity in the susceptibility derivative \cite{takahashi}. We already encountered this behavior in (\ref{XXXchi}), when studying the Heisenberg chain.

\section{Uni-axial Anti-ferromagnet: $\Delta <-1$}
\label{sec:XXZAFM}

In this regime, $\Re (\lambda) \in \left(-\pi,\pi\right)$ is mapped into $\Re (k) \in \left(-\pi,\pi\right)$.

For $h=0$, the ground state has zero magnetization and is given by $N/2$ real magnons satisfying the Bethe equation. Proceeding in the usual way, in the thermodynamic limit we have that the density of real rapidities satisfies the integral equation\index{Integral equation!linear} (\ref{XXZinteeq}) with $\Lambda = \pi$ and 
\be
   {\cal K} (\lambda) \equiv {\de \over \de \lambda} \; \theta_2 (\lambda) =
{\sinh \left( 2 \phi \right) \over \cosh \left( 2 \phi \right) - \cos \lambda} \; ,
\label{XXZinteqAFM1}
\ee
Due to periodicity, this integral equation can be solved through Fourier transform:
\bea
   \tilde{\theta}_n (j|\phi) & \equiv &  
   \int_{-\pi}^\pi {\de \lambda \over 2 \pi} \: \eu^{-\ii j \lambda} \:
   {\de \over \de \lambda} \theta_n (\lambda) 
   = \eu^{-n |j| \phi } \: , \\
   \tilde{\rho}_0 (j) & = & 
   {\tilde{\theta}_1 (j|\phi) \over 1 + \tilde{\theta}_2 (j|\phi)}
   = {1 \over 2 \cosh \left( j \phi \right)} \; , \\
   \rho_0 (\lambda) & = & 
   {1 \over 4 \pi} \sum_{j=-\infty}^\infty {\eu^{\ii j \lambda}  \over 
   	\cosh \left( j \phi \right)} 
   = {I (k) \over 2 \pi^2} \: {\rm dn}  \left( {I (k) \over \pi} \: \lambda,k \right) \: , \label{rho0XXZAFM}
\eea
where $I(k) \equiv \int_0^{\pi/2} {\de \theta \over \sqrt{1 - k^2 \sin^2 \theta}}$ 
is the {\it complete elliptic integral} of the first kind and ${\rm dn} \left( u,k\right)\equiv \sqrt{1 - k^2 \sin^2 \varphi} $ is one of the Jacobi elliptic functions, defined as the inverse of the {\it incomplete elliptic integral} of the first kind $u \equiv \int_0^{\varphi} {\de \theta \over \sqrt{1 - k^2 \sin^2 \theta}}$ \cite{lawden}.
The {\it elliptic modulus} $k$ is defined through $\phi = \pi {I \left( \sqrt{1-k^2} \right) \over I (k)}$.
It should be noted that in this phase the number of vacancies allowed for $N/2$ real rapidities is $N/2+1$: the ground state configuration generated by a symmetric distribution of quantum numbers is (nearly) degenerate with one in which these numbers are shifted by one unity. While the first has zero momentum, the latter's is $\pi$: in the $\Delta \to - \infty$ limit these two states become the (anti-)symmetric combination of the two Neel states $|\uparrow \downarrow \uparrow \downarrow \uparrow \dots \rangle \pm |\downarrow \uparrow \downarrow \uparrow \downarrow \dots \rangle$.

Low energy excitations are $2 \kappa$ spinons generated by removing $\kappa$ rapidities from the ground states. Each spinon\index{Spinon} contributes with energy \cite{samaj,takahashi}
\be
   \varepsilon (k) = h + J \;  {\sinh \phi \over \pi} I(k) \sqrt{1-k^2 \cos^2 k} \; .
   \label{AFMspinon}
\ee
The lowest energy state at $h=0$ has two spinons with vanishing momentum and finite energy gap $\varepsilon (0) = 2h_{\rm c}$ from (\ref{hcdef}) below: this means that this in an incompressible, gapped phase.

\subsection{Effect of the magnetic field on the anti-ferromagnet}

For small magnetic fields, the ground state configuration is still given be the same integral equation\index{Integral equation!linear}. The lowest field $h_{\rm c}$ necessary to induce a finite magnetization on the ground state can be found from (\ref{XXZhlambda}) by setting $\Lambda = \pi$. The dressed charge\index{Dressed!charge} equation can be solved through Fourier transform, yielding $Z(\lambda) = 1/2$, while from (\ref{rho0XXZAFM}) we have $\rho_0 (\pi) = {1 \over 2 \pi^2} \sqrt{1 - k^2} I(k) $
\be
   h_{\rm c} \equiv 2 \pi J \sqrt{\Delta^2 -1} \rho_0 (\pi) =
   J \;  {\sinh \phi \over \pi} \sqrt{1 - k^2} I(k)  \; .
   \label{hcdef}
\ee

For $h_{\rm c} < h < h_{\rm s}$ we reenter the paramagnetic phase, with the ground state magnetization increasing with $h$ until saturation. For $h>h_{\rm s}$ (\ref{hFM}) the ground state is fully polarized and the phase is ferromagnetic. The full phase diagram of the XXZ chain we have constructed is depicted in figure \ref{fig:XXZPhaseDiag}.

\setlength\extrarowheight{1pt}

\chapter{Algebraic Bethe Ansatz}
\label{chap:algebraic}

\abstract{
	The Algebraic Bethe Ansatz (ABA) approach is essentially a second quantization of the coordinate one we used so far. It uses the Yang-Baxter algebra of the transfer matrix to generate the wavefunctions by applying certain operators (which can be interpreted as quasi-particle creation operators) to a reference state (known as the {\it pseudo-vacuum}). The Bethe equations emerge as consistency conditions for these states to be eigenvectors of the transfer matrix.
	The ABA construction is one of the results in a long effort to understand the relation between seemingly different kinds of integrable systems. In fact, it is the quantum version of the Inverse Scattering Method (ISM): a construction that, through the Lax representation of classical integrable non-linear differential equations, has allowed a deeper understanding of these systems and, even most notably, the systematic construction of their soliton solutions. From another angle, the ABA is grounded on the relation between two-dimensional classical integrable statistical physics models and 1-D quantum ones. 
	One example of this connection is between the six-vertex model, whose solution is presented in Appendix \ref{app:2DClassical}, and the XXZ chain. After outlining the general ABA scheme in Sec. \ref{sec:ABAgeneralities}, we focus on this example. We revisit the structure of a scattering matrix in a lattice model in Sec. \ref{sec:ABAprel} and apply it to the construction of the transfer matrix of the XXZ chain in Sec. \ref{sec:Tconstruction}. In Sec. \ref{sec:ABAsol} and \ref{sec:QISM} we show how to construct the eigenstates and fundamental operators of the theory in the algebraic way and in Sec. \ref{sec:scalarp} we compute scalar products and norms of states. In Sec. \ref{sec:LAX} we introduce the Lax formalism to describe the Lieb-Liniger model. Finally, in Sec. \ref{sec:braid} and \ref{sec:Qgroups} we sketch the connection between the ABA and quantum groups.
    }

\section{Generalities on the algebraic approach}
\label{sec:ABAgeneralities}

The techniques developed to solve non-trivial integrable models pass through the enlargement of the physical space with the introduction of some auxiliary space or variable, to ``decouple'' the interaction so that the physical degrees of freedom do not interact among themselves, but only with the auxiliary space. This procedure simplifies the problem to the point of allowing an exact solution of this enlarged system. The original model can then be recovered by tracing over the auxiliary degrees of freedom \cite{solitonbook,ISM,baxterbook}.

In the ABA, one can think of this additional space as describing a new degree of freedom, a sort of a probe, that propagates inside the system (something like an unobservable gauge field that encodes the interaction between otherwise free particles in a gauge theory).  We denote the physical Hilbert space as ${\cal H}$ and the additional one as ${\cal V}_a$, where $a$ is a label used to distinguished these spaces, when more than one copy of them is needed. One introduces an operator called {\it monodromy matrix}\index{Monodromy matrix} $\bT_a (\lambda): {\cal H} \times {\cal V}_a \to {\cal H} \times {\cal V}_a$, acting in the enlarged space. This operator depends on the {\it spectral parameter} $\lambda$, which can be interpreted as the rapidity of the probe injected in the system. Tracing over the {\it ancillary space} yields the {\it transfer matrix}:\index{Transfer matrix} ${\bf T} (\lambda) \equiv \tr_a \bT_a (\lambda): {\cal H} \to {\cal H}$ and we seek to construct the eigenstates of the latter. The reason for this construction is that the spectral parameter defines a family of transfer matrices and the integrability of the model guarantees that all elements of this family commute with one another. In turns, this means that the transfer matrix is a generating function of a series of conserved charges in involution. Among them, one can identify the Hamiltonian of the quantum model and thus the diagonalization of the transfer matrix provides the eigenvectors of the Hamiltonian as well. 

The integrability of the model is encoded in the property that the order of application of two monodromy matrices\index{Monodromy matrix} is related by a similarity transformation, through an operator called {\it intertwiner}\index{Intertwiner operator} or {\it {\cal R}-matrix} ${\cal R}_{a,b} (\lambda): {\cal V}_a \times {\cal V}_b \to {\cal V}_a \times {\cal V}_b$:
\be
\bT_a (\lambda) \; \bT_b (\mu) \; {\cal R}_{a,b} (\mu - \lambda) =
{\cal R}_{a,b} (\mu-  \lambda) \; \bT_b (\mu) \; \bT_a (\lambda) \; ,
\label{TYB1}
\ee
The existence of such a relation is a consequence of the fact that the intertwiner\index{Intertwiner operator} satisfies itself a similar equation, called the {\it Yang-Baxter equation}\index{Yang-Baxter!equation} (YBE):
\be
{\cal R}_{1,2} (\lambda - \mu) \;
{\cal R}_{1,3} (\lambda - \nu) \; {\cal R}_{2,3} (\mu - \nu) =
{\cal R}_{2,3} (\mu - \nu) \;
{\cal R}_{1,3} (\lambda - \nu) \; {\cal R}_{1,2} (\lambda - \mu) \; ,
\label{RRRYBE-chap5}
\ee
which is depicted graphically in Fig. \ref{fig:YBE}. Here, $1, 2, 3$ indicate three different copies of ${\cal V}$, on which the ${\cal R}$-matrices acts (in pairs)\index{Intertwiner operator}.

\begin{figure}[t]
	\noindent\begin{minipage}[t]{11cm}
		\includegraphics[width=11cm]{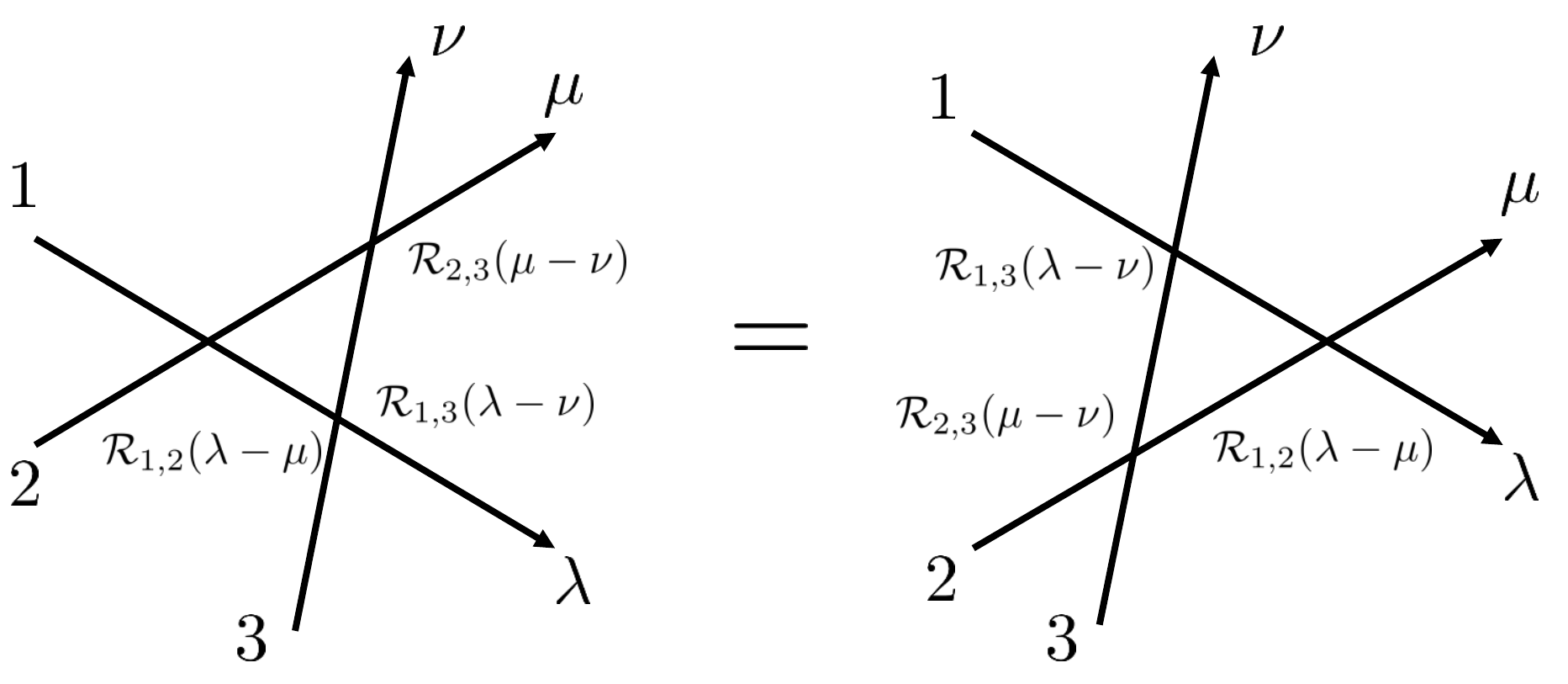}
	\end{minipage}
	\hfill
	\begin{minipage}[b]{5cm}
		\caption{Graphical representation of the Yang-Baxter equation (\ref{RRRYBE-chap5}). \vspace{3.7cm} $\quad$}
		\label{fig:YBE}
	\end{minipage}
\end{figure}

Eq. (\ref{RRRYBE-chap5}) is recognized as the heart of integrability, the equation from which the whole construction follows. In fact, each solution of (\ref{RRRYBE-chap5}) generates a family of integrable models. However, there is no prescription to reconstruct the intertwiner\index{Intertwiner operator} operator for a given integrable Hamiltonian. Thus, in practice, one looks for solutions of eq. (\ref{RRRYBE-chap5}) and, every time a ${\cal R}$-matrix\index{Intertwiner operator} is found, the Inverse Scattering Method (ISM) machinery is used to identify the model at hand. Once the ${\cal R}$-matrix is given, this identification is straightforward and direct, like calculating the derivative of a function. The inverse problem of finding the primitive of a function, however, is quite complicated and in general we do not have a systematic way to do so (even if we have a number of tricks and techniques to help), but we rely on our experience in taking derivatives. The situation is very similar to the problem of finding the ${\cal R}$-matrix of a given model\index{Intertwiner operator}. This is why the algebraic approach is referred to as the inverse problem.

Let us thus sketch the steps of the Quantum-ISM in generality and then proceed to present the explicit example of its application to the XXZ chain.
In general, one assumes that the physical Hilbert space can be decomposed as the direct product of the Hilbert spaces of each particle, or of each site ${\cal H}_j$: ${\cal H} = \otimes_{j=1}^N {\cal H}_j$. For the XXZ spin chain, the space at each site is $\mathbb{C}^2$, i.e. a two-dimensional complex vector corresponding to the probability space of having a spin-$1/2$ up or down. The ancillary space ${\cal V}$ in principle can be a different $\kappa$-dimensional space, but in the XXZ model $\kappa=2$ (in fact, ${\cal H}_j$ and ${\cal V}$ are isomorphic in this case):

\begin{itemize}
	
	\item For a given $\kappa$, one looks for a solution of the Yang-Baxter equation\index{Yang-Baxter!equation}  (\ref{RRRYBE-chap5}).
	
	\item Having a solution for the ${\cal R}$-operator\index{Intertwiner operator}, one looks for the ${\cal L}$-operator (or {\it Lax-operator})\index{Lax operator} ${\cal L}_{j;a}: {\cal H}_j \times {\cal V}_a \to {\cal H}_j \times {\cal V}_a$ which satisfies
	\be
	{\cal L}_{j,a} (\lambda) \; {\cal L}_{j,b} (\lambda') \;
	{\cal R}_{a,b} (\lambda' - \lambda) = {\cal R}_{a,b} (\lambda' - \lambda) \; {\cal L}_{j,b} (\lambda') \; {\cal L}_{j,a} (\lambda) \; .
	\label{LLR}
	\ee
	As explained in (\ref{YBEassociativity}), the existence of a solution to (\ref{LLR}) is guaranteed by (\ref{RRRYBE-chap5}) \cite{ISM,samaj}\footnote{The YBE (\ref{LLR}) satisfied by the ${\cal L}$-operator can be thought of as an algebra\index{Yang-Baxter!algebra}, whose structure factors are given by the ${\cal R}$-matrix\index{Intertwiner operator}. This algebra also has an adjoint representation, same as in traditional Lie-algebras, which is the YBE\index{Yang-Baxter!equation} (\ref{RRRYBE-chap5}) satisfied by the ${\cal R}$-matrix with itself.}. The ${\cal L}$-operator\index{Lax operator} can be thought of as the scattering matrix of the probe with a physical degree of freedom.
	
	\item The monodromy matrix\index{Monodromy matrix} is constructed as the product of the ${\cal L}$'s at\index{Lax operator} the $N$ different sites of a chain:
	\be
	\bT_a (\lambda) \equiv 
	{\cal L}_{N,a} (\lambda-\xi_N) \;
	{\cal L}_{N-1,a} (\lambda-\xi_{N-1}) \cdots {\cal L}_{1,a} (\lambda-\xi_1) \; ,
	\label{MonodromyMatrixdef}
	\ee
	where we allow each site to be endowed with a different spectral parameter $\xi_j$. If $\lambda$ identifies the rapidity of the probe injected into the system, $\xi_j$ is the rapidity of the degree of freedom sitting at site $j$, so that their scattering depends on $\lambda - \xi_j$.
	Note that, by construction, the monodromy matrix\index{Monodromy matrix} thus defined satisfies (\ref{TYB1}), as can be checked by writing it in terms of the ${\cal L}$-operators\index{Lax operator} and using repeatedly (\ref{LLR}) to shift the intertwiner\index{Intertwiner operator} through the chain (see also figure \ref{fig:TTR-YBE}).
	
	\item We now have to trace over the ancillary space. The monodromy operator\index{Monodromy matrix} can be represented as a $\kappa \times \kappa$ matrix, where each of the $\kappa^2$ entries are operators acting on ${\cal H}$. For instance, focusing on the $\kappa=2$ case, the monodromy matrix can be written as
	\be
	\bT_a (\lambda)
	= \left( \begin{array}{cc}
		{\bf A} (\lambda) & {\bf B} (\lambda) \cr
		{\bf C} (\lambda) & {\bf D} (\lambda) \cr
	\end{array} \right) \; .
	\label{MonodromyMatrix}
	\ee
	The transfer matrix\index{Transfer matrix} is then
	\be
	{\bf T} (\lambda) \equiv \tr_a \bT_a (\lambda) 
	= {\bf A} (\lambda) + {\bf D} (\lambda) \; .
	\label{TransferMatrix}
	\ee
	If $\dim {\cal H}_j=2$, the ${\bf A},{\bf B},{\bf C},{\bf D}$ operators can also be represented a $2^N \times 2^N$ matrices, whose explicit expressions are no simpler than the transfer matrix\index{Transfer matrix} and less telling than the Hamiltonian. However, when (\ref{TransferMatrix}) is inserted in (\ref{TYB1}), the Yang-Baxter equation\index{Yang-Baxter!equation} for the monodromy matrix\index{Monodromy matrix} provides a series of generalized commutation relations between the ${\bf A},{\bf B},{\bf C},{\bf D}$ at different spectral parameters, which can be exploited to generate the eigenstate of the system.
	
	\item One identifies a state $|0\rangle$ which is annihilated by ${\bf C}$:
	\be 
	   {\bf C} (\lambda) |0 \rangle = 0 \; .
	\ee
	This is a reference state and it is called {\it pseudo-vacuum}\index{Pseudo-vacuum} because it hosts no quasi-particle\index{Quasi-particle} excitation.
	
	\item States with $R$ quasi-particle excitations are constructed as
	\be
	   | \Psi \rangle = \prod_{j=1}^R {\bf B} (\lambda_j) |0 \rangle \; .
	   \label{eigenansatz}
	\ee
	The quasi-particle\index{Quasi-particle} rapidities are determined by the eigenvector condition
	\be
	   {\bf T} (\lambda) | \Psi \rangle 
	   = \big[ {\bf A} (\lambda) + {\bf D} (\lambda) \big] | \Psi \rangle
	   = \Lambda (\lambda) |\Psi \rangle \; .
	   \label{eigeneqT}
	\ee
	This condition can be worked out using the commutation relations between the ${\bf A},{\bf B},{\bf C},{\bf D}$ operators and is equivalent to the $R$ algebraic Bethe equations\index{Bethe!equations} for the $\lambda_j$ (equations that will not depend on $\lambda$).
	
	\item To calculate the expectation value of a given operator, one needs to express it through the ${\bf A},{\bf B},{\bf C},{\bf D}$ operators. Note that typically one is interested in local observables, while the operators appearing in the monodromy matrix\index{Monodromy matrix} are non-local. Thus, while in the ABA the states are expressed in a rather compact form, the observables are typically complicated objects. Nonetheless, the algebraic structure behind the construction allows for certain elegant manipulations and in some case to explicit results. One such explicit formula is the norm of (\ref{eigenansatz}), which is a fundamental ingredient to normalize all other correlation functions.	

\end{itemize}

\section{Preliminaries}
\label{sec:ABAprel}

Before we proceed with the algebraic construction, let us look once more at the two-body interaction from the scattering matrix point of view.
Let us write the two-body wavefunction as
\bea
   \Psi (x_1,x_2) & = & \sum_{\cP} \Psi ({\cal Q}|\cP) \, \eu^{\ii \sum_j x_{{\cal Q}j} k_{\cP j}}
   \\
   & = & \left\{ \begin{array}{ll}
                    \Psi(1,2|1,2) \, \eu^{\ii (x_1 k_1 + x_2 k_2)}
                    + \Psi (1,2|2,1) \, \eu^{\ii (x_1 k_2 + x_2 k_1)} \; , & \qquad x_1 < x_2 \cr
                    \Psi(2,1|1,2) \, \eu^{\ii (x_2 k_1 + x_1 k_2)}
                    + \Psi (2,1|2,1) \, \eu^{\ii (x_2 k_2 + x_1 k_1)} \; , & \qquad x_1 > x_2 \cr
                 \end{array} \right.
   \nonumber \\
   & = & \eu^{\ii X K} \left\{ \begin{array}{ll}
                    \Psi(1,2|1,2) \, \eu^{\ii x k} + \Psi (1,2|2,1) \, \eu^{- \ii x k} \; , 
                    & \qquad \qquad \qquad \: \, x < 0 \cr
                    \Psi(2,1|1,2) \, \eu^{- \ii x k} + \Psi (2,1|2,1) \, \eu^{\ii x k} \; , 
                    & \qquad \qquad \qquad \: \, x > 0 \cr
                 \end{array} \right. \; ,
\eea
where we used center-of-mass coordinates
\be
   X \equiv {x_1 + x_2 \over 2} \; , \qquad \quad
   x \equiv {x_1 - x_2 \over 2} \; , \qquad \quad
   K \equiv {k_1 + k_2} \; , \qquad \quad 
   k \equiv {k_1 - k_2} \; .
\ee
We explicitly wrote the dependence of the amplitudes $\Psi ({\cal Q}|\cP)$ on the order of particles (given by the ${\cal Q}$-permutation) and of the pairing with the different momenta (given by the $\cP$-permutation).

Let us now imagine a scattering experiment. We send in a beam from the left and we measure a reflected component on the left with amplitude $R(k)$ and a transmitted one to the right with amplitude $T(k)$:
\be
   \Psi (1,2|1,2) = R(k) \, \Psi (1,2|2,1) + T(k) \, \Psi (2,1|2,1) \; .
\ee
Similarly, if we start with an incident ray from the right we have
\be
   \Psi (2,1|1,2) = R(k) \, \Psi (2,1|2,1) + T(k) \, \Psi (1,2|2,1) \; .
\ee

We can cast these equations in matrix form, in several ways. We can write
\be
   \Psi^r (\cP') =  \left( \begin{array}{c} \Psi (1,2|1,2) \cr \Psi (2,1|1,2) \end{array} \right) =
   \left( \begin{array}{cc} R(k) & T(k) \cr T(k) & R(k) \cr \end{array} \right)
   \left( \begin{array}{c} \Psi (1,2|2,1) \cr \Psi (2,1|2,1) \cr \end{array} \right)
   = {\bf S}^r (k) \Psi (\cP) \; ,
\ee
where the identities of the particles are uncorrelated with the momenta. This representation is called {\it reflection-diagonal}. An alternative choice is the {\it transmission-diagonal representation}
\be
   \Psi^t (\cP') =  \left( \begin{array}{c} \Psi (2,1|1,2) \cr \Psi (1,2|1,2) \end{array} \right) =
   \left( \begin{array}{cc} T(k) & R(k) \cr R(k) & T(k) \cr \end{array} \right)
   \left( \begin{array}{c} \Psi (2,1|2,1) \cr \Psi (1,2|2,1) \cr \end{array} \right)
   = {\bf S}^t (k) \Psi (\cP) \; ,
\ee
where we identify each particle with the momentum it carries. Other representations are possible, but we will not use them. Since the particles are indistinguishable, we can equally think that after a scattering event a particle has gone through (transmission-diagonal) or has recoiled and released its momentum to the other particle (reflection-diagonal).

The reflection and transmission-diagonal representations are related by a matrix ${\bf \Pi} \equiv \begin{pmatrix}0 & 1 \\ 1 & 0 \end{pmatrix}$ that exchanges the particles:
\be
   \Psi^r (\cP) = {\bf \Pi} \, \Psi^t (\cP) \; , \qquad \qquad
   {\bf S}^r (k) = {\bf \Pi} \, {\bf S}^t (k) \;.
\ee
We can in fact write
\be
   {\bf S}^r (k) = R(k) + {\bf \Pi} \: T(k) \; , \qquad \qquad
   {\bf S}^t (k) = T(k) + {\bf \Pi} \: R(k) \; .
\ee

Using ${\bf \Pi}^2 = 1$, if the particle have bosonic/fermionic statistics, we have $\Pi = \pm 1$, thus
\be
   S^r (k) = R(k) \pm T(k) \; , \qquad \qquad S^t (k) = T(k) \pm R(k) = \pm S^r (k) \; .
\ee
The transmission and reflection coefficients are uniquely determined by the statistic of the particles and their scattering phase\index{Scattering phase} $S^r (k) = -\eu^{-\ii \theta_\pm (k)}$, where the $+/-$ sign refers to bosons/fermions. This is the scattering phase\index{Scattering phase} we calculated in  sections \ref{2bodysec}, \ref{sec:XXXBetheSol}, and \ref{sec:XXZBetheSol}.

In a lattice system, the scattering problem has to be supplemented with the information about the presence of a particle on the lattice site (equivalently, we can say that an additional quantum number is involved in the scattering event). Thus, the scattering matrix becomes a $4 \times 4$ matrix. We write the two-body interaction as a matrix connecting the $4$ possible states $|\uparrow \uparrow \rangle, |\uparrow \downarrow \rangle, |\downarrow \uparrow \rangle, |\downarrow \downarrow \rangle$ (we use the spin language, but we know we can equally replace a spin up/down with an empty/occupied site).
If the interaction conserves the magnetization/particle number, we have
\be
{\bf S}^r = \left( \begin{array}{cccc}
	\Theta & 0 & 0 & 0 \cr
	0 & R & T & 0 \cr
	0 & T & R & 0 \cr
	0 & 0 & 0 & \Theta \cr
\end{array} \right)
= \Theta (k) \left( \begin{array}{cccc}
	1 & 0 & 0 & 0 \cr
	0 & r & t & 0 \cr
	0 & t & r & 0 \cr
	0 & 0 & 0 & 1 \cr
\end{array} \right)
= \Theta (k) \: {\bf s}^r (k) \; ,
\ee
where we normalized by the amplitude for the aligned scattering and introduced the reduced reflection and transmission amplitudes $r \equiv R / \Theta$, $t \equiv T / \Theta$, with $\Theta (k) = - \eu^{-\ii \theta_\pm (k)}$. If the scattering event does not conserve particle number/magnetization, but only its parity,, the scattering matrix has two additional non-zero elements filling the whole anti-diagonal. This is the case for the XY\index{XY chain} or XYZ chain\index{XYZ chain}, but the algebraic construction for these models is significantly more involved and we will not pursue it further.

The exchange operator is represented as the $4 \times 4$ matrix
\be
{\bf \Pi} = \left( \begin{array}{cccc}
	\pm 1 & 0 & 0 & 0 \cr
	0 & 0 & 1 & 0 \cr
	0 & 1 & 0 & 0 \cr
	0 & 0 & 0 & \pm 1 \cr
\end{array} \right) \; ,
\label{Pidef}
\ee
where the $+$ sign applies to bosons and spin, while the $-$ is reserved for fermions. ${\bf \Pi}$ can be used to switch from the reflection-diagonal representation to the transmission-diagonal one: ${\bf S}^t = {\bf \Pi} {\bf S}^r = \Theta (k) {\bf s}^t (k)$. 

Notice that these matrices have a natural representation as a product of $2 \times 2$ Pauli matrices, for instance
\be
   {\bf s}^t (k) = {1 \over 2} \Big[ 1 + \sigma_z \sigma'_z + t(k) \left(1 - \sigma_z \sigma'_z \right)
   + r(k) \left( \sigma_x \sigma'_x + \sigma_y \sigma'_y \right) \Big] \; .
   \label{scatteringop}
\ee
Since $t(0)=0$, $r(0)=1$, we have
\be
  {\bf s}^t (0) = {\bf \Pi} = {1 + \vec{\sigma} \cdot \vec{\sigma}' \over 2} \; .
\ee

\section{Transfer Matrix for the XXZ chain}
\label{sec:Tconstruction}

Historically, the algebraic approach to the XXZ chain was developed following the transfer matrix\index{Transfer matrix} solution of the two-dimensional 6-vertex model\index{Vertex model} \cite{baxterbook}. It relies on the construction of certain operators that do not have a clear physical meaning within the quantum model, while they arise quite naturally in its classical counterpart and thus inherit their name from the latter construction. We discuss the solution of the 6-vertex\index{Vertex model} model in appendix \ref{app:2DClassical}: although it is not essential to understand the following derivations, we strongly recommend that the reader unfamiliar with these topics reads App. \ref{app:2DClassical} before proceeding, in order to provide context for the algebraic Bethe Ansatz approach.
Also, the solution of the classical model is somehow constructive, while, as we commented in the section \ref{sec:ABAgeneralities}, the ABA solution is quite indirect.

We start by looking for a solution of the Yang-Baxter equation\index{Yang-Baxter!equation} (\ref{RRRYBE-chap5}) for a $4 \times 4$ intertwiner\index{Intertwiner operator} matrix (that is, $\dim {\cal V}_a=2$). 
We take advantage of the work done in section \ref{sec:yangbaxter} to write the ${\cal R}$-operator\index{Intertwiner operator} as
\be
{\cal R}_{a,b} (\lambda) = a (\lambda) \; {1 + \tau_a^z \tau_b^z \over 2 } \; + \;
b (\lambda) \; {1 - \tau_a^z \tau_b^z \over 2} \; + \;
c (\lambda) \Big[ \tau_a^+ \tau_b^- + \tau_a^- \tau_b^+ \Big] \; ,
\label{Ransatz}
\ee
where $\tau_a^\alpha$ are Pauli matrices acting on ${\cal V}_a$ and the functions $a(\lambda), b(\lambda), c(\lambda)$ were determined in (\ref{BaxterParam}) as:
\bea
  &&
  a(\lambda) = \varphi (\lambda + \ii 2 \phi) \; , \qquad \qquad
  b(\lambda) = \varphi (\lambda ) \; , \qquad \qquad
  c(\lambda) = \varphi (\ii 2 \phi) \; , \\
  && \qquad 
   \varphi (\lambda) = \left\{ \begin{array}{lcll}
  	\sin \left( {\lambda \over 2} \right) &  \quad {\rm XXZ} \quad & \Delta >1 
  	& \quad\cosh \phi = \Delta \; , \cr
  	{\lambda \over 2} & \quad {\rm XXX} \quad & \Delta =1 
  	& \quad \phi=1 \; , \cr
  	\sinh \left( {\lambda \over 2} \right) & \quad {\rm XXZ} \quad  & |\Delta|<1 
  	& \quad \cos \phi =  \Delta \; .
  \end{array} \right.
  \label{varphidef}
\eea
Note that the parameter $\phi$ is kept fixed and common to every matrix satisfying (\ref{RRRYBE-chap5}). Compared to Baxter's parametrization of the six-vertex model\index{Vertex model} (\ref{BaxterParam}), in the quantum case we took the spectral parameter to the imaginary axis $\lambda \to \ii {\lambda \over 2}$, and rescaled it for later convenience. 

We now look for the {\it Lax operator} ${\cal L}_{j,a}$,\index{Lax operator} satisfying (\ref{LLR}). In general, the physical Hilbert space ${\cal H}_j$ does not have to be isomorphic to the ancillary one ${\cal V}_a$, they might have different dimensions and thus the matrix representation of the ${\cal L}$-operator\index{Lax operator} might be rectangular (while, by construction, the ${\cal R}$-operator\index{Intertwiner operator} is alway a $\kappa \times \kappa$ square matrix). However, as we found in section \ref{sec:yangbaxter}, for the XXZ chain the ${\cal L}$-operator\index{Lax operator} is also a $4 \times 4$ matrix functionally similar to (\ref{Ransatz}):
\be
   {\cal L}_{j,a} (\lambda) =
   {1 + \sigma_j^z \tau_a^z \over 2 } \;
   + \; t (\lambda) \; {1 - \sigma_j^z \tau_a^z \over 2} \;
   + \; r (\lambda) \; \left( \sigma_j^+ \tau_a^- + \sigma_j^- \tau_a^+ \right) \; ,
   \label{Lansatz}
\ee
where $\sigma_j^\alpha$ are Pauli matrices acting on the chain at the site $j$ and
\be
   t (\lambda) \equiv {b(\lambda) \over a (\lambda)}
   = {\varphi ( \lambda) \over \varphi (\lambda + \ii 2 \phi)} \; , \qquad \quad
   r (\lambda) \equiv {c (\lambda) \over a (\lambda) }
   = {\varphi (\ii 2 \phi) \over \varphi (\lambda + \ii 2 \phi)} \; .
   \label{trdef}
\ee
Note that $t(\lambda) t(-\lambda) + r (\lambda) r(-\lambda) =1$, $t(\lambda) r(-\lambda) + r (\lambda) t(-\lambda) =0$: ${\cal L}_{j,a} (\lambda)$ behaves like a scattering matrix.

Comparing the functional form of the ${\cal R}$\index{Intertwiner operator} and ${\cal L}$-operators\index{Lax operator} with (\ref{scatteringop}) we interpret $t(\lambda)$ and $r(\lambda)$ as the {\it transmission} and {\it reflection coefficients}, respectively.
We notice once more that, since $t(0)=0$ and $r(0)=1$, the Lax-operator\index{Lax operator} at $\lambda=0$ reduces to the permutation operator that exchanges the two spins/particles:
\be
   {\cal L}_{j,a} (0)
   = {1 + \sigma_j^z \tau_a^z \over 2 } \; + \;
   \left( \sigma_j^+ \tau_a^- + \sigma_j^- \tau_a^+ \right)
   = {1 \over 2} \left( {\cal I}_j \otimes {\cal I}_a +
   \overrightarrow{\sigma}_j \otimes \overrightarrow{\tau}_a \right)
   = {\bf \Pi}_{j,a} \; .
   \label{exchange}
\ee

We interpret the ${\cal L}$-operator\index{Lax operator} as the scattering matrix of the ancillary spin interacting with the physical spin on the chain.
It is convenient to consider the Lax-operator\index{Lax operator} as a $\kappa \times \kappa$ matrix (in the auxiliary space) with matrix elements given by operators in the physical Hilbert space
\be
\label{XXZL}
{\cal L}_{j,a} = \left( \begin{array}{cc}
	{1 + t (\lambda) \over 2} + {1 - t (\lambda) \over 2} \; \sigma_j^z & r (\lambda) \sigma_j^- \cr
	r (\lambda) \sigma_j^+ & {1 + t (\lambda) \over 2} - {1 - t (\lambda) \over 2} \; \sigma_j^z \cr
\end{array} \right) 
 = {1 \over \varphi (\lambda + \ii 2 \phi)} \left( \begin{array}{cc}
	\varphi \Big( \lambda + \ii (1 + \sigma_j^z) \: \phi  \Big) & 
	\varphi (\ii 2 \phi) \: \sigma_j^- \cr
	\varphi (\ii 2 \phi) \: \sigma_j^+  
	& \varphi \Big( \lambda + \ii (1 - \sigma_j^z) \: \phi  \Big) \cr
\end{array} \right) \; .
\ee

We can construct the {\it transition matrix} of the auxiliary spin as it moves across several sites:
\be
  \bT_a (n,m|\lambda) \equiv {\cal L}_{n,a} (\lambda-\xi_n) \;
  {\cal L}_{n-1,a} (\lambda-\xi_{n-1}) \cdots {\cal L}_{m,a} (\lambda-\xi_m) \; ,
  \qquad n \ge m \; .
  \label{TLLLL}
\ee
Here, the product between ${\cal L}$-operators\index{Lax operator} is a standard matrix product in ${\cal V}_a$ space, as they all act on the same auxiliary space. Thus the transition matrix remains a $\kappa \times \kappa$ matrix in ${\cal V}_a$, but each matrix element is an operator in the Hilbert space $\otimes_{j=m}^n {\cal H}_j$ of the physical sites involved.
In (\ref{TLLLL}) we allow each degree of freedom at site $j$ to carry a different rapidity $\xi_j$ which interacts with the rapidity $\lambda$ of the probe in ${\cal V}_a$ space. In intermediary steps, it is convenient to leave these {\it inhomogeneity parameters} free, to distinguish the different sites of the chain. However, at the end of the calculation one is generally interested in a translational invariant system and it is customary to set $\xi_j= \ii \phi/2$ for $j=1 \ldots N$ to restore the symmetry of the Bethe equations\index{Bethe!equations} (\ref{ABABE}).
The transition matrix across the entire chain is called the {\it monodromy matrix}\index{Monodromy matrix} (or {\it winding matrix}) $\bT_a (\lambda) \equiv \bT_a (N,1|\lambda)$. As an operator acting on ${\cal V}_a$, it is still a $\kappa \times \kappa$ matrix: for the XXZ chain it can be written as (\ref{TransferMatrix}), where ${\bf A}, {\bf B}, {\bf C}, {\bf D}$ are operators acting on ${\cal H}$, which can be each represented as $2^N \times 2^N$ matrices. However, their explicit representation will not be needed, as the ABA construction relies solely on their properties and the algebra they satisfy\index{Yang-Baxter!algebra}.

The monodromy matrix\index{Monodromy matrix} (\ref{MonodromyMatrix}) describes the scattering of a probe spin 
(living in the auxiliary vector space ${\cal V}_a$) with the whole XXZ chain, as the ghost spin propagates and interacts with each physical site.
Similarly with what we did for the ${\cal L}$-operator\index{Lax operator} (\ref{scatteringop}), we can write $\bT_a$ as
\be
   \bT_a (\lambda) =
   {1 \over 2} \left[ {\bf A} (\lambda) + {\bf D} (\lambda) \right] \: {\cal I}_a
   + {1 \over 2} \left[ {\bf A} (\lambda) - {\bf D} (\lambda) \right] \: \tau_a^z
   + {\bf B} (\lambda) \: \tau_a^+ + {\bf C} (\lambda) \: \tau_a^- \; .
   \label{MonodromyOperator}
\ee
Comparing this expression with (\ref{XXZL}), we notice that the ${\bf C}$ (${\bf B}$) operators act as spin raising (lowering) operator for the whole chain, respectively.

We consider two monodromy matrices\index{Monodromy matrix}, acting on different auxiliary space ${\cal V}_a$ and ${\cal V}_b$ and with different spectral parameters $\lambda$ and $\lambda'$. Remembering (\ref{MonodromyMatrixdef}) repeated use of (\ref{LLR}) shows that
\be
   {\cal R}_{a,b}^{-1} (\lambda' - \lambda) \; \bT_a (\lambda) \;
   \bT_b (\lambda') \; {\cal R}_{a,b} (\lambda' - \lambda) =
   \bT_b (\lambda') \; \bT_a (\lambda) \; ,
   \label{RTT}
\ee
as shown in figure \ref{fig:TTR-YBE}.
This is the Yang-Baxter equation\index{Yang-Baxter!equation} for the winding matrix and we will see later that, from an ABA point of view, it can be considered as a set of generalized commutation relations.
Physically, it means that it is equivalent to let two probes scatter through the physical chain, or to let these probe scatter on one another first, then propagate through the system and finally scatter again.

The {\it transfer matrix}\index{Transfer matrix} ${\bf T} (\lambda)$ (\ref{TransferMatrix}) is obtained by tracing over the ancillary space. This operation is equivalent to closing the system at infinity (with periodic boundary conditions) and thus requiring that the probe emerges from the interaction in the same state as it entered.
Taking the trace over ${\cal V}_a \otimes {\cal V}_b$ of (\ref{RTT}) and using the cyclic property of the trace we get
\be
  \left[ {\bf T} (\lambda), {\bf T} (\lambda') \right] = 0 \; .
  \label{TTcomm}
\ee
The Yang-Baxter equation\index{Yang-Baxter!equation} (\ref{RTT}) means that the entanglement process due to the probe propagation that generates the monodromy matrix\index{Monodromy matrix} can be factorized at the border and, by taking the trace over the ghost variables, the two chains can be disentangled. So, the transfer matrices generated by two ghost particles propagating with different parameters commute.

Eq. (\ref{TTcomm}) implies that the transfer matrix\index{Transfer matrix} is a generating function of  conserved quantities. Of course, the charges which can be generated expanding ${\bf T} (\lambda)$ depend on the expansion point. Moreover, it turns out that it is more convenient to expand the logarithm of the transfer matrix, since in this way one can construct (semi-)local operators. We can thus introduce a set of operators defined as
\be
   {\bf J}_{\{c\}} \equiv \sum_n \sum_j c_{n,j} \;
   \left. {\de^n \over \de \lambda^n} \;
   \ln {\bf T} (\lambda) \right|_{\lambda = \lambda_j} \; ,
   \label{Hc}
\ee
for certain $\lambda_j$, where $c_{n,j}$ are some coefficients.
It is clear that, using (\ref{TTcomm}) we have
\be
  \left[ {\bf J}_{\{c\}} , {\bf T} (\lambda) \right] = 0 \qquad \quad
  {\rm and} \qquad \quad
  \left[ {\bf J}_{\{c\}}, {\bf J}_{\{c'\}} \right] = 0 \; .
\ee
Expressions (\ref{Hc}) are known as {\it trace identities}\index{Trace identities} and define integral of motions in involutions, which can be used to characterize the state of the integrable system.

Let us now take the {\it homogeneous limit} in (\ref{MonodromyMatrixdef}): $\xi_j \to \xi$, for $j=1 \ldots N$ and let us consider, for instance, ${\bf T} (\xi) $. From (\ref{exchange}) we know that the monodromy matrix\index{Monodromy matrix} is composed by product of exchange operators. That is, the probe enters the system and exchanges its state with the first spin, then proceed to the next lattice site and exchanges its state, effectively leaving that spin with the state of the previous one and so on. After taking the trace and closing the chain, the net effect has been to shift every spin by one lattice site. To formalize this process, we remind the identity for the permutations
\be
   {\bf \Pi}_{j,a} \hat{X}_a = \hat{X}_j {\bf \Pi}_{j,a} \; ,
\ee
where $\hat{X}_a$ is some operator acting on the vector space $a$.
Thus, using (\ref{exchange}) we have (${\bf \Pi}_{j,l} = {\bf \Pi}_{l,j}$)
\be
   {\bf \Pi}_{j,a} {\bf \Pi}_{l,a} 
   = {\bf \Pi}_{j,l} {\bf \Pi}_{j,a} 
   = {\bf \Pi}_{l,a} {\bf \Pi}_{j,l} \; .
\ee
Then
\be
   \bT_a (\xi) 
   = {\bf \Pi}_{N,a} {\bf \Pi}_{N-1,a} \ldots {\bf \Pi}_{1,a}
   =  {\bf \Pi}_{1,a} {\bf \Pi}_{1,2} {\bf \Pi}_{2,3} \ldots {\bf \Pi}_{N-1,N} \; .
   \label{bTxi}
\ee
Since $\tr_a {\bf \Pi}_{j,a} = {\cal I}_j$, we get
\be
   {\bf T} (\xi) 
   ={\bf \Pi}_{1,2} {\bf \Pi}_{2,3} \ldots {\bf \Pi}_{N-1,N}
  =  \exp \left( \ii \hat{P} \right) \; ,
\ee
where $\hat{P}$ is the lattice momentum operator. Thus
\be
   \hat{P} = - \ii \ln {\bf T} (\xi) \; .
   \label{PTrel}
\ee

Next, let us look at the first logarithmic derivative of the transfer matrix\index{Transfer matrix}.
Using the properties of permutation operators
\bea
   \left. {\de \over \de \lambda} \bT_a (\lambda) \right|_{\lambda=\xi}
   & = & \sum_{j=1}^N {\bf \Pi}_{N,a} \ldots
   {\bf \Pi}_{j+1,a} {\cal L}'_{j,a} (0) {\bf \Pi}_{j-1,a} \ldots {\bf \Pi}_{1,a}
   \\
   & = & \sum_{j=1}^N {\bf \Pi}_{N,a} \ldots
   {\cal L}'_{j,j+1} (0) {\bf \Pi}_{j+1,a} {\bf \Pi}_{j-1,a} \ldots {\bf \Pi}_{1,a}
   \nonumber \\
   & = & \sum_{j=1}^N {\cal L}'_{j,j+1} (0)
   {\bf \Pi}_{1,2} {\bf \Pi}_{2,3} \ldots {\bf \Pi}_{j-1,j+1} \ldots {\bf \Pi}_{N-1,N} \Pi_{N,a} \; .
   \nonumber
\eea
Taking the trace over ${\cal V}_a$ we get
\be
   \left. {\de \over \de \lambda} {\bf T}_a (\lambda) \right|_{\lambda=\xi}
   = \sum_{j=1}^N {\cal L}'_{j,j+1} (0)
   {\bf \Pi}_{1,2} {\bf \Pi}_{2,3} \ldots {\bf \Pi}_{j-1,j+1} \ldots {\bf \Pi}_{N-1,N} \; .
\ee
Finally, multiplying by the inverse shift operator most of the permutation operators cancel out and we are left with
\be
   \left. {\de \over \de \lambda} \; \ln {\bf T} (\lambda) \right|_{\lambda=\xi} =
   \left. {\de \over \de \lambda} {\bf T}_a (\lambda) \right|_{\lambda=\xi} {\bf T}^{-1} (0)
   = \sum_{j=1}^N {\cal L}'_{j,j+1} (0) {\bf \Pi}_{j,j+1}
   = \sum_{j=1}^N \left. {\de \over \de \lambda} \; \ln {\cal L}_{j,j+1} (\lambda) \right|_{\lambda = 0} \; .
\ee

Now, we notice that
\be
   \left. {\de \over \de \lambda} {\cal L}_{j,j+1} (\lambda) \right|_{\lambda=0} =
   {1 \over \varphi (\ii 2 \phi)} \; \left[ {1 - \sigma_j^z \sigma_{j+1}^z \over 2 } \;
   - \; \varphi' (\ii 2 \phi) \left( \sigma_j^+ \sigma_{j+1}^- + \sigma_j^- \sigma_{j+1}^+ \right) \right]
\ee
and
\be
   {\cal L}'_{j,j+1} (0) {\bf \Pi}_{j,j+1} = {1 \over \varphi (\ii 2 \phi)} \; \left[
   - \varphi' (\ii 2 \phi) \; {1 - \sigma_j^z \sigma_{j+1}^z \over 2 } \; + \;
   \sigma_j^+ \sigma_{j+1}^- + \sigma_j^- \sigma_{j+1}^+ \right] \; .
\ee
The last identity is most easily checked in matrix form
\be
   {\cal L}'_{j,j+1} (0) {\bf \Pi}_{j,j+1} = {1 \over \varphi (\ii 2 \phi)}
   \left( \begin{array}{cccc}
                 0 & 0 & 0 & 0 \cr
                 0 & 1 & -\varphi' (\ii 2 \phi) & 0 \cr
                 0 & -\varphi' (\ii 2 \phi) & 1 & 0 \cr
                 0 & 0 & 0 & 0 \cr
          \end{array}  \right)
   \left( \begin{array}{cccc}
                 1 & 0 & 0 & 0 \cr
                 0 & 0 & 1 & 0 \cr
                 0 & 1 & 0 & 0 \cr
                 0 & 0 & 0 & 1 \cr
          \end{array}  \right)
   = {1 \over \varphi (\ii 2 \phi)}
   \left( \begin{array}{cccc}
                0 & 0 & 0 & 0 \cr
                0 & - \varphi' (\ii 2 \phi) & 1 & 0 \cr
                0 & 1 & - \varphi' (\ii 2 \phi) & 0 \cr
                0 & 0 & 0 & 0 \cr
   \end{array}  \right) \; .
\ee
We conclude that
\be
   \left. {\de \over \de \lambda} \; \ln {\bf T} (\lambda) \right|_{\lambda=\xi}
   = {1 \over \varphi (\ii 2 \phi)} \sum_{j=1}^N \left[ \varphi' (\ii 2 \phi) \; {\sigma_j^z \sigma_{j+1}^z - 1 \over 2 } \; + \;
   \sigma_j^+ \sigma_{j+1}^- + \sigma_j^- \sigma_{j+1}^+ \right] \; ,
\ee
where we recognize the form of the XXZ Hamiltonian (\ref{XXZham}), with no external field. From (\ref{varphidef}) we have $\varphi' (\ii 2 \phi) = \Delta$, thus
\be
   \hat{H}_{XXZ} = {\sqrt{\Delta^2 -1} \over 2}
   \left. {\de \over \de \lambda} \; \ln {\bf T} (\lambda) \right|_{\lambda=\xi} + {1 \over 4} \; \Delta \; N \; .
   \label{HTrel}
\ee
In this way, we proved that both the lattice momentum and the Hamiltonian of the XXZ chain are among the conserved quantities in convolution generated by the transfer matrix\index{Transfer matrix}.

Notice that only now we are able to identify the Hamiltonian of the model generated by the ansatz (\ref{Ransatz},\ref{Lansatz}). In general, one starts with a solution ${\cal R}$\index{Intertwiner operator} of the Yang-Baxter equation\index{Yang-Baxter!equation} (\ref{RRRYBE-chap5}), finds the Lax-operator\index{Lax operator} corresponding to a given Hilbert space for which the solution ${\cal R}$ acts as an intertwiner\index{Intertwiner operator} and with that constructs the monodromy\index{Monodromy matrix} and transfer matrices\index{Transfer matrix}. At this point, through the trace identities\index{Trace identities} it is possible to find out the Hamiltonian of the model that has just been solved. This is why this construction is called the (quantum) inverse scattering method. In our case, we made an educated ansatz for the ${\cal R}$-operator\index{Intertwiner operator} in (\ref{Ransatz}), thanks to the machinery developed for the 6-vertex model\index{Vertex model}.

\section{The ABA solution}
\label{sec:ABAsol}

The construction we developed so far can be compared with what we have done in the coordinate Bethe Ansatz approach: we factorized the interaction into a series of two-body scattering (although this time by introducing an auxiliary particle in the system) and we quantized it by closing the system at infinity with periodic boundary condition (in the ABA approach, by taking the trace over the auxiliary state).

To determine the eigenfunctions of the model, we look for eigenvectors of the transfer matrix\index{Transfer matrix} (\ref{TransferMatrix}). We construct them by injecting ghost particles from the auxiliary space ${\cal V}$ and constructing the eigenvalue equations using the Yang-Baxter equation\index{Yang-Baxter!equation}. These conditions will generate the same Bethe equations\index{Bethe!equations} we have found in the coordinate approach. But the advantage of the algebraic construction is that it provides a clearer characterization of the many-body eigenstates.

The algebraic approach exploits the similarity between (\ref{MonodromyOperator}) and (\ref{scatteringop}) to use the ${\bf B}$ (${\bf C}$) operators as creation/lowering (annihilation/raising) operators for the quasi-particle\index{Quasi-particle} excitations of the system. We also want to use the Yang-Baxter equation\index{Yang-Baxter!equation} for the monodromy matrix\index{Monodromy matrix} (\ref{TYB1}) to manipulate the eigenvector conditions formally without having to write out the states explicitly as we did in the coordinate approach. To this end, we seek the eigenstates of the transfer matrix\index{Transfer matrix} (\ref{eigeneqT}), and not directly of the Hamiltonian, knowing that anyway the two are diagonalized by the same states, see (\ref{HTrel}).

The ABA construction starts with the identification of a reference state $|0\rangle$, which we call {\it pseudo-vacuum}\index{Pseudo-vacuum}. This is a ``trivial'' eigenstate of the system which can be recognized by inspection and is specified by the requirements
\be
   A (\lambda) |0\rangle = \mathring{a} (\lambda) |0\rangle \; , \qquad
   D (\lambda) |0\rangle = \mathring{d} (\lambda) |0\rangle \; , \qquad
   C (\lambda) |0\rangle = 0 \; .
   \label{0statecond}
\ee
This identifies $|0\rangle$ as the highest weight state in the $SU(2)$ representation of (\ref{MonodromyOperator}).

To construct this pseudo-vacuum state\index{Pseudo-vacuum}, we look at (\ref{XXZL}) and notice that
\be
   {\cal L}_{j,a} (\lambda) |\uparrow_j \rangle =
   \left( \begin{array}{cc} 1 & 0 \cr 0 & t(\lambda) \cr \end{array} \right)
   |\uparrow_j \rangle +
   \left( \begin{array}{cc} 0 & r (\lambda) \cr 0 & 0 \cr \end{array} \right)
   |\downarrow_j \rangle \; ,
\ee
where $|\uparrow_j\rangle$ ($|\downarrow_j\rangle$) denotes the state with a spin up (down) at the j-th lattice site.
Thus, $|\uparrow_j\rangle$ makes the Lax-operator\index{Lax operator} upper-diagonal and the state
\be
   |0\rangle = \prod_{j=1}^N |\uparrow_j\rangle
   \label{pseudovacuum}
\ee
makes the monodromy matrix\index{Monodromy matrix} (\ref{MonodromyMatrix}) upper-diagonal as well and satisfies  (\ref{0statecond}) with
\be
  \mathring{a} (\lambda) = 1 \; , \qquad \qquad
  \mathring{d} (\lambda) = \prod_{j=1}^N t (\lambda-\xi_j) \; .
  \label{tildead}
\ee
Notice that these eigenvalues depend only on the form of the ${\cal L}$-operator\index{Lax operator} and thus on the inhomogeneity parameters $\xi_j$.

Comparing (\ref{scatteringop}, \ref{XXZL}) and (\ref{MonodromyOperator}) we see that, if we ``inject'' in the system a ghost particle (from the auxiliary space) with spin-down and we extract it in a spin-up state, the total spin of the chain decreases by one, meaning that one spin of the chain has made the opposite flip. This action is performed by the operator
\be
   \langle \uparrow_a | \bT_a (\lambda_j) | \downarrow_a \rangle
   = {\bf B} (\lambda_j) \; ,
\ee
which can be interpreted as a spin-flip operation that creates an excitation over the pseudo-vacuum $|0\rangle$\index{Pseudo-vacuum} with rapidity $\lambda_j$.
Thus, we construct a state with $R$ quasi-particle\index{Quasi-particle} excitations as (\ref{eigenansatz}) and we look under which conditions it satisfies the eigenstate equation (\ref{eigeneqT}).

The reason to use the spin-flip operator ${\bf B} (\lambda_j)$ instead of the local one $\sigma_j^-$ is that for ${\bf B}$ we can use the algebra\index{Yang-Baxter!algebra} of the Yang-Baxter equation\index{Yang-Baxter!equation} (\ref{RTT}) as defining a set of generalized commutation relations for the operators ${\bf A}, {\bf B}, {\bf  C}, {\bf D}$. These can be worked out by writing explicitly the matrix multiplications in (\ref{RTT}):
\bea
   \left[ {\bf A} (\lambda), {\bf A} (\mu) \right] & = &
   \left[ {\bf B} (\lambda), {\bf B} (\mu) \right] =
   \left[ {\bf C} (\lambda), {\bf C} (\mu) \right] =
   \left[ {\bf D} (\lambda), {\bf D} (\mu) \right] = 0 \; , 
   \label{BB} \\
   \left[ {\bf A} (\lambda), {\bf D} (\mu) \right] & = &
   g (\lambda - \mu) \left\{ {\bf C} (\lambda) {\bf B} (\mu)
   - {\bf C} (\mu) {\bf B} (\lambda) \right\} \; ,
   \label{AD} \\
   \left[ {\bf D} (\lambda), {\bf A} (\mu) \right] & = &
   g (\lambda - \mu) \left\{ {\bf B} (\lambda) {\bf C} (\mu)
   - {\bf B} (\mu) {\bf C} (\lambda) \right\} \; , 
   \label{DA} \\
   \left[ {\bf B} (\lambda), {\bf C} (\mu) \right] & = &
   g (\lambda - \mu) \left\{ {\bf D} (\lambda) {\bf A} (\mu)
   - {\bf D} (\mu) {\bf A} (\lambda) \right\} \; , 
   \label{BC}\\
   \left[ {\bf C} (\lambda), {\bf B} (\mu) \right] & = &
   g (\lambda - \mu) \left\{ {\bf A} (\lambda) {\bf D} (\mu)
   - {\bf A} (\mu) {\bf D} (\lambda) \right\} \; , 
   \label{CB} \\
   {\bf A} (\lambda) {\bf B} (\mu) & = & 
   f (\mu - \lambda) \; {\bf  B} (\mu) {\bf A} (\lambda)
   + g (\lambda - \mu) \; {\bf B} (\lambda) {\bf A} (\mu) \; ,
   \label{AB} \\
   {\bf B} (\lambda) {\bf A} (\mu) & = & 
   f (\mu - \lambda) \; {\bf  A} (\mu) {\bf B} (\lambda)
   + g (\lambda - \mu) \; {\bf A} (\lambda) {\bf B} (\mu) \; , 
   \label{BA} \\
   {\bf A} (\lambda) {\bf C} (\mu) & = & 
   f (\lambda - \mu) \; {\bf  C} (\mu) {\bf A} (\lambda)
   + g (\mu - \lambda) \; {\bf C} (\lambda) {\bf A} (\mu) \; ,
   \label{AC} \\
   {\bf C} (\lambda) {\bf A} (\mu) & = & 
   f (\lambda - \mu) \; {\bf  A} (\mu) {\bf C} (\lambda)
   + g (\mu - \lambda ) \; {\bf A} (\lambda) {\bf C} (\mu) \; , 
   \label{CA} \\
   {\bf D} (\lambda) {\bf B} (\mu) & = & 
   f (\lambda - \mu) \; {\bf B} (\mu) {\bf D} (\lambda)
   + g (\mu - \lambda) \; {\bf B} (\lambda) {\bf D} (\mu) \; ,
   \label{DB} \\
   {\bf B} (\lambda) {\bf D} (\mu) & = & 
   f (\lambda - \mu) \; {\bf D} (\mu) {\bf B} (\lambda)
   + g (\mu - \lambda) \; {\bf D} (\lambda) {\bf B} (\mu) \; ,
   \label{BD} \\
   {\bf D} (\lambda) {\bf C} (\mu) & = & 
   f (\mu - \lambda) \; {\bf C} (\mu) {\bf D} (\lambda)
   + g (\lambda - \mu) \; {\bf C} (\lambda) {\bf D} (\mu) \; ,
   \label{DC} \\
   {\bf C} (\lambda) {\bf D} (\mu) & = & 
   f (\mu - \lambda) \; {\bf D} (\mu) {\bf C} (\lambda)
   + g (\lambda - \mu) \; {\bf D} (\lambda) {\bf C} (\mu) \; ,
   \label{CD}
\eea
where we introduced
\be 
   f (\lambda) \equiv {1 \over t(\lambda)} 
   = {\varphi (\lambda + \ii 2 \phi) \over \varphi (\lambda)} \; , 
   \qquad \qquad \qquad
   g (\lambda) \equiv {r (\lambda) \over t (\lambda)}
   = {\varphi (\ii 2 \phi) \over \varphi (\lambda)} \; ,
   \label{fgdef}
\ee
and took advantage that $g(\lambda) = - g(-\lambda)$.
Eq. (\ref{BB}) establishes that in (\ref{eigenansatz}) the order in which we multiply the ${\bf B}$'s is not important, as we expect from the physical meaning of the Yang-Baxter\index{Yang-Baxter!equation}, i.e. that the order in which we inject the ghost interaction does not matter. Notice that the coefficients in (\ref{AD}--\ref{CD}) do not depend on the form of ${\cal L} (\lambda)$ and on the $\xi_j$, but only on the ${\cal R}$-matrix: the intertwiner\index{Intertwiner operator} provides the structure factors of the algebra\index{Yang-Baxter!algebra}.

We check (\ref{eigeneqT}) by progressively commute the ${\bf A}$ and ${\bf D}$ through the ${\bf B}$'s. This is physically equivalent to scattering the ghost particle with the excitations created by the ${\bf B}$'s through the whole system, giving rise to transmissions and reflections of the ghost.
For (\ref{eigenansatz}) to be an eigenstate it means that after each reflection, the ghost propagates in a way that the cumulative effect of all these processes interferes destructively, and thus at the end it has transmitted through the whole system keeping its degrees of freedom.

The commutation of ${\bf A} (\mu)$ and ${\bf D} (\mu)$ through the ${\bf B}$'s can be worked out by brute force, using (\ref{AB}, \ref{BA}, \ref{DB}, \ref{BD}). However, since from (\ref{BB}) all the ${\bf B}$'s are equivalent, out of this symmetry we conclude that at the end of the calculation we can only have
\bea
   {\bf A} (\mu) \prod_{j=1}^R {\bf B} (\lambda_j) 
   & = & \Xi \big( \mu;  \{\lambda_j \} \big)
    \left[ \prod_{j=1}^R {\bf B} (\lambda_j) \right] {\bf A} (\mu)
   + {\bf B} (\mu) \:  \sum_{l=1}^R \Xi_l \big( \mu;  \{\lambda_j \} \big)
   \left[ \prod_{j \ne l}^R {\bf B} (\lambda_j) \right] {\bf A} (\lambda_l) \; .
   \label{AprodB} \\
   {\bf D} (\mu) \prod_{j=1}^R {\bf B} (\lambda_j) 
   & = & \tilde{\Xi} \big( \mu;  \{\lambda_j \} \big)
   \left[ \prod_{j=1}^R {\bf B} (\lambda_j) \right] {\bf D} (\mu)
   + {\bf B} (\mu) \: \sum_{l=1}^R \tilde{\Xi}_l \big( \mu;  \{\lambda_j \} \big)
   \left[ \prod_{j \ne l}^R {\bf B} (\lambda_j) \right] {\bf D} (\lambda_l) \; .
   \label{DprodB}
\eea
Comparing these expressions with (\ref{AB},\ref{DB}), we understand that first term in (\ref{AprodB},\ref{DprodB}) is due to the first term on the RHS of (\ref{AB},\ref{DB}), which represents a sort of transmission, since the operators keep their rapidities. Thus, this first term is the result of a sequence of just transmissions through all the ${\bf B}$'s:
\be
   \Xi \big( \mu;  \{\lambda_j \} \big) = \prod_{j=1}^R f (\lambda_j- \mu) \; , \qquad \qquad
   \tilde{\Xi} \big( \mu;  \{\lambda_j \} \big) = \prod_{j=1}^R f (\mu - \lambda_j) \; .
   \label{Xidef}
\ee
The second term on the RHS of (\ref{AprodB},\ref{DprodB}) are the result of a single ``reflection'' (the second term in (\ref{AB}, \ref{DB})), followed by a sequence of transmissions, for instance, by singling out one of the ${\bf B}$ operators:
\be
   {\bf A} (\mu) {\bf B} (\lambda_l) 
   \prod_{j \ne l}^R {\bf B} (\lambda_j) 
   = \Big\{ f (\lambda_l -  \mu) {\bf B} (\lambda_l) \; {\bf A} (\mu)
   + g (\mu - \lambda_l ) \; {\bf B} (\mu) {\bf A} (\lambda_l) \Big\} 
   \prod_{j \ne l}^R {\bf B} (\lambda_j) \; .
   \label{ABex}
\ee
Additional reflections after the first generate terms which are odd under the exchange of the rapidities $\lambda_j$ and are thus not allowed by the symmetry of the ${\bf B}$'s (this means that at the end of the calculation they will cancel out as a result of destructing interference). Thus, only one reflection effectively takes place and we have 
\be
  \Xi_l \big( \mu;  \{\lambda_j \} \big)
  = g (\mu - \lambda_l) 
  \prod_{\stackrel{j=1}{j \ne l}}^R f(\lambda_j - \lambda_l) \; , 
  \qquad \qquad
  \tilde{\Xi_l} \big( \mu;  \{\lambda_j \} \big)
  = g (\lambda_j - \mu) 
  \prod_{\stackrel{j=1}{j \ne l}}^R f(\lambda_l - \lambda_j) \; .
\ee

Collecting these results, we conclude that
\bea
   {\bf T} (\mu) | \Psi \rangle =  \Big[ {\bf A} (\mu) + {\bf D} (\mu) \Big] 
   \prod_{j=1}^R {\bf B} (\lambda_j) | 0 \rangle
   & = & \Big[ \mathring{a} (\mu) \: \Xi \big( \mu;  \{\lambda_j \} \big)
   + \mathring{d} (\mu) \: \tilde{\Xi} \big( \mu;  \{\lambda_j \} \big) \Big] \Psi \rangle 
   \label{TPsi} \\
   &+&  {\bf B} (\mu) \sum_{l=1}^R 
   \Big[ \mathring{a} (\lambda_l) \: \Xi_l \big( \mu;  \{\lambda_j \} \big)
   + \mathring{d} (\lambda_l) \: \tilde{\Xi}_l \big( \mu;  \{\lambda_j \} \big)\Big] 
   \prod_{\stackrel{j=1}{j \ne l}}^R {\bf B} (\lambda_j) |0\rangle \; .
   \nonumber 
\eea
Thus, $|\Psi \rangle$ from (\ref{eigenansatz}) is an eigenvector of the transfer matrix\index{Transfer matrix} (\ref{eigeneqT}) with eigenvalue
\be
   \Lambda \big( \mu; \{ \lambda_j\} \big) =
   \mathring{a} (\mu) \prod_{j=1}^R f (\lambda_j -\mu) \;
   + \mathring{d} (\mu) \prod_{j=1}^R f (\mu - \lambda_j) 
   \label{Lambdamu}
\ee
if the off-diagonal terms vanish, i.e.
\be
   g (\lambda_l - \mu) \left[ \mathring{a} (\lambda_l)
   \prod_{\stackrel{j=1}{j \ne l}}^R f(\lambda_j - \lambda_l)
   - \mathring{d} (\lambda_l)
   \prod_{\stackrel{j=1}{j \ne l}}^R f(\lambda_l - \lambda_j) \right] = 0 \; , \qquad l=1,\ldots,R \; .
   \label{offdiagonal}
\ee
Notice that the dependence of (\ref{offdiagonal}) on the spectral parameter $\mu$ of the transfer matrix\index{Transfer matrix} has factorized (we know that transfer matrices commute for different spectral parameters and therefore the eigenstate conditions cannot depend on it). Thus, we can write the eigenstate condition in generality as
\be
   {\mathring{d} (\lambda_j) \over \mathring{a} (\lambda_j)}=
   \prod_{\stackrel{l=1}{j \ne j}}^R
   {f(\lambda_l - \lambda_j) \over f(\lambda_j - \lambda_l) } \; .
   \qquad \qquad j =1, \ldots, R \; ,
   \label{ABABE1}
\ee
Specializing with the definition of the structure function (\ref{fgdef}) and of the Lax-operator\index{Lax operator} (\ref{tildead}) for the XXZ chain (also taking the homogeneous limit $\xi_j = \xi$ for $j=1 \dots N$):
\be
   \left[ {\varphi ( \lambda_j - \xi ) \over \varphi (\lambda_j - \xi + \ii 2 \phi) } \right]^N =
   \prod_{l=1}^R {\varphi (\lambda_l - \lambda_j + \ii 2 \phi)
   \over \varphi (\lambda_j - \lambda_l + \ii 2 \phi) } \; ,
   \qquad j =1, \ldots, R \; .
   \label{ABABE}
\ee
Setting $\xi= \ii \phi$ we recognize in (\ref{ABABE}) the Bethe equations\index{Bethe!equations} for the XXZ model, see Table \ref{table:XXZparam}.

Note that from the transfer matrix\index{Transfer matrix} eigenvalues (\ref{Lambdamu}), through trace identities\index{Trace identities} such as (\ref{PTrel}, \ref{HTrel}), we can calculate the eigenvalues of conserved charges. For instance
\bea
P & = & - \ii \ln \Lambda \big( \xi ; \{\lambda_j \} \big)
= - \ii \sum_{j=1}^R \ln f (\lambda_j - \xi) 
= \ii \sum_{j=1}^R \ln { \varphi (\lambda_j - \xi + \ii 2 \phi) \over
	\varphi (\lambda_j - \xi)} \; , \\
E & = & {1 \over 4} \Delta N + {\sqrt{\Delta^2 -1} \over 2}
\left. {\de \over \de \mu} \ln  \Lambda \big( \mu ; \{\lambda_j \} \big) \right|_{\mu=\xi}
= {1 \over 4} \Delta N + {\sqrt{\Delta^2 - 1} \over 2} \sum_{j=1}^R
\left. {\de \over \de \mu} f (\lambda - \mu) \right|_{\mu=\xi}
\; ,
\eea
which coincide with the expressions we found in chapter \ref{chap:XXZmodel}, once we set $\xi = \ii \phi$.

Note that both the functions $f(\lambda)$ defining the ${\cal R}$-operator\index{Intertwiner operator} and $\mathring{a}(\lambda), \mathring{d}(\lambda)$, which depend on the parametrization of the ${\cal L}$-matrix\index{Lax operator}, enter into the Bethe equations\index{Bethe!equations} (\ref{ABABE1}) and into the transfer matrix\index{Transfer matrix} eigenvalues (\ref{Lambdamu}). It should be remarked that the Lax operator\index{Lax operator} is not uniquely defined by the requirement of satisfying the YBE\index{Yang-Baxter!equation} (\ref{LLR}). Different choices can correspond to different models (Hamiltonians). While these models will in general have a different ``bare'' scattering phase/momentum (see ${\mathring{d} (\lambda) \over \mathring{a}(\lambda)}$\index{Scattering phase} in (\ref{ABABE1})), they all get ``dressed'' by  the same kernel\index{Kernel}
\be
   {\cal K} (\lambda, \mu) \equiv \ii {\partial \over \partial \lambda}
   \ln {f(\lambda-\mu) \over f(\mu - \lambda)} \;.
   \label{Kfrel}
\ee
While the function $\mathring{\rho} (\lambda) \equiv {\mathring{d} (\lambda) \over \mathring{a}(\lambda)}$ on the LHS of (\ref{ABABE1}) can be chosen almost arbitrarily (for instance, we discuss in appendix \ref{app:2DClassical} how different parametrizations of the solution of the Yang-Baxter equation are equivalent), the intertwiner\index{Intertwiner operator} behind these different choices is always the same and thus identifies a kind of universality behind these integrable models.

In closing, we note that we can establish a direct parallel between the eigenvector condition just derived and the one employed in the coordinate approach, by repeatedly evaluating (\ref{TPsi}) at spectral parameters equal to one of the state rapidities: $\mu = \lambda_l$. In doing so, the probe scatters off the excitation created by a ${\bf D} (\lambda_l)$ and takes its ``identity'', since at vanishing rapidity difference the scattering matrix ${\cal R}$\index{Intertwiner operator} becomes the exchange operator (\ref{exchange}). Thus, in this way the excitation/probe is let scatter with the other excitations and the eigenvector condition asks that after this process it emerges unchanged (except for having developed some winding $2 \pi n$ of the scattering phase\index{Scattering phase}), exactly as we required in the coordinate construction. 

\section{Construction of the operators: the inverse scattering problem}
\label{sec:QISM}

In the coordinate approach, the eigenstates are written as the superposition of an exponentially large number of terms. This solution does not lend itself easily to the calculation of correlation functions. In the algebraic formulation, instead, the states are characterized directly in terms of the quasi-particle excitations\index{Quasi-particle}. Conservation of evil, however, implies that even simple operators appear complicated when expressed in the algebraic Bethe Ansatz way.

Let us consider, for instance, the representation of the spin operators in terms of the building blocks of the algebraic construction, namely ${\bf A}, {\bf B}, {\bf C}, {\bf D}$.
One of the reasons to keep general inhomogeneities $\xi_j$ in \ref{MonodromyMatrixdef} is to be able to manipulate individual sites before restoring translational invariance. For one point functions, however, we can take the homogeneous limit $\xi_j \to \xi$ and use (\ref{bTxi}) and (\ref{MonodromyMatrix}) to write \cite{gohman00}
\be
   \bT_a (\xi) 
   	= {\bf \Pi}_{1,a} \; {\bf T} (\xi) 
   	= { 1 + \vec{\sigma_1} \cdot \vec{\tau}_a \over 2} \; {\bf T} (\xi)
   	= \left( \begin{array}{cc}
   		{1 + \sigma_1^z \over 2 } & \sigma_1^- \cr
   		\sigma_1^+ & {1 - \sigma_1^z \over 2} \cr 
   	\end{array} \right) \exp\left( \ii \hat{P} \right) 
   	= \left( \begin{array}{cc}
   		{\bf A} (\xi) & {\bf B} (\xi) \cr
   		{\bf C} (\xi) & {\bf D} (\xi) \cr
   	\end{array} \right) \; ,
\ee 
which gives
\be
  \sigma_1^- = {\bf B} (\xi) \exp\left(- \ii \hat{P} \right) \; , \qquad \quad
  \sigma_1^+ = {\bf C} (\xi) \exp\left(- \ii \hat{P} \right) \; , \qquad \quad
  \sigma_1^z = \left[ {\bf A} (\xi) - {\bf D} (\xi) \right] \exp\left(- \ii \hat{P} \right) \; .
\ee
One can then use the lattice translation operator ${\bf T} (\xi) = \exp (\ii \hat{P} )$ to shift the operators at arbitrary sites:
\be
  \sigma_j^\alpha = 
  \exp\big[ \ii (j-1) \hat{P} \big] \sigma_1^\alpha \exp\big[\ii (1-j) \hat{P} \big]
  =  \left[ {\bf A} (\xi) + {\bf D} (\xi) \right]^{j-1} \;  \sigma_1^\alpha
  \; \left[ {\bf A} (\xi) + {\bf D} (\xi) \right]^{N-j} \; .
  \label{QISM0}
\ee
Although one can formally use the same procedure to write operators involving different sites, their expectation values result in general in indeterminate zero over zero expressions. Thus, it is better to postpone taking the homogeneous limit only after evaluating the correlators in the inhomogeneous case. Working with the inhomogeneities requires developing the appropriate representation of the shift operators for lattice site, as was introduced in \cite{kitanine99} using the so-called {\it F-basis}. We will not do it here, but the main result is that evaluating the transfer matrix\index{Transfer matrix} at one of the inhomogeneities corresponds to shifting the corresponding lattice site: the final expressions for the spin operators which can be used in computing correlation functions in the ABA approach are similar to (\ref{QISM0}) and are \cite{kitanine99,gohman00,maillet00}
\bea
     \sigma_j^- & = & 
     \left[ \prod_{l=1}^{j-1} \big( {\bf A} + {\bf D} \big) (\xi_l) \right] \; 
     {\bf B} (\xi_j) \; 
     \left[ \prod_{l=j+1}^N \big( {\bf A} + {\bf D} \big) (\xi_l) \right] \; , 
     \label{sigma-ABCD}\\
     \sigma_j^+ & = & 
     \left[ \prod_{l=1}^{j-1} \big( {\bf A} + {\bf D} \big) (\xi_l) \right] \;
     {\bf C} (\xi_j) \; 
     \left[ \prod_{l=j+1}^N \big( {\bf A} + {\bf D} \big) (\xi_l) \right] \; ,
     \label{sigma+ABCD} \\
     \sigma_j^z & = & 
     \left[ \prod_{l=1}^{j-1} \big( {\bf A} + {\bf D} \big) (\xi_l) \right] \; 
     \big( {\bf A} - {\bf D} \big) (\xi_j) \;
     \left[ \prod_{l=j+1}^N \big( {\bf A} + {\bf D} \big) (\xi_l) \right] \; .
     \label{sigmazABCD}
\eea

Form factors for the spin operators can then be calculated combining (\ref{sigma-ABCD}-\ref{sigmazABCD}) and (\ref{eigenansatz}) and using the Yang-Baxter relations\index{Yang-Baxter!equation} (\ref{BB}-\ref{CD}) as commutation relations as we did to prove the eigenvector condition \cite{kitanine99}. The homogeneous limit can then be taken at the end of the computation in a way that regularizes the resulting expression. Additional operators can be constructed starting from the basic building blocks (\ref{sigma-ABCD}-\ref{sigmazABCD}) to generate the different observables, thus reducing the computations of correlation functions to an algebraic problem. It is also clear from (\ref{sigma-ABCD}-\ref{sigmazABCD}) and (\ref{BB}-\ref{CD}) that the complexity of these problems can quickly become unmanageable as an increasing number of terms are generated in the process. Most of this construction was pioneered by the St. Petersburg school and culminated in \cite{ISM}. However, over the years it has become clear that the elegance of certain formulas emerging from the quantum inverse scattering is of limited practical help alone and needs further insights. For instance, often at a closer inspection of these formulations often reveals a structure that renders them amenable to explicit calculation, see, for instance, \cite{kitanineetal}. More recently, there has been progress due to alternative representations of the Bethe solution, known as {\it Separation of Variable approach} \cite{niccolietal} and {\it off-diagonal Bethe Ansatz} \cite{offdiagbethe,ODBA}.

\section{Scalar products and norms: Slavnov's and Gaudin's Formulas}
\label{sec:scalarp}

One of the cases in which it is possible to reach a compact expression out of the ABA construction is that of a scalar product between states. This results is important, for instance, in studying quantum system out of equilibrium, where one is interested in knowing the overlap between some initial state and one of the eigenstate of the system \cite{quenchaction}. As a particular case of scalar products, it is possible to calculate the norm of a Bethe state. The attentive reader might have noticed that so far we took good care of working with normalizable states (especially in presence of bound states\index{Bound state}), but never discussed how to calculate their norm, while, clearly, there is no point in calculating the expectation value of operators between certain states without knowing how the latter are normalized.

Defining the {\it dual pseudo-vacuum} $\langle 0|$\index{Pseudo-vacuum} as
\be
   \langle 0 | 0 \rangle = 1 \; , \qquad \quad
   \langle 0 | {\bf B} (\lambda) = 0 \; , \qquad \quad
   \langle 0 | {\bf A} (\lambda) = \mathring{a} (\lambda) \langle 0 | \; , \qquad \quad
   \langle 0 | {\bf D} (\lambda) = \mathring{d} (\lambda) \langle 0 | \; ,
\ee
we want to analyze the scalar product
\be
   S_R \Big( \{\mu_j \}, \{ \lambda_j \} \Big)  
   \equiv \langle 0 | \prod_{l=1}^R {\bf C} (\mu_l) 
   \prod_{j=1}^R {\bf B} (\lambda_j) | 0 \rangle \; .
   \label{Srdef}
\ee
It is easy to see that this quantity can be non-zero only if it contains the same number of ${\bf B}$ and ${\bf C}$ operators. For $R=1$, (\ref{Srdef}) can be readily calculated using (\ref{CB}):
\be
   S_1 (\mu,\lambda) = g(\mu - \lambda) \Big[ \mathring{a} (\mu) \mathring{d} (\lambda) - \mathring{a}(\lambda) \mathring{d}(\mu) \Big] 
    = g(\mu - \lambda) \; \mathring{d} (\lambda) \mathring{d} (\mu) 
    \Big[ {\mathring{a} (\mu) \over \mathring{d} (\mu)} - {\mathring{a}(\lambda) \over \mathring{d}(\lambda)} \Big]\; .
   \label{S1res}
\ee
It is possible to proceed similarly for higher $R$, by progressively commuting the operators, but this way of proceeding tends to hide the form of the final solution.

In fact, it seems that to arrive at a final, compact expression, one needs to assume that one of the sets of rapidities, say $\{ \lambda_j \}$, satisfies the Bethe equations\index{Bethe!equations} (\ref{ABABE}) \cite{slavnov89}. Under this assumption, we want to prove that 
\be
  \mathbb{S}_R \Big( \{\mu_j \}, \{ \lambda_j \} \Big) \equiv
  G_R \Big( \{\mu_j \}, \{ \lambda_j \} \Big) \: 
  \det_R {\bf M} \Big( \{\mu_j \}, \{ \lambda_j \} \Big) \; ,
  \label{bbSdef}
\ee
with
\bea
   G_R \Big( \{\mu_j \}, \{ \lambda_j \} \Big) & \equiv &
  	{ \prod_{j>l} g(\lambda_l - \lambda_j) g (\mu_j - \mu_l) \over 
  	\prod_{j=1}^R \prod_{l=1}^R g (\lambda_j - \mu_l) } \: 
   \prod_{j=1}^R \prod_{l=1}^R f (\lambda_j - \mu_l)  \; , \\
   M_{ab} \Big( \{\mu_j \}, \{ \lambda_j \} \Big) & \equiv & 
   {g^2 (\lambda_a - \mu_b) \over f(\lambda_a - \mu_b)} \; \mathring{d} (\lambda_a)
   \left[ \mathring{d}(\mu_b) - \mathring{a} (\mu_b) \prod_{j \ne a} {f (\mu_b - \lambda_j) \over f(\lambda_j - \mu_b)} \right] \; ,
\eea
is in fact equal to (\ref{Srdef}).

The proof relies on two key ingredients:
\begin{itemize}

 \item We note that scalar products, such as (\ref{S1res}), depend both on the functions appearing in the intertwiner\index{Intertwiner operator} ($f(\lambda)$ and $g(\lambda)$) and in the Lax operator\index{Lax operator} composing the monodromy matrix\index{Monodromy matrix}, through $\mathring{a} (\lambda)$ and $\mathring{d} (\lambda)$. It is convenient to disentangle the two contributions. In particular, we will consider the quantities $\mathring{a}_j^\lambda \equiv \mathring{a}(\lambda_j), \mathring{d}_j^\lambda \equiv \mathring{d} (\lambda_j), \mathring{a}_j^\mu \equiv \mathring{a}(\mu_j)$, and $\mathring{d}_j^\mu \equiv \mathring{d} (\mu_j)$ as parameters that we can vary independently from $\lambda_j$ and $\mu_j$. Using this point of view, we will consider (\ref{Srdef}) as a function of the $6R$ independent sets of parameters, but it is convenient to consider just a $4R$ dimensional subset of them
 \be
    S_R = 
    S_R \Big( \{ \mu_j \}, \{ \lambda_j \}, \{ \mathring{\sigma}_j \}, \{ \mathring{\rho}_j \} \Big) \; ,
 \ee
 where $\mathring{\sigma}_j \equiv \mathring{a}_j^\mu / \mathring{d}_j^\mu$ and $\mathring{\rho}_j \equiv \mathring{a}_j^\lambda / \mathring{d}_j^\lambda$.
 
  \item To establish that (\ref{Srdef}) and (\ref{bbSdef}) coincide, we will study them as analytical functions. We will show that they have the same poles and residue and thus are in fact the same function. These poles are located when one of the rapidities of the creation operators coincide with one of the annihilation operators: $\mu_j \to \lambda_l$. In physically relevant cases, this divergence is cured by the vanishing of the residue as $\mathring{\sigma}_j - \mathring{\rho}_l \to 0$. However, keeping the $\mathring{\sigma}_j$ and $\mathring{\rho}_l$ independent from $\mu_j$ and $\lambda_l$ allows us to keep this residue finite and thus to establish the analytical dependence of the scalar product on the rapidities alone.

\end{itemize}

Because of (\ref{BB}), the scalar products (\ref{Srdef}) are symmetric functions of the $\{\lambda_j \}$ and of the $\{ \mu_j \}$ separately. Let us single out two rapidities, for instance $\lambda_m$ and $\mu_R$ and extract the singular terms as $\mu_R \to \lambda_m$ (knowing that the behavior is the same for any pair of rapidities because of symmetry). We can place the two operators corresponding to these rapidities right next to each other in (\ref{Srdef}) and commute ${\bf C} (\mu_R) {\bf B} (\lambda_m)$ using (\ref{CB}). One term reverses their order and thus does not bring any singular contribution. The rest reads
\be
  S_R \stackrel{\mu_R \to \lambda_m}{\simeq}
  g(\mu_R - \lambda_m) \langle 0 | \prod_{l=1}^{R-1} {\bf C} (\mu_l) \Big[ 
  {\bf A} (\mu_R) {\bf D} (\lambda_m) - {\bf A} (\lambda_m) {\bf D} (\mu_R) \Big]
  \prod_{j\ne m} {\bf B} (\lambda_j) | 0 \rangle
  + \text{finite terms}
  \label{Sr1}
\ee
Now we progressively commute the ${\bf A}$'s toward the left through the ${\bf C}$'s and the ${\bf D}$'s toward the right through the ${\bf B}$'s. This procedure generates the same kind of terms as on the RHS of (\ref{AprodB}, \ref{DprodB}). For instance
\bea
   {\bf D} (\mu_R) \prod_{j \ne m} {\bf B} (\lambda_j) |0 \rangle
   & = & \mathring{d}^\mu_R \: \tilde{\Xi} \big( \mu_R;  \{\lambda_j \}_{j \ne m} \big)
   \: \prod_{j \ne m} {\bf B} (\lambda_j) |0 \rangle
   \nonumber \\
   && + \: \sum_{l \ne m} \mathring{d}^\lambda_l \: \tilde{\Xi}_l \big( \mu_R;  \{\lambda_j \}_{j \ne m} \big)
   \left[ \prod_{j \ne l,m} {\bf B} (\lambda_j) \right] {\bf B} (\mu_R) |0\rangle \; , \\
   {\bf D} (\lambda_m) \prod_{j \ne m} {\bf B} (\lambda_j) |0 \rangle
   & = & \mathring{d}^\lambda_m \: \tilde{\Xi} \big( \lambda_m;  \{\lambda_j \}_{j \ne m} \big) \:
   \prod_{j \ne m} {\bf B} (\lambda_j)  |0 \rangle
   \nonumber  \\
   && + \: \sum_{l \ne m} \mathring{d}^\lambda_l \: \tilde{\Xi}_l \big( \lambda_m;  \{\lambda_j \}_{j \ne m} \big)
   \left[ \prod_{j \ne l,m} {\bf B} (\lambda_j) \right] {\bf B} (\lambda_m) |0\rangle \; .
\eea
In the $\mu_R \to \lambda_m$ limit, the second terms in these expressions tend to one another. Thus, when inserted in (\ref{Sr1}) they compensate the pole in $g(\mu_R -\lambda_m)$ and contribute to the finite terms. Thus we have
\bea
   S_R \left( \{ \mu_j \}, \{ \lambda_j \}, \{ \mathring{\sigma}_j \}, \{ \mathring{\rho}_j \}\right)
   & \stackrel{\mu_R \to \lambda_m}{\simeq} &
   g(\mu_R - \lambda_m) \Big[ 
   \mathring{a}^\mu_R \: \mathring{d}^\lambda_m \: 
   \tilde{\Xi} \big( \mu_R;  \{\mu_j \}_{j<R} \big) 
   \tilde{\Xi} \big( \lambda_m;  \{\lambda_j \}_{j \ne m} \big)
   \nonumber \\
   && \qquad \qquad \quad
   - \mathring{a}^\lambda_m \: \mathring{d}^\mu_R \:
   \tilde{\Xi} \big( \lambda_m;  \{\mu_j \}_{j<R} \big)
   \tilde{\Xi} \big( \mu_R;  \{\lambda_j \}_{j \ne m} \big) \Big] \times
   \nonumber \\
   && \qquad \qquad  \times \quad
   \langle 0 | \prod_{l=1}^{R-1} {\bf C} (\mu_l) 
   \prod_{j \ne m} {\bf B} (\lambda_j) | 0 \rangle
   \: \: + \text{finite terms}
   \nonumber \\
   & = & g(\mu_R - \lambda_m) \mathring{d}^\lambda_m \: \mathring{d}^\mu_R  
   \Big[ \mathring{\sigma}_R - \mathring{\rho}_m \Big]
   \prod_{j=1}^{R-1} f (\mu_R - \mu_j) \prod_{j \ne m} f (\lambda_m - \lambda_j)
   \times \nonumber \\
   && \times 
   S_{R-1} \left( \{ \mu_j \}_{j<R}, \{ \lambda_j \}_{j \ne m}, \{ \mathring{\sigma}_j^{(R-1)} \}_{j<R}, \{ \mathring{\rho}_j^{(R-1)} \}_{j \ne m}\right)
   \: \: + \text{finite terms}
   \nonumber \\
   \label{Sr2}
\eea
where we took the $\mu_R \to \lambda_m$ limit in the expressions involving the $f(\lambda)$ function, but considered $\mathring{\sigma}_R$ and $\mathring{\rho}_m$ independent from this limit. In the final expression, $\mathring{\sigma}_j^{(R-1)} \equiv \mathring{\sigma}_j {f(\mu_j - \mu_R) \over f(\mu_R - \mu_j)}$, $\mathring{\rho}_j^{(R-1)} \equiv \mathring{\rho}_j {f(\lambda_j - \lambda_m) \over f(\lambda_m-\lambda_f)}$ redefine the eigenvalues of ${\bf A}$ and ${\bf D}$ on the vacuum to mimic the effect of the original rapidities $\lambda_m, \mu_R$ which have now been factorized out.

We have thus determined the analytic structure of the scalar product as a function of one of the rapidities: it has poles (only) as $\mu_R \to \lambda_m$, with residues given by (\ref{Sr2}) and it vanishes as $\mu_R \to \infty$.

To establish that (\ref{bbSdef}) coincide with (\ref{Srdef}), we need to assume that ${\mathring{a} (\lambda) \over \mathring{d} (\lambda)}$ and $\{\lambda_j \}$ are chosen such to satisfy the Bethe equations\index{Bethe!equations}, that is, $\mathring{\rho}_j \prod_{l=1}^R {f(\lambda_l - \lambda_j) \over f(\lambda_j - \lambda_l) } = 1$.
Note that $\{ \mathring{\sigma}_j , \mu_j \}$ are not bounded by a similar constraint.

We proceed by induction. For $R=1$, (\ref{bbSdef}) gives
\bea
   M_{11} (\mu, \lambda) & = &  {g^2 (\lambda - \mu) \over f(\lambda - \mu)} \: 
   \mathring{d} (\lambda) \Big[ \mathring{d} (\mu) - \mathring{a} (\mu) \Big] \; ,\\
   \mathbb{S}_1 \left( \mu , \lambda \right) 
   & = & {f(\lambda - \mu) \over g(\lambda - \mu)} H_{11} (\mu, \lambda) 
   = g(\mu - \lambda) \: \mathring{d}(\lambda) \: \Big[ \mathring{a} (\mu) - \mathring{d} (\mu) \Big] \; ,
\eea
which indeed agrees with (\ref{S1res}), once the Bethe equation\index{Bethe!equations} (\ref{ABABE1}) is satisfied, that is,  $\mathring{a} (\lambda) = \mathring{d} (\lambda)$.

Now we assume that $\mathbb{S}_{R-1} = S_{R-1}$ and want to prove it for $R$.
We have
\bea
   G_R \Big( \{\mu_j \}, \{ \lambda_j \} \Big) 
   & \stackrel{\mu_R \to \lambda_m}{\simeq} &
   (-1)^{R-m} \; { f (\lambda_m - \mu_R) \over g (\lambda_m - \mu_R) }
   \prod_{j \ne m} f (\lambda_j - \mu_R) \: \prod_{l=1}^{R-1} f (\lambda_m - \mu_l) 
   \times \nonumber \\
   && \quad \times 
   G_{R-1} \Big( \{\mu_j \}_{j<R}, \{ \lambda_j \}_{j \ne m} \Big) \; , 
   \label{GRrec}\\
   \det_R {\bf M} \Big( \{\mu_j \}, \{ \lambda_j \}, \{ \mathring{\sigma}_j \} \Big) 
   & \stackrel{\mu_R \to \lambda_m}{\simeq} & 
   (-1)^{R+m} {g^2 (\lambda_m - \mu_R) \over f(\lambda_m - \mu_R)} \; 
   \mathring{d}^\lambda_j 
   \left[ \mathring{d}^\mu_R - \mathring{a}^\mu_R \prod_{j \ne m} {f (\mu_R - \lambda_m) \over f(\lambda_m - \mu_R)} \right] 
   \times \nonumber \\
   && \quad \times 
   \det_{R-1} {\bf M} \Big( \{\mu_j \}_{j<R}, \{ \lambda_j \}_{j \ne m}, 
   \{ \mathring{\sigma}_j^{(R-1)} \}_{j<R} \Big)
   \: \: + \text{finite terms}
   \nonumber \\
   & = & (-1)^{R+m} {g^2 (\lambda_m - \mu_R) \over f(\lambda_m - \mu_R)} 
   \prod_{j \ne m} {f (\mu_R - \lambda_j) \over f(\lambda_j - \mu_R)} 
   \Big[ \mathring{a}^\lambda_m \: \mathring{d}^\mu_R - \mathring{a}^\mu_R \: \mathring{d}^\lambda_m \Big]
   \times \qquad \quad \nonumber \\
   && \quad \times 
   \det_{R-1} {\bf M} \Big( \{\mu_j \}_{j<R}, \{ \lambda_j \}_{j \ne m}, 
   \{ \mathring{\sigma}_j^{(R-1)} \}_{j<R} \Big)
   \: \: + \text{finite terms} \; .
   \label{detMRrec}
\eea
In the last line we used the Bethe equations\index{Bethe!equations} for the $\lambda$'s.
It should be noted that $G_R$ does not have poles as $\mu_R \to \lambda_m$ ($\lim_{\lambda \to 0} f(\lambda)/g (\lambda) = 1$). It does as $\mu_R \to \mu_j$ or $\lambda_m \to \lambda_j$, but in either of this limits two of the rows/columns of ${\bf M}$ become equal and thus $\det {\bf M}$ vanishes, rendering $\mathbb{S}_R$ finite. Collecting (\ref{GRrec}, \ref{detMRrec}):
\bea
   \mathbb{S}_R \Big( \{\mu_j \}, \{ \lambda_j \}, \{ \mathring{\sigma}_j \} \Big) 
   & \stackrel{\mu_R \to \lambda_m}{\simeq} & 
   g (\mu_R - \lambda_m)
   \prod_{j \ne m} f (\lambda_m - \lambda_j) \: \prod_{l=1}^{R-1} f (\mu_R - \mu_l)
  \Big[ \mathring{a}^\mu_R \: \mathring{d}^\lambda_m - \mathring{a}^\lambda_m \: \mathring{d}^\mu_R \Big]
  \times \nonumber \\
  && \qquad \times 
  S_{R-1} \Big( \{\mu_j \}_{j<R}, \{ \lambda_j \}_{j \ne m}, 
  \{ \mathring{\sigma}_j^{(R-1)} \}_{j<R} \Big) \: \: + \text{finite terms} \; .   
\eea
Thus, we see that $\mathbb{S}_R$ has the same poles and residues as $S_R$. This means that their difference $S_R - \mathbb{S}_R$ is bounded and thus by Liouville's theorem is a constant. Since both quantities vanish as one of the rapidities goes to infinity, this constant is zero and hence $S_R = \mathbb{S}_R$ as we set out to prove.

This proof is due to Slavnov \cite{slavnov89}, who built it on previous works and partial results \cite{korepin90,ISM} and first recognized that to make progress one of the sets of rapidities needed to satisfy the Bethe equations\index{Bethe!equations}.
It should be noted that comparing the algebraic construction of the quantum inverse scattering method with the transfer matrix\index{Transfer matrix} solution of the 6-vertex model\index{Vertex model} (appendix \ref{app:2DClassical}), the scalar product can be interpreted as the partition function of the classical model with domain wall boundary condition and Slavnov's fomula can thus be derived from the Izergin-Korepin formula for the latter \cite{foda12,kostov12}.

Pivotal in this construction has been the work by Gaudin \cite{gaudin}, who conjectured the analytical form of the norm of Bethe states from a careful analysis of numerical data. We can derive Gaudin's formula as the special case of Slavnov's result (\ref{bbSdef}) in which $\mu_j \to \lambda_j$ and the rapidities satisfy the Bethe equations\index{Bethe!equations} (\ref{ABABE1}):
\be
   \Omega_R \Big( \{ \lambda_j \} \Big) \equiv 
   \langle 0 | \prod_{l=1}^R {\bf C} (\lambda_l) 
   \prod_{j=1}^R {\bf B} (\lambda_j) | 0 \rangle 
   = \left. S_R \Big( \{\mu_j \}, \{ \lambda_j \} \Big)
   \right|_{\{\mu_j\} \to \{\lambda_j\}} \; .
   \label{OmegaRdef}
\ee
For $G_R$ and the off-diagonal terms of ${\bf M}$ the two sets of rapidities can be made equal without problems because, as we showed above, there are no singularities appearing in this limit:
\bea
   G_R \Big( \{ \lambda_j \}\Big) & = &
   \prod_{j \ne l} f (\lambda_j - \lambda_l) \; , \\
   M_{ab} \Big( \{ \lambda_j \}\Big) & \stackrel{a \ne b}{=} &
   \mathring{d} (\lambda_a) \mathring{d}(\lambda_b) \;
   {g^2 (\lambda_a - \lambda_b) \over f(\lambda_a - \lambda_b)}
   \left[ 1 - \mathring{\rho} (\lambda_b) \prod_{j \ne a} {f (\lambda_b - \lambda_j) \over f(\lambda_j - \lambda_b)} \right] 
   \nonumber \\
   & = &    \mathring{d} (\lambda_a) \mathring{d}(\lambda_b) \:
   {g^2 (\lambda_a - \lambda_b) \over f(\lambda_a - \lambda_b)} \; 
   \left[ 1 + {f (\lambda_a - \lambda_b) \over f(\lambda_b - \lambda_a)} \right] 
   \nonumber \\
   & = &    \mathring{d} (\lambda_a) \mathring{d}(\lambda_b) \: \varphi (\ii 2 \phi)
   \left[ {\partial_{\lambda_b} f(\lambda_a - \lambda_b) \over f(\lambda_a - \lambda_b)} 
   - {\partial_{\lambda_b} f(\lambda_b - \lambda_a) \over f(\lambda_b - \lambda_a)} \right] \: ,
\eea
where we used the fact that  $\lim_{\lambda \to 0} f(\lambda)/f(-\lambda) = -1$.
The diagonal terms of ${\bf M}$ have poles as the rapidities approach one another, which is compensated by the vanishing of $\lambda_m^\lambda \mathring{d}_R^\mu - \mathring{a}_R^\mu \mathring{d}_m^\lambda$ in (\ref{detMRrec}). We regularize this limit by setting $\mu_j = \lambda_j + \epsilon$ and work out:
\bea 
   M_{aa} \Big( \{\lambda_j + \epsilon \}, \{ \lambda_j \} \Big) & \equiv & 
   {g^2 (-\epsilon) \over f(-\epsilon)} \; 
   \mathring{d} (\lambda_a) \mathring{d}(\lambda_a + \epsilon)
   \left[ 1 - \mathring{\rho} (\lambda_a + \epsilon) \prod_{j \ne a} {f (\lambda_a - \lambda_j + \epsilon) \over f(\lambda_j - \lambda_a - \epsilon)} \right]
   \nonumber \\
   & \simeq & \mathring{d}^2 (\lambda_a) \; {\varphi (\ii 2 \phi) \over \epsilon}
   \left[ \mathring{\rho} (\lambda_a + \epsilon) \prod_{j \ne a} {f (\lambda_a - \lambda_j + \epsilon) \over f(\lambda_j - \lambda_a - \epsilon)} - 1 \right] 
   \nonumber \\
   & \stackrel{\epsilon \to 0}{\rightarrow} &
   \mathring{d}^2 (\lambda_a) \; \varphi (\ii 2 \phi) {\partial \over \partial \lambda_a}
   \mathring{\rho} (\lambda_a) \prod_{j \ne a} {f (\lambda_a - \lambda_j) \over f(\lambda_j - \lambda_a)} \; .
\eea   
Using the Bethe equations\index{Bethe!equations} as $\mathring{\rho} (\lambda_a) \prod_{j \ne a} {f (\lambda_a - \lambda_j) \over f(\lambda_j - \lambda_a)} =1$ we can write compactly the norm of a Bethe state as
\be
   \Omega_R \Big( \{ \lambda_j \} \Big) = \left[ - \ii \varphi (\ii 2 \phi) \right]^R
   \prod_{j=1}^R \mathring{d}^2 (\lambda_j) 
   \prod_{l \ne j} f (\lambda_j - \lambda_l) \:
   \det_R {\partial \over \partial \lambda_b} \Phi_a \Big( \{ \lambda_j \} \Big) \; ,
   \label{gaudin}
\ee
where
\be
   \Phi_a \Big( \{ \lambda_j \} \Big)  
   \equiv \ii \ln \left[ \mathring{\rho} (\lambda_a) \prod_{j \ne a} {f (\lambda_a - \lambda_j) 
   \over f(\lambda_j - \lambda_a)} \right]
   \equiv \Phi \Big( \lambda_a, \{ \lambda_j \}_{j \ne a} \Big)  \; .
\ee
This is the expression Gaudin conjectured \cite{gaudin} for the norm of a Bethe state. It has an interesting interpretation in terms of the action introduced by C.N. Yang and C.P. Yang in \cite{YangYang69}:
\be
   {\cal A} \equiv \sum_{j=1}^R \left[
   \int^{\lambda_j} \Phi \Big( \lambda, \{ \lambda_l \}_{l \ne j} \Big) \de \lambda - 2 \pi I_j \lambda_j \right] \; .
   \label{AYang}
\ee
Varying this action with respect to $\lambda_j$ looking for its minimum produces the Bethe equations\index{Bethe!equations} in logarithmic form
\be
   {\partial {\cal A} \over \partial \lambda_a} 
   = \Phi_a \Big( \{ \lambda_l \} \Big) - 2 \pi I_a =0 \; .
\ee
Thus, the determinant in (\ref{gaudin}) is a Jacobian that can be interpreted as the Hessian of the action (\ref{AYang}):
\be
   {\partial^2 {\cal A} \over \partial \lambda_a \partial \lambda_b}
   = {\partial \over \partial \lambda_b} \Phi_a \Big( \{ \lambda_j \} \Big)
   = \delta_{ab} \left[ z_a + \sum_{j=1}^R {\cal K} (\lambda_a -\lambda_j) \right]
   - {\cal K} (\lambda_a - \lambda_b)
\ee
where $z_j \equiv \ii \left. {\partial \ln \mathring{\rho} (\lambda) \over \partial \lambda} \right|_{\lambda = \lambda_j}$ and we used (\ref{Kfrel}). Hence, the normalizability of the Bethe state, i.e. the fact that the norm is positive and finite, implies that the Bethe solution is a minimum (stable) configuration of (\ref{AYang}) and can be used to prove the uniqueness of the ground state solution \cite{yangyangXXZ,ISM}\footnote{Note that in the paramagnetic phase of the XXZ chain at half-filling, this uniqueness is guaranteed only for $0<\Delta<1$ \cite{ISM}.}.

For instance, for the Lieb-Liniger model\index{Lieb-Liniger model} we have $\mathring{\rho} (\lambda) = \eu^{-\ii L \lambda}$ and the action is \cite{YangYang69}
\be
  {\cal A} = {L \over 2} \sum_{j=1}^N \lambda_j^2 - 2 \pi \sum_{j=1}^N I_j \lambda_j
  - {1 \over 2} \sum_{j,l}^N \Theta (\lambda_j - \lambda_l) \; ,
\ee 
where
\be
   \Theta (\lambda) \equiv
   \int^\lambda \theta (\lambda') \de \lambda'
   = c \ln \left( 1 + {\lambda^2 \over c^2} \right) 
   - 2 \lambda \arctan {\lambda \over c} \; .
\ee
The first variation of this action reproduces the Bethe equations\index{Bethe!equations} (\ref{betheeq}) and the second variation the norm of the states (note that in this case the factors before the Jacobian in (\ref{gaudin}) are equal to $1$).

Finally, we comment that using the aforementioned F-basis, it is possible to extract equivalent expressions of Slavnov's formula for the scalar product \cite{kitanine99}:
\be
   S_R \Big( \{\mu_j \}, \{ \lambda_j \} \Big) =
   {\det_R {\bf H} \Big( \{\mu_j \}, \{ \lambda_j \} \Big) \over 
   \prod_{j>l} \varphi (\lambda_l - \lambda_j) \varphi (\mu_j - \mu_l ) }	=
   { \det_R {\bf W} \Big( \{\mu_j \}, \{ \lambda_j \} \Big) \over
   	\det_R {\bf V} \Big( \{\mu_j \}, \{ \lambda_j \} \Big) } \; ,
\ee
where
\be
  H_{ab} \Big( \{\mu_j \}, \{ \lambda_j \} \Big) \equiv
  {\varphi (\ii 2 \phi) \over \varphi (\lambda_a - \mu_b)} \; 
  \mathring{d} (\lambda_a) 
  \left[ \mathring{a} (\mu_b) \; 
  \prod_{j \ne a} \varphi \left( \lambda_j - \mu_b + \ii 2 \phi \right) - \mathring{d} (\mu_b) \; 
  \prod_{j \ne a} \varphi \left( \lambda_j - \mu_b - \ii 2 \phi \right) \right] \; ,
\ee
and
\be
   W_{ab} \Big( \{\mu_j \}, \{ \lambda_j \} \Big) 
   \equiv {\partial \over \partial \lambda_a} 
   \Lambda \big( \mu_b ; \{ \lambda_j \} \big) \; ,
   \qquad \qquad
   V_{ab} \Big( \{\mu_j \}, \{ \lambda_j \} \Big)
   \equiv {1 \over \varphi (\mu_b - \lambda_a)} \; .
\ee
Here, $\Lambda \left( \mu_b ; \{ \lambda_j \} \right)$ are the eigenvalues of the transfer matrix\index{Transfer matrix} (\ref{Lambdamu}) and we remind that for Cauchy-like matrices like ${\bf V}$ we have
\be
   \det_R {\bf V} \Big( \{\mu_j \}, \{ \lambda_j \} \Big) =
   {\prod_{j>l} \varphi (\lambda_l - \lambda_j) \varphi (\mu_j - \mu_l) \over
   \prod_{j=1}^R \prod_{l=1}^R \varphi (\mu_j - \lambda_l)} \; .
\ee

\section{Algebraic approach to the Lieb-Liniger model: The Lax Representation}
\label{sec:LAX}

So far, in our presentation of thee ISM we have relied on the existence of a lattice over which the theory is defined. Let us discuss how to work with a continuous model such as the Lieb-Liniger\index{Lieb-Liniger model}. To do so, one considers the ``quantized'' version of the {\it Lax representation}, that allows to solve classical non-linear problems.
The main idea of the Lax method is to map the non-linear problem into a linear one, by adding an additional (auxiliary) degree of freedom. In its original form, it has been applied to the solution of classical integrable non-linear differential equations and allows for an explicit construction of (multi-)soliton solutions \cite{novikov1984,newell,solitonbook}, which characterize integrable PDE. The procedure also generates a family of commuting transfer matrices\index{Transfer matrix} which can be interpreted as the generating function of the conserved charges, including the Hamiltonian which produces the original non-linear problem.

The Lax representation of the Lieb-Liniger model\index{Lieb-Liniger model} is a straightforward quantization of that of the classical Non-linear Schr\"odinger equation (NLSE)\index{Non-Linear Schr\"odinger equation} (we remind that in the limit of weak interaction, in which the bosons form a quasi-condensate the long wavelengths properties of the quantum theory are well captured by the classical NLSE).
We introduce the two $\kappa \times \kappa$ matrices (here, as in thee rest of the chapter, $\kappa =2$):
\bea
   {\cal V} (x|\lambda) & \equiv & 
   \ii {\lambda \over 2} \: \sigma^z + {\bf \Upsilon} (x) \; , 
   \label{LLVdef} \\
   {\cal U} (x|\lambda) & \equiv & 
   \ii {\lambda^2 \over 2} \: \sigma^z + \lambda {\bf \Upsilon} (x)
   + \ii \sigma^z \left( \partial_x {\bf \Upsilon} + c \: \Psi^\dagger \Psi \right) \; ,
   \label{LLUdef}
\eea
with ${\bf \Upsilon} (x) \equiv \ii 
   \sqrt{c} \big[ \Psi^\dagger (x) \sigma^+ - \Psi (x) \sigma^- \big] 
   = \ii \sqrt{c} \begin{pmatrix}
   	0 & \Psi^\dagger (x) \cr - \Psi (x) & 0
   \end{pmatrix}$, where 
$\sigma^\alpha$ are Pauli matrices, $c$ is a constant and $\Psi (x)$ is a bosonic complex field with canonical commutation relation
\be
  \big[ \Psi (x), \Psi^\dagger (y) \big] = \delta (x-y) \; , \qquad \qquad
  \big[ \Psi (x), \Psi (y) \big] = \big[ \Psi^\dagger (x), \Psi^\dagger (y) \big] = 0 \; .
\ee
In the classical case $\Psi^\dagger (x) \to \Psi^* (x)$ and in the last equation one uses Poisson brackets.

The matrices ${\cal U}$ and ${\cal V}$ generate the temporal and spatial flow for a $\kappa$-dimensional vector field $\Phi (x,t)$:
\bea
   \partial_t \Phi (x,t) & = & \: \: {\cal U} (x|\lambda) \Phi (x,t) \; ,
   \nonumber \\
   \partial_x \Phi (x,t) & = & - {\cal V} (x|\lambda) \Phi (x,t) \; .
\eea
Since these equations are linear, they are easy to integrate separately, provided that the following consistency ({\it zero-curvature}) condition is satisfied
\be
   \big[ \partial_t - {\cal U} (x|\lambda),
   \partial_x + {\cal V} (x|\lambda) \big] = 0 \; ,
   \label{zerocurv}
\ee
for every $\lambda$ at each point $x$. Direct substitution shows that our choice for the {\it potential} ${\cal V} (x|\lambda)$ in (\ref{LLVdef}) and for the time evolution ${\cal U} (x|\lambda)$ in (\ref{LLUdef}) satisfy (\ref{zerocurv}) and thus decouple the dynamics. Two operators such as (\ref{LLVdef}, \ref{LLUdef}}), satisfying (\ref{zerocurv}), are called {\it Lax pair}.

The potential generates the infinitesimal translation. We can define the transition matrix $\bT (x,y|\lambda)$ which provides the evolution of the auxiliary field $\Phi$ from point $y$ to $x \ge y$ as the solution to the equation:
\be
\left[ \partial_x + {\cal V} (x|\lambda) \right] \bT (x,y|\lambda) = 0 \; ,
\ee
with the boundary condition $\bT (y,y|\lambda) = {\cal I}$. It has formal solution
\be
   \bT (x,y|\lambda)
   = : \eu^{ - \int_x^y {\cal V} (z|\lambda) \de z } : \; ,
   \label{TV}
\ee
where $: \ldots :$ indicates the normal ordering of placing creating operators $\Psi^\dagger$ to the left of the $\Psi$'s. Eq. (\ref{TV}) is the continuous analog of (\ref{TLLLL}) and satisfies
\be
\bT (x,z|\lambda) \; \bT (z,y|\lambda) = \bT (x,y|\lambda) \; , \qquad
x \ge z \ge y \; .
\ee
The monodromy matrix\index{Monodromy matrix} is the transition matrix for the whole chain: $\bT (\lambda) \equiv \bT (0,L|\lambda)$, and the transfer matrix\index{Transfer matrix} is obtained by tracing over the auxiliary $\kappa$-dimensional space: ${\bf T} (\lambda) \equiv \tr \bT (\lambda)$.

The trace identities\index{Trace identities} for this model are best evaluated at $\lambda \to \ii \infty$ and give \cite{ISM}:
\be
   \ln \left[ \eu^{\ii \lambda L/2} {\bf T} (\lambda) \right] 
   \stackrel{\lambda \to \ii \infty}{\rightarrow} \ii c \left\{
   {1 \over \lambda} \: \hat{J}_0 + 
   {1 \over \lambda^2} \left[ \hat{J}_1 - {\ii c \over 2} \: \hat{J}_0 \right] +
   {1 \over \lambda^3} \left[ \hat{J}_2 - \ii c \hat{J}_1 - {c^3 \over 3} \: \hat{J}_0 \right] + 
   \Ord \left( {1 \over \lambda^4} \right) \right\} \; ,
\ee
with the conserved charges 
\bea
   \hat{J}_0 & \equiv & \int \Psi^\dagger \Psi \de x 
   \hskip 4cm \text{Particle number} \; , \\
   \hat{J}_1 & \equiv & -  \ii \int \Psi^\dagger \partial_x \Psi \de x 
   \hskip 3.2cm \text{Momentum}\; , \\
   \hat{J}_2 & \equiv & \int \big[ \partial_x \Psi^\dagger \partial_x \Psi
   + c \: \Psi^\dagger \Psi^\dagger \Psi \Psi \big] \de x 
   \hskip 1cm \text{Lieb-Liniger Hamiltonian} \; , \\
   & \ldots \nonumber
\eea

The connection to lattice models is provided by introducing a lattice spacing $\delta$ which discretizes the positions $x_j = j \: \delta$. The evolution equations then read
\bea
   \partial_t \Phi (j,t) & = & {\cal U} (j|\lambda) \Phi (j,t) \; ,
   \nonumber \\
   \Phi (j+1,t) & = & {\cal L} (j|\lambda) \Phi (j,t) \; ,
\eea
where the spatial evolution is given by the {\it Lax operator}\index{Lax operator}
\be
  {\cal L} (j|\lambda) \equiv {\cal I} - {\cal V} (x_j|\lambda) + \Ord (\delta^2) 
  = \left( \begin{array}{ccc}
  	1 - \ii {\lambda \over 2} \: \delta & \quad & 
  	- \ii \sqrt{c} \: \Psi^\dagger_j \: \delta \cr
  	\ii \sqrt{c} \: \Psi_j \: \delta & \quad & 
  	1 + \ii {\lambda \over 2} \: \delta \cr
  \end{array} \right) + \Ord (\delta^2) \; ,
  \label{classicalLLL}
\ee
with $\Psi_j \equiv {1 \over \delta} \int_{x_{j-1}}^{x_j} \Psi (x) \de x $ $\left( \big[ \Psi_j , \Psi^\dagger_l \big] = {1 \over \delta} \delta_{j,l} \right)$ and ${\cal I}$ is the $\kappa \times \kappa$ unit matrix.

Both monodromy matrix\index{Monodromy matrix} (\ref{TV}) and Lax operator\index{Lax operator} (\ref{classicalLLL}) satisfy a YBE\index{Yang-Baxter!equation} (\ref{TYB1},\ref{LLR}) with intertwiner\index{Intertwiner operator}
\be 
  {\cal R}^{XXX}_{a,b} (\lambda) \equiv 
  {\lambda \over c} \; {\cal I}_{a,b} + \ii \; {\bf \Pi}_{a,b} \; ,
  \label{XXXR}
\ee
which is the ${\cal R}$-matrix\index{Intertwiner operator} of the XXX chain (\ref{Ransatz})\index{XXX chain}\index{Heisenberg chain}, with the rescaling $\lambda \to { 2 \lambda / c}$. Thus, the ABA proceeds writing the monodromy matrix\index{Monodromy matrix} as a $\kappa \times \kappa$ matrix, whose entries are operators satisfying the generalized commutation relations (\ref{BB}-\ref{CD}) with
\be
   f_{XXX} (\lambda) = 1 + \ii \; {c \over \lambda} \; , \qquad \quad
   g_{XXX} (\lambda) = \ii \; {c \over \lambda} \; .
   \label{XXXfg}
\ee

The pseudo-vacuum\index{Pseudo-vacuum} is just the Fock vacuum $|0 \rangle$ $\left( \Psi (x) |0 \rangle = 0 , \forall x \right)$. In a discretized setting, the eigenvalues of the Lax operator\index{Lax operator} on each lattice site $j$ are $\mathring{a}_j (\lambda) = 1 - \ii \; {\lambda \delta \over 2}$ and $\mathring{d}_j (\lambda) = 1 + \ii \; {\lambda \delta \over 2}$. For the whole chain, using $\lim_{N \to \infty} (1 + \ii \lambda \delta)^N = \eu^{\ii \lambda L}$, with $L = N \delta$, in the thermodynamic limit we have 
\be
   \mathring{a} (\lambda) = \eu^{-\ii {\lambda L \over 2}} \; , \qquad 
   \mathring{d} (\lambda) = \eu^{\ii {\lambda L \over 2}} \; , \qquad 
   \mathring{\rho} (\lambda) = \eu^{-\ii \lambda L } \; .
\ee
When plugged into the algebraic construction, for instance, (\ref{ABABE}) reproduces  the Bethe equations\index{Bethe!equations} (\ref{LLBetheEq}). In closing, we remark that the time evolution operator ${\cal U}$ can be extracted from the trace identities (of course, it includes the Hamiltonian, see \cite{ISM}) and the conserved charges evaluated on Bethe states are simple symmetric polynomials of the state rapidities: $J_n = \sum_{l=1}^R \lambda_l^n$.

\section{The Braid Limit}
\label{sec:braid}

\begin{figure}[b]
	\noindent\begin{minipage}[t]{5.5cm}
		\includegraphics[width=5.5cm]{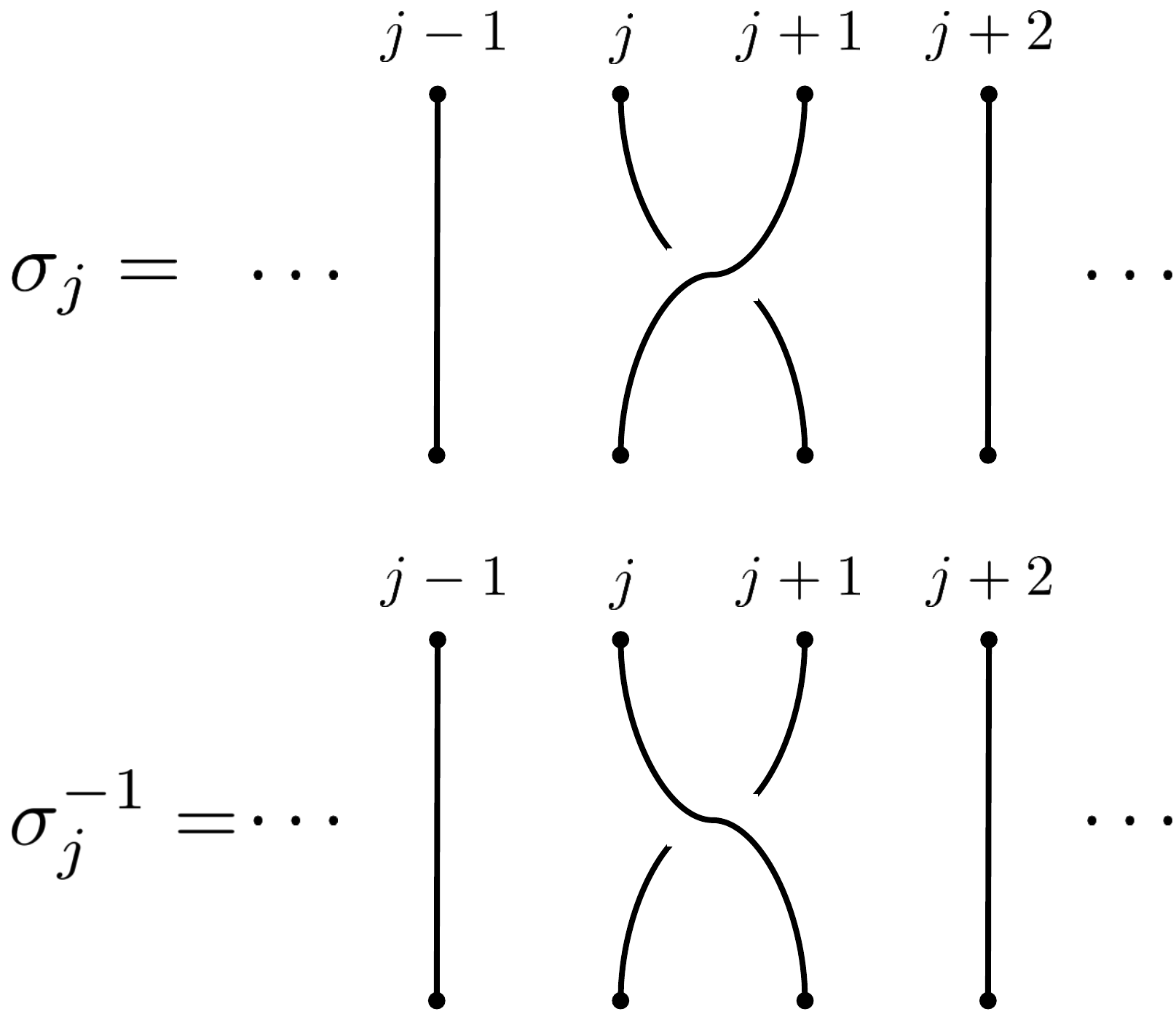}
	\end{minipage}
	\hfill
	\begin{minipage}[b]{10cm}
		$\qquad$\includegraphics[width=8cm]{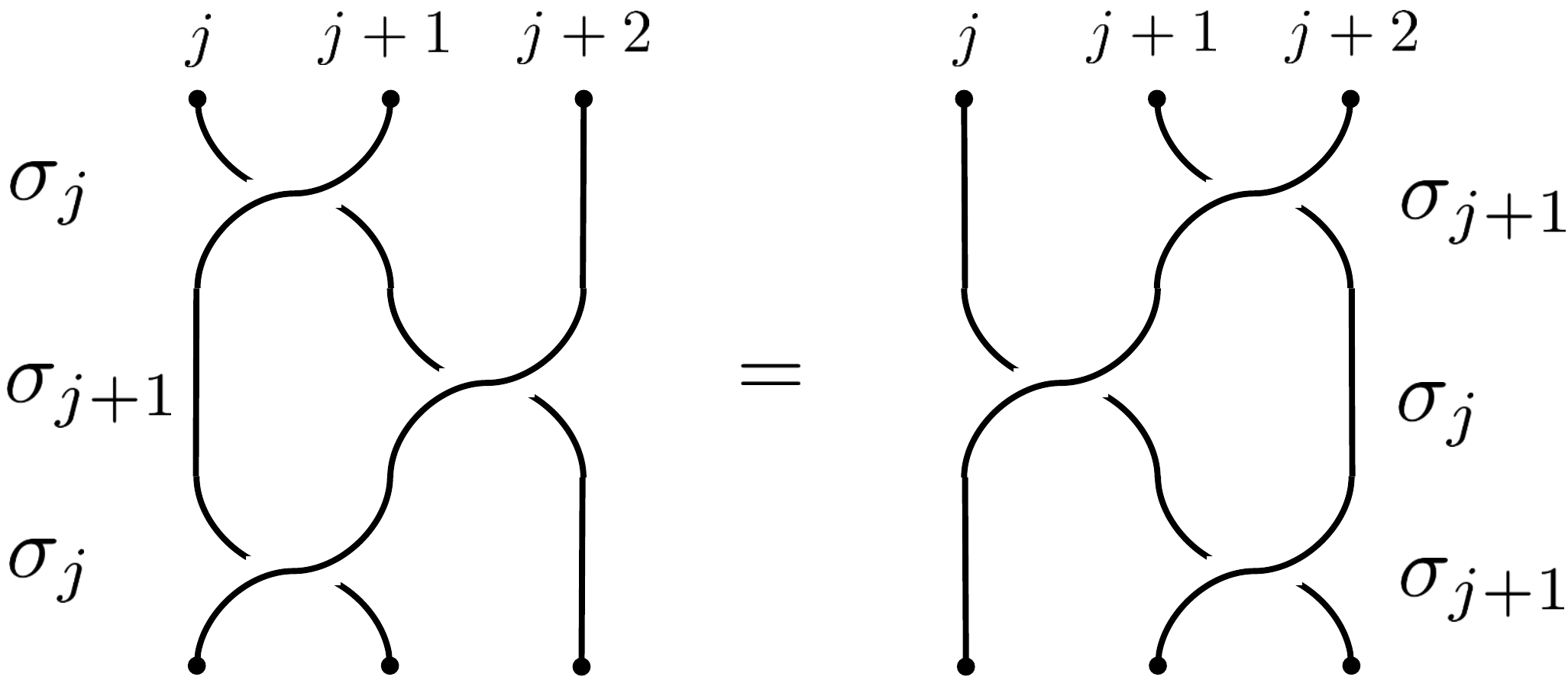}
		\caption{{\bf Left:} Graphical representation of the action of a braid element and of its inverse. {\bf Above:} Graphical representation of the first (fundamental) braid relation in (\ref{braidrel}).}
		\label{fig:braid}
	\end{minipage}
\end{figure}

The {\it braid group}\index{Braid!group} $B_N$ on $N$ strands is generated by $N-1$ elements $\sigma_j$ ($j=1,\ldots,N-1$) satisfying \cite{braid}
\be
   \sigma_j \sigma_{j+1} \sigma_j =  \sigma_{j+1}\sigma_j\sigma_{j+1} \; , \qquad\qquad
   \sigma_j \sigma_l = \sigma_l \sigma_j \text{ if } |l-j|\ge 2 \; , \qquad \qquad
   \sigma_j \sigma_j^{-1} = \sigma_j^{-1} \sigma_j = {\bf I} \; .
   \label{braidrel}
\ee
It is similar to the permutation group, but it keeps track of which wordline (strand) of the two permuted elements crosses above the other: the generator $\sigma_j$ braids the $j$-th strand under the $(j+1)$-th strand, while $\sigma_j^{-1}$ goes over, see Fig \ref{fig:braid}.
To explore the connection between the braid group\index{Braid!group} and the Yang-Baxter equation\index{Yang-Baxter!equation}, it is more convenient to switch to the reflection diagonal representation, defining
\be
  {\bf R} (\lambda) \equiv {\bf \Pi} \: {\cal R} (\lambda) 
  \qquad \qquad \Leftrightarrow \qquad \qquad 
  R_{\alpha \beta}^{\alpha' \beta'} (\lambda)
  \equiv {\cal R}_{\alpha \beta}^{\beta' \alpha'} (\lambda) \; .
\ee
This permuted intertwiner\index{Intertwiner operator} satisfies the YBE\index{Yang-Baxter!equation}:
\be
   \big( {\bf 1} \otimes {\bf R} (\lambda) \big) \:
   \big( {\bf R} (\lambda + \mu) \otimes {\bf 1} \big) \:
   \big( {\bf 1} \otimes {\bf R} (\mu) \big) =
   \big( {\bf R} (\mu) \otimes {\bf 1} \big) \:
   \big( {\bf 1} \otimes {\bf R} (\lambda + \mu) \big) \:
   \big( {\bf R} (\lambda) \otimes {\bf 1} \big) \; .  
\ee
Considering $N$ copies of the space ${\cal V}$, we can define a set of $N-1$ operators ${\bf R}_j$ on $\otimes_{j=1}^N {\cal V}_j$ as 
\be
  {\bf R}_j (\lambda) \equiv {\cal I} \otimes \cdots \otimes {\cal I} \otimes 
  {\bf R} (\lambda) \otimes {\cal I} \otimes \cdots \otimes {\cal I} \; ,
  \label{permRRR}
\ee
each of them acting non-trivially only on the $j$ and $j+1$ space.
Using this set of operators, the YBE\index{Yang-Baxter!equation} (\ref{permRRR}) can be written as
\be 
   {\bf R}_{j+1} (\lambda) \: {\bf R}_j (\lambda + \mu) \: {\bf R}_{j+1} (\mu) 
   = {\bf R}_j (\mu) \: {\bf R}_{j+1} (\lambda + \mu) \: {\bf R}_j (\lambda) \; .
   \label{BraidYBE}
\ee
This is the first of the defining equation (\ref{braidrel}), except for the role of the rapidities. This can be neutralized by setting $\lambda = \mu =0$, but in this limit the ${\bf R}$ operators are proportional to the identity and thus this is a trivial representation of the permutation group. Another possibility is to take $\lambda, \mu \to \pm \infty$. This is the {\it Braid Limit}, but, before taking it, it is convenient to rescale the operators as $\tilde{R}_{\alpha \beta}^{\beta' \alpha'} (\lambda) = \eu^{\lambda (\alpha - \alpha')/4} R_{\alpha \beta}^{\beta' \alpha'} (\lambda)$, which can be accomplished by rescaling the natural basis for the matrix representation \cite{quantumgroups}:
\be
   {\bf R}^{(\pm)} \equiv \pm 2 \lim_{\lambda \to \pm \infty} 
   \eu^{\mp{\lambda + \ii \phi \over 2}} \tilde{\bf R} (\lambda) \: :
   \quad 
   {\bf R}^{(+)} = {1 \over \sqrt{q}}
   \begin{pmatrix}
 	q & \quad & 0 & 0 & 0 \cr
	0 & \quad & 0 & 1 & 0 \cr
	0 & \quad & 1 & \: q-q^{-1} \: & 0 \cr
	0 & \quad & 0 & 0 & q \cr
   \end{pmatrix} \; , \: \:
   {\bf R}^{(-)} = \sqrt{q}
   \begin{pmatrix}
   	q^{-1} &  0 & 0 & \quad & 0 \cr
   	0 &  \: q^{-1} -q \: & 1 & \quad &0 \cr
   	0 &  1 & 0 & \quad & 0 \cr
   	0 &  0 & 0 & \quad & q^{-1} \cr
   \end{pmatrix} 
   \label{bfRpm}
\ee
where we introduced $q \equiv \eu^{\ii \phi}$. Note that ${\bf R}^{(+)} {\bf R}^{(-)} = {\cal I}$ and that we assume that $-1 < \Delta = \cos \phi < 1$. For $\Delta >1$, the braid limit\index{Braid!limit} is taken with $\lambda \to \pm \ii \infty$ and $q$ is analytically continued accordingly.

Each of this rescaled ${\bf R}$ matrices in the braid limit\index{Braid!limit} satisfies a YBE\index{Yang-Baxter!equation} like (\ref{BraidYBE}), but without spectral parameters. They form a representation of the braid group\index{Braid!group}, with ${\bf R}^{(+)}$ and ${\bf R}^{(-)}$ being the inverse of one another. We refer to \cite{quantumgroups} for a deeper discussion of this connection.

\section{A glimpse into Quantum Groups}
\label{sec:Qgroups}

The ${\bf R}$ operator of the Lieb-Liniger/XXX model (\ref{XXXR})\index{Lieb-Liniger model}\index{XXX chain}\index{Heisenberg chain} is the simplest (non-trivial) solution of the YBE\index{Yang-Baxter!equation}. In the braid limit\index{Braid!limit} it reduces directly to a permutation, constructed out of the $SL(2)$ algebra of Pauli and identity matrices (\ref{exchange}).
In the previous section we saw that the ${\bf R}$ operator for the whole XXZ chain constitutes a representation of the permutation/braid group\index{Braid!group}. We now want to argue that this results is the consequence of a deformation of the $SL(2)$ algebra, so that for every $\Delta$ the ${\bf R}$ operator can be seen as a deformed permutation operator.
This point of view allows for a classification of the solutions of the YBE\index{Yang-Baxter!equation}, as deformation of permutation operators for different dimensions $\kappa$.

The concept of deformation is at the heart of the {\it quantum group} construction. A proper account of this topic would require the introduction of the notion of Hopf algebra and of several connected structures, which can be found, for instance, in \cite{quantumgroups,evangelisti}. However, it is possible to explain the main ideas behind this construction by considering a quantum group just as the deformation of a traditional (classical) Lie algebra. 

We will do so by focusing on the relevant case of the $SL(2)$ group, which is the group of $2 \times 2$ matrices of unit determinant. It is also the group of linear transformation of $\mathbb{R}^2$ that preserves oriented areas. 
To define its deformation, we introduce the deformation parameter $q$ and consider the non-commutative, two-dimensional space spanned by a vector $(x,y)$ with the commutation property $xy = q \: yx$. While $q=1$ reproduces $\mathbb{R}^2$, for generic $q$ this non-commutative space has peculiar transformation properties. For instance:
\bea
  && \de = \de x \; \partial_x + \de y \; \partial_y \qquad \Rightarrow \qquad
  \de x \: \de y = - q^{-1} \: \de y \: \de x \;, \quad (\de x)^2 = (\de y)^2 = 0 \; , \quad \partial_x \partial_y = q^{-1} \partial_y \partial_x \; ,
  \label{noncomm}
  \\
  && \partial_x x = 1 + q^2 x \partial_x + (q^2-1)y \partial_y \; , \quad
  \partial_x y = q y \partial_x \; , \quad 
  \partial_y x = q x \partial_y \; , \quad
  \partial_y y = 1 + q^2 y  \partial_y \; , 
  \label{noncommcal1} \\
  && \partial_x \de x = q^{-2} \de x \: \partial_x \; , \quad
  \partial_x \de y = q^{-1} \de y \: \partial_x \: , \quad
  \partial_y \de x = q^{-1} \de x \: \partial_y \; , \quad
  \partial_y \de y = q^{-2} \de x \: \partial_y + (q^{-2} -1 ) \de x \: \partial_x \; ,
  \nonumber \\
  && x \: \de x = q^2 \de x \: x \; , \quad
  x \: \de y = q \de y \: x + (q^2 - 1) \de x \: y \; , \quad
  y \: \de x = q  \de x \: y \, \quad
  y \: \de y = q^2 \de y \: y \; .
  \label{noncommcal2}
\eea
For the linear transformation
\be
   \begin{pmatrix} x' \\ y' \end{pmatrix} = 
   {\bf T} \begin{pmatrix} x \\ y \end{pmatrix} \; , \qquad
   \begin{pmatrix} \partial_{x'} \\ \partial_{y'} \end{pmatrix} = 
   \left( {\bf T}^t \right)^{-1} \begin{pmatrix} \partial_x \\ \partial_y \end{pmatrix} \; , \qquad
   \begin{pmatrix} \de x' \\ \de y' \end{pmatrix} = 
   {\bf T} \begin{pmatrix} \de x \\ \de y \end{pmatrix} \; , \qquad \qquad
  {\bf T} \equiv \begin{pmatrix} a & b \\ c & d \\ \end{pmatrix} \; ,
\ee
to be consistent with commutation properties of the space (that is, for both prime and non-primed quantities to satisfy (\ref{noncomm})), the matrix entries have to satisfy:
\be
   ab = q \: ba \; , \quad ac = q \: ca \; , \quad bc = cb \; , \quad
   bd = q \: db \; , \quad cd = q \: dc \: , \quad 
   [a,d] = \left( q - q^{-1} \right) bc \; .
   \label{SLqentries}
\ee
Moreover, the quantity
\be
  {\det}_q {\bf T} = ad - q \: bc = da - q^{-1} bc 
  \label{qdetdef}
\ee
commutes with every entry of ${\bf T}$. to ensure the consistency of (\ref{noncommcal1}-\ref{noncommcal2}) (area preservation) we demand ${\det}_q {\bf T}={\bf I}$, where ${\bf I}$ is the unit operator. Note that 
\be
   {\bf T}^{-1} = \begin{pmatrix}
   	q & -q^{-1} b \\ -q \: c & a \\
   \end{pmatrix} \: , \qquad
   {\bf T} {\bf T}^{-1} = {\bf T}^{-1} {\bf T} = 
   \begin{pmatrix}
   	{\bf I} & 0 \\ 0 & {\bf I} \\
   \end{pmatrix} \; .
\ee
The group of $2 \times 2$ matrices, whose entries satisfy (\ref{SLqentries}) and with $q$-determinant (\ref{qdetdef}) equal to unity form the q-deformed $SL(2)$ group, called the quantum group $SL_q (2)$. Note that {\it ``quantum''} in this case refer to the fact that, parameterizing $q \equiv \eu^{-h}$, in the $h \to 0$ ($q \to 1$) limit one recovers the ``classical'' case.

Even before the concept of non-commutative space was proposed, a calculus with $q$-deformed objects was introduced, for instance to provide a non-standard measure that renders some integrals finite. For instance, the $q$-deformation of a number $x$ is defined as
\be
  [[x]]_q \equiv {q^x - q^{-x} \over q - q^{-1}} 
  \stackrel{q=\eu^{\phi}}{=}
  {\sinh (\phi x) \over \sinh \phi} = \prod_{n=-\infty}^\infty
  {x + \pi n \phi^{-1} \over 1 + \pi n \phi^{-1} } \; ,
  \label{xqdef}
\ee
where the last representation has the physical interpretation of a mapping of the complex plane to a strip via a multiplicative averaging, but it does not have a transparent limit for $\phi \to 0$.
It is also possible to define a generalization of the exponential function as
\be
  \eu_q^z \equiv \sum_{n=0}^\infty {z^n \over [n]_q!} \; , \qquad \qquad
  [n]_q! \equiv \prod_{j=1}^n {1-q^j \over 1 - q} \; , \quad [0]_q! \equiv 1 \; .
\ee
Notice that in $q \to 1$ limit all these quantities reduce to their classical counterparts (e.g $\lim_{q \to 1} [[x]]_q = x$).

From this point of view, the $SL_q (2)$ algebra is just the $q$-deformation of the classical $SL(2)$ one.
For instance, we can parametrized an element of the quantum group as 
\be
   {\bf T} = \begin{pmatrix}
   	\eu^{\alpha} & \eu^\alpha \: \beta \\
   	\gamma \: \eu^\alpha & \quad  \eu^{-\alpha} + \gamma \eu^\alpha \: \beta \\
   \end{pmatrix}
   = \eu_{q^{-2}}^{\gamma \mathscr{S}^-} \eu^{\alpha \mathscr{S}^z}
   \eu_{q^2}^{\beta \mathscr{S}^+}\; , \qquad \quad
   \left[\alpha, \beta \right] = \left( \ln q \right) \beta \; , \quad
   \left[\alpha, \gamma \right] = \left( \ln q \right) \gamma \; , \quad
   \left[\beta, \gamma \right] = 0 \; ,
\ee
where the generators $\mathscr{S}^\pm, \mathscr{S}^z$ obey the quantum $sl_q (2)$ algebra:
\be
  \left[ \mathscr{S}^z , \mathscr{S}^\pm \right] = \pm 2 \mathscr{S}^\pm \; , \qquad
  \left[ \mathscr{S}^+ , \mathscr{S}^- \right] 
  = \left[ \left[ \mathscr{S}^z \right]\right]_q
  = {q^{\mathscr{S}^z} - q^{-\mathscr{S}^z} \over q - q^{-1}} \; .
  \label{slq2S}
\ee
These are the defining equations of the quantum $sl_q(2)$ algebra, which reduces to the classical $sl(2)$ algebra in the $q \to 1$ limit. 
Notice that, unlike what happens for higher dimensional representations,the two-dimensional irreducible representation of the quantum algebra is provided directly by the spin-$1/2$ generators of $sl(2)$: $\mathscr{S}^\pm = \sigma^\pm$ and $\mathscr{S}^z = \sigma^z$, since $\left[ \left[ \sigma^z \right]\right]_q = \sigma^z$ because $\eu^{\alpha \sigma^z} = \left( \cosh \alpha \right) {\bf I} + \left( \sinh \alpha \right) \sigma^z$ (and thus ${q^{\sigma^z} - q^{-\sigma^z}} = (q - q^{-1}) \sigma^z$).
Notice, moreover, that for this representation $\eu_{q^{-2}}^{\gamma \mathscr{S}^-} = \eu^{\gamma \sigma^-}$ and $\eu_{q^2}^{\beta \mathscr{S}^+} = \eu^{\beta \sigma^+}$ because $\left( \sigma^\pm \right)^2 =0$. It should be stressed that our treatment of this quantum group does not do justice to its rich structure and that the equivalence between the quantum and classical representations is a consequence of these simplifications. In its full glory, $sl_q(2)$ is a bi-algebra and its 2D irreducible representation has six generators: ${\bf E}_0 \equiv \eu^\lambda \sigma^-$, ${\bf E}_1 \equiv \eu^\lambda \sigma^+$, ${\bf F}_0 \equiv \eu^{-\lambda} \sigma^+$, ${\bf F}_1 \equiv \eu^{-\lambda} \sigma^-$, ${\bf K}_0 \equiv q^{-\sigma^z}$, ${\bf K}_1 \equiv q^{\sigma^z}$, satisfying
\be
  {\bf K} \: {\bf E} = q^{2} \: {\bf E} \: {\bf K} \; , \qquad \qquad
  {\bf K} \: {\bf F} = q^{-2} \: {\bf F} \: {\bf K} \; , \qquad \qquad
  \left[ {\bf E} , {\bf F} \right] =
  { {\bf K} - {\bf K}^{-1} \over q - q^{-1} } \; ,
  \label{slq2EFK}
\ee
but we will not develop this representation further \cite{quantumgroups}.

If we write $q \equiv \eu^{\phi}$, the $q$-deformation of the Lax operator\index{Lax operator} for the Heisenberg chain\index{XXX chain}\index{Heisenberg chain} is
\bea
   \left[\left[ {\cal L}_{j,a}^{\rm XXX} (\lambda) \right]\right]_q & \propto &
   \big[ \big[ (\lambda + \ii) \: {\cal I}_{j,a} + 2 \ii \:
   \overrightarrow{\sigma_j} \cdot \overrightarrow{\tau_a} \big] \big]_q 
   \nonumber \\
   & = & {1 \over \sinh \phi} \left( \begin{array}{cc}
	x \; q^{{1 + \sigma_j^z \over 2}} - x^{-1} \; q^{-{1 + \sigma_j^z \over 2}} &
	(q - q^{-1}) \sigma_j^- \cr
	(q - q^{-1}) \sigma_j^+ &
	x \; q^{{1 - \sigma_j^z \over 2}} - x^{-1} \; q^{-{1 - \sigma_j^z \over 2}} \cr
    \end{array} \right) 
    \propto {\cal L}_{j,a}^{\rm XXZ} (\phi \lambda) \; ,
\eea
where $x \equiv \eu^{\phi \lambda/2}$ and where we dropped some (irrelevant) normalization factors in front of the Lax matrices\index{Lax operator}. Thus, we recognize that the deformation of the Lax operator of the Heisenberg chain has produced the one of the XXZ chain (\ref{XXZL}) with $\Delta = \cosh \phi >1$. Setting $q=\eu^{\ii \phi}$ yields the ${\cal L}$-matrix for $|\Delta|<1$..

In this notation, the Yang-Baxter equation\index{Yang-Baxter!equation} reads
\be
    {\cal L}_{j,a} (x) \; {\cal L}_{j,b} (y) \; {\cal R}_{a,b} (y/x) =
    {\cal R}_{a,b} (y/x) \; {\cal L}_{j,b} (y) \; {\cal L}_{j,a} (x) \; ,
    \label{multYBE}
\ee
and the entries of the intertwiner\index{Intertwiner operator} (\ref{Ransatz}) are written as
\be
   a \equiv q x - q^{-1} x^{-1} \; , \qquad
   b \equiv x - x^{-1} \; , \qquad
   c \equiv q - q^{-1} \; .
\ee

We now perform a similarity transformation which preserves the validity of the YBE (\ref{multYBE})\index{Yang-Baxter!equation}
\bea  
   \tilde{\cal L}_{j,a} (x) \equiv
   {\bf Q} (x) \: {\cal L}_{j,a} (x) \: {\bf Q}^{-1} (x) 
   & = & \left( x \sqrt{q} \right) \: {\cal L}_{j,a}^{(+)} - 
   \left( x \sqrt{q} \right)^{-1} \: {\cal L}_{j,a}^{(-)}\; \\
   \tilde{\cal R}_{a,b} (x/y) \equiv
   \left[ {\bf Q} (x) \otimes {\bf Q} (y) \right] {\cal R}_{a,b} (x/y)
   \left[ {\bf Q}^{-1} (x) \otimes {\bf Q}^{-1} (y) \right] 
   & = & \left( x \sqrt{q} \right) \: {\cal R}_{a,b}^{(+)} - 
   \left( x \sqrt{q} \right)^{-1} \: {\cal R}_{a,b}^{(-)}  \; ,
\eea
with $ {\bf Q} (x) \equiv \begin{pmatrix} x^{1/2} & 0 \\ 0 & x^{-1/2} \\
   \end{pmatrix}$,
and where ${\cal L}^\pm$, ${\cal R}^\pm$ are the permuted of the braid limit\index{Braid!limit} matrices (\ref{bfRpm}):
\be
    {\cal L}^{(+)} \equiv \begin{pmatrix}
    	q^{\sigma^z/2} & \left(q^{1/2} - q^{-3/2} \right) \sigma^- \\
    	0 & q^{-\sigma^z/2} \\
    \end{pmatrix} \; , \qquad \qquad
    {\cal L}^{(-)} \equiv \begin{pmatrix}
    	q^{-\sigma^z/2} & 0 \\
    	- \left( q^{3/2} - q^{-1/2} \right) \sigma^+ & q^{\sigma^z/2} \\
    \end{pmatrix} \; ,
\ee
and similarly for ${\cal R}^\pm$.

Through these transformations, the YBE (\ref{multYBE})\index{Yang-Baxter!equation} can be regarded as a polynomial in different powers of $x$ and $y$: the coefficients of each term of this polynomial contain different combinations of the braid matrices. Fulfillment of (\ref{multYBE}) means that each of these seven (rapidity-less) terms are equal to zero.

From an arithmetic point of view, the vanishing of these equations is equivalent to the fulfillment of the braid relations (\ref{BraidYBE}), but this is just a coincidence of this representation for the YBE\index{Yang-Baxter!equation}, where both Lax operator\index{Lax operator} and intertwiner\index{Intertwiner operator} have essentially the same functional form. In fact, the operators in (\ref{BraidYBE}) and (\ref{multYBE}) act on different space and thus the latter cannot be taken as braid relations. 

In generality, the vanishing of the coefficients for the different powers of $x$ and $y$ in (\ref{multYBE}) is a consequence of the algebra satisfied by the braid matrices ${\cal L}^\pm$, ${\cal R}^\pm$, as all these equations can be reduced to different combinations of (\ref{slq2S}). Thus, we conclude that the Yang-Baxter algebra\index{Yang-Baxter!algebra} of the six-vertex model\index{Vertex model} is a reflection of the underlying $sl_q(2)$ algebra (\ref{slq2S}, \ref{slq2EFK}), which provides the appropriate deformation of the fundamental solution generated by the permutation operator for the Heisenberg chain (\ref{XXXR})\index{XXX chain}\index{Heisenberg chain}.

It is possible to systematically deform any Lie group (possibly with more that one deformation parameter) and in this way to derive new, non-trivial, representations of the Braid group and, with a proper introduction of the rapidities, to construct the corresponding solutions of the Yang-Baxter equation\index{Yang-Baxter!equation} \cite{quantumgroups}.

\appendix

\chapter{Asymptotic behavior of Toeplitz Determinants}
\label{ToeplitzApp}

\abstract{
	As we showed in Chapter \ref{chap:XYModel}, virtually all the (static) correlation functions of the XY chain (as well as of other one-dimensional quadratic models) can be expressed in terms of Toeplitz matrices. For this reason and for their intriguing mathematical structure, these matrices have attracted the interest of mathematicians and of mathematical physicists. These efforts have granted an impressive control and understanding of these matrices. In this appendix we review some of the results concerning the asymptotic behavior of Toeplitz determinants and provide some references for additional analysis.
}

\section{Introduction}

The theory of Toeplitz determinants\index{Toeplitz!determinant} is intimately connected with the XY chain, since the pioneering works in \cite{LSM-1961,mccoy} for its spin-spin correlation functions. It is well known that the asymptotic behavior of the determinant of a Toeplitz matrix\index{Toeplitz!matrix} as the matrix size tends to infinity strongly depends upon the zeros and singularities of the generating function of the matrix.

Good reports on the subject have been recently compiled \cite{Ehrhardt-2001,deift13} and we refer to them for a more exhaustive review of what has been studied. Here we want to recapitulate what is known
about the determinant
\be
   D_n [\sigma] = \det ({\bf S_n}) =
   \det \left| s(j-k) \right|_{j,k=0}^n \; ,
\ee
of a $n+1 \times n+1$ Toeplitz matrix\index{Toeplitz!matrix}
\be
   {\bf S_n} = \begin{pmatrix}
   	                     s(0) & s(-1) & s(-2) & \ldots & s(-n) \\
                         s(1) & s(0)  & s(-1) & \ldots & s(1-n) \\
                         s(2) & s(1) & s(0) & \ldots & s(2-n) \\
                         \vdots & \vdots & \vdots & \ddots & \vdots \\
                         s(n) & s(n-1) & s(n-2) &\ldots & s(0) \
               \end{pmatrix} \; ,
\ee 
with entries generated by a function $\sigma(q)$:
\be
   s(l) \equiv \int_{-\pi}^{\pi} \sigma(q)
   \eu^{- \ii l q} {\de q \over 2 \pi} \: ,
\ee where the generating function $\sigma(q)$ is a periodic
(complex) function, i.e. $\sigma(q) = \sigma(2 \pi + q)$.

We will only consider generating functions with zero winding number, that is 
\be
   {\rm Ind } \, \sigma(q) \equiv
   \int_{-\pi}^{\pi} {\de q \over 2 \pi} \, {\de \over \de q}
   \log \sigma (q) = 0 \; ,
\ee
although this is not always the case even in the study of some correlators of the XY chain: in some regions of the phase diagram, \cite{mccoy} worked with generating function with non-zero winding number and we refer to it on how to reduce the problem to one with zero winding number, so that the theorems we present apply.

\section{The Strong Szeg\"o Theorem}

If $\sigma(q)$ is sufficiently smooth, \underline{non-zero} and
satisfies ${\rm Ind } \,\sigma(q) = 0$ (i.e., the winding number is $0$),
we can apply what is known as the {\it Strong Szeg\"o Limit Theorem}
(\cite{Hirschman}, \cite{mccoywu}), which states that the determinant
has a simple exponential asymptotic form
\be
   D_n [\sigma] \sim E[\sigma] \; F[\sigma]^n
   \qquad n \rightarrow \infty \; ,
   \label{szego}
\ee
where $F[\sigma]$ and $E[\sigma]$ are defined by
\be
  F[\sigma] \equiv \exp{\hat{\sigma}_0}, \qquad
  E[\sigma] \equiv
  \exp{\sum_{k=1}^\infty k \hat{\sigma}_k \hat{\sigma}_{-k}}
  \label{szegoexp}
\ee
and $\hat{\sigma}_k$ are the Fourier coefficients of the expansion of
the logarithm of $\sigma(q)$:
\be
  \log \sigma(q) \equiv
  \sum_{k=-\infty}^\infty \hat{\sigma}_k \eu^{\ii k q} \; .
  \label{sigmak}
\ee

\section{The Fisher-Hartwig Conjecture}

Over the years, Szeg\"o's Theorem has been extended to consider
broader classes of generating functions by relaxing the continuity
conditions which define a ``smooth function'', but it remained limited to
never-vanishing functions.
Therefore, some extensions have been proposed to the Strong Szeg\"o Theorem in
order to relax this latter hypothesis.
When the generating function has only point-wise singularities (or zeros), there exists a conjecture known as the Fisher-Hartwig Conjecture (FH)
\cite{FisherHartwig-1968}. The conjecture has been progressively proven over the years, by continuously extending and relaxing the conditions for its validity and by employing different methods.
For details over these steps, we refer to Ref. \cite{basor,widomsing,Ehrhardt-1997,deift11,krasovsky11}.

When $\sigma(q)$ has $R$ singularities at $q= \theta_r$ ($r=1,\dots,R$), we
decompose it as follows:
\be
   \sigma(q) = \tau(q) \prod_{r=1}^R
   \eu^{\ii \kappa_r [(q - \theta_r) \: {\rm mod} \: 2 \pi - \pi]}
   \left( 2 - 2 \cos (q - \theta_r) \right)^{\lambda_r} \; ,
   \label{fishdec}
\ee
so that $\tau(q)$ is a smooth function satisfying the conditions stated in the previous section.
Then according to FH the asymptotic formula for the determinant takes the form
\be
   D_n [\sigma] \sim 
   E \left[ \tau, \{ \kappa_a \}, \{ \lambda_a \}, \{ \theta_a \} \right] \; 
   n^{\sum_r \left( \lambda_r^2 - \kappa_r^2 \right)} \;
   F[\tau]^n \qquad n \rightarrow \infty \; ,
   \label{singexp}
\ee
where the constant prefactor is 
\bea
  E \left[ \tau, \{ \kappa_a \}, \{ \lambda_a \}, \{ \theta_a \} \right]
  \equiv & E[\tau] & \prod_{r=1}^R
  \tau_- \left( \eu^{\ii \theta_r} \right)^{-\kappa_r - \lambda_r}
  \tau_+ \left( \eu^{- \ii \theta_r} \right)^{\kappa_r - \lambda_r}
  \nonumber \\  & \times &
  \prod_{1 \le r \ne s \le R} \left( 1 - \eu^{\ii (\theta_s - \theta_r)}
  \right)^{(\kappa_r + \lambda_r) (\kappa_s - \lambda_s)}
  \nonumber \\ & \times &
  \prod_{r=1}^R { {\rm G} (1 + \kappa_r + \lambda_r)
  {\rm G} (1 - \kappa_r + \lambda_r) \over {\rm G} (1 + 2 \lambda_r) } \; .
  \label{fisherhartwig}
\eea
$E[\tau]$ and $F[\tau]$ are defined as in (\ref{szegoexp}) and
$\tau_{\pm}$ come from the decomposition
\be
   \tau(q) = \tau_- \left( \eu^{\ii q} \right) \;
   F[\tau] \; \tau_+ \left( \eu^{- \ii q} \right) \; ,
   \label{wienerhopf}
\ee
so that $\tau_+$ ($\tau_-$) is analytic and non-zero
inside (outside) the unit circle on which $\tau$ is defined. They also
satisfy the boundary conditions $\tau_+ (0) = \tau_- (\infty) = 1$.
${\rm G}(z)$ is the {\it Barnes G-function}, an analytic entire
function defined as
\be
   {\rm G}(z + 1) \equiv (2 \pi)^{z/2} \eu^{-[z + (\gamma_E + 1) z^2]/2}
   \prod_{n=1}^\infty \left( 1 + {z \over n} \right)^k
   \eu^{-z + {z^2 \over 2n}} \; ,
   \label{BGfun}
\ee
where $\gamma_E \sim 0.57721 \ldots$ is the Euler-Mascheroni constant.

In many simple cases it is possible to find  the
factorization of $\tau$ into the product of $\tau_+$ and $\tau_-$ by inspection.
More complicated examples require
a special technique to obtain this factorization, which is known as the
{\it Wiener-Hopf decomposition}\index{Wiener-Hopf method}, which we already mentioned in Sec. \ref{sec:XXXh} and \ref{sec:XXZparah}:
\bea
   \log \tau_+ (w) = \oint {\de z \over 2 \pi \ii}
   {\log \tau (z) \over z - w} & \qquad & |w| < 1,
   \nonumber \\
   \log \tau_- (w) = - \oint {\de z \over 2 \pi \ii}
   {\log \tau (z) \over z - w} & \qquad & |w| > 1,
   \label{wienint}
\eea
where the integral is taken counterclockwise over the unit circle.

In light of these formulas, it is useful to present the parametrization
(\ref{fishdec}) in a form which makes the analytical structure more
apparent.
Changing the variable dependence from $q$ to $z \equiv \eu^{\ii q}$, we
write
\be
   \sigma(z) = \tau(z) \prod_{r=1}^R
   \left( 1 - {z \over z_r} \right)^{\lambda_r + \kappa_r}
   \left( 1 - {z_r \over z_{} } \right)^{\lambda_r - \kappa_r},
\ee
where $z_r \equiv \eu^{\ii \theta_r}$.

\section{Generalized Fisher-Hartwig: Basor-Tracy Conjecture}
\label{gfhsec}

Despite the considerable success of the Fisher-Hartwig Conjecture, few
examples have been reported in the mathematical literature that do not
fit this result.
These examples share the characteristics that inequivalent
representations of the form (\ref{fishdec}) exist for the generating
function $\sigma(q)$.
Although no theorem has been proven concerning these cases, a
generalization of the Fisher-Hartwig Conjecture (gFH) has been suggested
by Basor and Tracy \cite{basor} and proven in \cite{deift14}.

If  more than one parametrization of the kind
(\ref{fishdec}) exists, we write them all as
\be
   \sigma(q) = \tau^i(q) \prod_{r=1}^R
   \eu^{\ii \kappa^i_r [(q - \theta_r) \: {\rm mod} \: 2 \pi - \pi]}
   \left( 2 - 2 \cos (q - \theta_r) \right)^{\lambda^i_r},
   \label{fishgendec}
\ee
where the index $i$ labels different parametrizations (for $R > 1$ there
can be only a countable number of different parametrizations
of this kind).
Then the asymptotic formula for the determinant is
\be
   D_n [\sigma] \sim \sum_{i \in \Upsilon}
   E \left[ \tau^i, \{ \kappa^i_a \}, \{ \lambda^i_a \}, \{ \theta_a \}
   \right] \; n^{\Omega(i)} \; F[\tau^i]^n \qquad n \rightarrow \infty,
   \label{singgenexp}
\ee
where
\be
   \Omega(i) \equiv
   \sum_{r=1}^R \left( \left( \lambda^i_r \right)^2 -
   \left( \kappa^i_r \right)^2 \right) \; ,
   \qquad \qquad \qquad
   \Upsilon =
   \left\{ i \: \| \: \R[\Omega(i)] = \max_j \R[\Omega(j)] \right\} \; .
 \label{Leading}
\ee

The generalization essentially gives the asymptotics of the Toeplitz
determinant\index{Toeplitz!determinant} as a sum of (FH) asymptotics calculated separately for
different leading (see Eq.~(\ref{Leading})) representations
(\ref{fishgendec}).

\section{Widom's Theorem}

If $\sigma(q)$ is supported only in the interval $\alpha \le q
\le 2 \pi - \alpha$, singularities are
no longer point-wise and one should apply Widom's Theorem
\cite{widomsupp}.
It states that the asymptotic behavior of the determinant in this case
is
\be
   D_n [\sigma] \sim 2^{1/12} \eu^{3 \zeta'(-1)}
   \left( \sin {\alpha \over 2} \right)^{-1/4} E[\rho]^2 \; n^{-1/4} \; F[\rho]^n
   \left( \cos {\alpha \over 2} \right)^{n^2},
   \label{widomasympt}
\ee
where $E$ and $F$ are defined in (\ref{szegoexp}) and
\be
   \rho(q) = \sigma
   \left( 2 \cos^{-1} \left[\cos {\alpha \over 2} \cos q\right] \right)
\ee
with the convention $0 \le \cos^{-1} x \le \pi$.

\chapter{Two-Dimensional Classical Integrable Systems}
\label{app:2DClassical}

\abstract{
	In chapter \ref{chap:algebraic} we present the algebraic formulation of the Bethe Ansatz solution. A central role in this construction is played by the {\it transfer} and {\it monodromy matrices}. These objects do not have an obvious interpretation within a one-dimensional quantum system. In fact, they are borrowed from the approach used in classical statistical models. In this appendix, we present Baxter's solution of the {\it 6-vertex model}: this is an exactly solvable 2-D model, whose Hamiltonian reduction reproduces the XXZ spin chain analyzed in chapter \ref{chap:XXZmodel}. The relation between these two models is similar to the one between a d-dimensional quantum field theory and a d+1-dimensional classical one. The integrability of the 6-vertex model is proven through the construction of an infinite set of conserved charges, whose generating function is the transfer matrix: this is the same operator as in the algebraic solution of the XXZ chain and one of the conserved charges is the Hamiltonian of the quantum model. Thus, the two models share the same solution and the construction we present in this chapter helps in understanding the content of chapter \ref{chap:algebraic}. 
	After outlining in Sec. \ref{sec:2Doverview} Baxter's philosophy for his construction, in Sec. \ref{sec:icemodels} we define the model and set up its partition function in terms of the transfer matrix, which is solved in Sec. \ref{sec:yangbaxter}. Finally, in Sec. \ref{sec:TQ} we introduce the $Q$-matrix, which allows for a direct diagonalization of the partition function.
	}

\section{Overview of the approach}
\label{sec:2Doverview}

In this chapter we consider a two-dimensional classical model, defined on a square lattice with $M$ horizontal rows and $N$ vertical lines, equipped with periodic boundary conditions in both directions.
A key role is played by the transfer matrix\index{Transfer matrix} ${\bf T}$: it is an operator that propagates a given configuration from one horizontal line to the next, i.e. it gives the weight in the partition function of a state with a given configuration on the $N$ sites of a horizontal line and another given configuration on the $N$ sites of the next line. If one knows the transfer matrix for a given model, the partition function can be found by repeatedly applying the transfer matrix $M$ times to propagate the bottom configuration to the top line and by taking the trace to close the system with periodic boundary conditions:
\be
   {\cal Z} = \Tr {\bf T}^M \; .
\ee
We are thus interested primarily in the eigenvalues $\Lambda_j$ of the transfer matrix, and in particular on its highest one $\Lambda_1$, since we can write
\be
   {\cal Z} = \Lambda_1^M \left[ 1 + \left( {\Lambda_2 \over \Lambda_1} \right)^M + \ldots \right] \; ,
\ee
where the terms in the brackets converge to $1$ in the thermodynamic limit $M \to \infty$. The other eigenvalues carry additional information: for instance, the second highest eigenvalue encodes the correlation length of the system \cite{baxterbook}.

To solve these models, instead of diagonalizing a single transfer matrix\index{Transfer matrix}, we will try to diagonalize a whole family of transfer matrices at the same time. We identify each member within a family by a parameter $\lambda$ (usually referred to as the {\it spectral parameter}), i.e. ${\bf T} (\lambda)$, and we try to diagonalize each ${\bf T} (\lambda)$ simultaneously for every $\lambda$.
This procedure might seem too ambitious at first, since one turns the hard problem of solving a system into the seemingly harder problem of solving a bunch of them. In fact, this technique brings out a deep structure due to the integrability. In particular, we will find that all transfer matrices within a family commute with one another and therefore they share the same eigenvectors. In turn, this means that the spectral parameter can be used to expand the transfer matrix\index{Transfer matrix} and generate an infinite set of integrals of motion in convolution. In spirit, for each $\lambda$ we can look for the ``easy'' eigenvectors of ${\bf T} (\lambda)$ knowing that they are eigenvectors of all other matrices. Then the rich algebraic structure allows to track down the eigenvalues of each vector for every value of the spectral parameter.

In order to uncover this rich structure, we need to consider, in addition to the Transfer matrix, also the {\it monodromy matrix}\index{Monodromy matrix} $\bT (\lambda)$, which is the operator that propagates an open horizontal line to the next, i.e. without imposing periodic boundary conditions at the end of the line. By definition, the monodromy matrix $\bT$ possesses an additional degree of freedom, compared to ${\bf T}$, corresponding to one of the $k$ allowed states at the beginning and end of the line. The tracing of this degree of freedom is equivalent to requiring the in and out state to coincide (i.e. imposing periodic boundary conditions) and thus reproduces the transfer matrix\index{Transfer matrix}.

The monodromy matrix\index{Monodromy matrix} is instrumental in proving the commutativity of the transfer matrices, because it satisfies the {\it intertwining relation}:
\be
\bT_j (\lambda) \; \bT_l (\mu) \; {\cal R}_{jl} (\mu - \lambda) =
{\cal R}_{jl} (\mu - \lambda ) \; \bT_l (\mu) \; \bT_j (\lambda) \; ,
\label{TYB}
\ee
where the ${\cal R}$-matrix ${\cal R}_{12} (\lambda, \mu)$, also known as the {\it intertwiner}, is a $k \times k$ matrix, where $k$ is the dimension of the Hilbert space in the horizontal direction. In (\ref{TYB}) the matrix product contracts only these horizontal degrees of freedom, that is, the monodromy matrix\index{Monodromy matrix} has to be considered as a $k \times k$ dimensional matrix, where each matrix element is an operator acting on the vertical space, and the subscript indicates that $\bT_j (\lambda)$ acts on the $j$-th row and that ${\cal R}_{jl} (\lambda)$ connects the horizontal degrees of freedom on the $j$-th and $l$-th row. Eq. (\ref{TYB}) is a first instance of the Yang-Baxter-like equation\index{Yang-Baxter!equation} and is represented pictorially in Fig. \ref{fig:TTR-YBE}.

By taking the trace over the horizontal space, we recover the original transfer matrix\index{Transfer matrix} ${\bf T}_j (\lambda) = \tr \bT_j (\lambda)$ (this is the partial trace only over the horizontal degree of freedom), and taking the trace of (\ref{TYB}) yields
\be
  \left[ {\bf T}_j (\lambda), {\bf T}_l (\mu) \right] = 0 \; .
\ee
Therefore, as we claimed before, transfer matrices\index{Transfer matrix} with different spectral parameters commute and thus share the same eigenvectors.

\section{Ice-type models}
\label{sec:icemodels}

The 6-vertex model\index{Vertex model} was originally introduced as a description of two dimensional ice. When water freezes, each oxygen atom is surrounded by four hydrogen ions: each of these ions will be closer to one of its neighboring oxygen than to the others, but always in a way such that each oxygen has two hydrogens closer to it and two further away. This is known as the {\it ice rule}.
This system is therefore modeled as a square lattice: each vertex is supposed to host an oxygen atom and each bond between two vertices is depicted with an arrow, indicating to which of the two oxygens the hydrogen ion is closer. What is important to us is that each bond has a degree of freedom that can take two values, which we can represent as $+$ and $-$, or $0$ and $1$, or with classical spin-$1/2$, etc.

\begin{figure}[t]
	\begin{center}
		\includegraphics[width=\columnwidth]{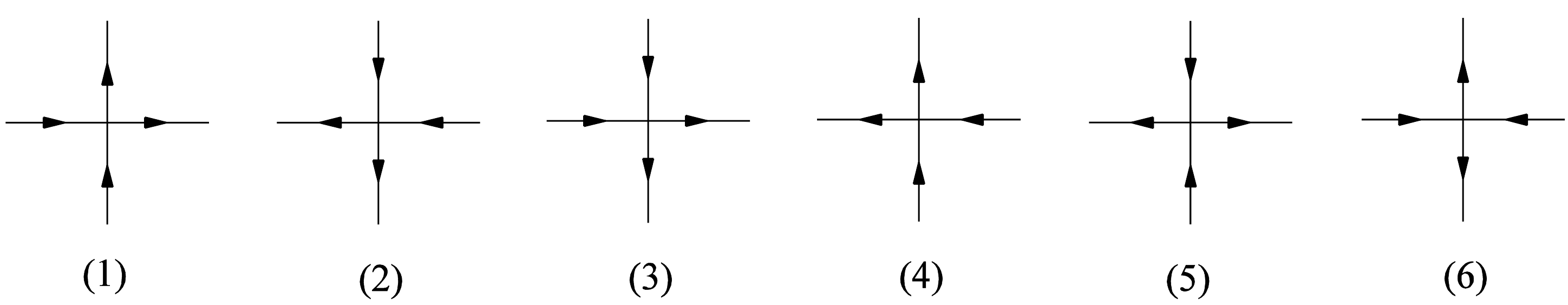}
	\end{center}
	\caption{The configurations allowed by the ice-rule, representing the six possible vertices.}
	\label{fig:6vertexarrows}
\end{figure}

Because of the ice rule, each vertex is surrounded by two arrows pointing towards it and two away: this constraint limits the number of possible vertex configurations to only $6$, which are listed in Fig. \ref{fig:6vertexarrows} and labeled from $1$ to $6$. This is the reason for which this model is also known as the {\it 6-vertex model}. There exist other integrable models similar to this and we should mention the $8$-vertex model\index{Vertex model}, where the ice rule is broken by adding two additional vertices, one with all the four arrows pointing towards the vertex and another with all arrows pointing away. Among the integrals of motion generated by the transfer matrix\index{Transfer matrix} of the $8$-vertex model one finds the Hamiltonian of the XYZ spin chain (and also of the XY model), thus its solution provides a solution of these quantum models as well (or vice-versa). The techniques needed to solve the systems connected to the 8-vertex model are more involved, because the additional vertices break the $U(1)$ symmetry to a $Z_2$. Also, one needs to work with elliptic functions, i.e. analytic functions that are (quasi-)periodic both in the real and imaginary direction. We will see that an entire parametrization (i.e. without branch-cuts) of the couplings of the $6$-vertex model is achieved using periodic (trigonometric) functions, which reduce to rational functions at the isotropic point ($\Delta =1$). For this reason, the $8$, $6$ and isotropic $6$-vertex models\index{Vertex model} are often refereed to as the elliptic, trigonometric and rational models, respectively.

The $6$-vertex model\index{Vertex model} is defined by the Boltzmann weights assigned to each vertex:
\be
   w_j = \eu^{-\beta \epsilon_j} \; , \qquad \quad
   j = 1 , \ldots, 6 \; ,
\ee
where $\beta = 1 / k_B T$ is the usual inverse temperature scaled by the Boltzmann's constant.
The partition function is given by the sum over all possible configurations of arrows on bonds, weighted by the above expressions:
\be
   {\cal Z} = \sum_{\{n_l\}} \exp \left[- \beta \left( n_1 \epsilon_1 +
   n_2 \epsilon_2 + n_3 \epsilon_3 + n_4 \epsilon_4 + n_5 \epsilon_5 +
   n_6 \epsilon_6 \right) \right]
\ee
where $\{ n_l \}$ are the number of vertices of type $l$ in the system and where configuration with vertices not satisfying the ice rule are given weight zero.

The periodic boundary conditions force the number of vertices of type $5$ and $6$ to be equal, since they act as sinks (sources) for horizontal (vertical) arrows, and vice-versa. This means that only the combination $n_5 (\epsilon_5 + \epsilon_6)$ appears in the partition function and we can chose $\epsilon_5 = \epsilon_6 = \epsilon_c$ with no loss of generality (since a different choice is unobservable).

If we further assume that the system is invariant under the simultaneous reversal of all arrows, then at equilibrium $n_1 = n_2$ and $n_3 = n_4$ and the partition function can be written as
\be
   {\cal Z} = \sum_{\{n_l\}} \exp \left\{- \beta \left[
   n_1 \left( \epsilon_1 + \epsilon_2 \right)
   + n_3 \left( \epsilon_3 + \epsilon_4 \right)
   + 2 \; n_5 \; \epsilon_c \right] \right\} \; ,
\ee
and for the same reasoning as before we can choose
\be
   \epsilon_1 = \epsilon_2 \equiv \epsilon_a \; , \qquad \qquad
   \epsilon_3 = \epsilon_4 \equiv \epsilon_b \; .
   \label{zerofieldcondition}
\ee
Condition (\ref{zerofieldcondition}), i.e. the invariance under arrow reversal, is known as the {\it zero-field condition}. In fact, if we add a field $E_y$ ($E_y$) in the vertical (horizontal) direction which couples to the arrows in each bond giving each up/down-pointing arrow an extra energy $\pm E_y$ and each right/left-pointing arrow the extra energy $\pm E_x$ we can break the degeneracy of the energies:
\bea
   \epsilon'_1 = \epsilon_a - E_x - E_y \; , \qquad && \qquad
   \epsilon'_2 = \epsilon_a + E_x + E_y \; ,
   \nonumber \\
   \epsilon'_3 = \epsilon_b - E_x + E_y \; , \qquad && \qquad
   \epsilon'_4 = \epsilon_b + E_x - E_y \; ,
\eea
and generates more vertices of one type or another. In fact, a finite $E_y$ does not spoil the integrability of the model and corresponds to a finite magnetic field in the XXZ chain. However, for the sake of simplicity, in the rest of this chapter we will always assume the zero-field condition $E_x = E_y = 0$.

If our lattice has $M$ rows and $N$ columns, we can write the partition function as a sum of contributions from each of the M rows
\be
   {\cal Z} = \sum_{r=1}^M \sum_{\{ m^r_l \}}
   a^{m^r_1 + m^r_2} \; b^{m^r_3 + m^r_4}  \; c^{m^r_5 + m^r_6} \; ,
\ee
where $\{m^r_l\}$ is the number of vertices of type $l$ in row $r$ and we introduce the parameters $a, b, c$ to identify the weights in the zero-field case:
\be
   a \equiv w_1 = w_2 \; , \qquad \quad
   b \equiv w_3 = w_4 \; , \qquad \quad
   c \equiv w_5 = w_6 \, .
   \label{abcws}
\ee

We can rewrite the contribution from each row by taking into account the configuration of arrows below and above it. If we denote by
\be
   \{ \gamma^r \} = \{\gamma_1^r, \gamma_2^r, \ldots, \gamma_N^r \}
\ee
the configuration of arrows immediately below row $r$ (since each arrow can assume two values --up or down-- for each row we have $2^N$ possible configurations spanned by $\{\gamma^r\}$) we can write the partition function in terms of the row-to-row transfer matrix\index{Transfer matrix} ${\bf T}_{\gamma^j}^{\gamma^{j+1}}$
\be
   {\cal Z} = \sum_{\gamma^1} \sum_{\gamma^2} \ldots \sum_{\gamma^M}
   {\bf T}_{\gamma^1}^{\gamma^2} {\bf T}_{\gamma^2}^{\gamma^3}
   \ldots {\bf T}_{\gamma^{M-1}}^{\gamma^M} {\bf T}_{\gamma^M}^{\gamma^1}
   = \tr {\bf T}^M \; ,
\ee
where ${\bf T}$ is a $2^N \times 2^N$ matrix with elements
\be
  {\bf T}_{\gamma}^{\gamma'} = \sum_{\{m_j\}}
  a^{m_1 + m_2} \; b^{m_3 + m_4}  \; c^{m_5 + m_6} \; ,
  \label{Tdefabc}
\ee
where the sum is over all vertex choices compatible with the ice-rule and the vertical configurations given by $\gamma$ and $\gamma'$, i.e. it is a sum over all possible configurations of horizontal arrows on the $N$ bonds of the row, modulo the ice-rule.

As a side note, we should remark that the number of down (up) arrows is conserved from one row to another (as a consequence of the toroidal boundary condition, on each row we must have the same number of sources and sinks, i.e. vertices of type $5$ and $6$). This means that the transfer matrix\index{Transfer matrix} has a block diagonal structure with blocks corresponding to configurations with the same number $R$ of down arrows entering and exiting the row, with $R=0, \ldots, N$. This structure is equivalent to the $U(1)$ symmetry we used within the Bethe Ansatz approach, that is, that there is no particle production and that states with $R$ particles scatter and evolve only into states with the same number of particles $R$. Because of this, the scattering matrix has the same block-diagonal structure as the transfer matrix\index{Transfer matrix} of the 6-vertex model\index{Vertex model}.

\section{The Transfer Matrix and the Yang-Baxter equations}
\label{sec:yangbaxter}

We now study the transfer matrix\index{Transfer matrix} in more detail. Let us consider a configuration where the arrows below the row are given by the configuration $\gamma = \{\gamma_1, \ldots, \gamma_N \}$ and the ones above are $\gamma' = \{\gamma'_1, \ldots, \gamma'_N \}$. We denote an up arrow by $\gamma_j = +1$ or $\gamma'_j = +1$ and a down arrow by $\gamma_j = -1$ or $\gamma'_j = -1$. We also denote the arrow on the horizontal bonds as $\alpha = \{\alpha_1, \ldots, \alpha_N \}$, with the convention that $\alpha_j = +1$ corresponds to a right-pointing arrow and a $\alpha_j = -1$ to a left-pointing one. With these notations in mind, we will refer to $\alpha_j$, $\gamma_j$, $\gamma'_j$ and so on as {\it spin variables} with spin up/down depending if they have value $+1/-1$.

We can write the transfer matrix\index{Transfer matrix} as
\be
   {\bf T}_{\gamma}^{\gamma'} = \sum_{\alpha_1} \ldots \sum_{\alpha_N}
   {\cal L}_{\alpha_1 \gamma_1}^{\alpha_2 \gamma'_1}
   {\cal L}_{\alpha_2 \gamma_2}^{\alpha_3 \gamma'_2}  \ldots
   {\cal L}_{\alpha_N \gamma_N}^{\alpha_1 \gamma'_N} \; ,
\ee
where ${\cal L}_{\alpha \; \gamma}^{\alpha' \gamma'}$ is a $4 \times 4$ matrix with entries given by the Boltzmann weights of the vertex configurations, i.e.
\begin{wrapfigure}{r}{6cm}
	  \vspace{-10pt}
	\begin{center}
		\includegraphics[width=5.5cm]{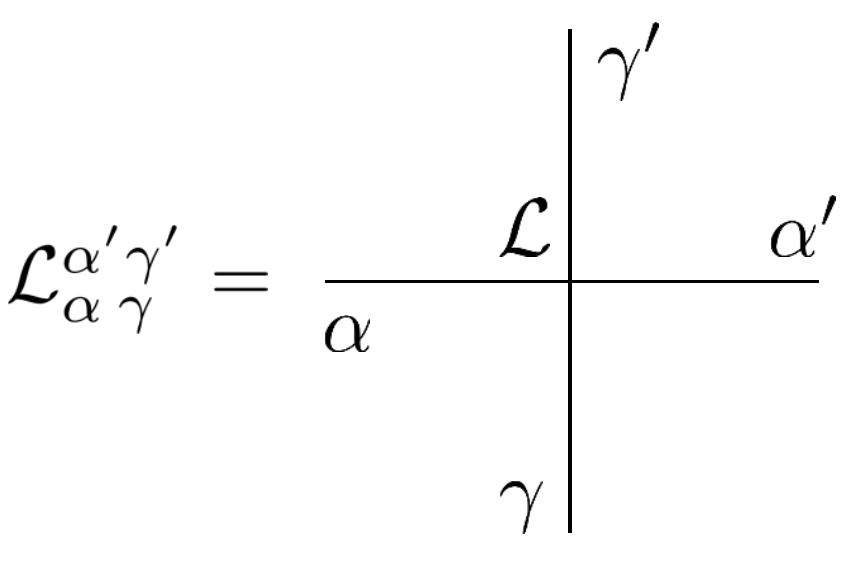}
	\end{center}
	\caption{Graphical representation of the ${\cal L}$-operator.}  
	\label{fig:L}
	  \vspace{-60pt}
\end{wrapfigure}
\bea
   {\cal L}_{+ +}^{+ +} = {\cal L}_{- -}^{- -} & = & a \; , \\
   {\cal L}_{+ -}^{+ -} = {\cal L}_{- +}^{- +} & = & b \; , \\
   {\cal L}_{+ -}^{- +} = {\cal L}_{- +}^{+ -} & = & c \; ,
\eea
with all other elements being zero due to the ice rule.
More explicitly, this ${\cal L}$-matrix can be written as
\be
   {\cal L} = \left( \begin{array}{cccc}
                              a & 0 & 0 & 0 \cr
                              0 & b & c & 0 \cr
                              0 & c & b & 0 \cr
                              0 & 0 & 0 & a \cr
                      \end{array} \right) \; ,
   \label{Lmatrix}
\ee
while fig. \ref{fig:L} shows its standard pictorial representation.

As we mentioned in the introduction, our strategy at this point is not to attempt to diagonalize directly the transfer matrix\index{Transfer matrix}, but instead to look under which conditions two transfer matrices with different parameters commute.
To this end, let us introduce a second transfer matrix ${\bf T}'$, defined as in (\ref{Tdefabc}), but with Boltzmann weights $a', b', c'$. Then\\
\begin{wrapfigure}{r}{6cm}
	\vspace{-30pt}
	\begin{center}
		\includegraphics[width=5.8cm]{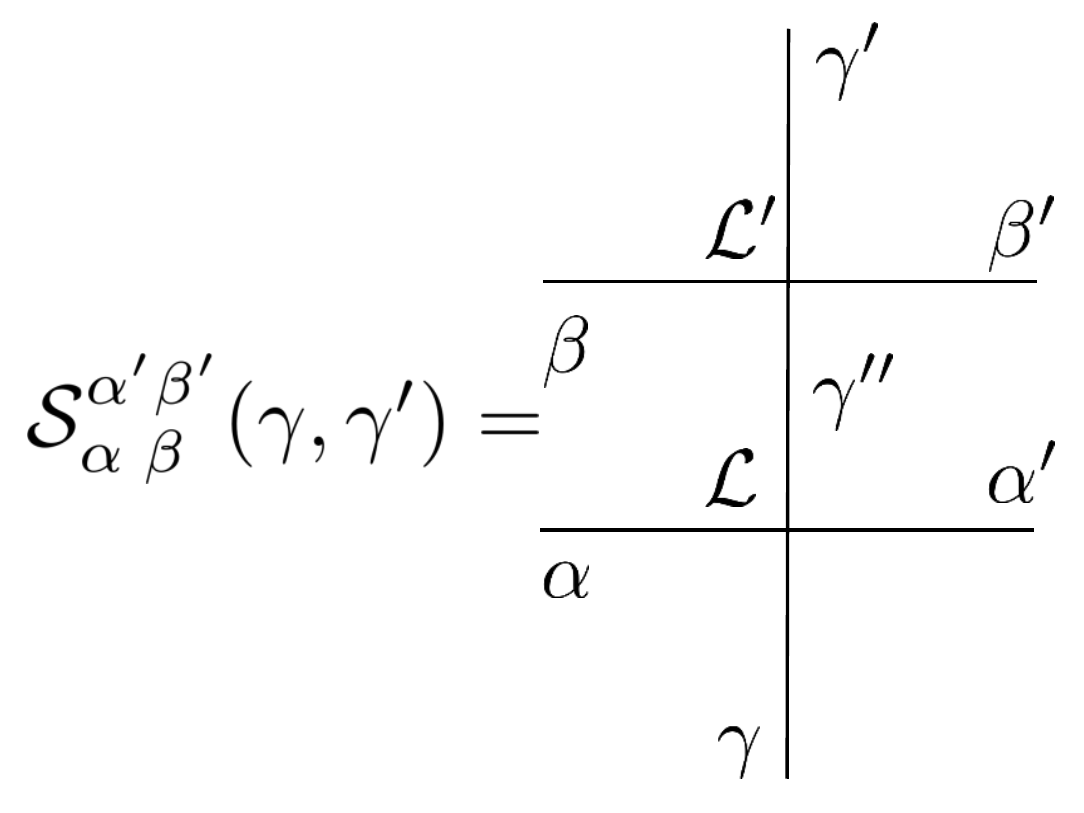}
	\end{center}
	\begin{center}
		\includegraphics[width=5.8cm]{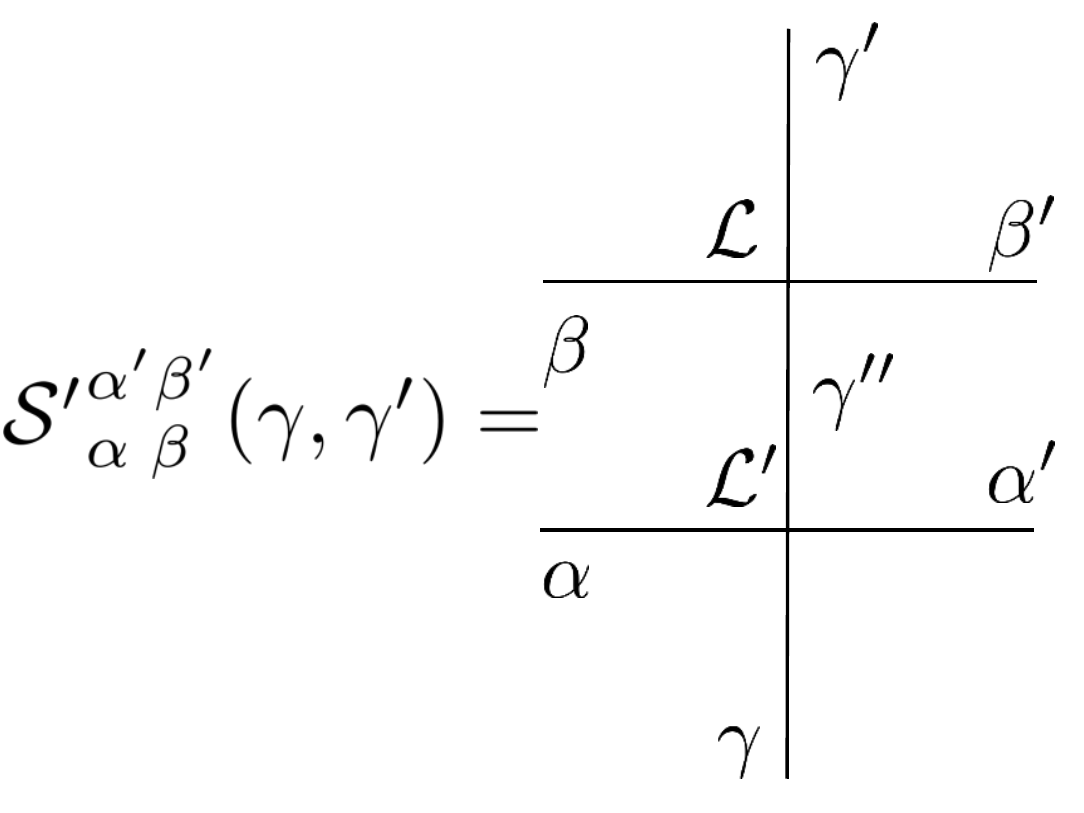}
	\end{center}
	\caption{Graphical representation of the ${\cal S}$-operators.}  
	\vspace{-20pt}
	\label{fig:S}
\end{wrapfigure}
\vspace{-.7cm}
\be
   \left( {\bf T} {\bf T}' \right)_{\gamma}^{\gamma'} =
   \sum_{\{\gamma''\}} {\bf T}_{\gamma}^{\gamma''} \; {\bf T}_{\: \gamma''}^{\prime \gamma'}
   = \sum_{\alpha_1, \ldots, \alpha_N} \sum_{\beta_1, \ldots, \beta_N}
   \prod_{j=1}^N {\cal S}_{\alpha_j \; \beta_j \; | \gamma_j}^{\alpha_{j+1} \beta_{j+1}| \gamma'_j} \; ,
   \label{TTprime}
\ee
where
\be
   {\cal S}_{\alpha \; \beta \; | \gamma}^{\alpha' \beta'| \gamma'} \equiv
   \sum_{\gamma''} {\cal L}_{\alpha \; \gamma}^{\alpha' \gamma''}
   {{\cal L}'}_{\beta \; \gamma''}^{\beta' \gamma'}
\ee
is the double-row transfer matrix\index{Transfer matrix} (i.e. the operator that propagates across two rows, with different weights for each row) and with the understanding that $\alpha_{N+1}=\alpha_1$ and $\beta_{N+1}=\beta_1$.
The operator ${\cal S}$ is a $8 \times 8$ matrix. If we keep the two vertical indices as fixed, we can write it as a $4 \times 4$ matrix as
\be
   {\cal S}_{\alpha \; \beta}^{\alpha' \beta'} ( \gamma, \gamma') \equiv
   {\cal S}_{\alpha \; \beta \; | \gamma}^{\alpha' \beta'| \gamma'} \; ,
\ee
and (\ref{TTprime}) as
\be
   \left( {\bf T} {\bf T}' \right)_{\gamma}^{\gamma'} = \tr
   {\bf {\cal S}} (\gamma_1, \gamma'_1) {\bf {\cal S}} (\gamma_2, \gamma'_2)
   \ldots {\bf {\cal S}} (\gamma_N, \gamma'_N) \; .
   \label{tautauprime}
\ee
We can also consider to invert the order of the two rows, but keeping the external legs fixed and write the resulting double-row transfer matrix\index{Transfer matrix} as
\be
   \left( {\bf T}' {\bf T} \right)_{\gamma}^{\gamma'} = \tr
   {\bf {\cal S}'} (\gamma_1, \gamma'_1) {\bf {\cal S}'} (\gamma_2, \gamma'_2)
   \ldots {\bf {\cal S}'} (\gamma_N, \gamma'_N) \; ,
   \label{tauprimetau}
\ee
where
\be
   {{\cal S}'}_{\alpha \; \beta}^{\alpha' \beta'} (\gamma, \gamma') \equiv
   \sum_{\gamma''} {{\cal L}'}_{\alpha \; \gamma}^{\alpha' \gamma''}
   {\cal L}_{\beta \; \gamma''}^{\beta' \gamma'} \; .
\ee

\begin{figure}[t]
	\noindent\begin{minipage}[t]{11cm}
		\includegraphics[width=10.5cm]{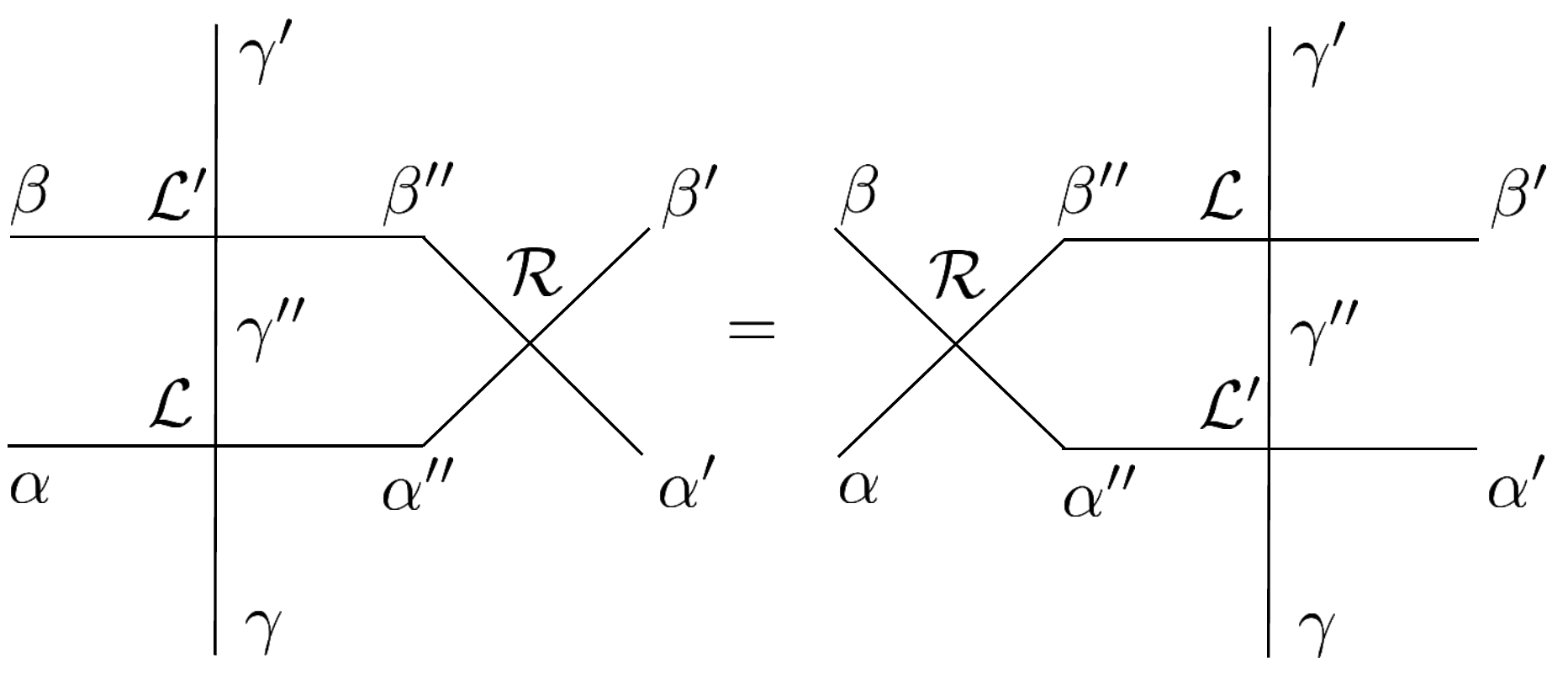}
	\end{minipage}
	\hfill
	\begin{minipage}[b]{5cm}
		\caption{Diagrammatic representation of the Yang-Baxter-like equation (\ref{SRCond},\ref{LLRYBE}).\vspace{3.2cm}$\qquad$}
		\label{fig:LLR-YBE}
	\end{minipage}
	\vspace{-.5cm}
\end{figure}

We want to find under which conditions these expressions commute, i.e. $ {\bf T} {\bf T}' = {\bf T}' {\bf T}$.
This will surely be true if there exists a $4 \times 4$ non-singular matrix ${\cal R}$ such that
\be
   {\cal S} (\gamma, \gamma') = {\cal R} \; {\cal S}' (\gamma, \gamma')
   {\cal R}^{-1} \; ,
   \label{SRCond}
\ee
where we remind that ${\cal S}$ is also a $4 \times 4$ matrix and $\gamma = \pm 1$ and $\gamma' = \pm 1$ are taken as parameters. If (\ref{SRCond}) is satisfied, then plugging it into (\ref{tautauprime}) and using the cyclic property of the trace we get (\ref{tauprimetau}) as we set to achieve.
Equation (\ref{SRCond}) can be written more explicitly as
\be
   \sum_{\alpha'', \beta'', \gamma''}
   {\cal L}_{\alpha \gamma}^{\alpha'' \gamma''} \;
   {{\cal L}'}_{\beta \gamma''}^{\beta'' \gamma'} \;
   {\cal R}_{\alpha'' \beta''}^{\alpha' \beta'}
   = \sum_{\alpha'', \beta'', \gamma''}
   {\cal R}_{\alpha \beta}^{\alpha'' \beta''} \;
   {{\cal L}'}_{\alpha'' \gamma}^{\alpha' \gamma''} \;
   {\cal L}_{\beta'' \gamma''}^{\beta' \gamma'} \; ,
   \label{LLRYBE}
\ee
and is the {\it Yang-Baxter} equation\index{Yang-Baxter!equation} for the ${\cal L}$-matrices. It can be represented pictorially as in Fig. \ref{fig:LLR-YBE}: we see that the ${\cal R}$-matrix acts as an intertwiner for the two ${\cal L}$-matrices since it connects the ``horizontal'' spins, but it does not act on the ``vertical'' ones.

At this point we make an ansatz, i.e. we assume that the ${\cal R}$-matrix has the same structure as an ${\cal L}$-matrix, i.e. that we can write it as in (\ref{Lmatrix}), but with different weights, namely $a''$, $b''$ and $c''$: ${\cal R} = {\cal L}''$. This is not to say that the ${\cal R}$-matrix can be identified with an ${\cal L}$-matrix (since they act on different spaces as operators), but only to assume that the ice-rules apply to ${\cal R}$ as well.

Then we can look for solutions of (\ref{LLRYBE}) by writing it as a system of 64 equations (coming from equating each component of the resulting matrix multiplication, or corresponding to all possible combination of the external spin variables). We take $a, b, c$ as given and we look for which choices of $a', b', c'$ and $a'', b'', c''$ (\ref{LLRYBE}) is satisfied. Notice that, since all equations are homogeneous, they do not fix the normalization of the matrices ${\cal L}'$ ${\cal R}$, so the parameters can be rescaled by a constant without violating (\ref{LLRYBE}), so only 4 out of the six parameters are meaningful to solve the Yang-Baxter equation\index{Yang-Baxter!equation}.

Of course there is one trivial solution:
\be
   {\cal L}' \propto {\cal L} \; , \qquad {\rm and} \qquad
   {\cal R}_{\alpha, \beta}^{\alpha' \beta'} = \delta_{\alpha \alpha'}
   \delta_{\beta \beta'} \; ,
\ee
but this amounts to say that the transfer matrix\index{Transfer matrix} commutes with simple multiples of itself (again, a rescaling of the parameters) and it is not interesting.
To look for non-trivial solutions, we notice that the ice-rule severely restricts the number of non-zero components of the ${\cal L}$ and ${\cal R}$-matrices (see \ref{Lmatrix}). In fact, ${\cal L}_{\alpha \gamma}^{\alpha' \gamma'} = 0$ unless $\alpha + \gamma = \alpha' + \gamma'$. This means that both sides of (\ref{LLRYBE}) are identically zero if $\alpha + \beta + \gamma \ne \alpha' + \beta' + \gamma'$ and this leaves only 20 non-trivial equations out of the 64.

Moreover, the zero-field condition implies that negating all the spin variables leaves the Boltzmann weights unchanged, so these 20 equations occur in 10 identical pairs. Finally, the symmetric structure of (\ref{LLRYBE}) under the reversal of spin pairs can be shown to lead to just these three inequivalent equations:
\bea
   a c' a'' & = & b c' b'' + c a' c'' \; ,
   \nonumber \\
   a b' c'' & = & b a' c'' + c c' b'' \; ,
   \label{YBconditions} \\
   c b' a'' & = & c a' b'' + b c' c'' \; .
   \nonumber
\eea
Thus, the matrix equation (\ref{LLRYBE}) is equivalent to three linear equations in three variables ($a,b,c$), with six parameters to be determined.
This is quite a miracle and is completely due to (and responsible for) the integrability of the model.
First, let us eliminate $a'', b'', c''$ from (\ref{YBconditions}): this leaves the single equation
\be
   {a^2 + b^2 - c^2 \over a b} = {{a'}^2 + {b'}^2 - {c'}^2 \over a' b'} \; .
   \label{deltadeltaprime}
\ee
This means that we can associate to each ${\cal L}$-matrix the quantity
\be
   \Delta \equiv  {a^2 + b^2 - c^2 \over 2 a b}
   \label{Deltaabc}
\ee
which has to remain invariant for each member of a family in order for the transfer matrices\index{Transfer matrix} to commute. In other words, ${\bf T}$ and ${\bf T}'$ can have different values of $a, b, c$, but they still commute as long as $\Delta = \Delta'$.

It is convenient to look for a parametrization of $a, b, c$ that will identically satisfy (\ref{deltadeltaprime}), by incorporating (\ref{Deltaabc}). An easy choice could be
\be
   a = a \; , \qquad
   b = a x \; , \qquad
   c = a \sqrt{1 - 2 \Delta x + x^2} \; .
\ee
The problem with this parametrization is that $c$ is not an entire function of $x$ and $\Delta$, due to the branch of the square root (an entire function does not have branch point or branch cuts). There are several possible choices for entire parametrizations: we will use 
\be
   a = \rho \; \sinh (\lambda + \phi) \; , \qquad
   b = \rho \; \sinh \lambda \; , \qquad
   c = \rho \; \sinh \phi \; , \qquad \qquad
   \Delta = \cosh \phi \; .
   \label{BaxterParam}
\ee
In conclusions, if the parameters of the ${\cal L}$-matrices are chosen according to (\ref{BaxterParam}) with different $\rho$ and $\lambda$, but the same $\phi$, then (\ref{deltadeltaprime}) is satisfied and the corresponding transfer matrices\index{Transfer matrix} commute. Since $\rho$ is just an irrelevant normalization constant, the transfer matrices belonging to a commuting family are denoted simply as ${\bf T} (\lambda)$, where $\lambda$ is called the {\it spectral parameter}. The dependence of the transfer matrix on $\phi = \cosh^{-1} \Delta$ is usually assumed and not explicitly written. Henceforth, we assume the same $\phi$ for all matrices.
As the ${\cal R}$-matrix has been chosen of the same form as the ${\cal L}$-matrix, it also has a parametrization like (\ref{BaxterParam}) with the same $\phi$ as the ${\cal L}$-matrices (this can be seen by eliminating the prime variables from (\ref{YBconditions}) to get $\Delta = \Delta''$).

Thus, two of the three equations in (\ref{YBconditions}) have given $\Delta = \Delta' = \Delta''$. Substituting our parametrization (\ref{BaxterParam}) for the unprimed, primed and double-primed variables in (\ref{YBconditions}) we see that the last equations gives
\be
   \lambda'' = \lambda' - \lambda \; .
\ee
Thus, from a given ${\cal L}$, we can construct a whole family of matrices ${\cal L} (\lambda)$ that satisfy the YBE\index{Yang-Baxter!equation} (\ref{LLRYBE}):
\be
    {\cal L}_{n,j} (\lambda) \; {\cal L}_{n,l} (\lambda') \;
    {\cal R}_{j,l} (\lambda' - \lambda) = {\cal R}_{j,l} (\lambda' - \lambda)
    {\cal L}_{n,l} (\lambda') \; {\cal L}_{n,j} (\lambda) \; ,
\ee
where we indicate explicitly the space on which each operator acts: ${\cal L}_{n,j}$ on the $n$-th column and $j$-th row; ${\cal R}_{j,l}$ on the $j$-th and $l$-th rows.

Summing over all configurations on a given row $j$ corresponds to taking the products of the ${\cal L}$-matrices at different sites: this defines the {\it monodromy matrix}\index{Monodromy matrix} $\bT_j (\lambda)$:
\be
   \bT_j (\lambda) \equiv {\cal L}_{N,j} (\lambda) \; {\cal L}_{N-1,j} (\lambda)
   \; \ldots \; {\cal L}_{1,j} (\lambda) \; .
\ee
\begin{figure}[t]
	\begin{center}
		\includegraphics[width=\textwidth]{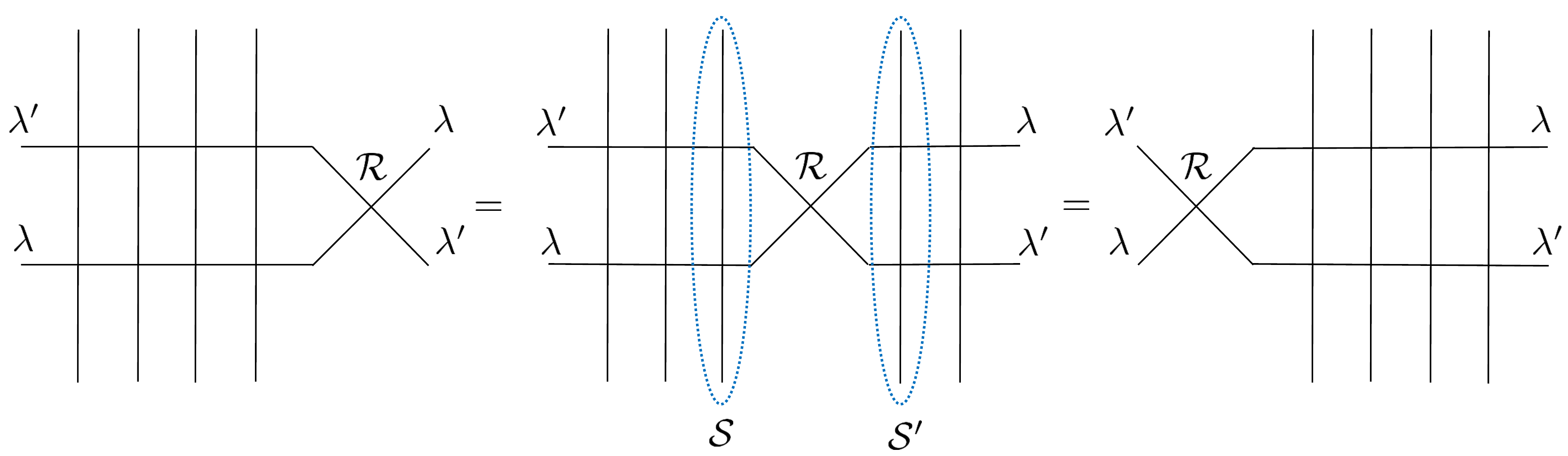}
		\caption{Diagrammatic representation of the Yang-Baxter-like equation for the monodromy matrix (\ref{TTRYBE}).}
		\label{fig:TTR-YBE}
	\end{center}
\vskip-.5cm
\end{figure}
This is a $2^{N+1} \times 2^{N+1}$ matrix that depends on the $N$ spin variables above and below the line and on the first and last horizontal spin.
If we consider two such monodromy matrices\index{Monodromy matrix}, acting on different rows and with different couplings, i.e. different spectral parameter, by using the Yang-Baxter equation\index{Yang-Baxter!equation} (\ref{LLRYBE}) for the ${\cal L}$-matrices, we can shift the intertwiner ${\cal R}$-matrix from one end to the other of the chain, see Fig. \ref{fig:TTR-YBE} to get
\be
   \bT_j (\lambda) \; \bT_l (\lambda') \;
   {\cal R}_{j l} (\lambda' - \lambda) =
   {\cal R}_{j l} (\lambda' - \lambda) \;
   \bT_l (\lambda') \; \bT_j (\lambda) \; ,
   \label{TTRYBE}
\ee
which is the Yang-Baxter\index{Yang-Baxter!equation} for the monodromy matrix\index{Monodromy matrix} (here we explicitly write the index $j$ and $l$ to remind us of the different spaces on which these operators act and the ${\cal R}$-matrix is intended to act only on the horizontal spin spaces). Taking the trace over the horizontal spins in (\ref{TTRYBE}) corresponds to closing the chain with periodic boundary conditions: since $\tr_j \bT_j (\lambda) = {\bf T}_j (\lambda)$, using the periodicity of the trace we find
\bea
   \left[ {\bf T}_j (\lambda) , {\bf T}_l (\lambda') \right] 
   & = & \tr_{j \otimes l} \Big( \bT_j (\lambda) \bT_l (\lambda') 
   - \bT_l (\lambda') \bT_j (\lambda) \Big)
   \nonumber \\
   & = & \tr_{j \otimes l} \Big( \bT_j (\lambda) \bT_l (\lambda') 
   - {\cal R}_{jl}^{-1} (\lambda' - \lambda) \bT_j (\lambda) \bT_l (\lambda')
   {\cal R}_{jl} (\lambda' - \lambda) \Big)
   = 0 \; .
   \label{tautau}
\eea

Let us remark that the proper Yang-Baxter equation\index{Yang-Baxter!equation} is a condition on the ${\cal R}$-matrix alone. To see this, let us consider the product of three monodromy matrices\index{Monodromy matrix} and notice that by applying (\ref{TTRYBE}) in different ways (order) we can get two different results:
\bea
    && \bT_j (\lambda) \; \bT_l (\mu) \; \bT_k (\nu) = 
    \label{YBEassociativity} \\
    && = {\cal R}^{-1}_{j l} (\lambda - \mu) \;
    {\cal R}^{-1}_{j k} (\lambda - \nu) \; {\cal R}^{-1}_{l k} (\mu - \nu)
    \times \bT_k (\nu) \; \bT_l (\mu) \; \bT_j (\lambda)
    \times  {\cal R}_{l k} (\mu - \nu) \;
    {\cal R}_{j k} (\lambda - \nu) \; {\cal R}_{j l} (\lambda - \mu)
    \nonumber \\
    && = {\cal R}^{-1}_{l k} (\mu - \nu) \;
    {\cal R}^{-1}_{j k} (\lambda - \nu) \; {\cal R}^{-1}_{j l} (\lambda - \mu) 
    \times \bT_k (\nu) \; \bT_l (\mu) \; \bT_j (\lambda)
    \times {\cal R}_{j l} (\lambda - \mu) \;
    {\cal R}_{j k} (\lambda - \nu) \; {\cal R}_{l k} (\mu - \nu) \; .
    \nonumber
\eea
Thus, in order to preserve associativity, we must require
\be
   {\cal R}_{l k} (\mu - \nu) \;
   {\cal R}_{j k} (\lambda - \nu) \; {\cal R}_{j l} (\lambda - \mu) =
   {\cal R}_{j l} (\lambda - \mu) \;
   {\cal R}_{j k} (\lambda - \nu) \; {\cal R}_{l k} (\mu - \nu)
   \label{RRRYBE}
\ee
which is the Yang-Baxter equation\index{Yang-Baxter!equation} for the ${\cal R}$-matrices. This is the fundamental equation defining an integrable model. It defines an algebra and finding solutions to (\ref{RRRYBE}) is in a sense equivalent to finding (adjoint) representations for the group. Every time a solution is identified for (\ref{RRRYBE}) in some $\kappa$-dimensional space, one can construct the corresponding ${\cal L}$-matrices and monodromy matrices\index{Monodromy matrix} that satisfy (\ref{LLRYBE}, \ref{TTRYBE}) and eventually identify the model one has just solved. In the case of the 6-vertex model\index{Vertex model}, we have found a trigonometric solution of (\ref{RRRYBE}) in terms of a $2^2 \times 2^2$ ($\kappa=2$) matrix, both for the ${\cal L}$ and ${\cal R}$ operators.\footnote{Note that, as the dimensions of the spaces on the horizontal and vertical bonds do not have to be the same, one can have systems for which the ${\cal L}$ matrix is rectangular, while the solutions of the Yang-Baxter equation\index{Yang-Baxter!equation} for the ${\cal R}$-operator are always square matrices.}.

The main advantage of having proven that transfer matrices\index{Transfer matrix} at different spectral parameters commute, is that we can now interpret the transfer matrix as a generator for the conserved charges of the theory (which are in infinite number, since the model is integrable). In practice, it is more convenient to consider the logarithm of the transfer matrix as the generating function of the integrals of motion, since in this way they turn out to be local operators with simple physical interpretation. We expand the logarithm of the generating function around $\lambda = 0$
\be
   \ln {\bf T} (\lambda) = \sum_{n=0}^\infty J_n \lambda^n \; .
\ee
Plugging this into (\ref{tautau}) we see that
\be
   \left[ J_n , J_m \right] = 0 \; ,
\ee
so that the coefficients of the expansions can be interpreted as conserved densities in involution with one another.
Let us look at these conserved quantities. If we set $\lambda =0$, we see that 
\be
   {\cal L}_{\alpha \gamma}^{\alpha' \gamma'} (\lambda = 0) = \rho \; \sinh \phi \; \delta_{\alpha \gamma'} \; \delta_{\alpha' \gamma} \; .
\ee
This means that the ${\cal L}$-operator transfers the in-horizontal spin to the out-vertical state and the in-vertical spin to the out-horizontal one. Successive application of this ${\cal L}$-operator, progressively shifts the in-vertical state in one column to the out-vertical spin in the next column. Taking the final trace over the first and last horizontal spin closes the chain and effectively moves the last vertical spin on the first column. Thus the net effect of the transfer matrix\index{Transfer matrix} at $\lambda=0$ is that of a shift by one lattice site, i.e.
\be
   {\bf T} (0) = \rho^N \; \sinh^N \phi \; \eu^{\ii \hat{P}} \; ,
\ee
where $\hat{P}$ is the lattice momentum operator.
Similarly, as we show in section \ref{sec:Tconstruction}, the first logarithmic derivative of the transfer matrix\index{Transfer matrix} at $\lambda = 0$ gives
\be
   \left. {\de \over \de \lambda} \ln {\bf T} (\lambda) \right|_{\lambda = 0}
   = {1 \over 2 \sinh \phi} \sum_{j=1}^N \left[
   \sigma_j^x \sigma_{j+1}^x + \sigma_j^y \sigma_{j+1}^y
   + \cosh \phi \left( 1 + \sigma_j^z \sigma_{j+1}^z \right) \right] \; ,
\ee
where $\sigma_j^\alpha$ are Pauli matrices, which emerge as matrix representations of Kronecker delta's. Thus, the logarithm of ${\bf T}$ at $\lambda = 0$ is proportional to the lattice momentum and its first logarithmic derivative gives an operator that is proportional to the Hamiltonian of the XXZ model (plus a constant). This shows the connection between the 6-vertex model\index{Vertex model} and the quantum spin chain and implies that all higher logarithmic derivatives of the transfer matrix\index{Transfer matrix} are also in convolution with the Hamiltonian. Thus, the transfer matrix and the XXZ chain share the same eigenvectors and the solution of one model translates into the other, although the natural questions one is interested in might differ between the two. Note that we already  encountered something similar, when we commented that the transfer matrix of the 2D classical Isingl model is the exponential of the Hamiltonian of the quantum Ising chain.

\section{T-Q relations}
\label{sec:TQ}

Finally, let us mention that it is possible to construct an additional operator ${\bf Q} (\lambda)$, called the Q-matrix, that allows for an easy derivation of the Bethe equations\index{Bethe!equations} and of the spectrum of the transfer matrix\index{Transfer matrix}. This construction was pioneered by Baxter \cite{baxterbook} and was instrumental in the development of the algebraic version of the thermodynamical Bethe Ansatz \cite{takahashi}\index{Thermodynamic Bethe Ansatz}.
The Q-matrix is defined as an operator that commutes with the transfer matrix\index{Transfer matrix}
\be
   \left[ {\bf T} (\lambda), {\bf Q} (\lambda') \right] =
   \left[ {\bf Q} (\lambda), {\bf Q} (\lambda') \right] = 0
   \label{TQcomm}
\ee
and satisfies the following equation
\be
   {\bf T} (\lambda) {\bf Q} (\lambda) = {\bf Q} (\lambda) {\bf T} (\lambda) =
   \sigma (\lambda - \phi) \; {\bf Q} (\lambda + 2 \phi) +
   \sigma (\lambda + \phi) \; {\bf Q} (\lambda - 2 \phi) \; ,
   \label{TQmatrix}
\ee
with
\be
   \sigma (\lambda) \equiv \left[ \rho \sinh \lambda \right]^N \; .
\ee
It can be proven \cite{baxterbook} that a Q-operator satisfying (\ref{TQcomm}, \ref{TQmatrix}) exists.

In the introduction we argued that the number of vertical spin up and down is conserved from one row to the next and thus that the transfer matrix\index{Transfer matrix} (and the partition function) of the 6-vertex model\index{Vertex model} have a block-diagonal structure, where each block corresponds and connect only configurations with a given number of spin down, say $R$. Since the Q-matrix commutes with the transfer matrix\index{Transfer matrix}, it shares the same eigenvectors and the same block-diagonal structure. Thus, we can diagonalize ${\bf T} (\lambda)$ and ${\bf Q} (\lambda)$ simultaneously, working in each individual block of dimension $R$.

For each eigenvector with $R$ down spins, the eigenvalue $Q (\lambda)$ of ${\bf Q} (\lambda)$ can be shown to be an entire function of $\lambda$ which vanishes at $R$ points $\lambda_j$, to be determined. The analytic structure of this function and the commutation (\ref{TQcomm}) of the Q-matrix for different spectral parameters imply that the eigenvalue can be written as 
\be
   Q (\lambda) = C \prod_{j=1}^R \sinh (\lambda - \lambda_j) \; ,
   \label{Qeigenvaues}
\ee
with some constant $C$.
Since ${\bf T}$ and ${\bf Q}$ commute, they can be simultaneously diagonalized in each block and the TQ-relation (\ref{TQmatrix}) can be written as a set of scalar equations
\be
   \Lambda (\lambda) Q (\lambda) =
   \sigma (\lambda - \phi) \; Q (\lambda + 2 \phi) +
   \sigma (\lambda + \phi) \; Q (\lambda - 2 \phi) \; .
   \label{TQscalar}
\ee
From (\ref{Qeigenvaues}) we see that $Q(\lambda)$ has $R$ zeros located at $\lambda = \lambda_j$ (i.e. there are $R$ values of $\lambda$ at which the Q-operator has vanishing determinant): evaluating (\ref{TQscalar}) at such zeros we get
\be
   \sigma (\lambda_j - \phi) \; Q (\lambda_j + 2 \phi) +
   \sigma (\lambda_j + \phi) \; Q (\lambda_j - 2 \phi) = 0 \; ,
   \label{TQZero}
\ee
i.e.
\be
   \left( {\sinh (\lambda_j + \phi) \over \sinh (\lambda_j - \phi) } \right)^N
   = - \prod_{l=1}^R {\sinh (\lambda_j - \lambda_l + 2 \phi) \over
   \sinh (\lambda_j - \lambda_l - 2 \phi) } \; , \qquad
   j=1, \ldots, R \; ,
\ee
which we recognize as the Bethe equations\index{Bethe!equations} for the XXZ model and which specify the parameters $\lambda_j$ in (\ref{Qeigenvaues}). So, in this construction, the Bethe equations arise as consistency equations for the TQ-relation. Having found the eigenvalues of ${\bf Q} (\lambda)$, we can substitute them into (\ref{TQscalar}) to find the spectrum of the transfer matrix\index{Transfer matrix}
\be
   \Lambda (\lambda) = \rho^N \left[ \sinh^N (\lambda - \phi) \prod_{j=1}^R
   {\sinh (\lambda - \lambda_j + 2 \phi) \over \sinh (\lambda - \lambda_j)}
   + \sinh^N (\lambda + \phi) \prod_{j=1}^R
   {\sinh (\lambda - \lambda_j - 2 \phi) \over \sinh (\lambda - \lambda_j)}
   \right] \; .
\ee

Thus, we accomplished what we set out to do: we determined the spectrum of the transfer matrices\index{Transfer matrix}, from which we can access all information contained in the partition function. To this end it was fundamental to extend the original problem of diagonalizing a single system to a whole family of commuting ones, since this gave us the freedom of choosing the most suitable $\lambda$ for each eigenvalue, namely the one for which the determinant of ${\bf Q} (\lambda)$ vanishes, see (\ref{TQZero}).

The TQ-construction has thus shown to be very helpful. Its limitation is that it does not give us direct access to the eigenvectors of the system. When we derived the Yang-Yang equation\index{Yang-Yang equation} for the thermodynamics of the Lieb-Liniger model we saw that one takes a similar point of view, focusing directly on the energy eigenvalues, instead of the eigenfunction. The operator generalization of the Yang-Yang equation\index{Yang-Yang equation} takes advantage of the TQ-relations to develop the {\it Thermodynamic Bethe Ansatz}\index{Thermodynamic Bethe Ansatz}, but this subject is not addressed in these notes, see \cite{samaj,tongeren16,AdSCFTreview}. The Algebraic Bethe Ansatz is a different way to use the transfer matrix\index{Transfer matrix} that starts from its eigenstates construction to characterize the system. This is the subject of the chapter \ref{chap:algebraic}.

\chapter{Field theory and finite size effects}
\label{app:CFT}

\abstract{
	While Bethe Ansatz techniques are quite efficient in providing the thermodynamics of an integrable system, the calculation of correlators requires the mastering of advanced methods related to the algebraic approach, as we briefly sketched at the end of chapter \ref{chap:algebraic}. In $1+1$ dimensions, however, we can rely on complementary analytical methods to access correlation functions, especially for gapless systems in the low-energy, long-distance limit: the {\it bosonization} approach and the {\it Conformal Field Theory} (CFT) description. They capture the universality of 1D systems, which do not conform to the standard paradigm of {\it Fermi liquids}, but are characterized by collective, emergent degrees of freedom. In Sec. \ref{BosSec} we introduce the simplest of this universality, known as {\it Luttinger liquid}, which is a $c=1$ CFT, to which both the Lieb-Liniger and the XXZ chain belong. We then show in Sec. \ref{sec:CFT} how Bethe Ansatz can be paired with these field theories, by considering finite size corrections to the thermodynamic limit. We introduce a controllable parameter (the system size) that allows to match the microscopic and the field theories, and show how the latter provides the asymptotics of correlation functions. Finally, in Sec. \ref{sec:LLbosonization} and \ref{sec:XXZbosonization} we provide the explicit field theory construction for the Lieb-Liniger and XXZ model.
	}

\section{Bosonization}
\label{BosSec}

In the solution of the Lieb-Liniger\index{Lieb-Liniger model} and XXZ chain\index{XXZ chain} we studied the low energy excitations and noticed that often they have remarkably different properties compared to the microscopic constituents of the system. This is true for Type I\index{Excitation!Type I} and II\index{Excitation!Type II} excitations of the LL and for the spinons\index{Spinon} of spin chains. This is a general feature observed also in non-integrable models, which reflects the fact that in low dimensions perturbations affect the entire system because scattering is unavoidable and thus excitations acquire a collective nature. 
	
Simple (i.e. one-component) critical (i.e. gapless) 1D systems are described by the {\it Luttinger liquid}\index{Luttinger!liquid} universality class: the low-energy behavior is captured by a free bosonic theory, which formalizes the observation that low-energy degrees of freedom have a sound wave (phononic) nature. From a CFT\index{CFT} point of view, the Luttinger liquid is a $c=1$ theory, and thus more information is needed to uniquely identify its operatorial content. 
The bosonization\index{Bosonization} procedure is the way in which, in principle, one extracts the collective behavior from the microscopic description. In practice, to close this derivation one would need to be able to exactly follow the renormalization group flow.  More complicated systems can have fractional central charges or $c>1$, and thus are generalizations of Luttinger liquids\index{Luttinger!liquid}, which are still described by a CFT\index{CFT}, possibly supplemented by a Kac-Moody algebra \cite{mussardobook,difrancescobook}.
In all these cases, Bethe Ansatz is useful in providing the non-perturbative results to determine the correct CFT\index{CFT} representation of integrable models in the scaling limit.

Let us start with a heuristic derivation of the bosonization description, which provides some physical intuition on this approach. We will use the example of a fermionic system, but in fact one can bosonize bosonic systems as well. For a pedagogical introduction to the latter approach we refer to \cite{giamarchibook} and we recommend \cite{bosonization} for detailed explanations on the bosonization techniques\index{Bosonization}.

Bosonization is a way to describe the dynamics of critical systems in terms of their collective behavior through a bosonic field. This is possible in one-dimension because the system is very much constrained: even if we try to excite an individual particle, all other particles have to rearrange to accommodate it, because there is no way for a particle to go around another without interacting. This kind of phenomenon is familiar to us already from our analysis of the excitations of integrable models using Bethe Ansatz.

Because of this collective nature, the description of the system in terms of its density of particles can be efficiently used to capture the whole dynamics, provided that the density field
\be
   \rho(x) \equiv \sum_j \delta (x - x_j) \; ,
\ee
(where $x_j$ is the position of the $j$-th particle) can be approximated with a smooth function. This amount to a hydrodynamic description for the system, where the field conjugated to the density is the velocity $v(x)$.
\be
   [\rho (x), v(y) ] = - \ii \delta' (x-y) \; .
\ee
A general structure for the evolution equations for such a system gives
\bea
   \dot{\rho} - \partial_x \left( \rho v \right) & = & 0 \; , \\
   \dot{v} - v \partial_x v + \partial_x F(\rho) & = & 0 \; ,
\eea
where the first is the continuity equation and the second is the proper dynamical Euler equation.

In general, these equations are non linear and very difficult to treat at the quantum level (moreover, there is no clear small-coupling expansion valid for all times). But they can be linearized around a classical solution and a linear hydrodynamics gives essentially a wave equation. This is to say that elementary (universal) excitations of a one-dimensional system are phonons.

Thus, under these general considerations we expect to be able to describe a 1-D system with a bosonic operator and a quadratic Hamiltonian. This description is called {\it bosonization}\index{Bosonization}, and we stress again that even bosonic theories can be bosonized, since this just means to give a linear-hydrodynamics formulation.

Let us describe how to bosonize a free fermionic theory, with microscopic Hamiltonian
\be
   {\cal H} =  - {1 \over 2 m} \; \Psi^\dagger (x) \partial_x^2 \Psi (x)
   = { k^2 \over 2 m} \:  \tilde{\Psi}^\dagger (k) \tilde{\Psi} (k) \; ,
   \label{FFH}
\ee
where $\partial_x \equiv \partial / \partial x$ and the last expression shows the  Fourier space representation.

CFT is a chiral theory, that is, the natural degrees of freedom are either right- or left-moving. Thus we will need to separate the fermions into their chiralities and apply the fundamental {\it bosonization identity}\index{Bosonization!identity}
\be
   \psi_\pm (x) \equiv {1 \over \sqrt{2 \pi}} \: : \eu^{\mp \ii \sqrt{4 \pi} \phi_\pm (x) } : \; ,
   \label{BosTrans}
\ee
where $\phi_\pm (x)$ are collective bosonic fields, $: {\cal O} : \equiv {\cal O} - \langle 0 | {\cal O} | 0 \rangle$ stands for the {\it normal ordering} and $+$ ($-$) refers to right-(left-)chirality. Note that the exponential mapping is periodic: the $\sqrt{4 \pi}$ factor determines the periodicity of the fields $\phi_\pm$ (also called the {\it compactification radius}) and for free fermions is equal for both chiral fields. The choice of $\sqrt{4 \pi}$ is convenient to ensure that the anti-commutation of the fermionic fields translate into canonical commutation relations for the bosonic ones. We will see that interactions change the compactification radii of the two chiral fields, in a way that preserves the commutation relation.

While the identity (\ref{BosTrans}) between a fermion and a boson holds in generality in one-dimension, the prescription for the normal ordering depends on the theory (and its ground state). This prescription is pivotal for the bosonization construction\index{Bosonization} to give meaningful results (and avoid spurious divergences) and it is not available in generality. However, if we concentrate only on low-energy excitations, we can linearize the spectrum and hence derive a clear and simple normal ordering rule.

\begin{wrapfigure}{r}{5.5cm}
	\vspace{-10pt}
	\begin{center}
		\includegraphics[width=5.3cm]{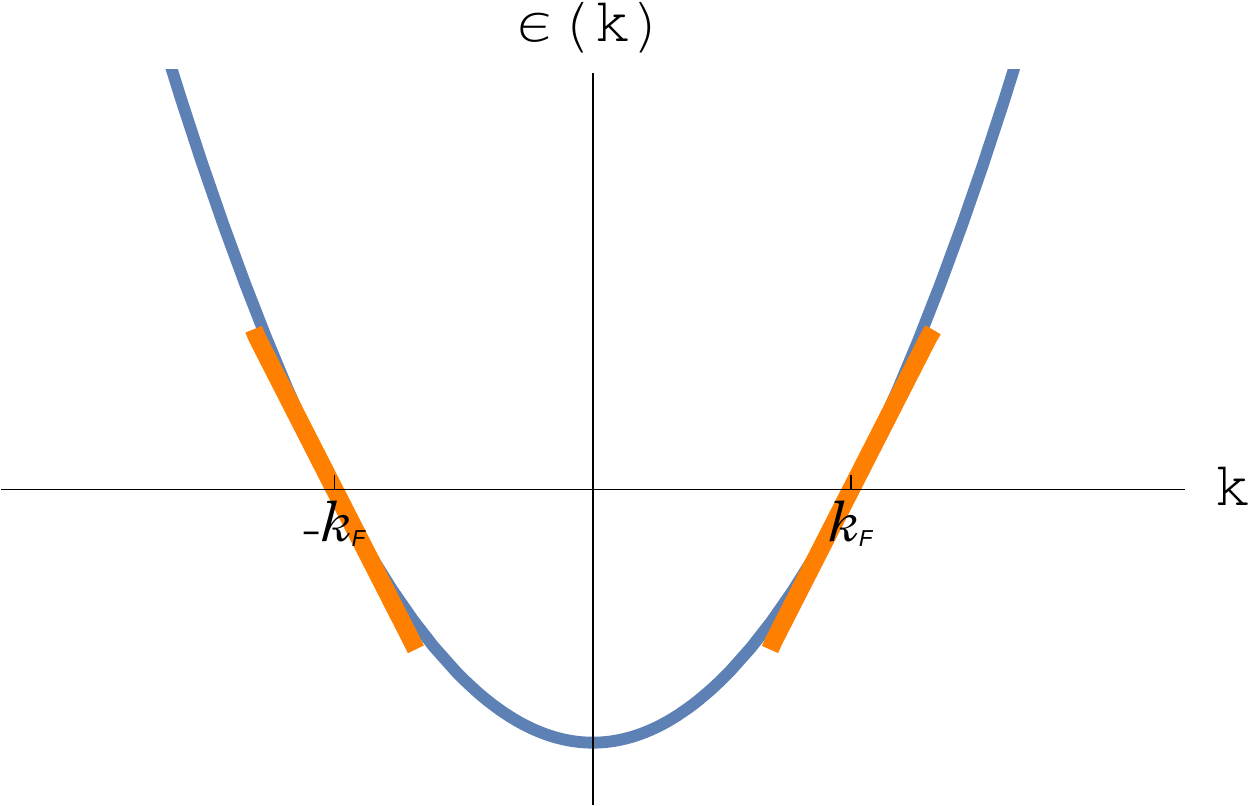}
	\end{center}
	\caption{Linearization of the spectrum around the Fermi points.}  
	\label{fig:linear}
	\vspace{-40pt}
\end{wrapfigure}
Thus, we write the free Hamiltonian (\ref{FFH}) as
\be
   {\cal H} = - {1 \over 2 m} \sum_{r=\pm}  \psi_r^\dagger \left( \partial_x + \ii r k_F \right)^2 \psi_r
\ee
and expand around the Fermi points (see Fig. \ref{fig:linear}) as
\be
   {\cal H} \simeq - {k_F^2 \over 2 m} \sum_{r=\pm} \psi_r^\dagger \psi_r
   - \ii {k_F \over m} \sum_{r=\pm} r \psi_r^\dagger \partial_x \psi_r + \ldots
   \label{linearspinH}
\ee
where the first term is interpreted as a chemical potential (which can be absorbed in a redefinition of the ground state energy), while the second term shows a linear spectrum for the excitations around the Fermi points $\pm k_F$. The left- and right-moving fields $\psi_\pm$ are obtained by expanding $\Psi(x)$ around the left/right Fermi points:
\be
   \psi_\pm (x) \equiv \int_{\pm k>0} {\de k \over 2 \pi} \: \eu^{\ii(k \mp k_F)x} \: \tilde{\Psi} (k)
   \quad \Rightarrow \quad
   \Psi (x) \simeq  : \eu^{\ii k_F x} \psi_+ (x) + \eu^{- \ii k_F x} \psi_- (x) :
   = \sum_{r=\pm} {\eu^{\ii r k_F x} \over \sqrt{2\pi}} \eu^{- \ii r \sqrt{4 \pi} \phi_r (x) } : \; .
   \label{psiRL} 
\ee
Most (i.e. low-energy) physical processes take place close to these points and thus this separation is a sensible approximation.

We can use the mapping (\ref{BosTrans}) to express various fermions bilinears in terms of the bosonic field. For instance, one can consider a quantity like
\bea
   : \psi_\pm^\dagger (x) \psi_\pm (x + \epsilon) : & = &
   \psi_\pm^\dagger (x) \psi_\pm (x + \epsilon) - \langle \psi_\pm^\dagger (x) \psi_\pm (x + \epsilon) \rangle
   \nonumber \\
   & = & {1 \over 2 \pi} \left[ : \eu^{\mp \ii \sqrt{4 \pi} \left( \phi_\pm (x + \epsilon) - \phi_\pm (x) \right) } :
   -1 \right] \eu^{4 \pi \langle \phi_\pm (x) \phi_\pm (x+\epsilon) \rangle}
   \nonumber \\
   & = & \pm { 1 \over 2 \ii \pi \epsilon }
   \left[ \eu^{\mp \ii \sqrt{4 \pi} \left( \phi_{L,R} (x + \epsilon) - \phi_{L,R} (x) \right) } - 1 \right]
   \label{bilinear}
\eea
where we used the identity
\be
   : \eu^A : \; : \eu^B : = : \eu^{A+B} : \; \eu^{\langle AB - {A^2 + B^2 \over 2} \rangle}
\ee
and the fact that
\be
   \langle \phi_\pm (0) \phi_\pm (x) - \phi_\pm^2 (0) \rangle =
   \lim_{\alpha \to 0} {1 \over 4 \pi} \ln {\alpha \over \alpha \pm \ii x} \; .
   \label{psicorr}
\ee
Here, $\alpha$ is a regulator that mimics a finite bandwidth and prevents the momentum from becoming too large (thus limiting the bandwidth to $\Lambda \sim 1 / \alpha$).

The prescription to calculate bilinears like (\ref{bilinear}) is known as {\it point splitting} and it takes into account that the square of a field in coordinate space is not defined and has to be regularized by discretizing the space. In practice, we saw in the second line of (\ref{bilinear}) that the normal ordering amounts to subtract $1/ \epsilon$ from the exponential, corresponding to the ground state contribution. Thus, from one side $\alpha$ in (\ref{psicorr}) captures the low-energy approximation, from the other $\epsilon$ in (\ref{bilinear}) is related to the underlying lattice of the microscopic theory.
We can expand (\ref{bilinear}) in powers of $\epsilon$
\be
   : \psi_\pm^\dagger (x) \psi_\pm (x + \epsilon) : =
   \sum_{n=0}^\infty {\epsilon^n \over n!} \psi_\pm^\dagger (x) \partial_x^n \psi_\pm (x)
   =  \pm { 1 \over 2 \ii \pi \epsilon }
   \left[ \eu^{\mp \ii \sqrt{4 \pi } \sum_{n=1}^\infty {\epsilon^n \over n!} \partial_x^n \phi_\pm (x) } - 1 \right] \; ,
   \label{bilexp}
\ee
which gives the generating function of the chiral fermionic currents
\be
   j^\pm_n (x) \equiv \psi_\pm^\dagger (x) \partial_x^n \psi_\pm (x)
\ee
in terms of the bosonic fields $\phi_\pm$.

By matching powers of $\epsilon$ in (\ref{bilexp}) we can write down these expressions. The density of fermion is
\be
   \rho_\pm = j^\pm_0 = \psi_\pm^\dagger (x) \psi_\pm (x) =
   - {1 \over \sqrt{ \pi} } \partial_x \phi_\pm (x),
   \label{rhoLR}
\ee
the current density is
\be
   j^\pm_1 = \psi_\pm^\dagger (x) \partial_x \psi_\pm (x) =
   \pm \ii \left( \partial_x \phi_\pm (x) \right)^2 - {1 \over \sqrt{4 \pi} } \partial_x^2 \phi_\pm \; .
\ee
The third term in the expansion is identified with the original quadratic Hamiltonian for the left/right movers
\be
   j^\pm_2 = - 2 m \: {\cal H}_\pm = \psi_\pm^\dagger (x) \partial^2_x \psi_\pm (x)
   = { 4 \over 3} \sqrt{\pi} \left( \partial_x \phi_\pm (x) \right)^3
   \pm \ii \left( \partial_x \phi_\pm \right) \left( \partial_x^2 \phi_\pm \right)
   - {1 \over 3 \sqrt{ \pi} } \partial_x^3 \phi_\pm \; .
   \label{cubicHam}
\ee
While in terms of fermions it is a well defined Hamiltonian operator, its bosonic form shows dangerous cubic terms (note that the last term is a total derivative and thus contributes only as a boundary term). Thus, we see that the regularization prescription employed for the normal ordering maps the free system into an unstable bosonic theory, with a cubic potential that is not bounded from below and thus cannot sustain a stable quantum vacuum. Physically, this failure originates from the separation of the fermionic field into left and right movers, since this separation breaks down moving closer to the bottom of the band. Mathematically, to take into account this effect we need to modify (\ref{psicorr}), which is valid for relativistic theories. In certain cases (for instance for certain integrable models) it is possible to find a suitable prescription to write a non-linear bosonization \cite{abanov09}\index{Bosonization!Non-linear} or to incorporate the corrections to go beyond the Luttinger liquid\index{Luttinger!liquid} and to take into account the curvature of the spectrum \cite{khodas07,imambekov09,imambekov12}.

As long as we are interested in low energy physics, however, we can exploit the fact that the Fermi momentum $k_F$ is a large parameter and thus the terms neglected in (\ref{linearspinH}) are suppressed. We retain the linearized version of the free fermionic theory and use it as our Hamiltonian, which is bosonized through the expressions found above as
\be
   {\cal H} \sim - \ii {k_F \over m} \left[ j_1^+ - j_1^- \right] + \ldots
   = {k_F \over m} \left[ \left( \partial_x \phi_+ \right)^2
   + \left( \partial_x \phi_- \right)^2 \right] + \ldots
   \label{bosHam}
\ee

Out of the two chiral fields we can define a bosonic field and its dual
\be
   \phi (x) \equiv \phi_+ (x) + \phi_- (x) \; , \qquad \qquad
   \theta (x) \equiv \phi_+ (x) - \phi_- (x) \; .
   \label{phithetadef}
\ee
Using (\ref{psiRL}) and the fermionic commutation relation, one can prove that these bosonic fields satisfy the commutation relation
\be
   [\phi(x), \theta (y) ] = \ii \vartheta_H (y-x) \; ,
\ee
where $\vartheta_H (x)$ denotes the Heaviside step function. By differentiating we have
\be
   [\phi(x), \partial_y \theta(y) ] = [\theta(x), \partial_y \phi(y)] = \ii \delta(x-y) \; ,
\ee
which means that we can identify the derivative of the dual field $\theta (x)$ as the conjugate of $\phi (x)$ (or viceversa):
\be
  \Pi (x) \equiv {1 \over v_0} \partial_t \phi (x) = \partial_x \theta (x) \; ,
\ee
where $v_0 \equiv k_F /m$ is the sound velocity of the free system.

Thus, the linearized free fermionic theory is mapped into a free bosonic theory
\be
   {\cal H} = {v_0 \over 2} \; \int \left[ \big( \Pi (x) \big)^2
   + \big( \partial_x \phi (x) \big)^2 \right] \de x \; .
   \label{FBH}
\ee
Physically, the bosonic field is the {\it displacement field} and one should notice the similarity between the bosonization identity\index{Bosonization!identity} (\ref{BosTrans}) and the Jordan-Wigner transformation\index{Jordan-Wigner!transformation} (\ref{JordanWigner}). In fact, $\phi (x)$ counts the number of particles to the left of $x$ and its derivative gives the particle density, see (\ref{rhoLR}).
In particular we have
\bea
   \rho (x) =  \Psi^\dagger (x) \Psi (x)
   & = & \rho_0 + \psi_+^\dagger (x) \psi_+ (x) + \psi_-^\dagger (x) \psi_- (x)
   + \eu^{- \ii 2 k_F x} \psi_+^\dagger (x) \psi_- (x) + \eu^{\ii 2 k_F x} \psi_-^\dagger (x) \psi_+ (x)
   \nonumber \\
   & = & \rho_0 - {1 \over \sqrt{\pi}} \partial_x \phi (x)
   + {1 \over \pi} \cos \left[ \sqrt{ 4 \pi} \phi (x) - 2 k_F x \right] \; ,
   \label{rhophi}
\eea
so that we identify the bosonic field with a density wave ($\rho_0$ is the constant, background, density of particles).

We have shown that low-energy excitations of the free fermions Hamiltonian (\ref{FFH}) can be described in terms of a simple quadratic boson, corresponding to a quantum sound wave. What is remarkable is that the operators appearing in (\ref{FBH}) are the only marginal operators in this bosonic theory \cite{difrancescobook}. This means that any interaction term added to the free fermionic theory, as long as it does not open a gap (i.e., drives the system away from criticality), once bosonized, possibly using the mapping (\ref{BosTrans}) and the point-splitting prescription, results in a series of irrelevant operators and a combination of $\big( \Pi(x) \big)^2$ and $\big( \partial_x \phi (x) \big)^2$. Thus, this renormalization group argument implies that all one-dimensional critical fermionic theories are mapped by bosonization\index{Bosonization} into a quadratic theory like 
\be
   {\cal H} = {v_S \over \pi} \int \left[ K \left( \Pi (x) \right)^2
   + {1 \over K} \left( \nabla \phi (x) \right)^2 \right] \de x \; ,
   \label{bosonizationH}
\ee
where $v_S$ has the dimension of a velocity and can be interpreted as the (renormalized) Fermi velocity of the interacting system and $K$ is a dimensionless parameter that is related to the compactification radius of the theory, or to the exclusion statistic area occupied by a particle in phase-space. Interactions which open a gap result in relevant operators in the bosonic theory, usually sine or cosine terms in the field and/or its dual. A single term of this kind gives a ``simple'' Sine-Gordon theory, additional terms can make the resulting field theory difficult to analyze, but it is often the case that one of them is dominant (in the RG sense): thus close to criticality one can usually extract the behavior of the system using a suitable Sine-Gordon model \cite{bosonization}.

To recap, the low-energy excitations of any one-dimensional gapless system can be mapped using the bosonization\index{Bosonization} procedure into a bosonic Gaussian theory (\ref{bosonizationH}), where all the interaction effects are captured by just two parameters: $v_S$ and $K$.
Notice that the {\it Luttinger parameter}\index{Luttinger!parameter} $K$ can be removed from the Hamiltonian (\ref{bosonizationH}) by a rescaling of the fields
\be
   \phi (x) \to {1 \over \sqrt{K}} \; \phi (x) \; , \qquad \qquad
   \theta (x) \to \sqrt{K} \; \theta (x) \; .
\ee
This corresponds to a redefinition of the compactification radius of the chiral fields. Using (\ref{phithetadef})
\be
   \phi_\pm = {1 \over 2 \sqrt{K}} \big[ \phi (x) \pm K \; \theta (x) \big] \; .
   \label{phipmK}
\ee
In general, $K=1$ corresponds to free fermions; $K>1$ encodes attractive fermions and $0<K<1$ repulsive fermions. Free bosons are not stable in one dimension and they would correspond to $K \to \infty$. Thus, any finite $K$ corresponds to repulsive bosons all the way to the $K=1$ limit of perfectly repulsive bosons (the so-called Tonks-Girardeau limit, i.e. $c \to \infty$ of the Lieb-Liniger model\index{Lieb-Liniger model}). Bosonic systems with $K<1$ can be reached in the super-Tonks–Girardeau regime \cite{STG}.

One of the fundamental advantages of having mapped an interacting system to a Gaussian theory like (\ref{bosonizationH}) is that the correlation functions are easily obtainable. For instance, see (\ref{psicorr}) and (\ref{phipmK}), we have
\bea
   \langle \left[ \phi (x,\tau) - \phi (0,0) \right]^2 \rangle & = &
   \lim_{\alpha \to 0} {K \over 2 \pi} \ln {x^2 + (v_S \tau + \alpha)^2 \over \alpha^2 } \; , \\
   \langle \left[ \theta (x,\tau) - \theta (0,0) \right]^2 \rangle & = &
   \lim_{\alpha \to 0} {1 \over 2 \pi K} \ln {x^2 + (v_S \tau + \alpha)^2 \over \alpha^2 } \; ,
\eea
where $\tau \equiv \ii t$ is the Euclidean time.

The principal operators of the theory are vertex operators of the form
\be
   V(\beta, z) 
   \equiv \eu^{ \ii \beta \phi_+ (z= v_S \tau - \ii x)} \; ,
   \qquad
   \bar{V}(\bar{\beta}, \bar{z}) 
   \equiv \eu^{ \ii \bar{\beta} \phi_- (\bar{z} = v_S \tau + \ii x)} \; .
\ee
Correlation functions of vertex operators can be calculated using the power of a Gaussian theory:
\be
   \Big\langle \eu^{\ii \sum_j \left[ \beta_j \phi_+ (z_j) + \bar{\beta}_j \phi_- (\bar{z}_j) \right]} \Big\rangle
   = \eu^{{1 \over 2} \big\langle \left[ \sum_j \beta_j \phi_+ (z_j) + \bar{\beta}_j \phi_- (\bar{z}_j) \right]^2 \big\rangle} \; ,
\ee
which is non-zero only if $\sum_j \beta_j = \sum_j \bar{\beta}_j = 0$. 
In general, these correlation functions decay like power-law $\langle {\cal O} \rangle \sim r^{-2 \Delta}$, with a characteristic exponent $\Delta$. If $\Delta<2$ the corresponding operator is {\it relevant} in an RG sense; if $\Delta>2$ it is {\it irrelevant}, while $\Delta=2$ corresponds to the marginal case \cite{difrancescobook}.
Using this machinery and (\ref{BosTrans}) one can calculate the asymptotic behavior of physical correlators. For instance
\be
  \langle \rho (x,\tau) \rho(0,0) \rangle \simeq {K^2 \over 2 \pi^2} \;
  {1 \over (x^2 + v_S^2 \tau^2)^2} + B {\cos 2 k_F x \over (x^2 + v_S^2 \tau^2)^{2K} }
  + \ldots \; .
  \label{rhorhocorr}
\ee

Finally, let us mention that the bosonization construction\index{Bosonization} is very general and applicable to any one-dimensional critical system. Even if we showed the construction explicitly only for a microscopic fermionic theory, it can be generalized to any model. The approximation to linear spectrum (low-energy modes) is pivotal to ensure that the resulting theory is just quadratic. To bosonize a spin system, one can first perform a Jordan-Wigner transformation\index{Jordan-Wigner!transformation} to map it into a fermionic theory and then bosonize these fermions (note that a spin chain at half filling -i.e. zero magnetization- has $k_F = \pi/2$, which corresponds to having a smooth and a staggered component in the spin density, see (\ref{rhophi}) and section \ref{sec:XXZbosonization}). It is also possible to bosonize a bosonic theory \cite{giamarchibook}, in that the mapping does not have to do with the statistics of the particle, but with the fact that fundamental excitations are collective.
With systems with additional degrees of freedom, like the Hubbard model or various spin ladders, one can bosonize each degree of freedom and study their interaction (and competition) in the collective description. However, these systems often acquire additional symmetries for which graded CFTs can provide a more powerful description \cite{bosonization}.

\section{Conformal Field Theory parameters from Bethe Ansatz}
\label{sec:CFT}

The physical ideas behind bosonization\index{Bosonization} were pioneered in a seminal paper by Haldane in \cite{Haldane-1981}, building over previous works. Over the years, it has been understood that the success of these ideas is rooted on the universality of Conformal Field Theory (CFT)\index{CFT}, which in $1+1$-dimensions is particularly powerful.

At a critical point there are no relevant length scales and the theory is invariant under rescaling. In a relativistic theory, the group responsible for this invariance is the conformal group. In $1+1$-dimensions, this symmetry is enhanced to an infinite number of generators and becomes powerful enough to constrain the structure of the theory and of the correlation functions in a significant way.
We expect the reader to be familiar with the basic ideas behind CFT\index{CFT} and refer to \cite{difrancescobook} for an exhaustive treatment of the subject. Nonetheless, let us introduce few basic concepts for the sake of completeness.

Conformal Field Theory, being two-dimensional, is best represented in terms of complex variables
\be
   z \equiv - \ii ( x - v_S \; t) = v_S \; \tau - \ii x \; , 
   \qquad \qquad \qquad
   \bar{z} \equiv \ii ( x + v_S \; t ) = v_S \; \tau + \ii x  \; ,
\ee
where $v_S$ is the sound (light) velocity and $\tau \equiv \ii t$. CFT\index{CFT} assumes Lorentz invariance and thus all massless excitations move with the same velocity $v_S$. CFT is also a chiral theory, therefore the left and right moving sectors tend to be independent from one another.

The quantum generators of the conformal transformations are called {\it Virasoro operators} $L_n, \bar{L}_n$ and satisfy the algebra (independent for the holomorphic and antiholomorphic sectors)
\bea
   \left[ L_n, L_m \right] & = & (n-m) L_{n+m} + {c \over 12} \; n (n^2-1) \delta_{n,-m} \; , 
   \label{virasoro1} \\
   \left[ \bar{L}_ n, \bar{L}_m \right] & = & (n-m) \bar{L}_{n+m} + {\bar{c} \over 12} \; n (n^2-1) \delta_{n,-m} \; ,
   \label{virasoro2}
\eea
where $c, \bar{c}$ is the {\it central charge}, or {\it conformal anomaly}. The $L_n, \bar{L}_n$ are nothing but the coefficients in a Laurent expansion of the stress tensor in powers of $z, \bar{z}$:
\be
   T (z) = \sum_{n=-\infty}^\infty {L_n \over z^{n+2}} \; , \qquad
   \bar{T} (\bar{z}) = \sum_{n=-\infty}^\infty {\bar{L}_n \over \bar{z}^{n+2}} \; .
\ee

Under a conformal transformation $z = z(w)$, $\bar{z} = \bar{z} (\bar{w})$ a primary field $\phi(z, \bar{z})$ transforms as
\be
   \phi (w , \bar{w}) =
   \left( {\partial z \over \partial w } \right)^{\Delta^+}
   \left( {\partial \bar{z} \over \partial \bar{w} } \right)^{\Delta^-}
   \phi \left( z \left( w \right), \bar{z} \left( \bar{w} \right) \right) \; ,
\ee
where the {\it conformal dimensions} $\Delta^\pm$ characterize the field and specify the two-point correlation function
\be
  \langle \phi (z_1,\bar{z}_1) \phi (z_2, \bar{z}_2) \rangle =
  (z_1 - z_2)^{-2\Delta^+} (\bar{z}_1 - \bar{z}_2)^{-2\Delta^-} \; .
  \label{phiphi}
\ee

At this point, the strategy to identify the parameters of the CFT\index{CFT} is to exploit scale invariance to bring the system to a cylinder geometry, that is, we apply periodic boundary condition in the space direction. In doing so, from one side we connect to the setting employed in Bethe Ansatz, but most of all we introduce a scale $L$ in the model, which opens a finite-size energy gap. The conformal mapping to a cylinder is
\be
   z = \eu^{2 \pi w / L} \; , \qquad \qquad 
   w = v_S \; \tilde{\tau} - \ii \; \tilde{x} \; , \quad
   0 \le \tilde{x} < L \; .
\ee
In this geometry, the asymptotic behavior of the 2-point function (\ref{phiphi}) becomes
\be
  \langle \phi (w_1,\bar{w}_1) \phi (w_2, \bar{w}_2) \rangle_L \sim
  \eu^{{2 \pi \Delta^+ \over L} \left[ \ii (\tilde{x}_1 - \tilde{x}_2) - v_S (\tilde{\tau}_1 - \tilde{\tau}_2) \right]} \;
  \eu^{{2 \pi \Delta^- \over L} \left[ -\ii (\tilde{x}_1 - \tilde{x}_2) - v_S (\tilde{\tau}_1 - \tilde{\tau}_2)
  \right]} \; ,
  \label{phiphiw}
\ee
where we see that finite size effects turned a power-law into an exponential behavior.
This expression can be compared with a standard spectral decomposition ($\tilde{\tau}_2>\tilde{\tau}_1$)
\be
  \langle \phi (w_1,\bar{w}_1) \phi (w_2, \bar{w}_2) \rangle_L =
  \sum_Q \left| \langle 0 | \phi (0,0) | Q \rangle \right|^2
  \eu^{-(\tilde{\tau}_1 - \tilde{\tau}_2)(E_Q - E_0) + \ii (\tilde{x}_1 - \tilde{x}_2)(P_Q - P_0)}
  \label{phiseries}
\ee
where $E_0$, $P_0$ are the energy and momentum of the ground state, while $E_Q$, $P_Q$ are the energy and momentum of one of the intermediate states $Q$, which constitute a complete set.

Matching the leading term of this expansion with (\ref{phiphiw}) gives
\bea
   E_Q - E_0 & = & {2 \pi v_S \over L} (\Delta^+ + \Delta^-)
   \label{EQ} \\
   P_Q - P_0 & = & {2 \pi \over L} (\Delta^+ - \Delta^-)
   \label{PQ}
\eea
Comparing the energy and momentum of the different low-energy states as obtained from Bethe Ansatz with (\ref{EQ}, \ref{PQ}) we can identify the scaling dimensions of the operators corresponding to these states.

Having determined the primary fields, to identify the CFT\index{CFT} we need the central charge in (\ref{virasoro1},\ref{virasoro2}). Once more, finite size effects help, because for a CFT\index{CFT}, the energy of the system goes as
\be
   E \simeq L \; e - c \: {\pi \over 6 L} \: v_S + \Ord (L^{-2}) \; .
   \label{CFTE}
\ee
Thus, we can determine $c$ for a gapless solvable model, by studying the finite size behavior of the ground state energy and by knowing the speed of low-energy excitations $v_S$.
Let us now show how we can determine, using the Bethe Ansatz solution, the parameters of the field theory. In order, we will extract the velocity of low energy modes (velocity of sound), the central charge and the scaling dimensions/Luttinger parameter \index{Luttinger!parameter}of the fields corresponding to the Bethe states.

\subsection{Sound velocity}
\label{sec:soundvel}

We employ the microscopical definition of the Fermi velocity as the derivative of the dressed energy\index{Dressed!energy} by the dressed momentum\index{Dressed!momentum} at the Fermi point:
\be
   v_S \equiv 
   \left. {\partial \varepsilon (\lambda) \over \partial k(\lambda)} \right|_{\lambda=\Lambda}
   = \left( {\partial \varepsilon (\lambda) \over \partial \lambda} \right)
   \left/ \left( {\partial k (\lambda) \over \partial \lambda} \right) \right|_{\lambda=\Lambda} \; .
   \label{vSdef}
\ee
The dressed functions satisfy the dressing equation with the bare quantity as a source:
\bea
   \rho (\lambda)
   + {1 \over 2 \pi} \int_{\Lambda}^\Lambda {\cal K} (\lambda - \mu) \; \rho (\mu) \; \de \mu 
   & = & {1 \over 2 \pi} p_0^\prime (\lambda) \; , 
   \label{rhoint3} \\
   \varepsilon (\lambda)
   + {1 \over 2 \pi} \int_{-\Lambda}^\Lambda {\cal K} (\lambda - \mu) \; \varepsilon (\mu) \; \de \mu 
   & = & \epsilon_0 (\lambda) \; ,
   \label{epsilonint3}
\eea 
while the dressed momentum is given by
\be
   k (\lambda) = p_0 (\lambda)
   -  \int_{-\Lambda}^\Lambda  \theta (\lambda -\mu) \; \rho (\mu) \; \de \mu  \; .
   \label{kint3}
\ee
Comparing (\ref{rhoint3}) and (\ref{kint3}) we notice
\be
{\partial k (\lambda) \over \partial \lambda} = 2 \pi \rho (\lambda) \; ,
\ee
and thus
\be
   v_S = \left. {1 \over 2 \pi \rho (\Lambda) }
{\partial \varepsilon (\lambda) \over \partial \lambda} \right|_{\lambda =\Lambda} \; .
\label{vF1}
\ee

It is also possible to take a more macroscopic approach and define $v_S$ as the derivative of the pressure ${\cal P}$ (which at zero temperature equals the negative of the ground state energy, see (\ref{T0P})) with respect to the density $n$. This definition can be proven equivalent to the one we employ and additional identities can be derived through formal manipulation of the integral equation\index{Integral equation!linear}. We refer the interested reader to \cite{ISM}.

\subsection{Central Charge}
\label{sec:centralcharge}

We already anticipated that both the Lieb-Liniger\index{Lieb-Liniger model} and the XXZ\index{XXZ chain} for $|\Delta|<1$ are described in the scaling limit by a $c=1$ CFT\index{CFT}.
To confirm this statement, we compare (\ref{CFTE}) with the finite-size corrections obtained for the Bethe Ansatz solution. 

We start recalling the Euler-Maclaurin formula which captures how an integral approximates a sum:
\be
   \sum_{j=1}^N f (x_j) = \int_{a}^{b} f(x) \de x + \left. {f \over 2} \right|_a^b
   - \left. {b_2 \over 2} {\de f \over \de x} \right|_a^b + \ldots \; .
   \label{EMFormula}
\ee
Here $b_2 = {1 \over 6}$ is the second Bernoulli number, $x_1 = a$ and $x_N = b$ and the additional terms, which we do not need, are known in terms of higher Bernoulli numbers and higher derivatives at the boundaries.

The energy of the ground state is given by $E = \sum_{j=1}^N \epsilon_0 (\lambda_j)$, where the $\lambda_j$ are the ground state solution of the Bethe equations (with quantum numbers $I_j$ symmetrically distributed around $0$). As $N \to \infty$, the distance between consecutive $\lambda$'s is of the order of $1 / N$. We define a function $\lambda (x)$ as $\lambda \left( I_j / L \right) = \lambda_j$.
We use (\ref{EMFormula}) to write:
\be
   E = L \int_{-N/(2L)}^{N/(2L)} \epsilon_0 \big( \lambda(x) \big) \de x
   - \left. {1 \over 24 L} {\partial \epsilon_0 \over \partial x}
   \right|_{x=-N/2L}^{x=N/2L} + \ldots
   = L \int_{-N/(2L)}^{N/(2L)} \epsilon_0 \big( \lambda(x) \big) \de x
   - \left. {1 \over 24 L \rho(\Lambda)} {\partial \epsilon_0 \over \partial \lambda}
   \right|_{\lambda=-\Lambda}^{\lambda=\Lambda} + \ldots \; ,
   \label{EME1}
\ee 
where we used ${\de \lambda / \de x} = {1 / \rho(\lambda)}$.
We also need to account for the finite size corrections to the Bethe equations in going from (\ref{betheeq}) to (\ref{rhoint3}):
\be
   \rho_l (\lambda) + + {1 \over 2 \pi} 
   \int_{-\Lambda}^\Lambda {\cal K}(\lambda - \mu) \rho_L (\mu) \de \mu =
   {1 \over 2 \pi} \left\{ p_0^\prime (\lambda) + {1 \over 48 \pi L^2 \rho (\Lambda) }
   \left[ {\cal K}'(\lambda - \Lambda) - {\cal K}'(\lambda + \Lambda) \right] \right\} \; .
\ee
We write the solution as
\be 
   \rho_L (\lambda) = \rho (\lambda) + \rho^{(1)} (\lambda)
\ee
where $\rho (\lambda)$ is the solution of the infinite size integral equation (\ref{rhoint3})\index{Integral equation!linear} and $\rho^{(1)} (\lambda)$ accounts for the finite size corrections
and has formal solution
\be
   \rho^{(1)} (\lambda) = {1 \over 48 \pi L^2 \rho (\Lambda) }
   \int_{-\Lambda}^\Lambda {\cal U}_\Lambda (\lambda,\mu) 
   \Big[ {\cal K}'(\mu - \Lambda) - {\cal K}'(\mu + \Lambda) \Big] \de \mu
\ee
in terms of the Green's function (\ref{Greenf})\index{Green's function}.
We now use the density of rapidities in (\ref{EME1})
\bea
   E & = &
    L \int_{-\Lambda}^\Lambda \epsilon_0 (\lambda) \rho_L (\lambda) \de \lambda
   - {1 \over 12 L} {\epsilon'_0 (\Lambda) \over \rho(\Lambda)} + \ldots
   \nonumber \\
   & = & L \int_{-\Lambda}^\Lambda \epsilon_0 (\lambda) \rho (\lambda) \de \lambda
   + {1 \over 48 \pi L \rho(\Lambda)} \int_{-\Lambda}^\Lambda \epsilon_0 (\lambda)
   \: {\cal U}_\Lambda (\lambda,\mu) 
   \Big[ {\cal K}'(\mu - \Lambda) - {\cal K}'(\mu + \Lambda) \Big] \de \lambda \: \de \mu
   - {1 \over 12 L} {\epsilon'_0 (\Lambda) \over \rho(\Lambda)} + \ldots
   \nonumber \\
   & = & L \int_{-\Lambda}^\Lambda \epsilon_0 (\lambda) \rho (\lambda) \de \lambda
   + {1 \over 48 \pi L \rho(\Lambda)} \int_{-\Lambda}^\Lambda \varepsilon (\mu)
   \Big[ {\cal K}'(\mu - \Lambda) - {\cal K}'(\mu + \Lambda) \Big] \de \mu
   - {1 \over 12 L} {\epsilon'_0 (\Lambda) \over \rho(\Lambda)} + \ldots
   \nonumber \\ 
   & = & L \int_{-\Lambda}^\Lambda \epsilon_0 (\lambda) \rho (\lambda) \de \lambda
  - {\pi \over 6 L} {\varepsilon' (\Lambda) \over 2 \pi \rho(\Lambda)} + \ldots
  \label{finiteE2}
\eea
where the self-consistent (to zeroth order) limit of integration $\Lambda \simeq \lambda_N + {1 \over 2 L \rho(\lambda_N)}$ has been taken into account in the first line and in the last line we used the derivative of (\ref{epsilonint3}) for the  dressed energy\index{Dressed!energy} function.

Comparing (\ref{finiteE2}) with (\ref{CFTE}) and remembering the expression we found for the sound velocity (\ref{vF1}), we conclude that $c=1$ as anticipated.

\subsection{Conformal dimensions from finite size}
\label{sec:confdim}

To evaluate the conformal dimensions of the primary fields, we use
(\ref{EQ},\ref{PQ}) and we need the momentum and energy gap of the lowest
excitations of the theory. 
In chapters \ref{chap:LLmodel}, \ref{chap:XXXmodel}, and \ref{chap:XXZmodel} we separately studied the low energy excitations of the different models. To provide a unified account for their contributions, a central role is played by the {\it dressed charge} \index{Dressed!charge} function $Z(\lambda)$. We recall that this function is defined as the solution of the integral equation\index{Integral equation!linear} (\ref{Zdef1})
\be
Z(\lambda) + {1 \over 2 \pi} \int_{-\Lambda}^\Lambda {\cal K} (\lambda - \nu) Z(\nu) \de \nu = 1  \; ,
\label{Zdef3}
\ee
and shares interesting relations with other thermodynamic quantities.

For starters, we note that the chemical potential $h$ enters linearly in the bare energy $\epsilon_0 (\lambda;h)=\epsilon_0 (\lambda;0) \pm h$\footnote{For the Lieb-Liniger model we have a minus sign, while the XXZ chain has a plus.}. Thus, comparison with (\ref{epsilonint3}) shows that
\be
{\partial \varepsilon \over \partial h} = \pm Z(\lambda) \; .
\label{epsilonZrel}
\ee
By explicitly writing the dependence of the energy on the function's support we notice that the condition $\varepsilon (\lambda|\Lambda) = 0$ means 
\be
  \left. {\partial \varepsilon (\lambda|\Lambda) \over \partial \lambda} \right|_{\lambda=\Lambda}
  + \left. {\partial \varepsilon (\lambda|\Lambda) \over \partial \Lambda} \right|_{\lambda=\Lambda} = 0
  \: .
\ee
Thus, (\ref{epsilonZrel}) can be rewritten as
\be
   {\partial \Lambda \over \partial h} \cdot 
   \left. {\partial \varepsilon \over \partial \Lambda} \right|_{\lambda=\Lambda} 
   = - {\partial \Lambda \over \partial h} \cdot 
   \left. {\partial \varepsilon \over \partial \lambda} \right|_{\lambda=\Lambda} 
   = \pm Z (\lambda) \; ,   
\ee
and 
\be
   {\partial h \over \partial \Lambda} = 
   \mp {\varepsilon' (\Lambda) \over Z (\Lambda)} \; .
   \label{phpLambda}
\ee

Next, we consider the density of particles $n = \int_{-\Lambda}^\Lambda \rho (\lambda) \de \lambda$ and compute
\be
  {\partial n \over \partial \Lambda} = 
  \rho (\Lambda) + \rho (-\Lambda ) + 
  \int_{-\Lambda}^\Lambda {\partial \rho(\lambda) \over \partial \Lambda} \: \de \lambda \; .
  \label{pnpLambda0}
\ee
Using (\ref{rhoint3}) we have
\be
   {\partial \rho(\lambda) \over \partial \Lambda} + {1 \over 2 \pi}
   \int_{-\Lambda}^\Lambda {\cal K} (\lambda - \mu) \; {\partial \rho (\mu) \over \partial \Lambda} \; \de \mu =
   - {1 \over 2 \pi} \; \rho(\Lambda) \big[ {\cal K} (\lambda - \Lambda) + {\cal K} (\lambda + \Lambda) \big] \; ,
\ee
which, using the Green's function\index{Green's function} (\ref{Greenf}) has formal solution 
\be
   {\partial \rho (\lambda) \over \partial \Lambda} 
   = - {1 \over 2 \pi} \; \rho (\Lambda) \int_{-\Lambda}^{\Lambda} {\cal U}_\Lambda (\lambda, \mu) \big[ {\cal K} (\mu - \Lambda) + {\cal K} (\mu + \Lambda) \big] \de \mu\; .
\ee
Integrating and noting that (\ref{Zdef3}) also has formal solution $\int_{-\Lambda}^\Lambda {\cal U}_\Lambda (\lambda, \mu) \de \lambda = Z(\mu)$ we have
\bea
   \int_{-\Lambda}^\Lambda {\partial \rho (\lambda) \over \partial \Lambda} \; \de \lambda & = &
   - {1 \over 2 \pi} \; \rho (\Lambda) \int_{-\Lambda}^{\Lambda} \de \lambda
   \int_{-\Lambda}^{\Lambda} {\cal U}_\Lambda (\lambda, \mu) \big[ {\cal K} (\mu - \Lambda) + {\cal K} (\mu + \Lambda) \big] \de \mu
   \nonumber \\ 
   & = & - {1 \over 2 \pi} \; \rho (\Lambda) \int_{-\Lambda}^{\Lambda} 
   Z (\mu) \big[ {\cal K} (\mu - \Lambda) + {\cal K} (\mu + \Lambda) \big] \de \mu 
   \nonumber \\
   & = & - \rho (\Lambda) \left[ 2 - Z (\Lambda) - Z (-\Lambda) \right] \; .
\eea
Inserting this into (\ref{pnpLambda0}) and remembering that both $\rho(\lambda)$ and $Z(\lambda)$ are even functions we obtain
\be
   {\partial n \over \partial \Lambda} = 2 \; \rho (\Lambda) \; Z (\Lambda) \; .
   \label{pnpLambda}
\ee

We are now ready to classify the contributions from the different types of low energy excitations, whose origin can be traced to three fundamental processes in the Bethe Ansatz construction \cite{ISM}
\begin{enumerate}
\item Particles at the Fermi points $\pm \Lambda$ can be boosted: the quantum numbers $I_1$ and $I_N$ are changed by a finite amount $N^-$ (at $-\Lambda$: $I_1 \to I_1 - N^-$) or $N^+$ (at $\Lambda$: $I_N \to I_N + N^+$);
\item A number of particles $\Delta N$ can be added (or subtracted) to (from) the system and placed (removed) around the Fermi points;
\item Some particles (let say $d$) can backscatter, i.e. transfered from one Fermi point to the other. This process is equivalent to shifting all quantum numbers $\{ I_j \}$ by $d$, i.e. to a state with $\{ I_j + d \}$.
\end{enumerate}

Let us review these processes and their contributions to the energy and momentum of the system. For clarity, in these manipulations we have in mind the Lieb-Liniger model, but everything is valid for the XXZ chain as well (for instance, substituting $L$ with $N$ for the system's size).

\subsubsection{Boosting: $N^\pm$}

This excitation can be thought of as the creation of a particle/hole pair and the energy and momentum change can be expressed through the dressed quantities, like in (\ref{kint3}, \ref{epsilonint3}), as
\bea 
   \Delta P = k \left( \lambda_p \right) - k \left( \lambda_h \right) & \simeq & 
   \left. {\partial k \over \partial \lambda} \right|_{\lambda = \Lambda}
   \left( \lambda_p - \lambda_h \right) \; ,
   \label{KN} \\
   \Delta E = \varepsilon  \left( \lambda_p \right) - \varepsilon \left( \lambda_h \right) & \simeq & 
   \left. {\partial \varepsilon \over \partial \lambda} \right|_{\lambda = \Lambda}
   \left( \lambda_p - \lambda_h \right) \; .
   \label{EN}
\eea
We also know that by moving the last Bethe number by an integer $N^+$ we give the state the momentum $\Delta P ={2 \pi \over L} N^+$. Combining this with (\ref{KN}) we have $\lambda_p - \lambda_h = {2 \pi N^+ \over L k' (\Lambda)}$, which, substituted in (\ref{EN}) gives
\be
   \Delta E = {2 \pi \over L} {\varepsilon' (\Lambda) \over k' (\Lambda)} N^+
   = {2 \pi \over L} \: v_S \; N^+ \; ,
   \qquad \qquad \qquad
   \Delta P = {2 \pi \over L} \; N^+ \; ,
   \label{NpmDelta}
\ee
where we used the definition of the sound velocity (\ref{vSdef}). Similar expressions apply for $N^-$.

\subsubsection{Creation/Annihilation: $\Delta N$}

Using the linear dependence of the energy on the chemical potential, we write
\be
  E = L
  \int_{-\Lambda}^\Lambda \epsilon_0 (\lambda|h) \; \rho (\lambda) \; \de \lambda
  = L \int_{-\Lambda}^\Lambda \epsilon_0 (\lambda|0) \; \rho (\lambda) \; \de \lambda
  \pm L \; h \; n 
  \equiv L \left[ e_0 (n) \pm h \; n \right] \; ,
\ee
where $n$ is the particle density. Adding $\Delta N$ particles changes the density by a small amount $n \to n + {\Delta N \over L}$ and the energy accordingly
\be
  \Delta E = L \left[ e_0 \left( n + {\Delta N \over L } \right) - e_0 (n) 
  \pm h \; {\Delta N \over L} \right]
  \simeq L \left[ \left( {\partial e_0 (n) \over n } \pm h \right) {\Delta N \over L}
  + {1 \over 2} {\partial^2 e_0 (n) \over \partial n^2} \left( {\Delta N \over L} \right)^2 + \ldots \right] \; .
\ee
The ground state/equilibrium condition ensures that the linear term has to vanish. Thus $\partial e_0/ \partial n = \mp h$ and
\be
   \Delta E \simeq \mp { (\Delta N)^2 \over 2 L} {\partial h \over \partial n} 
   = \mp { (\Delta N)^2 \over 2 L} 
   \left( {\partial n \over \partial \Lambda} \right)^{-1}
   {\partial h \over \partial \Lambda} =
   {2 \pi v_s \over L} \left( {\Delta N \over 2 Z (\Lambda)} \right)^2 \; , 
   \qquad \quad \Delta P =0 \; ,
   \label{DeltaNDelta}
\ee
where we used (\ref{phpLambda},\ref{pnpLambda}) and the expression (\ref{vF1}) for the sound velocity.

\subsubsection{Backscattering: $d$}

Shifting all Bethe numbers by $d$ produces an analogous shift in the support of the rapidity density by $\delta$. Repeating the derivations of Sec. \ref{sec:LLexcitations} one finds that the integral equation for the back-flow \index{Back-flow} for this process is
\be
   \left( \hat{\cal I} + {1 \over 2 \pi} \hat{{\cal K}}_\Lambda \right) J = - d \; ,
\ee
which, compared to (\ref{Zdef3}), means $J(\lambda) = - Z(\lambda) d$. At the same time, the microscopical definition of the back-flow (\ref{Jdef}) gives $J(\Lambda) = {- \delta \over \lambda_N - \lambda_{N-1}} = - L \rho(\Lambda) \delta$, from which we extract the relation between $d$ and $\delta$ as
\be
   \delta = {Z (\Lambda) \over L \rho (\Lambda)} \; d \; .
   \label{deltadrel}
\ee
After the shift, the energy of the system is
\be
  E (\delta) = L \int_{-\Lambda + \delta}^{\Lambda + \delta} \epsilon_0 (\lambda) \rho_\delta (\lambda) \de \lambda = {L \over 2 \pi} \int_{-\Lambda + \delta}^{\Lambda + \delta} \varepsilon_\delta (\lambda) p_0^\prime  (\lambda) \de \lambda \; ,
  \label{Edelta}
\ee
where, in analogy with (\ref{rhoint3},\ref{epsilonint3}),
\bea
   \rho_\delta (\lambda) + {1 \over 2 \pi} 
   \int_{-\Lambda + \delta}^{\Lambda + \delta} {\cal K} (\lambda - \mu) \; \rho_\delta (\mu) \; \de \mu & = & {1 \over 2 \pi} \; p_0^\prime (\lambda) \; ,
   \label{rhodeltaint} \\
   \varepsilon_\delta (\lambda) + {1 \over 2 \pi} 
   \int_{-\Lambda + \delta}^{\Lambda + \delta} {\cal K} (\lambda - \mu) \; \varepsilon_\delta (\mu) \; \de \mu & = & \epsilon_0 (\lambda) \; ,
   \label{epsilondeltaint}
\eea
and where the identity between the two expressions in (\ref{Edelta}) is readily established by using the formal solution of the integral equation in terms of the Green's function\index{Green's function} (\ref{Greenf}). 

To calculate the energy change, first we expand the dressed energy as:
\be
   \varepsilon_\delta (\lambda) = \varepsilon (\lambda) +
   \left. {\partial \varepsilon_\delta (\lambda) \over \partial \delta} \right|_{\delta = 0} \delta 
   + {1 \over 2} \left. {\partial^2 \varepsilon_\delta (\lambda) \over \partial \delta^2} \right|_{\delta = 0} \delta^2 + \ldots 
\ee
and evaluate the individual terms with the help of (\ref{epsilondeltaint}). Taking a first derivative of (\ref{epsilondeltaint}) w.r.t $\delta$ we have
\be
   {\partial \varepsilon_\delta (\lambda) \over \partial \delta} + {1 \over 2 \pi}   \int_{-\Lambda + \delta}^{\Lambda + \delta} {\cal K} (\lambda - \mu) \:
   {\partial \varepsilon_\delta (\mu) \over \partial \delta} \; \de \mu
   = {1 \over 2 \pi} \left[ 
   {\cal K} (\lambda + \Lambda - \delta) \;
   \varepsilon_\delta (-\Lambda + \delta) -
   {\cal K} (\lambda - \Lambda - \delta) \;
   \varepsilon_\delta (\Lambda + \delta)  \right] \; .
   \label{1stepsilondelta}
\ee
The RHS vanishes in the $\delta \to 0$ limit because the dressed energy vanishes at the boundaries and thus $\left. {\partial \varepsilon_\delta (\lambda) \over \partial \delta} \right|_{\delta =0} = 0$.
Taking an additional derivative of (\ref{epsilondeltaint}) and evaluating it at $\delta =0$ yields
\bea
    && {\partial^2 \varepsilon_\delta (\lambda) \over \partial \delta^2} + {1 \over 2 \pi}   \int_{-\Lambda}^{\Lambda} {\cal K} (\lambda - \mu) \:
    {\partial^2 \varepsilon_\delta (\mu) \over \partial \delta^2} \; \de \mu
    = - {1 \over \pi} \; \varepsilon^\prime (\Lambda) 
    \big[ {\cal K} (\lambda - \Lambda) + {\cal K} (\lambda + \Lambda) \big] 
    \nonumber \\
    & \Rightarrow &
    {\partial^2 \varepsilon_\delta (\lambda) \over \partial \delta^2} =
    - {1 \over \pi} \; \varepsilon^\prime (\Lambda) \int_{-\Lambda}^\Lambda
    {\cal U}_\Lambda (\lambda, \mu) \; \big[ 
    {\cal K} (\mu - \Lambda) + {\cal K} (\mu + \Lambda) \big] \de \mu \; ,
    \label{d2epsilondelta}
\eea
where $\varepsilon^\prime (\Lambda) = \left. {\partial \varepsilon \over \partial \lambda} \right|_{\lambda = \Lambda} = \lim_{\delta \to 0} \varepsilon_\delta^\prime (\Lambda)$.

We perform a similar expansion for the energy (\ref{Edelta}):
\be
   E(\delta) = E(0) + 
   \left. {\partial E \over \partial \delta} \right|_{\delta =0} \delta
   + {1 \over 2} \left. {\partial^2 E \over \partial \delta^2} \right|_{\delta =0} \delta^2 + \ldots
\ee
For the first order we have
\be
    {\partial E \over \partial \delta} = {L \over 2 \pi} \left[
    \int_{-\Lambda + \delta}^{\Lambda + \delta} {\partial \varepsilon_\delta (\lambda) \over \partial \delta} \; p_0^\prime  (\lambda) \de \lambda 
    + \varepsilon_\delta (\Lambda + \delta) p_0^\prime (\Lambda + \delta)
    - \varepsilon_\delta (-\Lambda + \delta) p_0^\prime (-\Lambda + \delta)
    \right] \stackrel{\delta \to 0}{\rightarrow} 0 \; ,
\ee
because of (\ref{1stepsilondelta}) and of $\varepsilon (\Lambda)=\varepsilon(-\Lambda)=0$.
The next order, at $\delta = 0$ gives
\bea  
   {\partial^2 E \over \partial \delta^2} & = &
   {L \over 2 \pi} \left[ 
   4 \; \varepsilon' (\Lambda) \; p_0^\prime (\Lambda) + \int_{-\Lambda}^\Lambda
   {\partial^2 \varepsilon (\lambda) \over \partial \delta^2} \; p_0^\prime (\lambda)
   \; \de \lambda \right]
   \nonumber \\
   & = & {L \over 2 \pi} \left\{ 
   4 \; \varepsilon' (\Lambda) \; p_0^\prime (\Lambda) 
   - {1 \over \pi} \; \varepsilon^\prime (\Lambda)
   \int_{-\Lambda}^\Lambda \int_{-\Lambda}^\Lambda
   {\cal U}_\Lambda (\lambda, \mu) \; 
   \big[ {\cal K} (\mu - \Lambda) + {\cal K} (\mu + \Lambda) \big] 
   p_0^\prime (\lambda) \; \de \mu \; \de \lambda \right\}
   \nonumber \\
   & = & {L \over \pi} \; \varepsilon' (\Lambda) \left\{2 \; p_0^\prime (\Lambda) 
   - \int_{-\Lambda}^\Lambda \big[ {\cal K} (\mu - \Lambda) + {\cal K} (\mu + \Lambda) \big] \rho (\mu) \; \de \mu \right\}
   = 2 L \; \varepsilon' (\Lambda) \; \rho(\Lambda) \; .
\eea
Thus, combining this result and (\ref{deltadrel}), the change in energy is
\be 
  \Delta E = E(\delta) - E(0) \simeq {1 \over 2} 
  \left. {\partial^2 E \over \partial \delta^2} \right|_{\delta =0} \delta^2 
  = L \; \varepsilon' (\Lambda) \; \rho(\Lambda) 
  \left( {Z (\Lambda) \over L \rho (\Lambda)} \; d \right)^2
  = {2 \pi \over L} v_S \; Z^2 (\Lambda) d^2 \; .
  \label{dDeltaE}
\ee

From physical considerations, moving an excitation from the left Fermi point to the right one yields a change of momentum $\Delta P = 2 k_F$ (and thus moving $d$ particles corresponds to $\Delta P = 2 k_F d$). Similarly to what we did in Sec. \ref{sec:LLexcitations}, we can derive this result from using Bethe Ansatz. From (\ref{kint3}) we have for a single backscattering process:
\bea
  \Delta P = k(\Lambda) - k (-\Lambda) & = & p_0 (\Lambda) - p_0 (-\Lambda) 
  - \int_{-\Lambda}^\Lambda \big[ \theta (\Lambda - \mu) - \theta (-\Lambda - \mu) \big] \rho (\mu) \; \de \mu 
  \nonumber \\
  & = & \int_{-\Lambda}^\Lambda \de \lambda \left[ p_0^\prime (\lambda) 
  - \int_{-\Lambda}^\Lambda {\cal K} (\lambda - \mu) \; \rho (\mu) \; \de \mu \right]
  \nonumber \\
  & = & 2 \pi \int_{-\Lambda}^\Lambda \rho (\lambda) \; \de \lambda = 2 \pi \; n =
  {2 \pi \over L} \; N \; ,
  \label{kFcalc}
\eea
which shows that $k_F = \pi N/L$ as for free fermions. Notice that, in conjunction with the $\Delta N$ process analyzed in the previous subsection,
\be
   \Delta P = {2 \pi \over L} \; \left( N + \Delta N \right) d \; .
   \label{dDeltaP}
\ee
  
\subsubsection*{Summary}

Collecting (\ref{NpmDelta},\ref{DeltaNDelta},\ref{dDeltaE},\ref{dDeltaP}) we have
\bea
   \Delta E & = & {2 \pi v_S \over L} \left[
   \left( {\Delta N \over 2 {\cal Z} } \right)^2 +
   \left( {\cal Z} d \right)^2 + N^+ + N^- \right] \; , 
   \label{deltaE3} \\
   \Delta P & = & 2 k_F \; d + {2 \pi \over L}
   \left( N^+ - N^- + \Delta N \; d \right) \; ,
   \label{deltaP3}
\eea
where ${\cal Z} = Z(\Lambda) = Z(-\Lambda)$ is the value of the {\it dressed charge} \index{Dressed!charge} function $Z(\lambda)$ at the Fermi boundary and $k_F = \pi N /L$. 

Comparing with (\ref{psiRL}), we identify the first term in the momentum (corresponding to the backscattering process) as arising from the expansion of the operators around the Fermi momenta $\pm k_F$: as it connects the two chirality sectors, this momentum contribution is external with respect to the CFT\index{CFT} description. Comparing the other terms in (\ref{deltaE3}, \ref{deltaP3}) with (\ref{EQ}, \ref{PQ}), the conformal dimensions of the operators corresponding to these elementary excitations are
\be
   \Delta^\pm = 
   {1 \over 2} \left( {\Delta N \over 2 {\cal Z} }
   \pm {\cal Z} \; d \right)^2 + N^\pm \; .
   \label{Deltapm}
\ee
In the conformal language, $N^\pm$ describes the level of the descendants
and $\Delta N$ is a characteristic of the local field $\phi(x,t)$.
Comparison of (\ref{Deltapm}) with (\ref{phipmK}) hints at the identification $K= {\cal Z}^2$.

\section{Bosonization of the Lieb-Liniger model}
\label{sec:LLbosonization}

The bosonization\index{Bosonization} of the Lieb-Liniger model using Bethe Ansatz can be found in \cite{Haldane-1981,haldane81,korepin84,bogoliubov86}.
Comparing (\ref{Zdef3}) with (\ref{intbetheeq}, \ref{epsilonen}), for the Lieb-Liniger model\index{Lieb-Liniger model} we have
\be
   Z(\lambda) = 2 \pi \rho (\lambda) = - {\partial \varepsilon (\lambda) \over \partial h} \; .
   \label{Zrhorel}
\ee
Moreover, thanks to Galilean invariance, as argued in \cite{haldane81}, we have
\be
   {\cal Z}^2 = {2 \pi n \over v_S} \; .
   \label{Z2nu}
\ee
To prove this, we consider the integral equation\index{Integral equation!linear} satisfied by the derivatives of the quasi-momenta density and of the dressed energy\index{Dressed!energy}, obtained by differentiating (\ref{intbetheeq}, \ref{epsilonen}), using $\partial_\lambda {\cal K} (\lambda - \nu)= - \partial_\nu {\cal K} (\lambda - \nu)$ and integrating by parts:
\bea
   \rho' (\lambda) 
   + {1 \over 2 \pi} \int_{-\Lambda}^{\Lambda} {\cal K} (\lambda -\nu) \rho' (\nu) \de \nu
   & = & {\rho (\Lambda) \over 2 \pi} 
   \Big[ {\cal K} (\lambda - \Lambda) - {\cal K} (\lambda + \Lambda) \Big] \; , \\
   \varepsilon' (\lambda)
   + {1 \over 2 \pi} \int_{-\Lambda}^\Lambda {\cal K} (\lambda - \nu) \; \varepsilon' (\nu) \; \de \nu
    & =&  2 \lambda \; ,
\eea
where we used that $\varepsilon (\Lambda) = \varepsilon (-\Lambda) = 0$ and that the solution of the integral equation\index{Integral equation!linear} are even function.
These equations have formal solution in terms of the Green's function\index{Green's function} (\ref{Greenf}):
\be
   \rho'(\lambda) = {\rho (\Lambda) \over 2 \pi} 
   \int_{-\Lambda}^\Lambda {\cal U}_\Lambda (\lambda,\nu)
   \Big[ {\cal K} (\nu - \Lambda) - {\cal K} (\nu + \Lambda) \Big] \de \nu \; , 
   \qquad \qquad
   \epsilon' (\lambda) = 2 \int_{-\Lambda}^\Lambda {\cal U}_\Lambda (\lambda,\nu)
   \: \nu \: \de \nu \; .
   \label{epsrhoprime}
\ee
Armed with these identities we compute
\bea
\int_{-\Lambda}^{\Lambda} \lambda \; \rho' \: (\lambda) \de \lambda 
& = & {\rho (\Lambda) \over 2 \pi} \int_{-\Lambda}^{\Lambda} \lambda
\int_{-\Lambda}^\Lambda {\cal U}_\Lambda (\lambda,\nu)
\Big[ {\cal K} (\nu - \Lambda) - {\cal K} (\nu + \Lambda) \Big] \de \nu \: \de \lambda
\label{secondid1} \\
& = & - \rho (\Lambda) \int_{-\Lambda}^{\Lambda} \lambda
\Big[ {\cal U}_\Lambda (\lambda, \Lambda) - {\cal U}_\Lambda (\lambda, -\Lambda)
-\delta(\lambda - \Lambda) + \delta (\lambda + \Lambda) \Big] \de \lambda
\label{secondid2} \\
& = & 2 \Lambda \rho (\Lambda) -  \rho (\Lambda) \epsilon' (\Lambda) \; ,
\label{secondid3}
\eea
which should be compared with what one gets by directly integrating by parts:
\be
\int_{-\Lambda}^{\Lambda} \lambda \; \rho' \: (\lambda) \de \lambda 
= 2 \Lambda \rho (\Lambda) - n \; .
\label{firstid1}
\ee
The RHS of (\ref{secondid1}) is obtained by plugging the expression for $\rho'(\lambda)$ from (\ref{epsrhoprime}), while to get (\ref{secondid2}) we recalled the definition of the Green's function\index{Green's function} (\ref{Greenf}). Finally, in (\ref{secondid3}) we used $\epsilon' (\lambda)$ from (\ref{epsrhoprime}). Comparing (\ref{secondid3}) and (\ref{firstid1}) we get
\be
\varepsilon' (\Lambda) \rho (\Lambda) = n \; .
\label{epsprimerho}
\ee
Note that in this derivation the choice of the LHS of (\ref{secondid1},\ref{firstid1}) was pivotal and its form is a consequence of the Galilean invariance of the model, that is $\epsilon_0 (\lambda) = \lambda^2 - h$.
Using the first identity of (\ref{Zrhorel}), (\ref{epsprimerho}), and (\ref{vF1}) we have (\ref{Z2nu}), which relates the dressed charge\index{Dressed!charge} to thermodynamic, macroscopic, observables.

To obtain, for instance, the field correlator $\langle \Psi (x,\tau) \Psi^\dagger (0,0) \rangle$ of the Lieb-Liniger\index{Lieb-Liniger model}, one sets $\Delta N =1$. To obtain the leading term, we further set $N^\pm = d = 0$:
\be
   \langle \Psi (x,\tau) \Psi^\dagger (0,0) \rangle \simeq A |x + \ii v_S \tau|^{-1/(2 {\cal Z}^2)} \; .
\ee
Higher terms are obtained in a series (\ref{phiseries})
\be
   \langle \Psi (x,\tau) \Psi^\dagger (0,0) \rangle = \sum_{d, N^\pm}
   {A(d,N^\pm) \eu^{- 2 \ii d k_F x} \over (x - \ii v_S \tau)^{2 \Delta^+}
   (x + \ii v_S \tau)^{2 \Delta^-} }\; ,
   \label{LLpsipsidag}
\ee
where $\Delta^\pm$ are given by (\ref{Deltapm}), with $\Delta N=1$ and $d$ and $N^\pm$ integers.

Similarly, for the density correlator we set $\Delta N = 0$:
\bea
  \langle \rho (x,\tau) \rho (0,0) \rangle - \langle \rho (0,0) \rangle^2 
  & = & {B_1 \over (x + \ii v_S \tau)^2} + {B_1 \over (x - \ii v_S \tau)^2} 
  + B_2 {\cos 2 k_F x \over |x + \ii v_S \tau|^{2 {\cal Z}^2}} 
  \label{LLrhorho} \\
  & = & 2 B_1 {x^2 - (v_S \tau)^2 \over [x^2 + (v_S \tau)^2]^2} 
  + B_2 {\cos 2 k_F x \over |x + \ii v_S \tau|^{2 {\cal Z}^2}}  \; ,
\eea
where the first term in (\ref{LLrhorho}) corresponds to $d=N^+=0, N^-=1$, the second to $d=N^-=0, N^+=1$, and the third to $d= \pm 1, N^\pm = 0$.
Note that these series are consistent with the Luttinger liquid\index{Luttinger!liquid} universality and with the result of bosonization\index{Bosonization}. In particular, comparing (\ref{rhorhocorr}) with the asymptotic behavior of the density correlators one can extract the Luttinger parameter\index{Luttinger!parameter}: 
\be
   K = {\cal Z}^2 \; .
   \label{KZ2}
\ee

In this way, we have identified the CFT\index{CFT} describing the low energy properties of the Lieb-Liniger model\index{Lieb-Liniger model}, through its parameters. As it is usually the case with this model, the finiteness of the interval over which the integral equations\index{Integral equation!linear} are defined means that their exact solution is obtained only approximately. Quite accurate results can be easily derived analytically in the asymptotic regimes of strong and weak interaction.
This analysis is best performed in terms of the universal parameter $\gamma$ introduced in (\ref{gammaLL}) and of the rescaled setting we introduced starting eq. (\ref{LLunivrescaling}), that is $\lambda \equiv \Lambda \: x$, $c \equiv \Lambda \: g$.

In the Tonks-Girardeau regime of strong interaction ($\gamma \gg 1$), the Bose system behaves like a free fermionic one \cite{TG}. The kernel vanishes\index{Kernel} in the $g \to \infty$ limit and can be approximated by a constant ${\cal K} (x - y) \simeq 2/g$ for large $g$. Thus to first order in $1/g$ we can approximate $\rho(x)$ with a constant, yielding
\be
   \rho (x) =  {g \over 2 \pi g - 4} \; , \qquad G(g) = {g \over \pi g - 2} 
   \qquad \qquad \Rightarrow \qquad \qquad  \gamma = \pi g -2 \;.
\ee
Using (\ref{Zrhorel}), (\ref{KZ2}) and (\ref{Z2nu}) we have
\be
{\cal Z} = {2 \pi g \over 2 \pi g - 4} 
 \qquad \quad \Rightarrow  \qquad \quad
K =  1 + {4 \over \gamma} + \Ord\left( {1 \over \gamma^2 }\right) \; ,
\quad
v_S = 2 \pi n \left[ 1 - {4 \over \gamma} + \Ord\left( {1 \over \gamma^2 }\right) \right] \; .
\label{TTLL}
\ee

\begin{wrapfigure}{r}{8cm}
	\vspace{-10pt}
	\begin{center}
		\includegraphics[width=7.8cm]{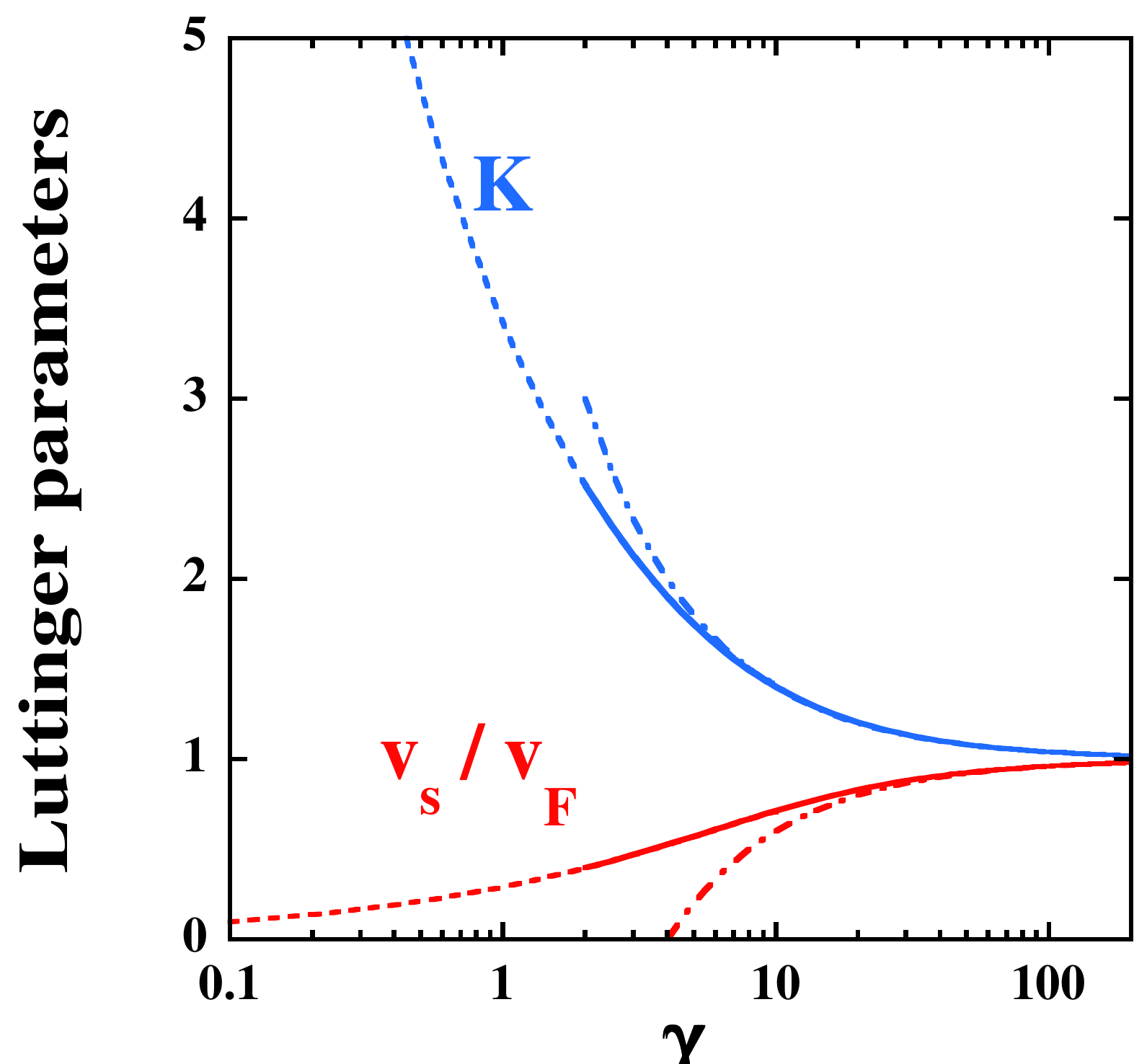}
	\end{center}
	\caption{Comparison between the asymptotic behaviors (\ref{TTLL}, \ref{weaklyLL}) and the numerical solution for the Luttinger parameters. This plot is taken from \cite{cazalilla04}, where $v_F$ is the sound velocity of a free system with the same density. In our notation $v_F = 2\pi n$.}
	\label{fig:LLvsLL}
	\vspace{-70pt}
\end{wrapfigure}

In the weakly interacting limit ($\gamma \ll 1$), one might want to use the asymptotic solution (\ref{weakLLrho}). Unfortunately, these expressions are not sufficiently accurate at the integration boundaries. A careful analysis of the small coupling solution of these integral equation\index{Integral equation!linear} can be found in \cite{hutson62} and could be applied to our problem as well. However, as pointed out already in \cite{LL63}, in this regime one can use Bogoliubov theory for weakly interacting gases \cite{stringari} to extract the sound velocity and thus the Luttinger parameter\index{Luttinger!parameter} through (\ref{Z2nu}):
\bea
  v_S & = & 2 \pi n {\sqrt{\gamma} \over \pi} \left( 1- {\sqrt{\gamma} \over 2 \pi}\right)^{1/2} \; , 
  \nonumber \\
  K & = & {\pi \over \sqrt{\gamma}} \left( 1- {\sqrt{\gamma} \over 2 \pi}\right)^{-1/2} \; .
  \label{weaklyLL}
\eea

The good agreement of these asymptotic behaviors with numerical solutions was already noticed in \cite{LL63}. The comparison for the Luttinger parameters\index{Luttinger!parameter} was done in \cite{cazalilla04}, from which Fig. \ref{fig:LLvsLL} is taken, and shows that the asymptotic regimes can be used safely for $\gamma <1$ and $\gamma>10$.

\section{Bosonization of the XXZ model}
\label{sec:XXZbosonization}

It is instructive to repeat the bosonization\index{Bosonization} procedure for the XXZ chain\index{XXZ chain}, to see it at work in a prototypical example.
We will see that the scaling limit of this model corresponds to a sine-Gordon theory \cite{bosonization}. We can start by writing the spin model using spin-less fermions, using the Jordan-Wigner transformation (\ref{JordanWigner})\index{Jordan-Wigner!transformation}:
\be
   S_n^z = : \psi_n^\dagger \psi_n : = \psi_n^\dagger \psi_n - {1 \over 2} \; .
\ee
We expand the fermionic field around the Fermi points, in terms of the chiral fields:
\be
  \psi_n \to \sqrt{a} \left[ (-\ii)^n \psi_+ (x) + \ii^n \psi_- (x) \right] \; ,
\ee
where $a$ is the lattice spacing, $x = a n$ and we took the system at half filling ($k_F = \pi/2$), i.e. at zero magnetization. The spin density, written in terms of chiral fields, decomposes into the sum of a smooth and oscillating (staggered) component \cite{lukyanov98}:
\bea
    S^z (x) & = & \rho (x) + (-1)^n M(x) \; ,\\
    \rho (x) & = & : \psi_+^\dagger (x) \psi_+ (x) : + : \psi_-^\dagger (x) \psi_- (x) : \; , \\
    M (x) & = & : \psi_+^\dagger (x) \psi_- (x) : + : \psi_-^\dagger (x) \psi_+ (x) : \; ,
\eea
where $S_n^z \to a S^z (x)$. The XXZ Hamiltonian (\ref{XXZham}) can be written as 
\be
   {\cal H} = 
   - \sum_n \left[ {1 \over 2} \left( S^+_n S^-_{n+1} + S^-_n S^+_{n+1} \right)
   + \Delta S^z_n S^z_{n+1} \right] 
   \simeq {\cal H}_0 + {\cal H}_{int} \; ,
\ee
where the first two terms are the kinetic part of a free theory, which, in the linear approximation, give
\be
   {\cal H}_0 = - \ii v_0 \int \de x \left[ \psi_+^\dagger \partial_x \psi_+ - \psi_-^\dagger \partial_x \psi_- \right] \; ,
\ee
and the interaction term can be written as
\be
   {\cal H}_{int} = v_0 \Delta \int \de x \left[ : \rho (x) \rho (x + a): - M(x) M(x+a) \right] \; .
\ee

The bosonization\index{Bosonization} of the kinetic term gives (\ref{FBH}).
For the interaction terms we have
\bea
   \rho(x) & = & {1 \over \sqrt{\pi}} \partial_x \phi (x) \; , \\
   M (x) & \simeq & - {1 \over \pi a} : \sin \sqrt{4 \pi} \phi(x) : \; , \\
   \lim_{a \to 0} M(x) M(x+a) & = & -{1 \over (\pi a)^2} \cos \sqrt{16 \pi} \phi(x)
   - {1 \over \pi} \left( \partial_x \phi \right)^2 + {\rm const} \; .
\eea
The cosine term originates from the sa-called Umklapp processes $\psi_+^\dagger (x) \psi_+^\dagger (x+a) \psi_- (x+a) \psi_- (x) + h.c.$ where two particles are removed from one Fermi point and added at the other. This scattering event corresponds to a transfer of momentum $4k_f$ and it is possible only when the Fermi point is such to allow the lattice to recoil and absorb this excess momentum, as it happens for $k_F = \pi/2$.

Putting these contributions together, the continuous version of the XXZ Hamiltonian\index{XXZ chain} reads
\be
   {\cal H} = \int \de x \left\{ {v_0 \over 2} \left[ \Pi^2
   + \left( 1 + {4 \Delta \over \pi} \right) \left( \partial_x \phi \right)^2 \right]
   + {v_0 \Delta \over (\pi a)^2} \; : \cos \sqrt{16 \pi} \phi : \right\} \; .
   \label{XXZHbos}
\ee
This is clearly a na\"ive analysis, since higher oder terms and fusion rules renormalize the coefficients in this Hamiltonian. Nonetheless, the three terms in (\ref{XXZHbos}) are sufficient to capture the scaling limit of the\index{XXZ chain} XXZ chain\footnote{An anisotropy between the $x$ and $y$ component would generate an additional cosine in the dual field $\theta (x)$, which produces a competition between fields in the scaling limit of the XYZ chain\index{XYZ chain} and thus additional challenges.}. It is also customary to normalize the fields so to absorb the $\Delta$-dependent coefficient of the cosine: doing so we rescale the energy scale through the speed of sound and transfer the effect of the interaction into the compactification radius of the bosons \cite{bosonization}. Studying the conformal dimension of the cosine terms, one sees that it is irrelevant for $|\Delta|<1$. At $\Delta =-1$ (Heisenberg AFM) the chiral symmetry gets broken by the Umklapp term and the cosine term turns relevant and opens a gap toward the uni-axial AFM phase \cite{lukyanov98}. At $\Delta=1$ the low energy excitations are magnons\index{Magnon} with quadratic dispersion relation and thus the bosonization\index{Bosonization} prescription breaks down due to the restoration of Galilean symmetry, instead of the Lorentz one. We concentrate now only on the paramagnetic phase, where cosine terms can be neglected and the effective Hamiltonian is (\ref{bosonizationH}).

The bosonization\index{Bosonization} of the Heisengerg/XXZ chain through Bethe Ansatz was done originally in \cite{haldanespin81,izergin85,affleck89,woynarovich89}.
The sound velocity can be calculated using (\ref{vF1}), but we already determined it in (\ref{dispeXXZ}) as 
\be
   v_S = {\pi \sin \gamma \over 2 \gamma} \; ,
\ee
where $\Delta = - \cos \gamma$.
The Luttinger parameter\index{Luttinger!parameter} can be extracted from the fractional charge (\ref{Zdef3}) as $K={\cal Z}^2$. At zero magnetic field, the support of the integral equation is over the whole real axis and can be solved by Fourier transform, yielding a constant fractional charge $ Z(\lambda) = {\pi \over 2 (\pi - \gamma)}$ (\ref{ZLambdaas}). This is not the value to be used for the Luttinger parameter\index{Luttinger!parameter}, as can be checked by considering a finite, but small, magnetic field (large $\Lambda$). The calculation can be done perturbatively through the Wiener-Hopf method\index{Wiener-Hopf method} as shown in Sec. \ref{sec:XXZparah}, which gives to first order a different constant value for the dressed charge\index{Dressed charge} at the boundary, which can be continued analytically for $\Lambda \to \infty$ to give the value at the boundary at infinity \cite{yangyangXXZ,ISM}:
\be
  \lim_{\Lambda \to \infty} Z(\Lambda) = \sqrt{ \lim_{\Lambda \to \infty} Z(0)} =
  \sqrt{\pi \over 2 (\pi - \gamma)} \qquad \qquad \Rightarrow \qquad \qquad
  K = {\pi \over 2 (\pi - \gamma)} \; .
\ee
This result agrees with the na\"ive (perturbative) answer one can derive from (\ref{XXZHbos}) at small $\Delta$, $K \simeq 1 - {2 \Delta \over \pi} + \Ord (\Delta^2)$, but deviates from it as one moves away from $\Delta=0$.
While $\Delta =0$ corresponds to free fermions ($K=1$), $\Delta >0$ gives repulsive fermions ($K<1$) and $\Delta <0$ attractive ones ($K>1$). The Heisenberg AFM chain\index{XXX chain}\index{Heisenberg chain} ($\Delta =1$) has $K = {1 \over 2}$ and $v_S = {\pi \over 2}$, while the ferromagnetic point $\Delta =-1$ is not conformal ($v_S=0$).
For $h \ll 1$, next orders in the Wiener-Hopf solution give 
\be
   K \stackrel{h << 1}{\simeq} {\pi \over 2 (\pi - \gamma)} \times \left\{ 
   \begin{array}{lcl}
   	1 + {1 \over 2 \ln {h_0\over h}}
   	& \qquad \gamma =0 \; ,
   	& \qquad \left( \Delta = -1 \right) \cr
   	1 + \alpha_1 \; h^{4 \gamma \over \pi - \gamma} \; , 
   	& \qquad 0 < \gamma < {\pi \over 3}
   	& \qquad \left( -1 < \Delta < -.5 \right) \cr
   	1 + \alpha_2 \; h^2 \; , 
   	& \qquad {\pi \over 3} < \gamma < \pi 
   	& \qquad \left(-.5<\Delta<1\right) \cr
   \end{array}
   \right.
\ee
which show that the $h=0$ value is approached through different exponents, which vary continuously for $-1<\Delta<-0.5$ and stay  constant for $\Delta>-0.5$ (the constants $h_0,\alpha_{1,2}$ can be found, for instance in \cite{ISM}).

The final bosonized representations for the spin operators are \cite{bosonization,lukyanov98}
\bea
S^z (x) & = & \sqrt{ K \over 2 \pi} \partial_x \phi (x) - \text{const} \; (-1)^j \sin \sqrt{4 \pi K} \phi (x) \; , \\
S^\pm (x) & = & \text{const} \; (-1)^j \eu^{\pm \ii \sqrt{\pi \over K} \theta (x)} \; .
\eea 

As we discussed in Sec. \ref{sec:XXZparah}, the XXZ chain remains critical for magnetic fields smaller than $h_{\rm s}$ (\ref{hFM}). 
Close to saturation the support of the integral equations shrink toward zero: we can use the perturbative result (\ref{rhoZsat}) together with (\ref{Lambdasat}) to find:
\be
   Z(\Lambda) \stackrel{h \to h^-_{\rm s}}{\simeq}  
   1 - {\Lambda \over \pi} \; {\cal K} (0) = 
   1 - {2 \over \pi} {\tan \gamma /2 \over \tan \gamma} \sqrt{h_{\rm s} - h} \; , 
   \qquad \Rightarrow \qquad
   K \stackrel{h \to h^-_{\rm s}}{\simeq}  
   1 + {4 \over \pi} {\Delta \over 1 - \Delta} \sqrt{h_{\rm s}-h} \; ,
\ee
which corresponds to $K \simeq 1 - {2 \over \pi} \sqrt{h_{\rm s}-h}$ for the AFM Heisenberg chain\index{XXX chain}\index{Heisenberg chain} close to the saturation point.

The spin flip correlation function in the paramagnetic regime (even at finite $h<h_{\rm s}$) corresponds to $\Delta N=1$ (and $d=N^\pm=0$ to leading order)
\be
   \langle S^- (x,\tau) S^+ (0,0) \rangle \simeq A \left| x + \ii v_S \tau \right|^{-1/(2 {\cal Z})} \; ,
\ee
which is the same as (\ref{LLpsipsidag}). Similarly, 
\be
  \langle S^z(x,t) S^z(0,0)\rangle = 
  C_1 {x^2 - (v_S \tau)^2 \over [x^2 + (v_S \tau)^2]^2} 
  + C_2 {\cos 2 k_F x \over |x + \ii v_S \tau|^{2 {\cal Z}^2}} \; ,
 \qquad 0\le h \le h_{\rm s} \; ,
\ee
which looks like (\ref{LLrhorho}). Here, see the discussion around (\ref{kFcalc}), $k_F = \pi \int \rho_0 (\lambda) \de \lambda$, which equals $k_F={\pi \over 2}$ at $h=0$. A deviation from this behavior was reported in \cite{affleck89} for the AFM isotropic Heisenberg chain ($\Delta=1$) at $h=0$, due to the enhanced symmetry of this point:
\be
   \langle S^z(z) S^z(0)\rangle \simeq \tilde{A} \: (-1)^{|z|} \; |z|^{-1} \; \sqrt{\ln |z|} \; , \qquad \qquad \text{at }h=0 \;, \Delta=1.
\ee

\backmatter

\printindex




\end{document}